\documentclass[12pt,a4paper]{article}\usepackage[top=20mm,left=12mm]{geometry}\textwidth18.2cm\textheight25.0cm
\usepackage{amsmath,amssymb,amsthm,amsfonts}
\usepackage{lineno,url}\usepackage{float}
\floatstyle{plaintop}\restylefloat{table}
\usepackage[tableposition=top]{caption}
\usepackage{graphicx,epsfig}
\usepackage{multirow}\usepackage{pstricks}\usepackage{fancyvrb}
\usepackage{listings,paralist, longtable}
\usepackage{setspace}
\newtheoremstyle{myplain}
  {-\baselineskip\topsep}   
  {\topsep}   
  {\itshape\setstretch{1.05}}  
  {0pt}       
  {\bfseries} 
  {.}         
  {5pt plus 1pt minus 1pt} 
  {}       
\newtheoremstyle{mydef}
  {-\baselineskip\topsep}   
  {\topsep}   
  {\setstretch{1.05}}  
  {0pt}       
  {\bfseries} 
  {.}         
  {5pt plus 1pt minus 1pt} 
  {}       
\theoremstyle{myplain}
\newtheorem{theorem}{Theorem}[section]\newtheorem{lemma}[theorem]{Lemma}
\newtheorem{definition}[theorem]{Definition}
\newtheorem{corollary}[theorem]{Corollary}
\theoremstyle{mydef}
\newtheorem{example}[theorem]{Example}
\newtheorem{remark}[theorem]{Remark}
\newtheorem{exercise}[]{}[subsection]
\def\bexe{\begin{exercise}}\def\eexe{\eex\end{exercise}}
\def\bsol{\begin{solution}}\def\esol{\eex\end{solution}}
\def\bexa{\begin{example}}\def\eexa{\end{example}}
\def\brem{\begin{remark}}\def\erem{\end{remark}}
\def\bthm{\begin{theorem}}\def\ethm{\end{theorem}}
\def\blem{\begin{lemma}}\def\elem{\end{lemma}}
\def\bcor{\begin{corollary}}\def\ecor{\end{corollary}}
\def\bdefi{\begin{definition}}\def\edefi{\end{definition}}
\newcommand{\IDEA}{\textbf{Idea of the Proof.} }

\def\bmip{\begin{minipage}{\textwidth}}\def\emip{\end{minipage}}
\def\huga#1{\begin{gather} #1 \end{gather}}
\def\hugast#1{\begin{gather*} #1 \end{gather*}}
\def\hual#1{\begin{align} #1 \end{align}}
\def\hualst#1{\begin{align*} #1 \end{align*}}

\newcommand{\R}{{\mathbb R}}
\newcommand{\N}{{\mathbb N}}

\def\CK{{\cal K}}\def\CT{{\cal T}}
\def\CE{{\cal E}}\def\CD{{\cal D}}  
\def\CF{{\cal F}}\def\CG{{\cal G}}
\def\CO{{\cal O}}\def\CS{{\cal S}}
\def\CM{{\cal M}}

\def\ee{{\rm e}}\def\ii{{\rm i}}

\def\uti{\tilde{u}}
\def\xt{\tilde{x}}

\def\ga{\gamma}\def\om{\omega}
\def\noi{\noindent}\def\ds{\displaystyle}
\def\vt{\vartheta}\def\pa{{\partial}}\def\lam{\lambda}
\newcommand{\bi}{\begin{itemize}}\newcommand{\ei}{\end{itemize}}
\newcommand{\ben}{\begin{enumerate}}\newcommand{\een}{\end{enumerate}}
\newcommand{\bce}{\begin{center}}\newcommand{\ece}{\end{center}}
\newcommand{\bci}{\begin{compactitem}}\newcommand{\eci}{\end{compactitem}}
\newcommand{\bcen}{\begin{compactenum}}\newcommand{\ecen}{\end{compactenum}}
\newcommand{\bcena}{\begin{compactenum}[(a)]}
\newcommand{\reff}[1]{(\ref{#1})}
\newcommand{\ov}[1]{{\overline {#1}}}

\newcommand{\spr}[1]{\left\langle #1 \right\rangle}
\newcommand{\hs}[1]{{\hspace{#1}}}\newcommand{\vs}[1]{{\vspace{#1}}}

\def\eps{\varepsilon}\def\aqui{\Leftrightarrow}
\def\ra{\rightarrow}

\newcommand{\barr}{\begin{array}}\newcommand{\earr}{\end{array}}
\newcommand{\bpm}{\begin{pmatrix}}\newcommand{\epm}{\end{pmatrix}}
\newcommand{\bsm}{\left(\begin{smallmatrix}}
\newcommand{\esm}{\end{smallmatrix}\right)}
\newcommand{\ba}{\begin{array}}\newcommand{\ea}{\end{array}}
\def\dd{\, {\rm d}}\def\ri{{\rm i}}
\def\rds{{\rm ds}}
\def\cc{{\rm c.c.}}\def\er{{\rm e}}
\def\re{{\rm Re}}\def\im{{\rm Im}}
\def\om{\omega}\def\Om{\Omega}
\def\hot{{\rm h.o.t}}\def\ddt{\frac{\rm d}{{\rm d}t}}
\def\del{\delta}

\def\nab{\nabla}\def\eex{\hfill\mbox{$\rfloor$}}

\def\Del{\Delta}

\def\sig{\sigma}
\def\al{\alpha}

\def\Ga{\Gamma}
\def\kap{\kappa}

\def\bd{\begin{displaymath}} \def\ed{\end{displaymath}}
\def\ba{\begin{array}} \def\ea{\end{array}}
  
\def\eps{\varepsilon}

\def\pdep{{\tt pde2path}}

\def\p{\tt p} 
\def\oop{{\tt OOPDE}}

\def\mlab{{\tt Matlab}}\def\ptool{{\tt pdetoolbox}}

\def\dhome{/hh/path/pde2path/demos/acpbc}

\definecolor{codegreen}{rgb}{0,0.6,0}
\definecolor{codegray}{rgb}{0.5,0.5,0.5}
\definecolor{codepurple}{rgb}{0.58,0,0.82}
\definecolor{backcolour}{rgb}{0.95,0.95,0.92}
 
\lstdefinestyle{mystyle}{
    backgroundcolor=\color{backcolour},   
    commentstyle=\color{codegreen},
    keywordstyle=\color{black},
    numberstyle=\small\color{codegray},
    stringstyle=\color{codepurple},
    basicstyle=\footnotesize\ttfamily,
    breakatwhitespace=false,         
    breaklines=true,                 
    captionpos=b,                    
    keepspaces=true,                 
    numbers=left,                    
    numbersep=5pt,                  
    showspaces=false,                
    showstringspaces=false,
    showtabs=false,                  
    tabsize=2, 
  xleftmargin=4mm,
}
 
\newlength{\tew}

\def\rds{{\rm ds}}
\def\ig{\includegraphics}\def\ds{{\rm d}s}\def\dst{\displaystyle}

\def\medskip{}\def\bigskip{}
\def\ssmatrix#1{\bigl(\begin{smallmatrix}#1 \end{smallmatrix} \bigr)}

\def\lin{{\rm span}}
\def\nab{\nabla}\def\DLB{\Del_{{\rm LB}}}\def\DelS{\Del_{\CS_R}}
\def\DelT{\Del_{\CT_{R,\rho}}}
 \def\xti{\tilde x}\def\yti{\tilde y}\def\zti{\tilde z}\def\uti{\tilde u}

\def\trulle{{\tt trullekrul}}\def\ma{mesh adaptation} 

\newfloat{Algorithm}{ht}{lop}[section]
\usepackage{hyperref}
\renewcommand{\arraystretch}{1.15}\renewcommand{\baselinestretch}{1.1}
\def\taskip{\renewcommand{\arraystretch}{1}\renewcommand{\baselinestretch}{1}}
\def\teskip{\renewcommand{\arraystretch}{1.1}
\renewcommand{\baselinestretch}{1.25}}
\def\hulst#1#2{\taskip\lstinputlisting[#1]{#2}\teskip}
\def\hutab#1#2{\taskip\begin{\table}#1\end{table}\teskip}

\def\sm{\small}
\newcommand{\btab}[2]{\begin{tabular}{#1}#2\end{tabular}}

\begin{document}
\text{}\vspace{0mm}
\begin{center}\Large
Pattern formation with \pdep\ -- a tutorial \\[4mm]
\normalsize 
Hannes Uecker\\[2mm]
\footnotesize
Institut f\"ur Mathematik, Universit\"at Oldenburg, D26111 Oldenburg \\
hannes.uecker@uni-oldenburg.de \\
\normalsize
\today
\end{center}
\begin{abstract} 
We explain some \pdep\ setups for pattern formation in 1D, 2D and 3D. 
A focus is on 
steady bifurcation points of higher multiplicity, typically due 
to discrete symmetries, but we also review 
general concepts of pattern formation and their handling in \pdep, 
including localized patterns and homoclinic snaking, again in 1D, 2D and 3D, 
based on the demo {\tt sh} (Swift--Hohenberg equation). 
Next, the demos {\tt schnakpat} (a Schnakenberg reaction--diffusion system) 
and {\tt chemtax} (a quasilinear RD system with cross--diffusion 
from chemotaxis) simplify and unify 
previous results in a simple and concise way, {\tt CH} (Cahn-Hilliard) 
and {\tt fCH} (functionalized Cahn--Hilliard) 
deal with mass constraints,  {\tt hexex} deals with 
(multiple) branch points of higher degeneracy in a scalar problem 
on a hexagonal domain, and {\tt shgc} illustrates some global coupling. 
The demos {\tt acS, actor,schnakS} and {\tt schnaktor} (the Allen--Cahn 
and Schnakenberg models on spheres and tori) consider pattern formation on curved surfaces, {\tt cpol} considers a problem of cell polarization 
described by bulk--surface coupling, and 
{\tt bruosc} (Brusselator) explains how 
to augment autonomous systems by a time periodic forcing. 
Along the way we also comment on the choice of meshes, 
on time integration,  and we 
give some examples of branch point continuation and Hopf point continuation 
to approximate stability boundaries. 
\end{abstract}

\noi

\tableofcontents
\section{Introduction}
The \mlab\ bifurcation and continuation package \pdep\ \cite{p2p, p2phome} 
can be used to study solution branches and bifurcations in 
pattern forming systems (PFS), in particular reaction diffusion  
systems of the form 
\begin{equation} \label{rd1} M_d\pa_t u=D\Delta u+f(u)=:-G(u,\lam), 
\quad  u=u(x,t) \in \R^N,\ t \geq 0,\ x\in \Om, 
\end{equation}
where $\Om\subset\R^d $  is a bounded domain, $d=1,2,3$ (1D, 2D and 3D case, respectively),  $D\in\R^{N\times N}$ is a positive
(semi-)definite diffusion matrix, $\Delta=\pa_{x_1}^2+\ldots+\pa_{x_d}^2$, 
where the ``reaction part'' $f$ is a smooth function, where $\lam$ in $G(u,\lam)$ stands 
for one or several parameters present, and where \reff{rd1} can be completed 
by various kinds of boundary conditions (BCs). Moreover, the dynamical 
mass matrix $M_d\in 
\R^{N\times N}$ in \reff{rd1} may be singular, allowing to also write 
elliptic--parabolic systems in the form \reff{rd1}, or to rewrite some 
4th order problems such as the Swift--Hohenberg equation in the form 
\reff{rd1} in a consistent way. See, e.g., 
\cite[\S4.2]{p2p} and \cite{uwsnak14, U16, w16, BGUY17, ZUFM17, hotheo} for 
examples, mostly related to pattern formation and Turing bifurcations 
\cite{Mur}. 

In applications (discrete) symmetries of the domain often enforce higher multiplicities of BPs. For instance, for Turing bifurcations over square 
domains with Neumann BCs 
we have ``stripes in $x_1$'' and ``stripes in $x_2$'' as two 
kernel vectors, and altogether we obtain three (modulo discrete spatial shifts)  bifurcating branches, namely 
stripes (twice) and spots as a superposition of stripes. 
In the following, we always use 
\huga{\label{mdef}
m=\dim{N(G_u(u_0,\lam_0))} 
}
to denote the dimension of the kernel of $G_u(u_0,\lam_0)$, and call this $m$ 
 the {\em multiplicity of the BP} $(u_0,\lam_0)$. 

The (analytically) higher multiplicity $m\ge 2$ of BPs 
in situations as above can be circumvented by some tricks, which for instance have been used in \cite{p2p,uwsnak14}. 
Essentially we can exploit the fact that even on ideal domains, 
the discretization breaks up multiple BPs, and/or we can strengthen 
this breakup by slightly distorting the domain. 
However, besides the lack of elegance, using these tricks  has some serious disadvantages: 
(a) The localization of close together simple BPs (obtained from the 
breakup of multiple BPs) is quite inefficient. 
(b) The branching behavior at the (artificially) simple BPs is in general quite 
different from that at the originally multiple BP. For instance, 
two simple stripes may hide the spots also present. 
This then requires further 
tricks/analytical understanding to relate the numerics to the true analytical 
situation. 

Algorithms for branch switching at steady BPs of higher multiplicity, aimed particularly at pattern 
formation in $d\ge 2$ space dimensions,  have been 
implemented in \pdep\ since 2018. 
Here we explain these in a somewhat wider context and review in a tutorial style 
some general ideas of applying \pdep\ to PFS in 1D, 2D and 3D. 
To make the tutorial somewhat self-contained, in \S\ref{pfssec} we briefly review some basics of PFS, in particular those related to amplitude equations 
and symmetries,  using 
the Swift--Hohenberg (SH) equation as an example problem. 

Table \ref{demotab} lists the demos discussed in this tutorial. 
In \S\ref{shsec} and \S\ref{schnaknum} we 
explain the \pdep\ demos {\tt sh}  and {\tt schnakpat}, which 
implement the SH equation and the Schnakenberg reaction diffusion system, 
respectively, over various 1D, 2D and 3D domains, mostly with 
homogeneous Neumann BCs (NBCs). In particular, the demo {\tt sh} 
also explains how to use $M$ in \reff{rd1} to rewrite the 4th order SH equation as a 2--component 2nd order system in a consistent way, 
and {\tt schnakpat} simplifies and unifies in a concise way many of the results from \cite[\S4.2]{p2p} and \cite{uwsnak14}. 
Additionally, we also consider some periodic BCs (pBCs) to illustrate 
how to deal with the interplay of discrete and continuous  symmetries.

\begin{table}[ht]\taskip
\caption{Subdirectories of {\tt /demos/pftut}. \label{demotab}}
\bce\vs{-4mm}
{\small 
\begin{tabular}{l|p{0.8\textwidth}}\hline
sh&The Swift--Hohenberg (SH) equation on 1D, 2D and 3D boxes with homogeneous Neumann BCs, 'main' demo directory, \S\ref{shsec}\\
shpbc&The SH in 2D with pBCs, \S\ref{sy1sec}\\
shEck&SH, 1D, with branch--point continuation to approximate 
the Eckhaus instability, \S\ref{shecksec} \\
shgc&SH with global coupling, to illustrate customized linear system solvers, \S\ref{gcsec}\\
schnakpat&A Schnakenberg 2-component reaction diffusion model, \S\ref{schnaknum}\\
CH&Cahn-Hilliard (CH) model, to illustrate a mass constraint setup, \S\ref{chsec}\\
fCH&a 'functionalized' CH model, \S\ref{fchsec}\\
hexex&A scalar problem on a hexagonal domain, with multiple BPs of higher 
degeneracy, \S\ref{hexsec}\\
chemtax&A quasilinear RD system modeling chemotaxis, \S\ref{chemsec}\\
actor, acS&Allen--Cahn equations on tori and spheres, based 
on Laplace--Beltrami operators, \S\ref{act-sec} and \S\ref{acs-sec}, mainly 
as a preparation for schnakS and schnaktor\\
schnakS/tor&The Schnakenberg model on spheres and tori, \S\ref{ss-sec} and \S\ref{st-sec}\\
accyl&An Allen-Cahn eqn on a cylinder with boundary coupling to 
a Poisson eqn, \S\ref{accylsec}\\
cpol&a model for cell--polarization with  bulk--surface coupling, \S\ref{cpolsec}\\
bruosc&The Brusselator, with oscillating Turing patterns and period doubling, \S\ref{brusec}\\
\hline
\end{tabular}
}
\ece
\end{table}
\teskip

In \S\ref{intersec} we collect some shorter demos. These deal for instance 
with approximation of the Eckhaus instability curve (in the SH equation)
via  BP continuation, with mass constraints 
(in the Cahn--Hilliard (CH) problem and a 'functionalized' CH problem), 
with multiple branch--points of higher order 
indeterminacy (in a scalar problem on a hexagonal domain), with 
a quasilinear chemotaxis problem, and with global coupling (again in a SH 
equation), which requires some customized linear system and eigenvalue solvers. 

In \S\ref{pfsurf} we consider pattern formation on 
curved surfaces. The case of the sphere is in particular interesting from 
a symmetry point of view, yielding BPs of rather high multiplicity, 
and again requires to deal with both discrete and continuous symmetries at bifurcation. Moreover, we give an example how to patch together 
problems living on different domains and only coupled via a common 
boundary, and we consider a simple model for cell--polarization with 
a bulk--surface coupling. 

While in \S\ref{shsec}-\S\ref{pfsurf} we restrict to steady patterns, 
in \S\ref{brusec} we give an outlook on oscillatory patterns, including secondary bifurcations of periodic orbits 
such as period doubling, and explain a trick how to consider time periodic 
forcing. Along the way we also comment on the choice of meshes in 2D and 3D (\S\ref{msec}), 
and on tips and tricks (\S\ref{tntsec}, including deflation and 
time--integration, aka direct numerical simulation (DNS)) 
how to deal with problems which are characterized by a high multiplicity of 
solutions. The software \pdep, including 
all the demo directories and a number of further tutorials can be downloaded 
at \cite{p2phome}. 

\section{Some theory: pattern formation in the Swift-Hohenberg equation}
\label{pfssec}
Consider the (quadratic-cubic) Swift-Hohenberg (SH) equation 
\huga{\label{swiho} \pa_t u = -(1+\Delta)^2 u + \lam u  +\nu u^2-u^3, 
\quad u=u(x,t) \in \R,\ x\in\Om\subset \R^d, 
}
with instability parameter $\lam\in\R$,  second parameter $\nu\in\R$, 
and BCs $\pa_n u|_{\pa\Om}=\pa_n (\Delta u)|_{\pa\Om}=0$.
The original (cubic) SH model \cite{sh77} corresponds to $\nu=0$, 
while the case $f(u)=\nu u^3-u^5$ instead of $f(u)=\nu u^2-u^3$ 
is called the cubic-quintic SH equation. Swift--Hohenberg equations of this 
type are canonical and much studied model 
problems for pattern formation in dissipative system 
\cite{CH93, pismen06, SU17}. 
For later comparison with the numerics, we start with some 
theory for \reff{swiho}, 
already using numerical results from the \pdep\ demo directory 
{\tt sh} for illustration, but conversely no problem specific analytical 
results (except of symmetries) are used in the numerics. 

For us, the main advantage of the SH equation compared to RD systems 
of type \reff{rd1}, which may show exactly the same type of (Turing) 
instabilities, is that the SH equation 
allows much simpler and explicit computation of the amplitude equations 
on the center manifold at bifurcation from the trivial branch. 
Additionally, \reff{swiho} is a gradient system $\pa_t u=-\nabla \CE(u)$  
wrt the energy 
\huga{\label{en1}
\CE(u)=\int_\Om \frac 1 2 ((1+\Delta)u)^2-\frac 1 2 \lam u^2-F(u)\dd x, 
\quad F(u)=\int_0^uf(v)\dd v,
} 
where either $\Om=\R^d$ or $\Om$ a bounded domain and as above we assume 
the homogeneous Neumann BCs $\pa_x u|_{\pa\Om}=\pa_x\Delta u|_{\pa\Om}=0$. 
In particular, local minima of $\CE$ are stable stationary solutions of \reff{swiho}, and \reff{swiho} does not have time--periodic solutions (with 
finite energy). 
Moreover, the translational invariance of $\CE$ yields the 
existence of a spatially conserved quantity for steady solutions, 
a Hamiltonian, cf., e.g., \mbox{\cite[Proposition 1]{strsnake}.} 
If for instance we consider the steady problem in a spatial dynamics formulation in 1D, i.e., 
$U=(u_1,u_2,u_3,u_4):=(u,\pa_x u,\pa_x^2 u,\pa_x^3u)$ such that 
$$
\frac{{\rm d}}{{\rm d}x} U=\bigl(u_2, u_3, u_4,-2u_2-(1-\lam)u_1+f(u_1)\bigr)^T, 
$$  
then the Hamiltonian, written as a function of $u$, 
\hual{
H(u)&=\pa_xu\pa_x^3u-\frac 1 2 (\pa_x^2u)^2+(\pa_x u)^2+\frac 1 2 (1-\lam)u^2-F(u),\quad F(u)=\int_0^uf(v)\dd v, \label{ham1d}
}
is conserved, i.e., $\frac{{\rm d}}{{\rm d}x}H(u(x))=0$. 
While we do not make use of the energy \reff{en1}, 
$H$ can be used to discuss the location (in parameter space) of 
localized patterns, see \S\ref{shsec}. A similar Hamiltonian also 
exist in 2D, see, e.g., \cite{strsnake}. 

For all $\lam\in\R$, 
 \reff{swiho} has
the spatially homogeneous state $u^* \equiv 0$ (trivial branch). 
For $\Om=\R^d$, the linearization $\pa_t v=-(1+\Delta)^2v+\lam v$  
at $u^* \equiv 0$ has the solutions $v(x,t)
= \ee^{ \ii k\cdot x+ \mu(k) t }$, $k\in\R^d$, where
\begin{equation}
  \mu(k, \lam) =- (1-|k|^2)^2+  \lam, \quad |k|^2:=k_1^2+\ldots+k_d^2.  
\end{equation}
Thus, $u^*\equiv 0$ is asymptotically stable for $\lam<0$, unstable 
for $\lam>0$ with respect to periodic waves with wave vector $k$ with $|k|=k_c = 1 $, 
and in 1D we expect a pitchfork  bifurcation of spatially $2\pi$ periodic patterns  at $\lam=0$, if permitted by the domain and the BCs. 

\brem\label{nurem} 
a) In the following we rather briefly describe the computations of the 
pertinent  amplitude equations. More details can be found in, e.g., \cite{ampsys}, where we 
moreover describe the \pdep\ tool {\tt ampsys}, which can be used to 
{\em automatically} compute the  amplitude equations 
with minimal user input, in 1D, 2D and 3D, 
and for SH type of equations and RD systems.\\
b) Since \reff{swiho} with $f(u)=\nu u^2-u^3$ has the equivariance 
$(u,\nu)\mapsto (-u,-\nu)$ it is sufficient to restrict to $\nu\ge 0$. 
\eex
\erem

\subsection{1D} 
\begin{figure}[ht]
\bce 
\begin{tabular}{p{0.25\textwidth}l}
(a) spectral curve, and admissible $k$;&\quad(b) modes for $k=1,3/4,5/4$. \\
\ig[width=0.23\textwidth]{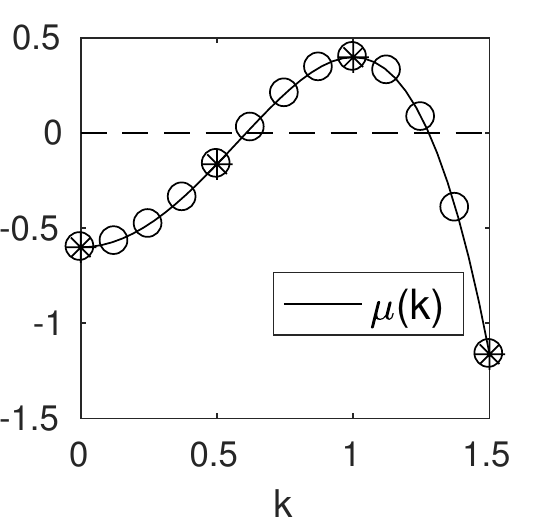}&\quad 
\raisebox{4mm}{\ig[width=0.5\textwidth]{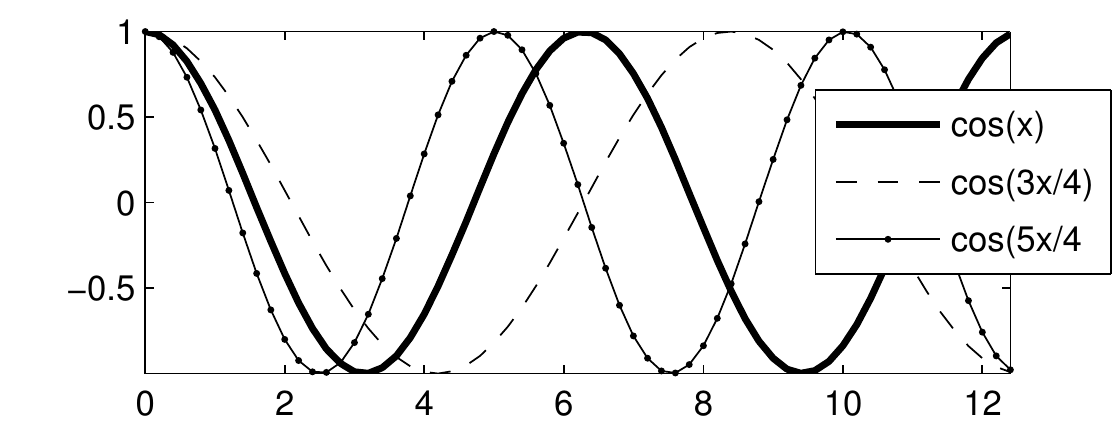}}
\end{tabular}
\ece
\vs{-5mm}
   \caption{{\small (a) Dispersion relation with admissible $k$ for $\Om=(0,\pi)$ (*) and $\Om=(0,4\pi)$ ($\circ$). (b) First 3 modes for $\Om=(0,8\pi)$. 
  \label{shf77a}}}
\end{figure}

Over $\R$ we have two bands of unstable wave numbers $k$ around $\pm 1$, i.e., 
\huga{\label{ba1}
\text{$\CK_u=\left\{k\in\R: |k|\in\left(\sqrt{1-\sqrt{\lam}},\sqrt{1+\sqrt{\lam}}
\right)\right\}$. }
}
If $\Om=(-l\pi/2,l\pi/2)$), then the admissible wave numbers 
are $k\in\frac 1 {2l}\N$, and for large $l$ we have many bifurcation 
points for small $\lam> 0$. The first bifurcation at $\lam_1=0$ has $k_1=1$, 
then $k_{2,3}=1\pm 1/(2l)$, $k_{4,5}=1\pm 1/l, \ldots$, which are usually called sidebands of $k=1$. See Fig.~\ref{shf77a} for how the sidebands are filled for increasing $l$. Still, generically, BPs are simple. 
For $l\ra\infty$ the center 
manifold becomes smaller and smaller, and in the limit the bifurcating 
solutions must be described by the Ginzburg--Landau equation as an 
amplitude equation, see, e.g., \cite{mie99} and \cite[Chapter 10]{SU17}. 

For simplicity we first restrict to the primary bifurcation at $\lam=0, k=1$. 
To compute the amplitude equation on the center manifold
we make the ansatz $\lam=\mu\eps^2$, where $\mu=1$ or $\mu=-1$, and 
\huga{\label{ba2}
u(t,x)=\eps A_1(T)e_1+\eps^2\left[\frac 1 2 A_0(T)+A_2(T)e_2\right]+\cc+\hot, \quad e_j=\er^{\ri j x}, 
}
with complex coefficients $A_j=A_j(T)$, which depend on the slow time scale 
$T=\eps^2 t$. Furthermore, $\hot$ denotes higher order terms which are not relevant for the present computation, and $\cc$ stands for the complex conjugate 
of the preceding terms, to obtain real valued $u$. 
The $\cc$ of, e.g., $A_1e_1$ is also conveniently written as $A_{-1}e_{-1}$. Here the BCs enforce $\im(A_j)=0$ such that $A_{-j}=A_j$ and for instance 
$|A_j|^2=A_j^2$, but for the sake of generality we pretend that the $A_j$ are genuinely complex 
for $j\ne 0$, which, e.g., is the case for periodic BCs or homogeneous 
Dirichlet BCs. 

Plugging \reff{ba2} into \reff{swiho} we first obtain the $\CO(\eps^2)$ terms 
$
0=-A_0e_0-9A_2e_2+\nu(2|A_1|^2 e_0+2A_2^2e_2)+\cc, 
$ 
and solving for $A_0=2\nu|A_1|^2$ and $A_2=\frac 2 9 A_1^2$,  and collecting terms at $\CO(\eps^3e_1)$ yields 
\huga{
\dot A_1=A_1(\mu-c_1|A_1|^2)\text{ with } c_1=3-\frac{38}9\nu^2.
}
Thus, for $\nu^2<\nu_0^2:=\frac{27}{38}$ ($\nu^2>\nu_0^2$) we obtain a 
supercritical (subcritical) pitchfork bifurcation of $2\pi$ periodic solutions. 
In \S\ref{shsec} we first verify this numerically, and then focus on the case 
$\nu>\nu_0$. On large domains the subcritical bifurcation then yields interesting secondary bifurcation to snaking branches of localized patterns, 
see \S\ref{shsec}. 

\subsection{2D}
 Over $\R^2$, for $\lam>0$ we have an annulus $\CK_u(\lam):=\{k\in\R^2: |k|\in[\sqrt{1-\sqrt{\lam}},\sqrt{1+\sqrt{\lam}}]\}$ of unstable wave vectors. 
On a bounded box, its side-lengths determine which 
discrete wave vectors fall into $\CK_u(\lam)$,  respectively onto 
$\pa \CK_u(\lam)$, which in turn determines the sequence of bifurcation points, 
and in particular the dimension of the kernel. 
\subsubsection{A square domain.}\label{shsqd} 
\begin{figure}[ht]
\bce 
\begin{tabular}{lp{6cm}p{6cm}}
{\small (a) Admissible $(k_1,k_2)$}&{\small (b) 
'natural' planforms, and numerical eigenvectors at $\lam=0$.}&
{\small (c) planforms, and numerical eigenvectors at $\lam=1/16^2$.}\\
\ig[width=0.27\textwidth]{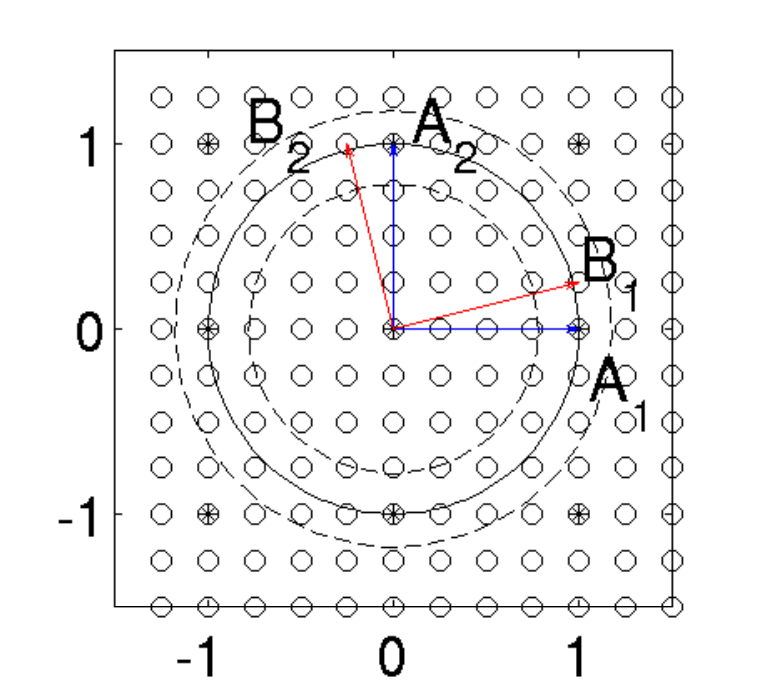}&
\hs{-4mm}\raisebox{25mm}{\begin{tabular}{l}
\ig[width=0.18\textwidth]{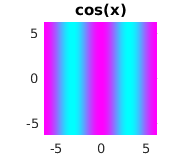}\\
\ig[width=0.18\textwidth]{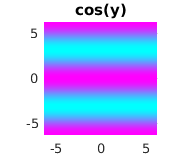}
\end{tabular}
\hs{-8mm}
\begin{tabular}{l}
\ig[width=0.16\textwidth]{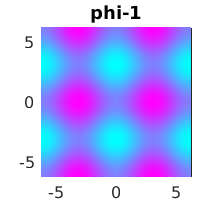}\\
\ig[width=0.16\textwidth]{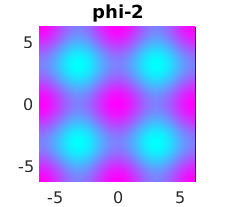}
\end{tabular}}
&
\hs{-4mm}\raisebox{25mm}{\begin{tabular}{l}
\ig[width=0.14\textwidth]{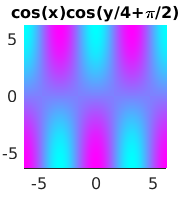}\\
\ig[width=0.14\textwidth]{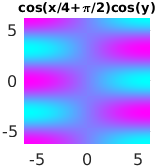}
\end{tabular}
\hs{-3mm}
\begin{tabular}{l}
\ig[width=0.14\textwidth]{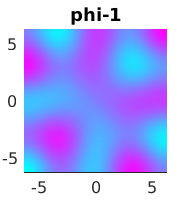}\\
\ig[width=0.14\textwidth]{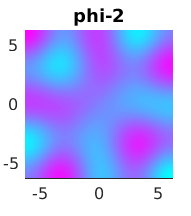}
\end{tabular}}
\end{tabular}
\ece
\vs{-5mm}
   \caption{{\small Spectral situation for the linearization of the SH equation around $u\equiv 0$ over square domains. (a) Admissible wave vectors $k$ for $\Om=(-\pi/2,\pi/2)^2$ (*) and 
$\Om=(-2\pi,2\pi)^2$ ($\circ$), respectively, with Neumann BCs. The amplitudes 
$A_{1,2}$ and $B_{1,2}$ are used in the amplitude equations below. 
(b,c) Kernels at 
the first two bifurcations. In (b), the left column shows the 'natural' planforms of stripes, corresponding to $A_1$ and $A_2$, and the right column the kernel vectors obtained 
numerically. From these the bifurcation directions must be obtained 
from the solution of the CBE \reff{cbe} below. Here it is rather easy to see 
that $\cos(x), \cos(y)$ correspond to $\phi_1+\phi_2$ and $\phi_2-\phi_1$. 
However, this already becomes slightly more difficult at the second BP in (c), 
where again the left column shows the planforms corresponding to $B_1$ and $B_2$. 
  \label{shf77b}}}
\end{figure}

We first let $\Om=(-l_1\pi,l_1\pi)\times 
(-l_2\pi, l_2\pi)$, $l_1,l_2\in\N/2$, such that $\mu_1=0$ at $\lam=0$ is double with 
$k^{(1)}=(1,0)$, $k^{(2)}=(0,1)$. 
The 'natural' associated planforms $u_1=\cos(x)$ and $u_2=\cos(y)$ are called 
stripes. However, any linear combination of these 
vertical and horizontal stripes are also in the kernel.  In particular, 
combinations of type $u_1+u_2$ yield spots, and to see what (if any) patterns bifurcate we should compute the amplitude equations. This has 
for instance been carried out in general form in \cite{bard91}. These computations can greatly benefit from 
symmetry considerations, which yield that the reduced system must 
always be of the form \reff{sqae} below, and that the only possible 
bifurcating branches are stripes and (regular) spots. 
Finally, numerical kernel 
computations just yield two (orthogonal) kernel vectors, not knowing a 
'natural' base, see Fig.~\ref{shf77b} for some examples.

Here we briefly go through the amplitude equation computations, in a rather ad hoc way, see 
\cite{GoS2002,hoyle} for background on symmetry considerations. 
We let $\lam=\mu\eps^2$, $\mu=\pm 1$, and $e_{m,n}=\er^{\ri(mx+ny)}$, 
and make the ansatz 
\huga{
u=\eps(A_1 e_{1,0}+A_2e_{0,1})+\eps^2(\frac 1 2 A_0+A_{1,1}e_{1,1}+A_{-1,1}e_{-1,1}+A_{2,0}e_{2,0}+A_{0,2}e_{0,2})+\cc, 
\label{shcm1}
}
where again $\cc$ stands for the complex conjugate since we look for real 
solutions. 
The BCs enforce that the $A_k$ and $A_{k_1,k_2}$ are all real, and we could 
as well use $e_1=\cos(x_1)$ and $e_2=\cos(x_2)$ and similar 
for $e_{k_1,k_2}$, but the (here formally) complex calculus is 
more general (i.e., also applies to periodic BCs, where $\im A_k$ may 
be non zero), and, moreover, is typically more convenient. 
Similarly, for convenience we use $A_1$ and $A_2$ instead of the more consistent
notations $A_{1,0}$ and $A_{0,1}$, respectively. Finally,  $A_k, A_{k_1,k_2}$ 
in \reff{shcm1} are again functions of the slow time $T=\eps^2 t$. 
\allowdisplaybreaks
From \reff{shcm1} we obtain 
\hualst{
u^2=&\eps^2\biggl[A_1^2e_{2,0}{+}A_2^2e_{0,2}{+}2(|A_1|^2{+}|A_2|^2)e_0{+}2A_1A_2e_{1,1}{+}
2A_{-1}A_2e_{-1,1}\biggr]\\
&{+}2\eps^3\biggl[A_0(A_1e_1{+}A_2e_2){+}(A_{2,0}A_{-1}{+}A_{1,1}A_{-2}{+}A_{1,-1}A_2)e_1
{+}(A_{0,2}A_{-2}{+}A_{1,1}A_{-1}{+}A_{-1,1}A_1)e_2\biggr]\\
&+\cc+\hot,\\
u^3=&3\eps^3\bigg[(|A_1|^2{+}2|A_2|^2)A_1e_1{+}(2|A_1|^2{+}|A_2|^2)A_2e_2)\biggr]
+\cc+\hot, \\
\pa_t u=&\eps^3(\dot A_1e_1+\dot A_2e_2)+\cc+\hot, 
}
where $\hot$ stands for both, terms of higher order in $\eps$ and terms 
that at $\CO(\eps^3)$ 
fall onto stable wave vectors and are hence irrelevant for the 
further computations. Collecting terms at $\CO(\eps^2)$ and solving for 
$A_0, A_{2,0}, A_{0,2}, A_{1,1}$ and $A_{-1,1}$ yields 
\hualst{
&A_0=2\nu(|A_1|^2{+}|A_2|^2),\quad A_{2,0}=\frac {\nu} 9 |A_1|^2, \quad 
A_{0,2}=\frac {\nu} 9 |A_2|^2, \quad A_{1,1}=2\nu A_1A_2,\quad 
A_{-1,1}{=}2\nu A_{-1}A_2, 
} 
and the complex conjugate equations for $A_{-2,0},\ldots,A_{1,-1}$. 
Now collecting terms at $\CO(\eps^3 e_1)$ and $\CO(\eps^3 e_2)$ yields 
the amplitude equations 
\huga{\label{sqae} \frac{d}{dT}\bpm A_1\\ A_2\epm=\bpm A_1(\mu-c_1 |A_1|^2-c_2|A_2|^2)\\
A_2(\mu-c_1|A_2|^2-c_2|A_1|^2)\epm, \quad c_1=3-\frac{38}9\nu^2, \quad c_2=6-12\nu^2. 
}
The amplitude equations (truncated at third order) for the bifurcations from $u\equiv u_0$ on a square with 
two dimensional kernel {\em always} take the form \reff{sqae}, see 
\cite{bard91,GoS2002} and \cite[\S4.3.1, \S5.3]{hoyle}, and the specifics 
of the system condense in the coefficients $c_1,c_2$. 

From \reff{sqae} we find that the bifurcation problem at $(u,\lam)=(0,0)$ on the square is 3-determined except if $c_1=0$ or $|c_1|=|c_2|$. Here, 
a problem is called $k$--determined if the Taylor expansion 
up to order $k$ is sufficient to uniquely determine all small solutions, 
i.e., if any small perturbation of order 
$k+1$ does not qualitatively change the set of (small) solutions, 
see \cite{mbiftut} and \cite[\S6.7]{mei2000} for further discussion. 
If $c_2=c_1$, then, 
returning to real notation $A_{1,2}\in\R$ and wlog assuming that $\mu,c_1>0$, 
\reff{sqae} has the circle $A_1^2+A_2^2=\mu/c_1$ of nontrivial solutions. 
For $c_2=-c_1$, we have 'vertical branches' of spots $\mu=0$ and $(A_1,A_2)=s(1,\pm1)$, $s\in\R$, and for $c_1=0$ 
we have vertical branches $\mu=0$ of stripes $(A_1, A_2)=s(0,1), (A_1, A_2)=s(1,0), s\in\R$. In all these cases, 
the bifurcating branches would be determined at fifth order. 

Our particular problem \reff{sqae} at the first BP is thus 
3-determined except if $\nu\in\{\nu_1,\nu_2,\nu_3\}$, where 
$$
\text{$\nu_1:=\sqrt{\frac{27}{70}}$\ \ ($c_1=c_2>0$), \quad 
$\nu_2:=\sqrt{\frac{81}{146}}$\ \ ($c_1=-c_2>0$), \quad 
$\nu_3=\sqrt{\frac{27}{38}}$\ \ ($c_1=0$).}
$$
For $\nu\not\in\{\nu_1,\nu_2,\nu_3\}$ we have the nontrivial solutions 
\huga{\label{sqsol}
\begin{split}
&A_1=A_2=\pm\sqrt{\mu/(c_1+c_2)}\text{ (spots),  } 
A_1=\pm \sqrt{\mu/c_1}, A_2=0\text{ (or $A_1,A_2$ interchanged, stripes),}
\end{split}
}
where we assume the right sign of $\mu$ for the respective solutions 
to exist (sub-or supercritically).  Moreover, also the stability 
of the nontrivial solutions can immediately be evaluated, 
see  \cite[Theorem]{bard91}, \cite[Fig.~4.10]{hoyle}. 

\begin{figure}[ht]
\bce 
\ig[width=0.6\textwidth]{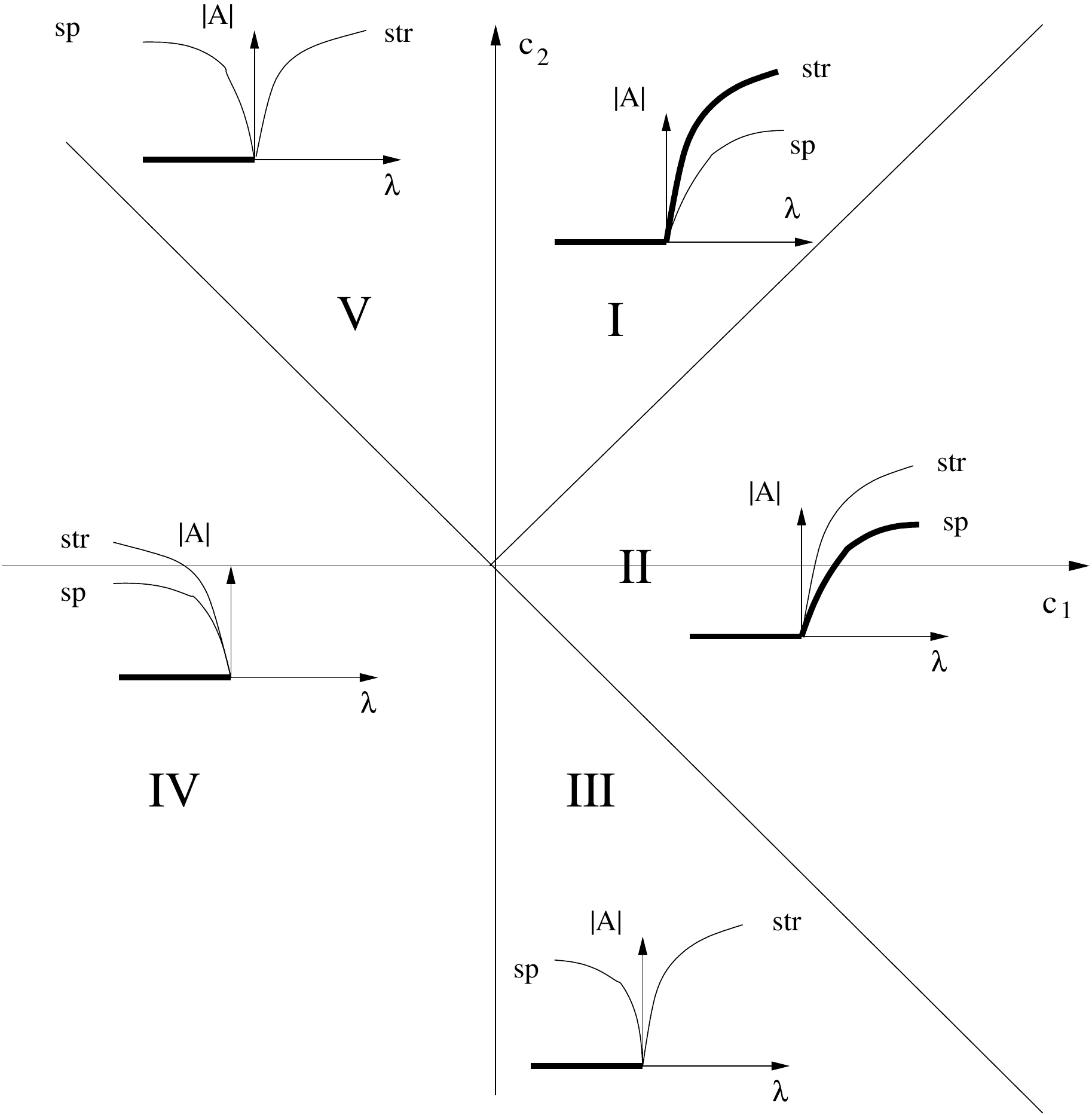}\qquad 
\raisebox{50mm}{
\begin{tabular}{l}
\ig[width=0.22\textwidth,height=30mm]{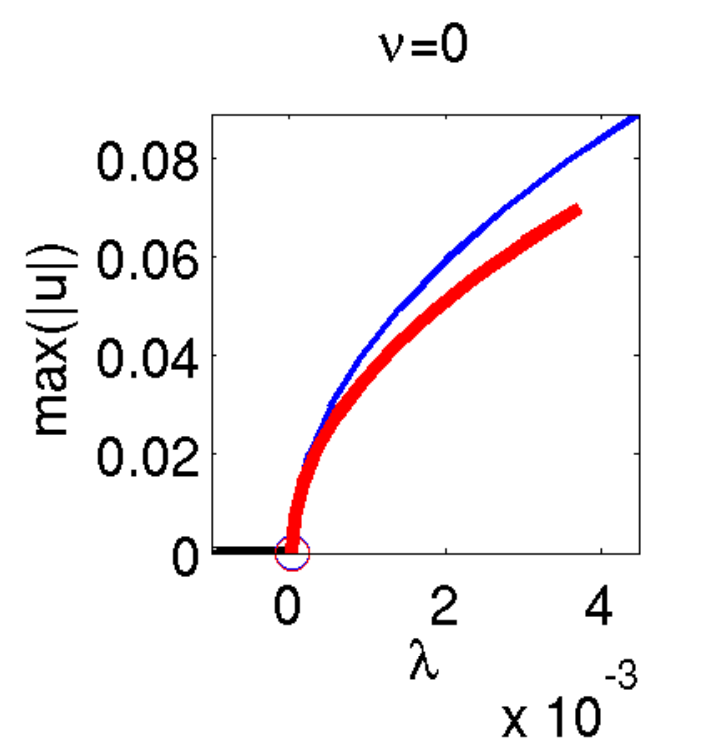}\\
\ig[width=0.22\textwidth,height=30mm]{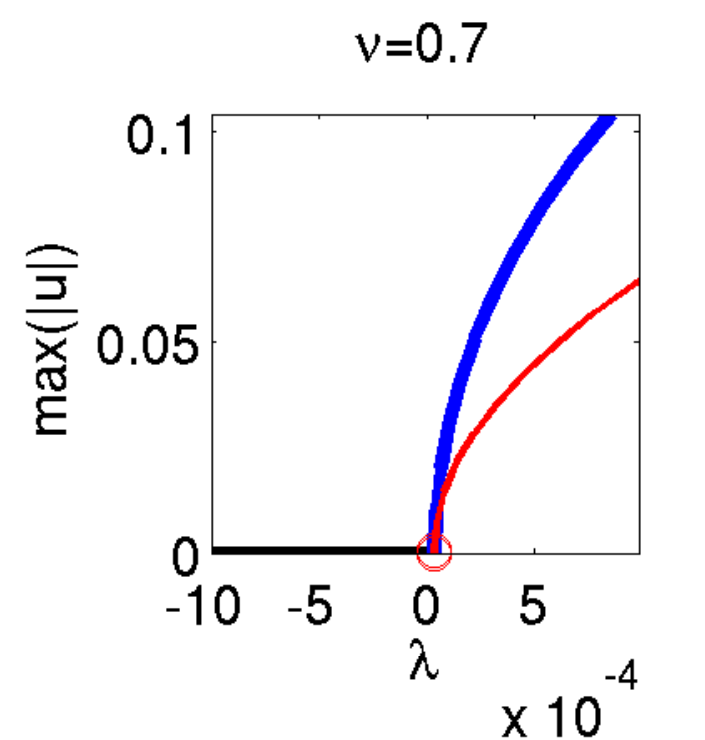}\\
\ig[width=0.22\textwidth,height=30mm]{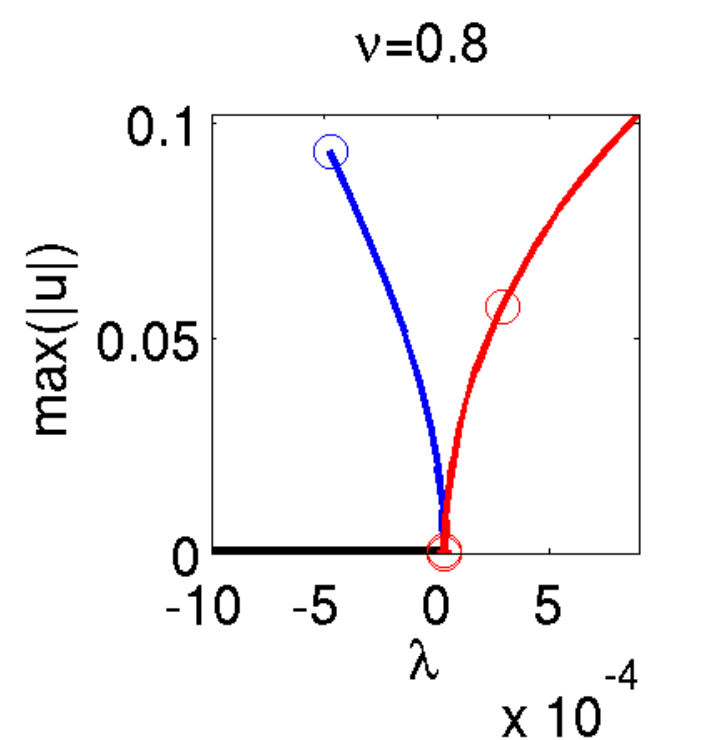}\\
\ig[width=0.22\textwidth,height=30mm]{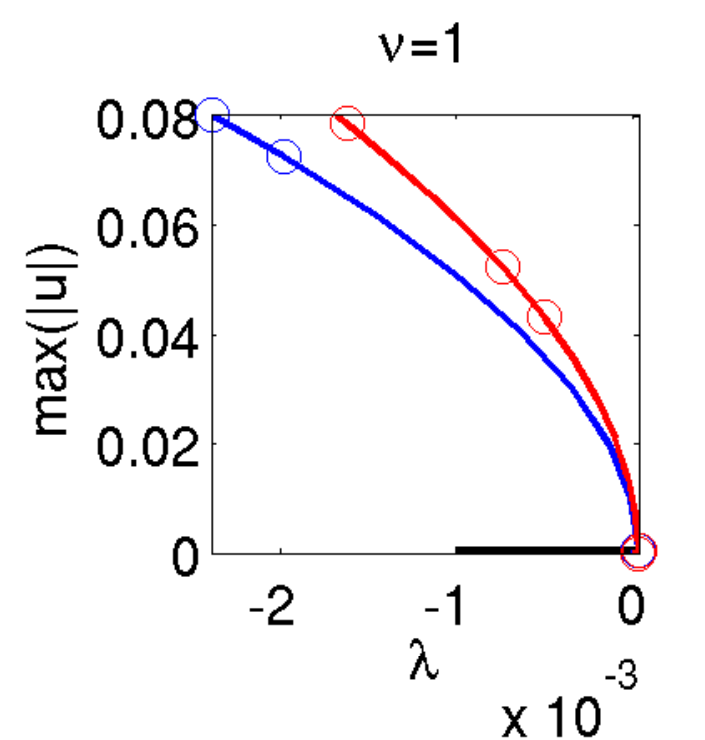}
\end{tabular}}
\ece
\vs{-2mm}
   \caption{{\small Left: Bifurcation (and stability) diagram 
for \reff{sqae} in the $c_1$-$c_2$ plane. Right: corresponding 
numerical bifurcation diagrams for \reff{swiho} in dependence of $\nu$, 
stripes=red, spots=blue, thick lines=stable solutions, thin lines=unstable 
solutions. 
  \label{shf77b2}}}
\end{figure}

\blem\label{sqstablem} The stripes are stable if  
$0<c_1<c_2$. The spots are stable if $0<|c_2|<c_1$. 
\elem 
The bifurcation behavior is illustrated in Fig.~\ref{shf77b2}. 
On the boundaries of the sectors 
we would need higher order terms in \reff{sqae} to discuss solutions. 
However, if we ignore these boundaries, then the two statements in 
Lemma \ref{sqstablem} read 'if and only if'. An interesting immediate consequence of this is that (close to bifurcation) spots and stripes are 
mutually exclusive as stable patterns. For \reff{swiho}, the sectors for 
$c_1,c_2$ from \reff{sqae} as a function of $\nu$ are given in \reff{nutab}, 
and the right column of Fig.~\ref{shf77b2} confirms the predictions 
of the amplitude equations via \pdep, see \S\ref{shsec}. 
\huga{\label{nutab}
\text{
\begin{tabular}{c|cccc|}
$I_\nu$&$\left[0,\nu_1\right)$&$(\nu_1,\nu_2)$&
$(\nu_2, \nu_3)$&$(\nu_3,\infty)$\\
sector&I&II&III&IV
\end{tabular}\ .
}
}

\brem\label{sh2nd}
The second bifurcation point is also double, and Fig.~\ref{shf77b}(c) illustrates the kernel over $\Om=(-2\pi,2\pi)^2$, spanned by, e.g., 
$e_1=\cos(x)\cos(y/4+\pi/2)$ and $e_2=\cos(x/4+\pi/2)\cos(y)$. If 
we make an ansatz 
$u=\eps(B_1\er^{\ri(x+y/4)}+B_2\er^{\ri(-x/4+y)})+\cc+\hot$, 
then an analogous computation as above yields the 
same amplitude equations \reff{sqae}. 
The nontrivial branches are unstable close to bifurcation as they bifurcate where the trivial branch is already unstable. However, 
such initially unstable branches may stabilize at larger amplitude, 
and in fact they do for \reff{swiho}, which 
is one reason why they may also be interesting. 
\eex \erem

\subsubsection{A rectangle with a hexagonal dual grid}\label{shhexd}
\begin{figure}[ht]
\bce 
\begin{tabular}{ll}
\ig[width=0.25\textwidth
]{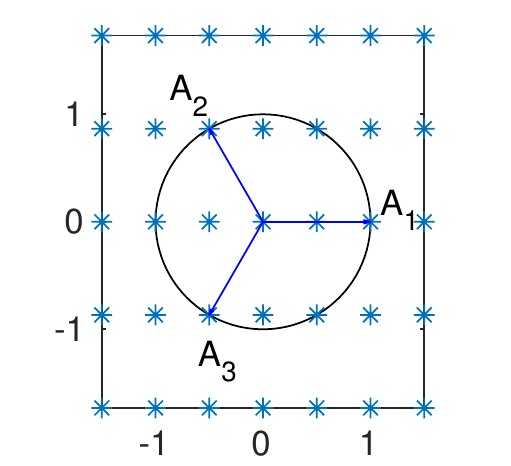}&\ig[width=0.25\textwidth]{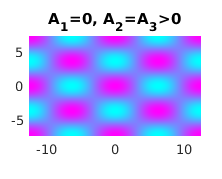}\\
\ig[width=0.25\textwidth]{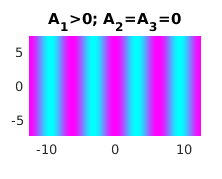}&
\ig[width=0.25\textwidth]{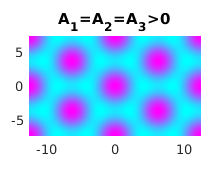}
\end{tabular}
\ece
\vs{-8mm}
   \caption{{\small Admissible wave vectors for the domain 
$\Om=(-\pi,\pi)\times(-\pi/\sqrt{3},\pi/\sqrt{3})$, and associated planforms 
at bifurcation, for clarity plotted over $\Om=(-4\pi,4\pi)\times(-4\pi/\sqrt{3},4\pi/\sqrt{3})$. 
  \label{shf77c}}}
\end{figure}

A special situation occurs for problems with quadratic terms over domains which allow resonant wave vector triads, i.e., critical wave vectors $k^{(1)}, k^{(2)}$ and $ k^{(3)}$ such that any $k^{(j)}$ is a linear combination of the other two. As these lie on the circle $|k|=k_c=1$, 
the angle between them is $2\pi/3$, and on a rectangular domain this 
is compatible with the BCs for  $\Om=(-l_1\pi,l_1\pi)\times (-l_2\pi/\sqrt{3}, l_2\pi/\sqrt{3})$, such that $\mu_1=0$ at $\lam=0$ is double with 
$k^{(1)}=(1,0)$, $\phi_1=\cos(x)$, $k^{(2)}=(-1/2,\sqrt{3}/2)$, and 
$k^{(3)}=-(1/2,\sqrt{3}/2)$, and, e.g., $\phi_2(x,y)=e_2+e_3=\cos(x/2)\cos(\sqrt{3}y/2)$, see Fig.~\ref{shf77c}. 

The ansatz 
\huga{\label{hexa} 
u(x,t)=A_1(t)e_1+A_2(t)e_2(t)+A_3(t)e_3+\hot
}
and the pertinent symmetry considerations yield the 
amplitude equations (in complex notation, although again for Neumann BCs 
the $A_j$ are real, and, moreover $A_3=A_2$), truncated at third order, 
\huga{\label{hexeq} 
\begin{split}
\dot A_1&=\lam A_1+\ga\ov{A_2A_3}-c_1 |A_1|^2A_1-c_2(|A_2|^2+|A_3|^2)A_1,\\
\dot A_2&=\lam A_2+\ga\ov{A_1A_3}-c_1 |A_2|^2A_3-c_2(|A_1|^2+|A_3|^2)A_2,\\
\dot A_3&=\lam A_3+\ga\ov{A_1A_2}-c_1 |A_3|^2A_3-c_2(|A_1|^2+|A_2|^2)A_3. 
\end{split}
}
The $\eps$--scaling from \reff{shcm1} is omitted in \reff{hexa} because 
the derivation of \reff{hexeq} assumes that the quadratic terms in 
the original system are small (and in particular $\ga=0$ if the 
quadratic terms vanish, i.e., if the original system has the symmetry $u\mapsto -u$). 
For \reff{swiho} with $f(u)=\nu u^2-u^3$ we obtain (recalling that 
we treat $|\nu|$ as small, and ) 
\huga{\label{hexcoeff} 
\ga=2\nu+\CO(|\lam\nu|), \quad c_1=3+\CO(|\lam|+\nu^2),\quad 
c_2=6+\CO(|\lam|+\nu^2).
} 
The problem has similar (non-generic) indeterminacies as \reff{sqae}, but is generically 3-determined, and for $\nu{=}0$ we have 
we have three bifurcating branches: 
\hualst{
\text{stripes:}\quad&A_1=\pm\sqrt{\mu/c_2},\quad A_2=A_3=0,\\
\text{hexagons:}\quad&A_1=A_2=A_3=A,\quad A=\frac{\ga}{2(c_1+2c_2)}\pm\sqrt{\frac {\ga^2}{4(c_1+2c_2)^2}+\frac \mu{c_1+2c_2}}\\
\text{patchwork quilt:}\quad&A_1=0,\quad A_2=A_3=\pm\sqrt{\frac\mu{c_1+c_2}} 
\text{ \quad ({\em only} if $\ga=0$).}
}
For $\ga\ne 0$, the hexagons bifurcate transcritically. Those with $A>0$ ($A<0$) are 
also called 'up' ('down') hexagons or 'spots' ('gaps'), and the spots 
have a fold at $\mu=-\frac{\ga^2}{2(c_1+2c_2)}$. Analogously, the stripes 
with $A_1>0$ ($A_1<0$) could be called 'up' and 'down', but these 
are related via shifting by $\pi$ in $x$. 
The pq branch only 
bifurcates from the trivial branch if $\ga=0$, and turns into two 
secondary 'mixed-modes' (or 'rectangle') 
branches for $\ga\ne 0$, which connect  
the stripes and the up hexagons, and the 'down' stripes and the down 
hexagons (respectively), 
namely
\hualst{
\text{mixed modes:}\quad&A_1=-\frac{\ga}{c_1-c_2},\quad A_2=A_3=
\pm\sqrt{\frac 1{c_1+c_2}(\mu-\frac{\ga^2c_1}{(c_1-c_2)^2})}. 
}

The stability of these branches obtained from \reff{hexeq} is as follows. 
If $\ga=0$, then the stripe and pq-branches are stable and the hex branch 
is unstable. For $\ga\ne 0$ the stripes are unstable at bifurcation, and 
become stable at $\mu=\ga c_1/(c_1-c_2)^2$. The up hexagons are stable 
after the fold, until $\mu=\ga^2(2c_1+c_2)/(c_1-c_2)^2$, and 
the mixed modes for $\ga\ne 0$ are never stable. 
These results from the amplitude equations \reff{hexeq} are 
confirmed and illustrated 
by \pdep\ numerics over $\Om=(-2\pi,2\pi)\times (-2\pi/\sqrt{3},2\pi/\sqrt{3})$ 
in Fig.~\ref{shf10} in \S\ref{sh2dnum}

\subsection{3D} \label{sh3dsec}
In 3D, the situation naturally becomes more complex. 
We now have a spherical shell $S(\lam) 
:=\{k\in\R^3: |k|\in(\sqrt{1-\sqrt{\lam}},\sqrt{1+\sqrt{\lam}}\}$
of unstable wave vectors, and 
the determination of the branching behavior from 
the trivial branch is a complicated problem which in general requires lengthy 
analysis based on results from (symmetry) group theory. 

The simplest situation is the so called simple cubic (SC) lattice, spanned 
by the wave vectors $k_1=(1,0,0), k_2=(0,1,0), k_3=(0,0,1)$, 
where wlog we focus on the first bifurcation such that $k_c=1$. 
This corresponds to a cube of side-lengths $l=2j\pi$, $j\in\N$, e.g.~$\Om=(-\pi,\pi)^3$, with periodic 
BCs, while Dirichlet or Neumann BCs as above 
reduce the problem to smaller solution sets, which we therefore call a 
sublattice problem. The ansatz for the amplitude equations reads $\lam=\eps^2\mu, \mu=\pm 1, 0<\eps\ll 1$, and 
\allowdisplaybreaks[0]
\hual{
u=&\eps(A_{100}+A_{010}+A_{001})\notag\\
&+\eps^2(\frac 1 2 A_{000}+A_{110}+A_{101}+A_{011}+A_{200}
+A_{020}+A_{002}+A_{1-10}+A_{10-1}+A_{01-1})+\cc,
\label{shcm3d1}
}
where $A_{100}, A_{111}$ etc are shorthands for $\tilde{A}_{lnm}(T)e_{lnm}(x)$, 
i.e., coefficient function $\tilde{A}_{lnm}$ and mode $e_{lnm}=\exp(\ri(lx_1+nx_2+mx_3))$. 
Then going through the analysis (where again symmetry theory is very helpful),  
at $\CO(\eps^3 e_{100}), \CO(\eps^3 e_{010})$, and $\CO(\eps^3 e_{001})$ we 
obtain the amplitude equations 
\huga{\label{sqae3D} 
\frac{d}{dT}\bpm A_1\\ A_2\\A_3\epm=
\bpm 
A_1(\mu-c_1|A_1|^2-c_2(|A_2|^2+|A_3|^2))\\
A_2(\mu-c_1|A_2|^2-c_2(|A_1|^2+|A_3|^2))\\
A_3(\mu-c_1|A_3|^2-c_2(|A_1|^2+|A_2|^2))
\epm, 
}
where again $c_1=3-\frac{38}9\nu^2$ and $c_2=6-12\nu^2$. 
Naturally, this contains the system \reff{sqae} 
as a subsystem with $A_3=0$ (or $A_1=0$ or $A_2=0$), 
and the stripes and spots of the 2D problem are now classified as 
{\em lamellas} $A_1=\pm \sqrt{\mu/c_1}, A_2{=}A_3{=}0$, and {\em tubes}  $A_1=A_2=\pm\sqrt{\mu/(c_1+c_2)}, A_3=0$, respectively. 
Clearly $A_1,A_2,A_3$ can be permuted,  
giving different orientations of the lamellas and tubes. Additionally 
we have the {\em rhombs}  $A_1=A_2=A_3=\pm\sqrt{\mu/(c_1+2c_2)}$, where again depending on $c_1,c_2$ 
we assume the right sign of $\mu$. 

Moreover, \reff{sqae3D} is 3--determined except if $c_1=0, 
|c_1|=|c_2|$, and  additionally if $|c_1|=2|c_2|$. For $c_1=0, 
|c_1|=|c_2|$ we have non-isolated solutions as above with $A_3=0$. 
For $c_1=2c_2$ we have the sphere $A_1^2+A_2^2+A_3^2=\mu/c_1$ of 
non-isolated solutions, and for $c_1=-2c_2$ we have vertical branches $\mu=0$, $(A_1,A_2,A_3)=s(1,\pm1,\pm1)$, $s\in\R$, of rhombs. The additional exceptional 
values of $\nu$ are $\nu_4=\sqrt{\frac{81}{178}}$ 
($c_1=2c_2$) and $\nu_5=\sqrt{\frac{243}{252}}$ 
($c_1=-2c_2$). 
The stabilities of the nontrivial branches on the amplitude equations 
level can efficiently be computed using symmetry, see, e.g., \cite{CKnob99,CKnob99,CKnob01}, 
which inter alia yields that the tubes are always unstable close to 
bifurcation, and either the rhombs or the lamellas can be stable, but 
not both, see also \cite[Theorem 1]{alberetal05}. 

In \S\ref{3dshnumsec} we illustrate some of these results for the 
SH equation on the cube $\Om=(-\pi,\pi)^3$, and additionally present 
results for a so called BCC (body-centered cubic) (sub-)lattice problem 
on the  cube $\Om=(-\sqrt{2}\pi,\sqrt{2}\pi)^3$. %
\section{Numerics for the Swift--Hohenberg equation}\label{shsec}
\def\dhome{./pftut/sh}\def\dname{sh}
\subsection{Overview}
Besides the connection to the analytical results 
from \S\ref{pfssec}, the SH equation gives an opportunity to show how 
to rewrite this 4th order (in space) equation as a 2--component 2nd order 
system in a consistent way. Recall 
that \pdep\ uses the finite element method (FEM) to convert a system of (2nd order in space) PDEs 
of the form \reff{rd1} into a system of ODEs 
\def\ru{{\rm u}}
\huga{\label{ode1}
\CM\ddt\ru=-G(\ru)
}
for the unknown nodal values $\ru\in\R^{n_u}$, where 
$\CM\in \R^{n_u\times n_u}$ 
denotes the so called mass matrix.%
\footnote{See, e.g., \cite{actut,hotheo} for 
details on the general classes of systems of 
PDEs that \pdep\ can treat.} 
Thus, let $(u_1,u_2)=(u,\Delta u)$ to obtain the 2nd order system 
\huga{\label{shsys}
\bpm 1&0\\0&0\epm \pa_t \bpm u_1\\ u_2\epm=
\bpm -\Delta u_2-2u_2-(1-\lam)u_1+f(u_1)\\-\Delta u_1+u_2\epm, 
}
which immediately translates into the FEM formulation 
(dropping the notational distinction between the function $u$ and the 
nodal values $\ru$) 
\huga{\label{shfem}
\CM\dot u=-(\CK u-F(u)), \\ 
\CM=\bpm M&0\\0&0\epm, \ \CK=\bpm 0&-K\\K&M\epm, 
\ F(u)=\CM\bpm (\lam{-}1)u_1{-}2u_2{+}f(u_1)\\ 0\epm, 
\notag 
}
where $K$ and $M$ correspond to the scalar stiffness 
and mass matrices, i.e., $M^{-1}K$ corresponds to $-\Delta$. 
\brem{\label{mrem} 
(a) 
The formulation \reff{shsys}, resp.~\reff{shfem} with the singular $\CM$ on the lhs 
yields the correct eigenvalues and hence stability information, and, moreover, 
can be used for time integration via {\tt tint}. See also 
\cite{hotheo} for the analogous construction for the Kuramoto-Sivashinsky 
equation, where it is used to compute Hopf bifurcations and periodic 
orbits. 

(b) Using the same $\CM$ on the rhs of \reff{shfem} as on the lhs is merely for convenience; --we 
could also implement $F$ in some other way. In particular, 
we use a 'simplified FEM' setup, where we do not interpolate 
$u$ from the nodal values to the element centers and then 
evaluate the nonlinearity and subsequently the pertinent integrals 
over elements. See \cite[Remark 1.1]{actut} for further comments.  
}\eex \erem  

See Table \ref{shdirtab} for an overview of files used to 
implement \reff{shfem} in 1D, 2D and 3D.  
In {\tt oosetfemops} we essentially preassemble matrices ${\tt p.mat.M}=\CM$ and ${\tt p.mat.K}=\CK$ , and then set up the rhs 
in {\tt sG.m} (and the Jacobian in {\tt sGjac})  in a standard way, 
see Listing \ref{shl2}. 
Also the init routine is completely standard, and thus below we restrict 
to brief remarks on the script files. 

\begin{table}[ht]\taskip
\caption{Main scripts and functions in {\tt pftut/sh}. 
\label{shdirtab}}
\bce\vs{-4mm}
{\small 
\begin{tabular}{l|p{0.73\textwidth}}
script/function&purpose, remarks\\
\hline
cmds1d, cmds1dhplot&scripts for 1D, essentially yielding Fig.~\ref{shf1}. \\%
cmds2dsq&script for 2D square domain, essentially yielding Fig.~\ref{shf77b}.\\%
cmds2dhex&script for 2D rectangular domain for hexagons, essentially yielding 
Fig.~\ref{shf10}.\\
cmds2dhexfro&script for localized hex patterns on a 2D long rectangular domain, 
see Fig.~\ref{shf11}\\
cmds2dhexfroada&\ma\ for localized hex patterns, 
see Fig.~\ref{shf11a}\\
cmds2dhexb&similar to cmds2dhex, but on domain twice as large; meant to 
illustrate tips and tricks, for problems with 'too many' patterns, essentially yielding Fig.~\ref{shf11b}.\\
cmds2dtint&script for patterns from initial guesses and 
time integration, see Fig.~\ref{shtf}.\\
cmds3dSC&script for simple cube (SC) 3D patterns, see Fig.~\ref{shf12}.\\
cmds3dBCC&script for body centered cube (BCC) 3D patterns, see Fig.~\ref{shf12b}.\\
cmds3dBCC&\ma\ for BCC 3D patterns\\ 
cmdsBCClong&script for localized BCC patterns obtained from initial 
guesses, see Fig.~\ref{shtf2}.\\
cmdsBCClongada&\ma\ for localized BCC patterns, see Fig.~\ref{shtf3}.\\
\hline 
shinit&initialization\\
oosetfemops&set FEM matrices (stiffness K,  and two mass matrices M, M0) \\
sG,nodalf,sGjac&encodes $G$ with 'nonlinearity' in nodalf, and Jacobian\\
spjac&Jacobian for fold continuation\\
shbra1d&modification of {\tt stanbra} for putting the Hamiltonian $H$ on 
the branch\\
geth&function to compute the Hamiltonian for the spatial dynamics formulation\\
e2rs&Element2RefineSelection function, used for mesh adaption, here just ad-hoc\\
hfplot,spl,spplots&convenience functions to plot solutions, Fourier transforms,  and planforms
\end{tabular}
}
\ece
\end{table}\teskip

\hulst{caption={},
label=shl1,language=matlab,stepnumber=10, linerange=1-10}{\dhome/oosetfemops.m}

\hulst{caption={},
label=shl111,language=matlab,stepnumber=5, linerange=1-10}{\dhome/nodalf.m}
\hulst{caption={{\small  {\tt oosetfemops.m}, {\tt nodalf.m} and {\tt sG.m} from {\tt pftut/sh}. Here, the 1st component of {\tt nodalf} contains 
``everything but diffusion'', including the linear terms 
$(\lam-1)u_1-2u_2$, while the 2nd component of {\tt nodalf} is $0$  
as we implement the 2nd equation from \reff{shsys} via $\CK$. 
}}, 
label=shl2, language=matlab,stepnumber=5, firstnumber=1}{\dhome/sG.m}

\subsection{1D} \label{sh1Dn}
\begin{figure}[H]
\bce 
\begin{tabular}{p{0.33\textwidth}}
{\small (a1) Primary branch (black) and two 
branches of localized patterns}\\
\ig[width=0.3\textwidth,height=65mm]{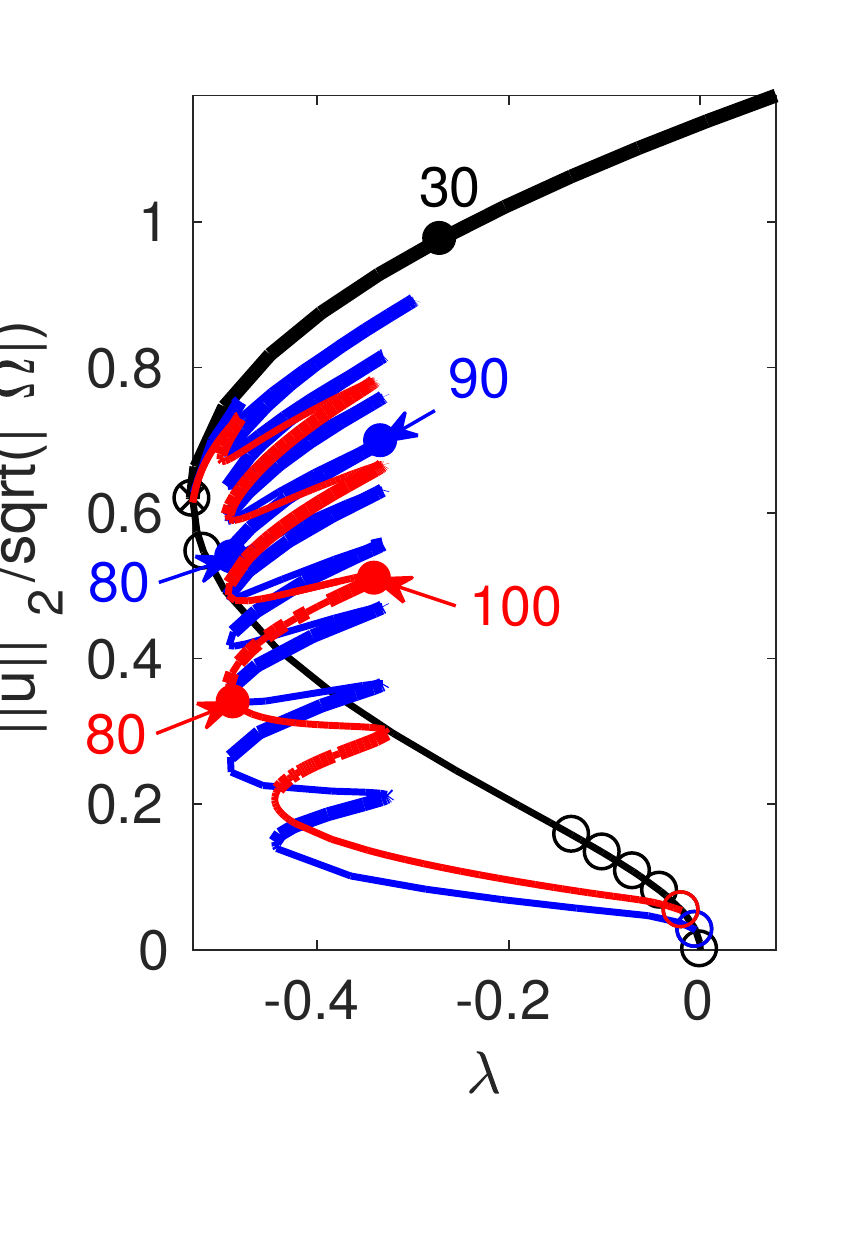}\\[-8mm]
{\sm (a2) blow-up from (a1)}\\
\ig[width=0.28\textwidth]{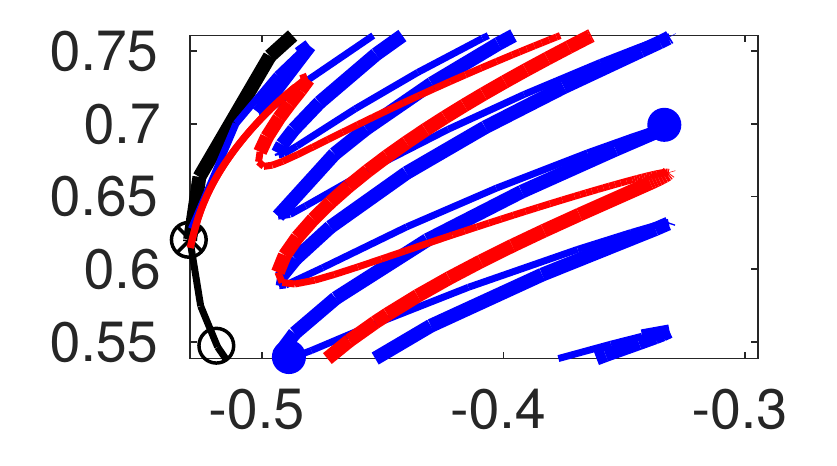}\\
{\sm (a3) $H$ on the first four periodic branches, $\Om=(-20\pi,20\pi)$}\\ 
\ig[width=0.28\textwidth]{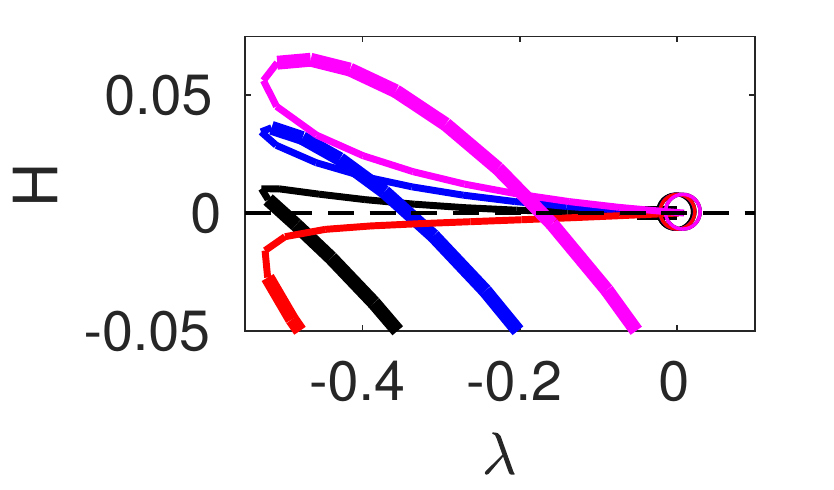}
\end{tabular}
\begin{tabular}{p{0.3\textwidth}}
{\sm (b) solution plots, including $H$}\\
\ig[width=0.3\textwidth]{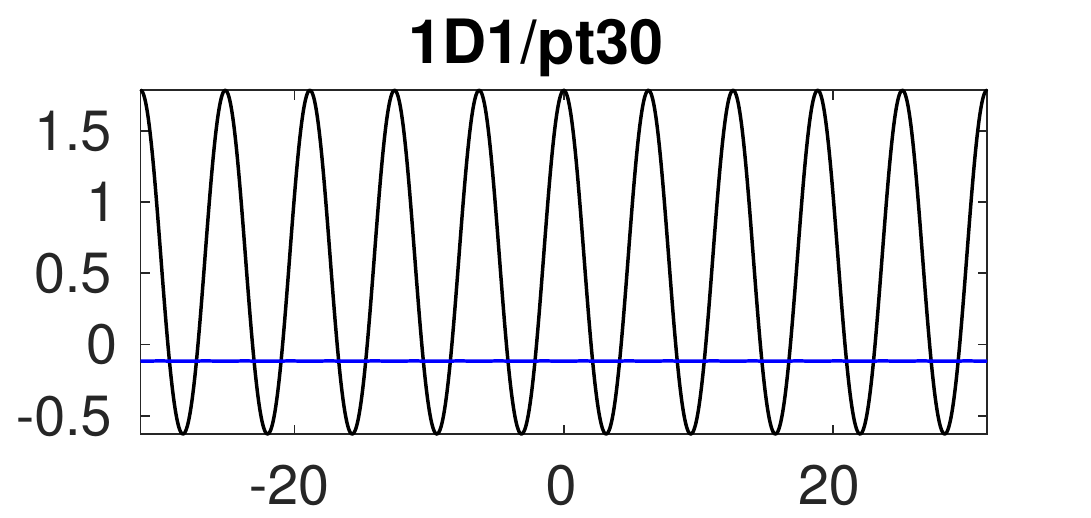}\\
\ig[width=0.3\textwidth]{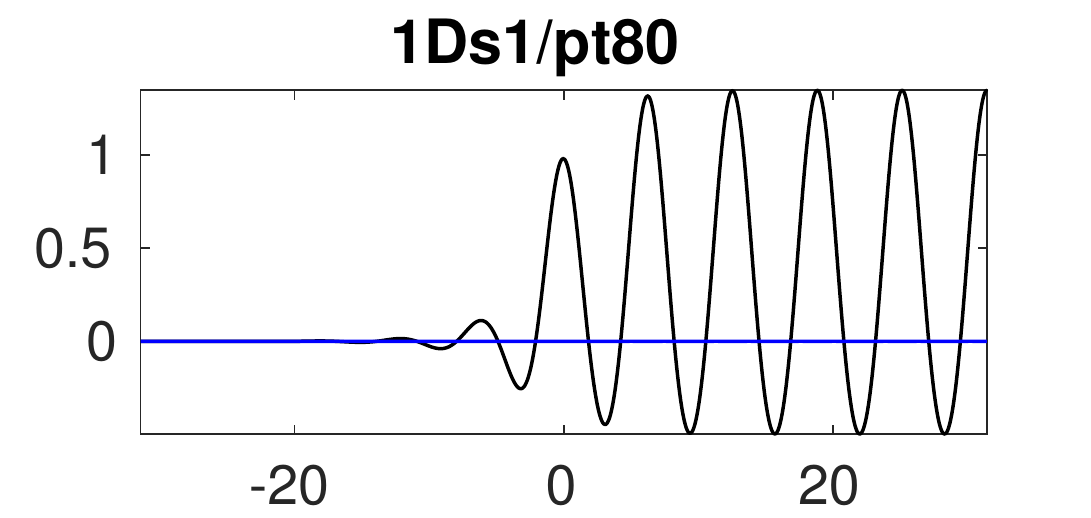}\\
\ig[width=0.3\textwidth]{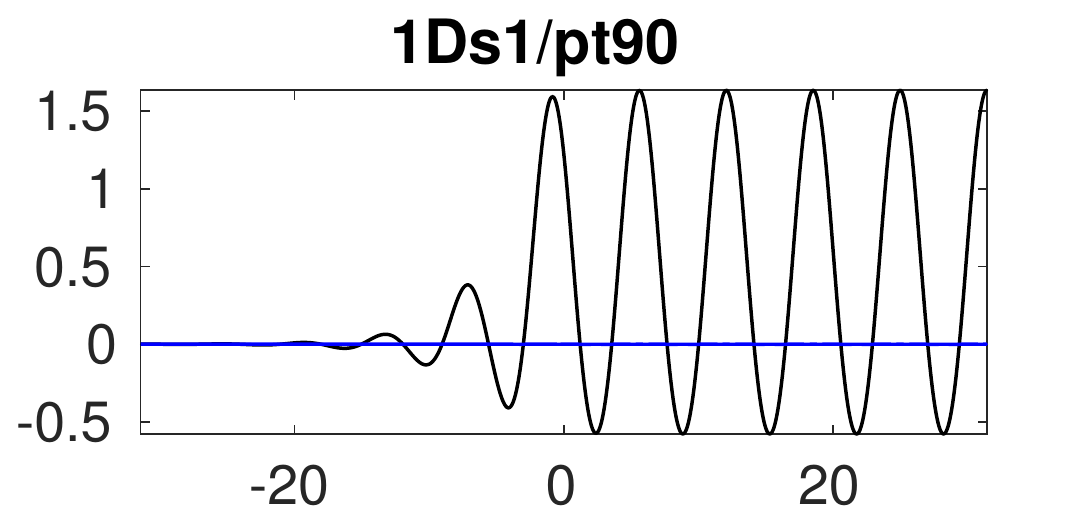}\\
\ig[width=0.3\textwidth]{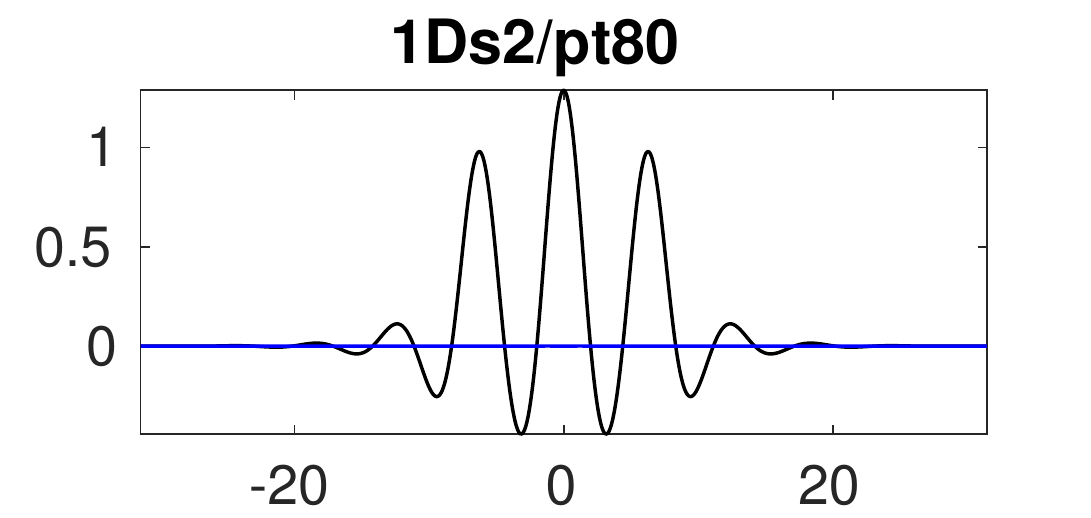}\\
\ig[width=0.3\textwidth]{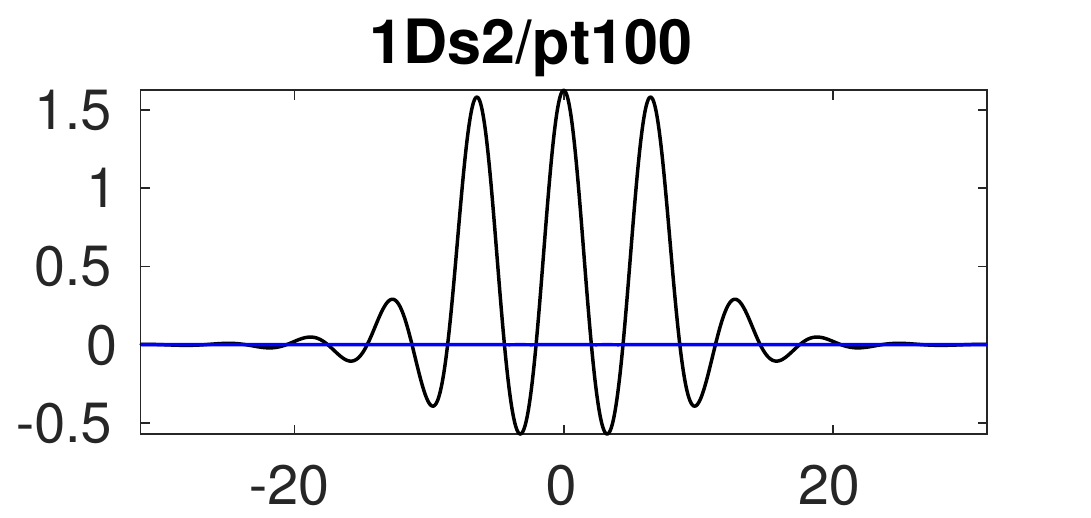}
\end{tabular}
\begin{tabular}{p{0.27\textwidth}}
{\sm (c) $|\CF(u)(k)|$ from (b)}\\
\ig[width=0.3\textwidth]{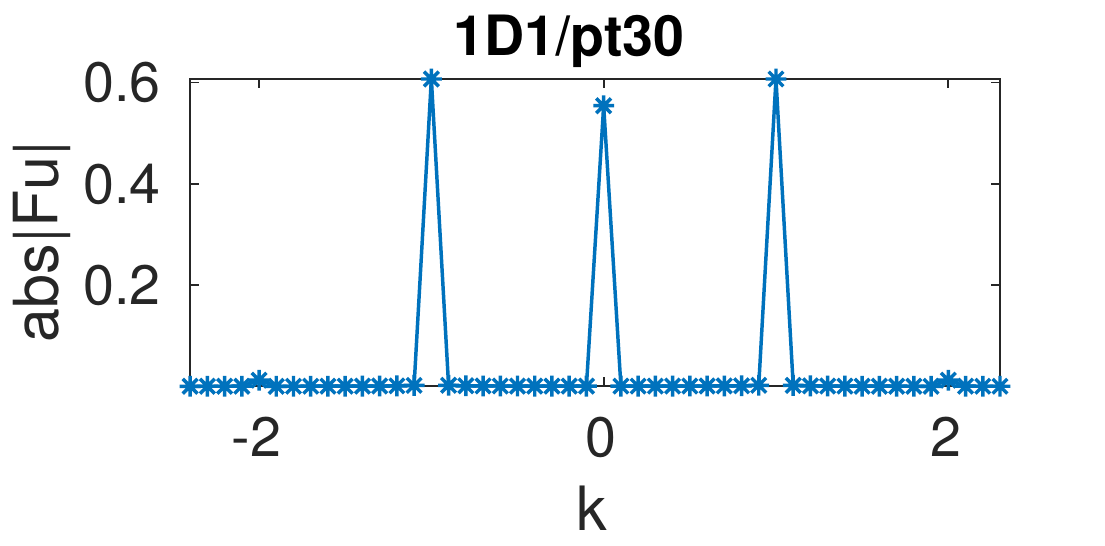}\\
\ig[width=0.3\textwidth]{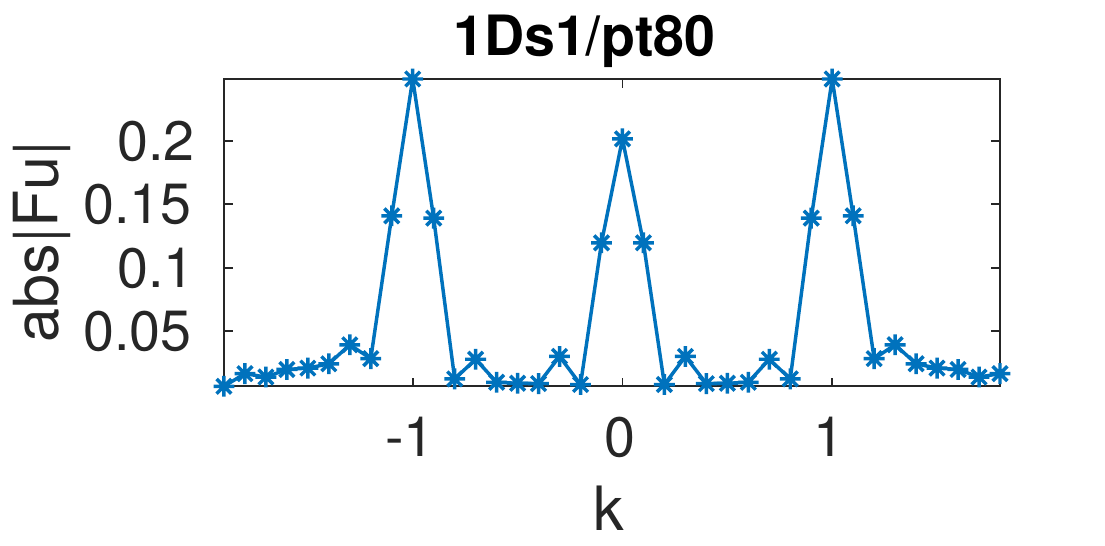}\\
\ig[width=0.3\textwidth]{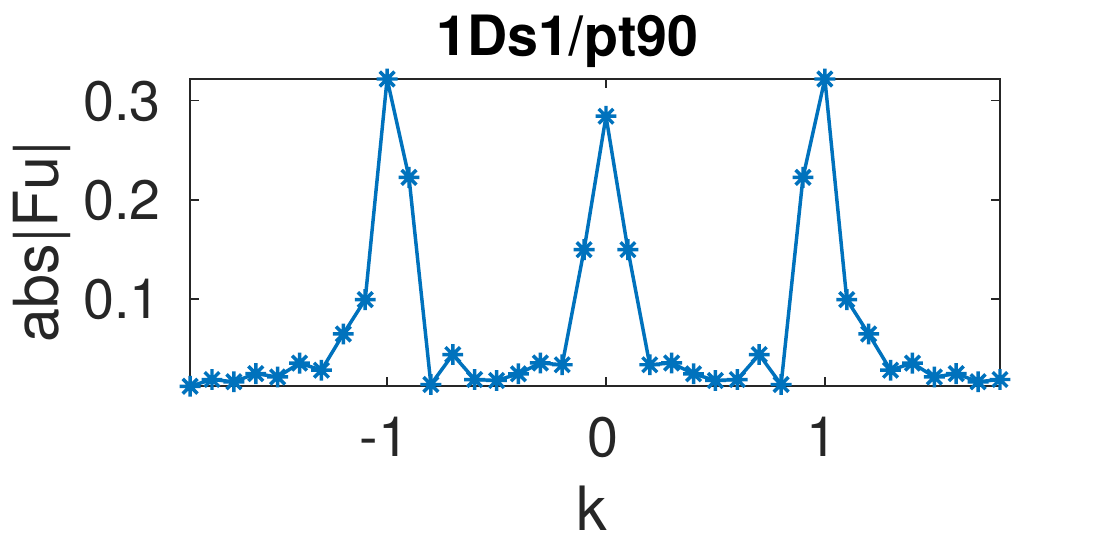}\\
\ig[width=0.3\textwidth]{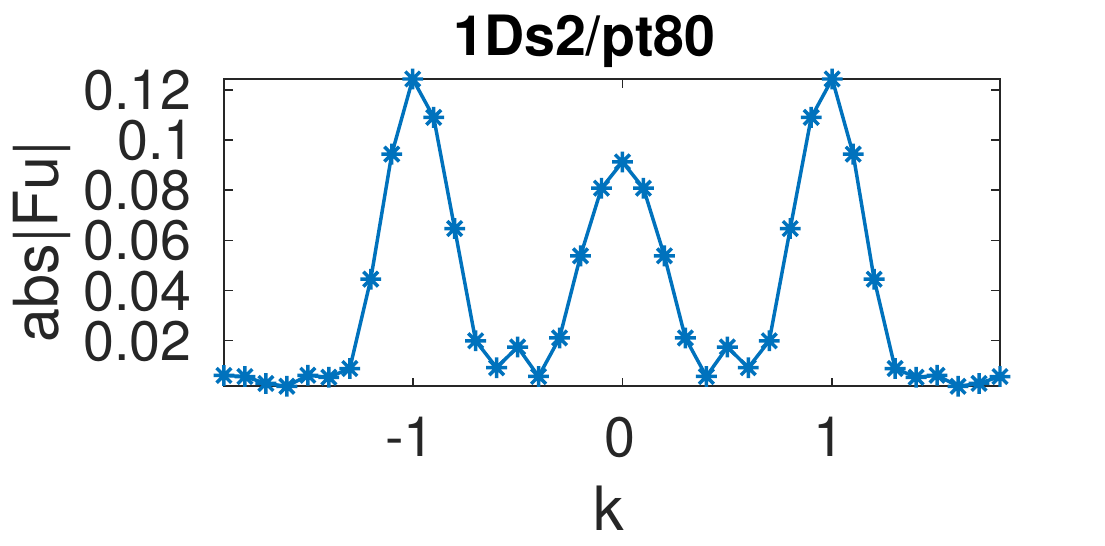}\\
\ig[width=0.3\textwidth]{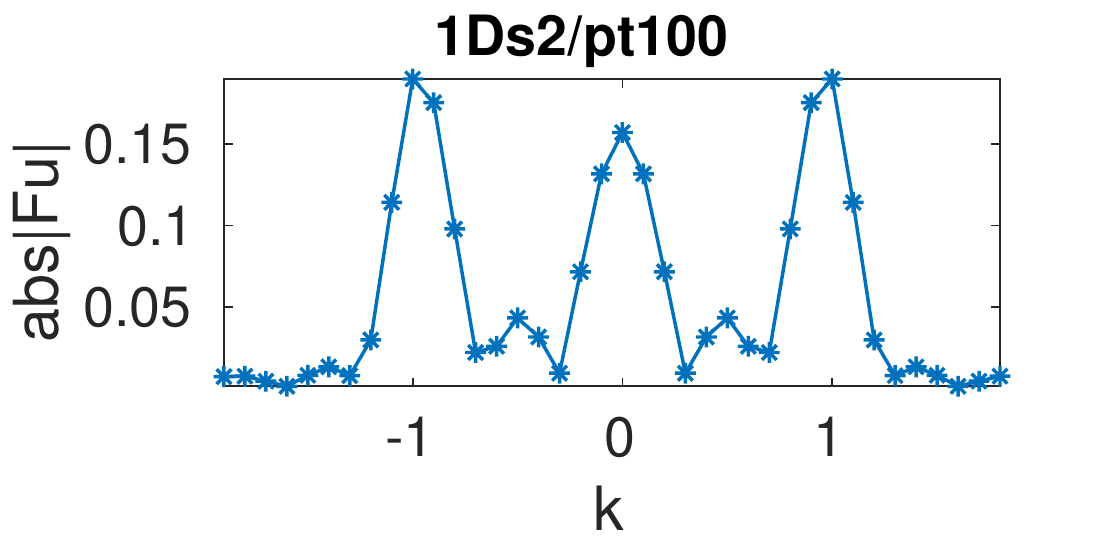}
\end{tabular}
\ece
\vs{-5mm}
   \caption{{\small (a1) Subcritical bifurcation of primary periodic patterns 
(black branch) in the SH equation with $\nu=2$, $\Om=(-10\pi,10\pi)$ and secondary bifurcations of snaking branches of a front (blue, {\tt 1Ds1}) and a localized pattern (red {\tt 1Ds2}). 
(a2) blow-up from (a1) showing how the snaking branches reconnect to the 
primary periodic branch. (a3) The Hamiltonian $H$ on the first four 
bifurcating branches of periodic patterns on $\Om=(-20\pi,20\pi)$, hence corresponding 
to wave numbers $k=1$ (black), $k=1-\frac 1 {80\pi}$ (blue), 
$k=1+\frac 1 {80\pi}$ (red) and $k=1-\frac 1 {40\pi}$ (magenta). (b,c) Solution 
plots, see main text for further comments. 
  \label{shf1}}}
\end{figure}

In 1D, the bifurcations of spatially periodic solution branches are 
simple, and for $\nu>\nu_3{=}\sqrt{27/38}$ the primary bifurcation 
at $\lam=0$ is 
subcritical. An interesting consequence are 
secondary bifurcations to steady (approximate) fronts 
between $u{\equiv}0$ and periodic patterns, and to localized 
patterns, and the ``snaking'' of the associated branches. 
This is illustrated in Fig.~\ref{shf1}(a1,a2) for \reff{swiho} 
over $\Om{=}(-10\pi,10\pi)$.

Snaking branches of localized patterns have attracted much interest in 
recent years \cite{burke,kno2008,BKLS09,hokno2009,strsnake,KC13,deWitt19,KUW19}. 
Since the SH equation has the spatial Hamiltonian \reff{ham1d}, i.e., 
$
H(u)=\pa_xu\pa_x^3u-\frac 1 2 (\pa_x^2u)^2+(\pa_x u)^2+\frac 1 2 (1-\lam)u^2-F(u), 
$
$F(u)=\int_0^uf(v)\dd v$, and since $H(0)=0$, a front between $u=0$ and 
\def\uper{u_{{\rm per}}}
a periodic pattern must connect to a pattern $\uper=\uper(\lam)$ 
with $H(\uper(\lam),\lam)=0$. Over $\Om=\R$ we have a continuum of 
periodic patterns with wave-numbers near $k=1$. Over finite domains, the admissible wave numbers of the periodic patterns are discrete, but for localized patterns the 'local' wave numbers 
in the patterns are free again. Thus, the local wave numbers can and must 
vary with $\lam$ in the snake. 
For simplicity, i.e., by mirroring 
the solutions over the right boundary, 
we also call the (approximate) fronts 'localized patterns'. 
In Fig.~\ref{shf1}(a3) we plot $H(\uper(\lam),\lam)$ 
for the first four periodic branches over $\Om=(-20\pi,20\pi)$. Comparison 
with (a1) shows that the wave numbers in the snake should roughly vary 
between $k=1-\frac 1 {80\pi}$ and $k=1$, which is confirmed by the 
solution plots and Fourier plots in (b,c). At the left folds in the snake ($\lam\approx -0.49$), the wave-number is very close to $1$, while at the right folds, corresponding to the intersection 
of the blue branch in (a3) with $H=0$, it is close to $k=1-\frac 1 {80\pi}$. The Fourier plots, however are slightly under-resolved to truly show the shifts of the 
maxima around $k=1$ between the left and right folds. Also note that while 
the $0$ mode ($k=0$) is clearly visible, the second harmonic 
($k=2$) in, e.g., {\tt 1D1/pt30} is very small. 

Regarding the implementation, the main additional function is {\tt geth} 
to compute $H$, which is then put on the branch for plotting in {\tt shbra1d}. 
Additional to the above results, at the end of {\tt cmds1d} we 
continue the fold for illustration of fold continuation in a system, 
cf.~\cite{sftut}, 
with a straightforward implementation of {\tt spjac}. Moreover, in \S\ref{gcsec} 
(demo {\tt gcsh}) we add a global coupling to \reff{swiho} and illustrate how this modifies 
the branches in Fig.~\ref{shf1}. 

\subsection{Intermezzo: branch switching at BPs of higher multiplicity} \label{bssec} 
As discussed in \S\ref{pfssec}, for 'natural' (i.e.~highly symmetric) 
choices of domains in 2D and 3D, bifurcation points $(u_0,\lam_0)$ on homogeneous branches often have a multiplicity 
\huga{\label{mdef2}
m=\dim N(G_u(u_0,\lam_0))\ge 2. 
} 
Here we briefly review the algorithm 
for branch switching at multiple  bifurcation 
points from \cite{mbiftut}. 
Let $(u_0,\lam_0)$ be a bifurcation 
point of multiplicity $m\ge 2$, and let 
\hual{\label{sbp1}
N(G_u^0)=\lin\{\phi_1,\ldots,\phi_m\}, \ 
N(G^{0T}_u)=\lin\{\psi_1,\ldots,\psi_m\}, \ \spr{\phi_i,\psi_j}=\del_{ij}, 
\text{ and }   G_\lam^0\in R(G_u^0), 
}
where $(G_u^0, G_\lam^0)=(G_u(u_0,\lam_0),G_\lam(u_0,\lam_0))$. 
Then there exists  a unique $\phi_0\in N(G^0_u)^\perp$ such that 
$G_u^0\phi_0+G_\lam^0=0$ and $\spr{\phi_0,\psi_j}=0$, $j=1,\ldots,m$. 
The ansatz 
\allowdisplaybreaks
\huga{\label{qans}
u'(s_0)=\sum_{j=0}^m \al_j\phi_j,\quad 
\al_0=\lam'(s_0), 
} 
where $\al_j=\spr{\phi_j,\dot u(s_0)}, 1\le j\le m$, and 
differentiating $G(u(s),\lam(s))=0$ twice and evaluating at $s_0$ 
yields the quadratic bifurcation equations (QBE) 
\hual{
&B(\al_0,\al)=0\in\R^m, \label{qeqa} \\
&B_i(\al_0,\al)=\sum_{j=1}^m\sum_{k=1}^m a_{ijk}\al_j\al_k+2\sum_{j=1}^m b_{ij}\al_j\al_0
+c_i\al_0^2, \quad 1\le i\le m, \notag\\
&a_{ijk}=\spr{\psi_i,G_{uu}^0[\phi_j,\phi_k]}, \ \ b_{ij}=\spr{\psi_i, 
G_{uu}^0[\phi_0,\phi_j]+G_{u\lam}^0\phi_j}, \ \ 
c_i=\spr{\psi_i,G_{uu}^0[\phi_0,\phi_0]+2G_{u\lam}^0\phi_0+G_{\lam\lam}^0}.
\notag
}

 The QBE are quadratic homogeneous in $(\al_0,\al)$, 
and hence solutions are only determined up to a factor $\ga$. 
They are necessary conditions for bifurcating branches. 
Conversely, each distinct {\em isolated} zero $(\al_0,\al)$ gives 
a distinct solution branch of $G(u,\lam)$ \cite{KL72}. Here 
$(\al_0,\al^*)$ is called isolated if for fixed $\al_0$ and some $\del>0$ 
the only solution in $U_\del^{\R^m}(\al^*)$ is $\al^*$. 
By the implicit function theorem, a sufficient 
condition for this is that $J(\al)=\pa_\al B(\al_0,\al)$ is non-singular. 
Without loss of generality we may fix $\al_0=0$ (if $m=1$) or $\al_0=1$, but for scaling 
reasons (relative to $\al$) it turns out that some small $\al_0$ is 
more suitable, and our default choice for solving the QBE by a 
Newton loop in $\al$ is $\al_0=0.001$, and initial guesses for 
$\al$ as all tuples $\al\ne 0$ with $\al_i\in\{0,\pm 1\}$, $i=1,\ldots,m$. 

For $m=1$, if $(\al_0,\al_1)$ is one solution (from the already given branch) 
of the QBE with $a_{111}\al_1+b_{11}\al_0\ne 0$, 
then $(u_0,\lam_0)$ is a bifurcation point. Moreover, we can solve 
the QBE explicitly, and this is done in the \pdep\ simple BP branch switching routine {\tt swibra}, see \cite[\S2]{p2p}, or \cite[Algorithm 2.1]{mbiftut}. 

The case $m\ge 2$ is more difficult,  and the QBE \reff{qeqa} may 
(and typically will) not yield all bifurcating branches, but only 
those that are $2$--determined, see \cite{mbiftut}.  
Pitchfork bifurcations are at best 3--determined, and in this case 
we use the ansatz 
\hual{&u(s)=u_0+s\sum_{i=1}^m \al_i\phi_i+s^2w, \quad \lam(s)=\lam_0+\beta s^2, 
\label{cbeans}
}
with unknowns $\al\in\R^m$, $\beta\in\R$ and $w\in N(G_u)^\perp={\rm span}\{\phi_1,\ldots,\phi_m\}^\perp, \text{ i.e. } 
\spr{\psi_i,w}=0, i=1,\ldots,m$. Differentiating twice, solving for $w$ at $s=0$, and differentiating once more and evaluating at $s=0$ yields the system 
\hual{\label{cbe}&C(\al,\beta)=0\in\R^m
}
of $m$ cubic bifurcation equations (CBE), see \cite{mbiftut}. 
Again wlog we can fix $\beta=\pm 1$ (but numerically use $\beta=\pm\beta_0$ 
with default choice $\beta_0=0.001$), and then each isolated solution $\al$ 
of \reff{cbe} gives a tangent $(u'(0),\lam'(0))=(\sum_{i=1}^m \al_i\phi_i,0)$ to a distinct bifurcating branch. Alternatively, we may choose a small $s$ in \reff{cbeans} 
and use $(u_0,\lam_0)+(u(s),\lam(s))$  as a predictor for the bifurcating  branch. 

The functions {\tt p0=qswibra(dir,bpt)} ({\tt p0=cswibra(dir,bpt)}) 
attempt to solve 
the QBE (CBE) (unless {\tt aux.besw=0}, see below), and store the computed tangents in the fields 
{\tt p0.mat.qtau} (/{\tt ctau}), respectively, and 
store the kernel vectors $\phi_1,\ldots,\phi_m$ in {\tt p0.mat.ker}. 
Additionally, {\tt cswibra} also stores the predictors $s\sum_{i=1}^m \al_i\phi_i+s^2w$ in {\tt p0.mat.pred}. 
Subsequently we can choose a tangent $\tau{=}(u'(0),\lam'(0))$ via 
{\tt p=seltau(p0,nr, newdir,sw)}, where depending on {\tt sw=2}, {\tt sw=3} 
and {\tt sw=4} we select vector {\tt nr} from {\tt p0.mat.qtau}, {\tt p0.mat.ctau} or {\tt p0.mat.pred}, respectively. 
Alternatively, and as a fallback for problems only determined at higher order, 
we can generate a guess for a tangent to a new branch according to 
$\tau=\sum_{i} \ga_i{\tt p.mat.ker}(i)$ via {\tt p=gentau(p0,ga)}
where the sum runs from 1 to {\tt length(ga)}. 
In  Algorithm \ref{mswiba} we summarize the approach, and 
in Table \ref{swauxtab} we collect auxiliary arguments for fine 
tuning of {\tt qswibra} and {\tt cswibra}, but otherwise we refer to 
\cite{mbiftut} for further mathematical comments. Some additional care 
must be taken if, e.g., the system has continuous symmetries. In this 
case, {\tt aux.ali} can be used to choose 'active' kernel vectors 
for setting up and solving the quadratic or cubic bifurcation equations,  
see \S\ref{sy1sec}.

\begin{Algorithm}[ht]
\fbox{\parbox{0.98\textwidth}{\parbox{0.96\textwidth}{
\bci
\item[1.] Call {\tt p0=qswibra(dir,fname)} to find nontrivial solutions of 
the QBE \reff{qeqa} and to store these in {\tt p.mat.qtau}. 
Additionally, store a base of the kernel 
of $G_u$ in {\tt p.mat.ker}.  
\item[2a.] If 1 yields nontrivial solutions of the QBE: use 
{\tt p=seltau(p0,nr,newdir,2)} to choose tangent {\tt nr} as 
a predictor to the new branch, to be stored in {\tt newdir}. 
\item[2b.] Use {\tt cont} to continue the new branch, 
return to 2a to follow more branches.  \\[-2mm]
\item[3.] Subsequently/alternatively (if the absence of transcritical 
branches is known) to 1,2, call 
{\tt p0=cswibra(dir,fname)}  to find nontrivial solutions of 
the CBE \reff{cbe}. The tangents are then stored in {\tt p.mat.ctau}, 
and 'effective' predictors $(u,\lam)$ are computed from \reff{cbeans} 
with $s={\tt ds}$, normalized and stored in {\tt p.mat.pred}. 
\item[4.] Proceed as in 2, i.e.: Call {\tt p=seltau(p0,nr,newdir,3)} for 
choosing tangent {\tt nr} as predictor, or {\tt p=seltau(p0,nr,newdir,4)} 
to choose the quadratic predictor {\tt nr}. Afterwards call 
{\tt cont}.  \\[-2mm]
\item[5.] For (possible) branches additional to those found in 
1.--4.: use {\tt p=gentau(p0,v,newdir)} to generate 
guesses for tangents to new branches according to $\tau=\sum_{i} v(i){\tt p.mat.ker}(i)$, 
where the sum runs from 1 to {\tt length(v)}. Afterwards call 
{\tt cont}. 
\item[6.] If {\tt cont} fails after branch-switching, try, e.g., different {\tt ds} 
(for the quadratic predictor {\tt p=seltau(p0,nr,newdir,4)} this theoretically requires 
a new call to cswibra). 
\eci 
}}}
\caption{{\tt qswibra, cswibra}, and subsequent {\tt seltau, gentau} 
 for branch-switching at multiple bifurcation points. The arguments 
{\tt dir, bpt} stand for the \pdep\ setting that the pertinent 
branch point has filename {\tt fname} in directory {\tt dir}. 
The function {\tt qcswibra} first calls {\tt qswibra}, then {\tt cswibra}. If 
{\tt q(c)swibra} is called at a BP with 1D kernel ($m=1$), then 
it directly calls {\tt swibra}. Use {\tt q(c)swibra(dir,fname,aux)} 
to pass optional arguments {\tt aux} listed 
in Table \ref{swauxtab} to {\tt qswibra/cswibra}. 
\label{mswiba} }
\end{Algorithm}


\begin{table}[ht]\taskip
\caption{Entries in the auxiliary argument {\tt aux} of {\tt qswibra} and {\tt cswibra}. We assume the call {\tt p0=qswibra(dir,pt,aux)} 
such first a BP {\tt p0} is loaded from {\tt dir/pt}. Since we do not 
save data stored in {\tt p.mat} to disk, in this case we first need to recompute 
the kernel {\tt p0.mat.ker}, and essentially to avoid this we also allow 
the call {\tt p0=qswibra(p0,aux)}. Similar for {\tt cswibra}. See  
\cite{mbiftut} for the (scaling) purposes of $\al_0$ and $\beta_0$, 
and further comments. 
\label{swauxtab}}
\bce\vs{-4mm}
{\small 
\begin{tabular}{l|p{15cm}}
field&purpose, remarks\\
\hline
soltol&tolerance (default $10^{-10}$) to solve the QBE/CBE, i.e., $|F(\al)|<{\tt soltol}$. \\
isotol&tolerance (default $0.1$) to classify 
solutions $\al$ of the QBE/CBE as isolated if 
$|\det \pa_\al F(\al)|>{\tt isotol}\max|\pa_\al F(\al)|$.\\
mu2&to override p.nc.mu2 where an eigenvalue $\mu$ is considered to be zero 
if $|\mu|<{\tt mu2}$.\\
m&to explicitly give the dimension of the kernel, instead of $m=\sharp\{\mu:
|\mu|<{\tt mu2}\}$\\
al0v&to override $\al_0{=}0.001$ for {\tt qswibra}; can be a 
vector $\al_0{=}(\al_0(1),\ldots,\al_0(j))$\\
bet0&to override $\beta_0{=}0.001$ for {\tt cswibra}. \\
alc&to override the initial guesses for the Newton loop for $\al$.\\
ral&use random initial guesses for $\al$ if {\tt ral=1}\\
ds&to override the steplength selection {\tt p.nc.dsmax/10}; also used for computing the quadratic 
predictor $(u,\lam)$ from \reff{cbeans}.\\
besw&if {\tt besw=0}, then q(c)swibra only compute and store the kernel vectors; useful if subsequently only these are used as (approximate) predictors 
via {\tt gentau}. Default {\tt bews=1}. \\
ali&active list of kernel vectors; useful in case of continuous symmetries, 
see \S\ref{sy1sec}. Default {\tt ali=[]}, which uses all kernel vectors $\phi_j$, 
$j=1,\ldots,m$. \\
hasker&for a subsequent call to {\tt qswibra} or {\tt cswibra} in the syntax, 
e.g., {\tt p0=cswibra(p0,aux)}; if hasker=1, then the kernel is taken from 
{\tt p0.mat.ker} and not recomputed; useful for experimenting with 
parameters, and for instance used in {\tt qcswibra}. \\
keeplss&if 1, then the linear system solver {\tt p0.fuha.lss} is used for 
solving the linear systems inside {\tt cswibra}. Otherwise, and as default 
setting, {\tt lsslu} is used. \\
\end{tabular}
}
\ece
\end{table}\teskip

\subsection{Patterns in 2D}
\label{sh2dnum}
In 2D we can use the same basic setup ({\tt oosetfemops} and {\tt sG})  
as in 1D, but now need to deal with the multiplicity $m\ge 2$ of BPs over domains that (by symmetry) generate higher dimensional kernels. As indicated in 
Algorithm \ref{mswiba} the idea is to find all bifurcating branches via 
numerical solution of the associated QBE and CBE by Newton loops for 
different fixed $\al_0, \beta$, with different initial guesses $\al$.  
For this, the two key \pdep\ functions {\tt p0=qswibra(dir,pt,aux)} and 
{\tt p0=cswibra(dir,pt,aux)} can and sometimes must be fine tuned via the auxiliary argument {\tt aux}, which can 
have the fields from Table \ref{swauxtab}. 

\subsubsection{A square domain}\label{2dsqnsec}
We proceed by example and consider in Listing \ref{shl4} the script {\tt cmds2dsq.m} used to generate the branch plots in Fig.~\ref{shf77b2}. 
Here we know a priori that all bifurcations are pitchforks, and hence 
can restrict to {\tt cswibra}. 

\hulst{caption={{\small (Selection from) {\tt \dname/cmds2dsq.m}. Script 
for the 2D SH equation on the square domain $\Om=(-2\pi,2\pi)^2$, which yields a 
double branch points at $\lam=0$. {\tt cswibra} in line 5 finds 4 bifurcation 
directions, falling into the two classes of spots and stripes. Thus, 
in lines 10,11 we select one spot and one stripe branch and continue these. 
In line 14 we have a subsequent call (hence set {\tt aux.hasker=1}) to {\tt cswibra} for the indetermiante case $\nu=\nu_1$, cf.~\reff{nutab}, where {\tt cswibra} correctly finds non-isolated solutions. 
The remainder of {\tt cmds2dsq} deals with the other 
cases for $\nu$ (see Fig.~\ref{shf77b2}), plotting, and the 2nd bifurcation 
point, cf.~Remark \ref{sh2nd}. 
 }}, 
label=shl4, language=matlab,stepnumber=5, linerange=3-19}{\dhome/cmds2dsq.m}

\subsubsection{$D_4$ symmetry on a non--square domain}\label{d4sec} 
\begin{figure}[ht]
\bce 
\btab{l}{
\btab{llll}{{\sm (a)}&{\sm (b)}&{\sm (c)}&{\sm (d)}\\
\hs{-5mm}\raisebox{30mm}{\btab{l}{
\ig[width=0.13\tew]{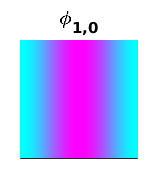}\hs{-4mm}
\ig[width=0.13\tew]{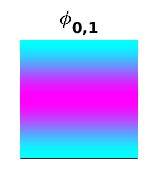}\\
\ig[width=0.13\tew]{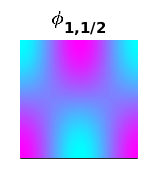}\hs{-4mm}
\ig[width=0.13\tew]{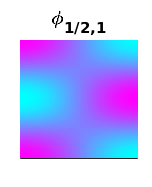}}}
&\hs{-5mm}\raisebox{30mm}{\btab{l}{
\ig[width=0.13\tew]{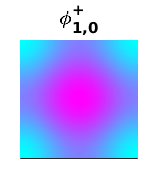}\hs{-4mm}
\ig[width=0.13\tew]{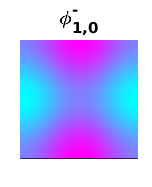}\\
\ig[width=0.13\tew]{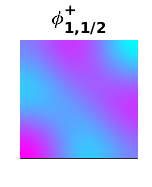}\hs{-4mm}
\ig[width=0.13\tew]{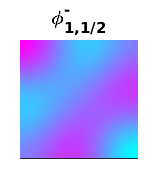}}}
&\hs{-2mm}\ig[width=0.27\tew]{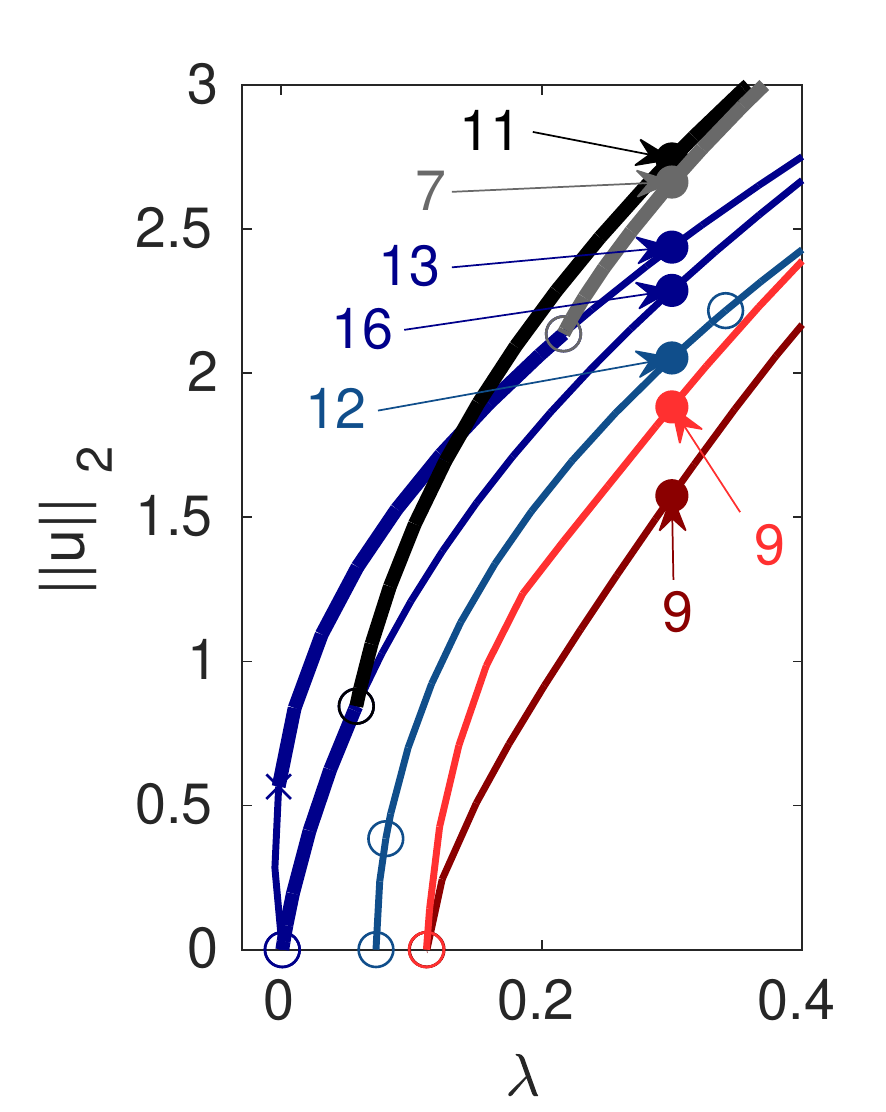}
&\hs{-5mm}\raisebox{30mm}{\btab{l}{
\ig[width=0.2\tew]{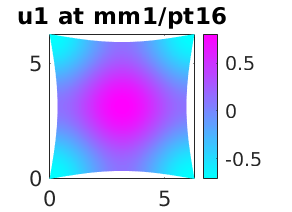}\\
\ig[width=0.2\tew]{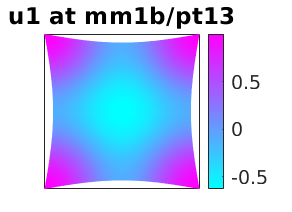}}}}
\\
{\sm (e)}\\
\ig[width=0.2\tew]{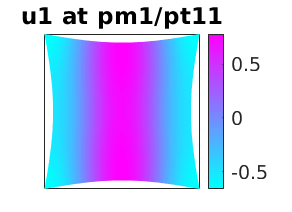}
\ig[width=0.2\tew]{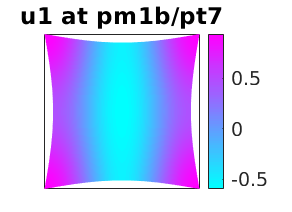}
\ig[width=0.2\tew]{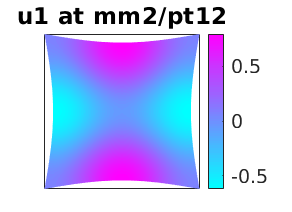}
\ig[width=0.2\tew]{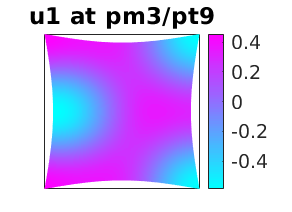}
\ig[width=0.2\tew]{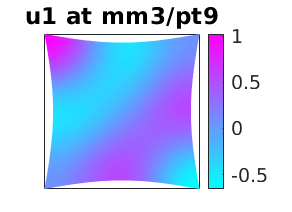}}
\ece 

\vs{-5mm}
\caption{{\small The SH equation on a distorted square with $D_4$ symmetry. 
(a) Pure modes at the first two BPs on the square $\Om=(0,2\pi)^2$. (b) 
Associated mixed modes.  
(c,d) BD and sample solutions over $\Om_{0.35}$ with $\nu=0.5$, from 
{\tt cmds2dpsq}.  \label{shfpsq}}}
\end{figure}

The double multiplicity 
of the BPs in Fig.~\ref{shf77b} and of many further BPs follows from 
$\Om$ being a square domain and the Neumann BCs. However, the 
generic form of the amplitude equations \reff{sqae} (in the case of a double BP) 
follows from the $D_4$ equivariance. Here we briefly discuss an example 
of the SH equation on a domain which is not a square (and hence we have 
no explicit kernel vectors and the multiplicity of BPs is not clear a priori) 
but which is $D_4$ invariant (such that in case of double BPs the amplitude 
equations still have the form \reff{sqae}). Essentially, we apply 
results from \cite{Craw91} which explain why some double BPs break up while 
other stay double, and what primary bifurcations occur. 
We consider \reff{swiho} on the perturbed $\Om_\del$, where we perturb 
the edges 
$\Ga_1=\{(t,0)\},  \Ga_2=\{(2\pi,t)\}, 
\Ga_3=\{(2\pi{-}t,2\pi)\},  \Ga_4=\{(0,2\pi{-}t)\}$
where $t\in[0,2\pi)$,  to  
\hugast{\Ga_1^\del=\{(t,\del\sin(\frac t 2))\}, 
\Ga_2^\del=\{(2\pi{-}\del\sin(\frac t 2),t)\}, 
\Ga_3^\del=\{(2\pi{-}t,2\pi{-}\sin(\frac t 2))\}, 
\Ga_4^\del=\{(\del\sin(\frac t 2),2\pi{-}t)\},}
see Fig.~\ref{shfpsq}(d) for the shape of $\Om_\del$. 
In Fig.~\ref{shfpsq}(a) we recall the 'pure modes' 
\huga{\text{$\phi_{m,n}(x,y)=\cos(mx)\cos(ny)$, $n,m\in\N/2$,}}
at the first two branch points $\lam=0$ 
and $\lam=1/16$ on the (unperturbed) square $\Om=(0,2\pi)^2$ with Neumann BCs, 
where we take the horizontal reflection 
$\ga_1:(x,y)\mapsto (\pi-x,y)$ and the diagonal reflection $\ga_2:(x,y)\mapsto(y,x)$ as the generators of $D_4$. Then $\ga_1\phi_{m,n}=(-1)^{2m}\phi_{m,n}$ 
and $\ga_{2}\phi_{m,n}=\phi_{m,n}$. 
In \cite{Craw91} it is explained that the 'mixed' modes 
$\phi_{m,n}^\pm=\phi_{m,n}\pm\phi_{n,m}$, see Fig.~\ref{shfpsq}(b), 
are more natural basis functions of the kernels. 
These fulfill 
\hual{
\ga_2\phi_{m,n}^\pm&=\pm\phi_{m,n}^\pm, \text{ and }\\
\ga_1\phi_{m,n}^\pm&=(-1)^{2m}\times\left\{
\barr{ll}\phi_{m,n}^\pm&2(m+n)\text{ even (parity)}, \\
\phi_{m,n}^\mp&2(m+n)\text{ odd (parity)}.\earr\right.
} 
Using these symmetries, the effect of the perturbation $\del\sin(t/2)$ 
of the square is further analyzed in \cite{Craw91}. 
For odd parity, the perturbation analysis yields that the 
double eigenvalues stay double, with the simultaneous bifurcation 
of branches with pure and mixed mode symmetry. However, 
the even parity modes break 
into two BPs where only branches of mixed mode symmetry bifurcate,  
while the previously primary pure modes now become 
secondary bifurcation. Precisely this is illustrated in Fig.~\ref{shfpsq}(c,d) 
for the first two BPs, for $\del=0.35$. Additionally, the mixed mode 
branch {\tt mm1} now becomes transcritical as the hidden symmetry 
of a shift by half a period in $x$ or $y$ is lost, while all other 
primary branches must remain pitchforks due to the presence of 
their alternative orientations via $\ga_2$. 

Concerning the implementation we refer to the script {\tt cmds2dpsq.m}, 
and to {\tt shinitpsq.m} for the initialization of the perturbed square, 
which uses {\tt freegeompdeo.m} to contruct the perturbed square 
PDE object.

\begin{figure}[ht]
\bce 
\begin{tabular}{l}
{\small (a) $\nu=0$ (cubic case), str and pq stable at bifurcation, hex unstable; BD, zoom, and example solutions}\\
\ig[width=0.24\textwidth, height=50mm]{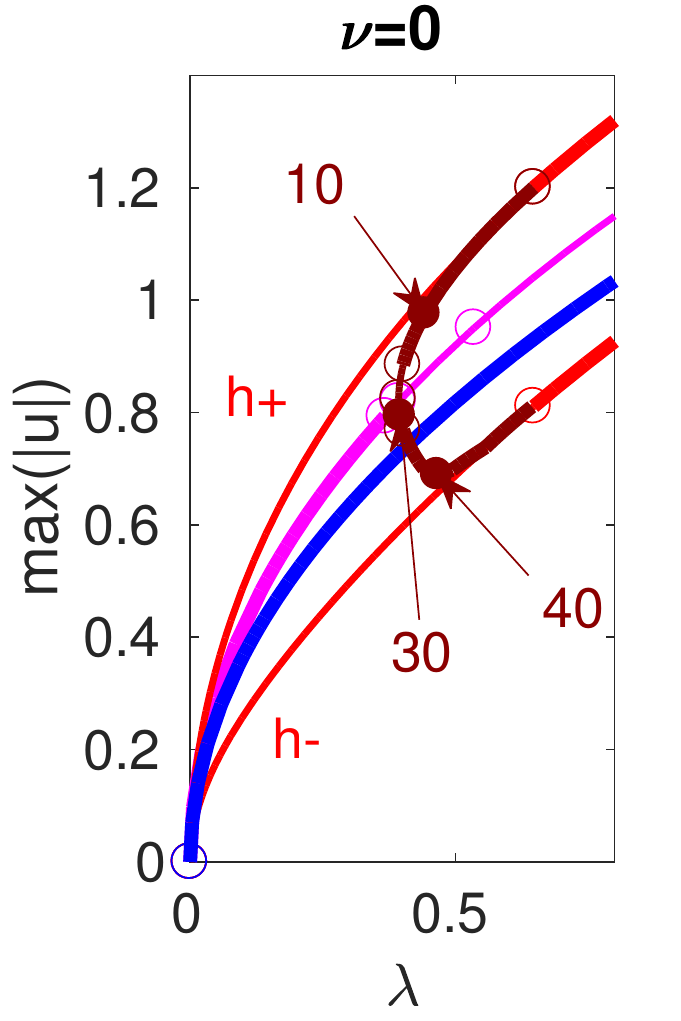}
\ig[width=0.2\textwidth]{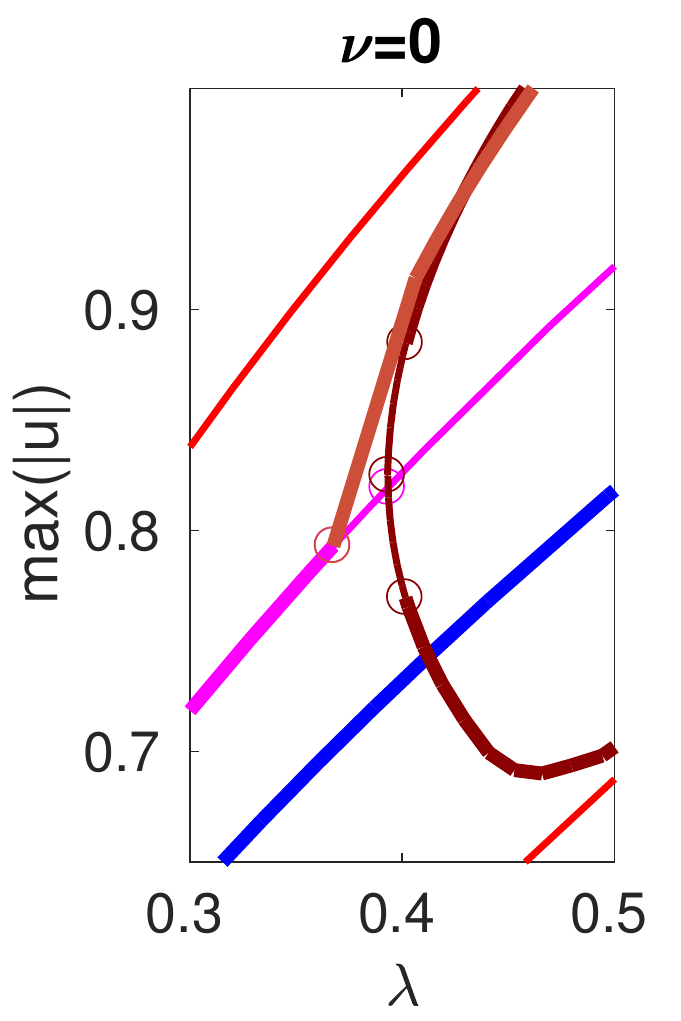}
\raisebox{25mm}{\begin{tabular}{ll}
\ig[width=0.24\textwidth]{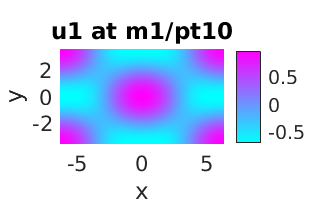}&
\hs{-2mm}\ig[width=0.24\textwidth]{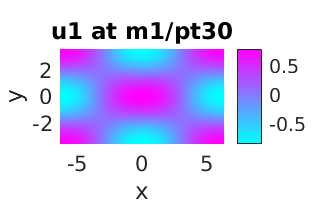}\\
\ig[width=0.24\textwidth]{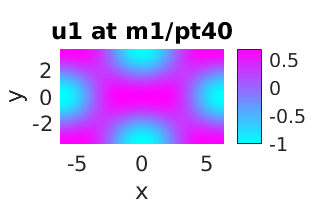}&
\hs{-2mm}\ig[width=0.24\textwidth]{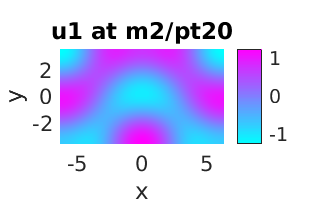}
\end{tabular}}\\
{\small (b) $\nu=1.3$, BD and example solutions}\\[0mm]
\ig[width=0.26\textwidth,height=55mm]{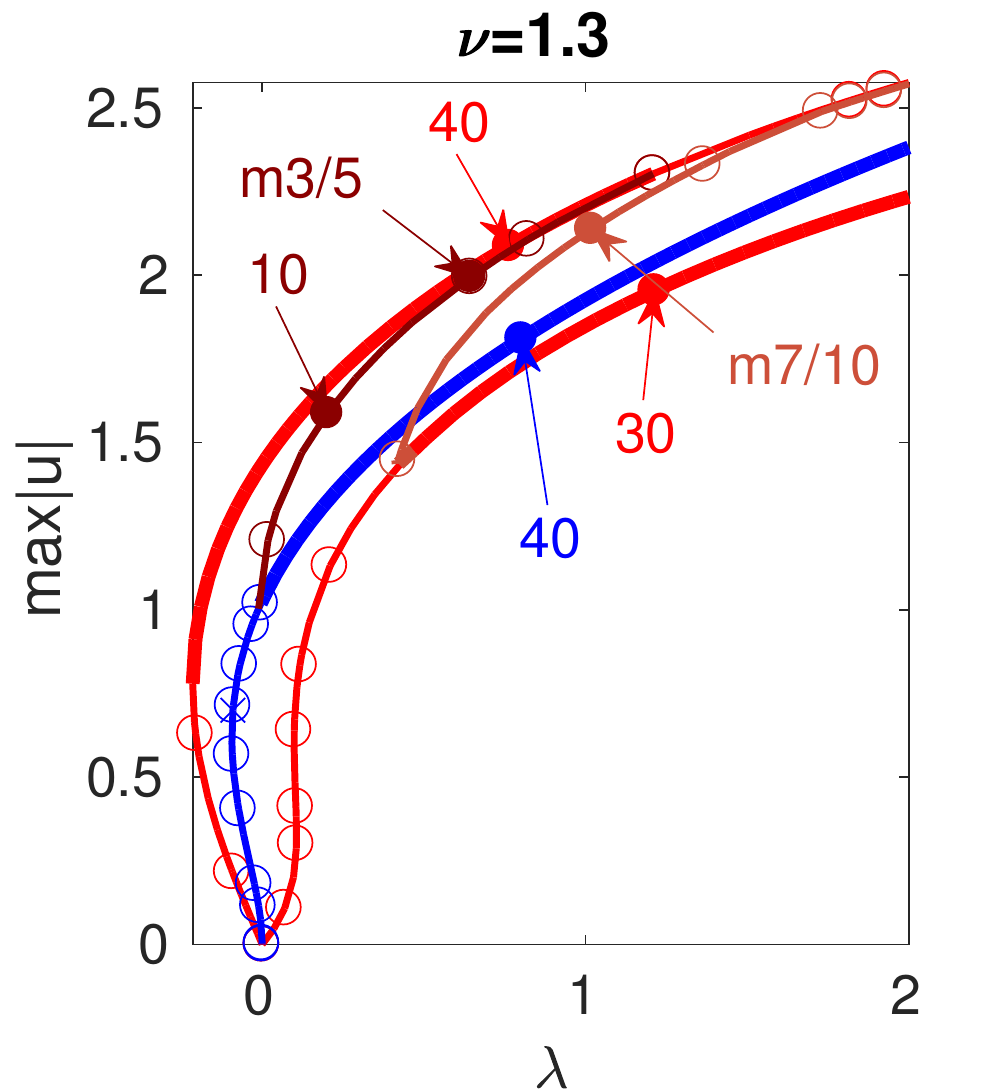}
\hs{-2mm}\raisebox{25mm}{\begin{tabular}{lll}
\ig[width=0.24\textwidth]{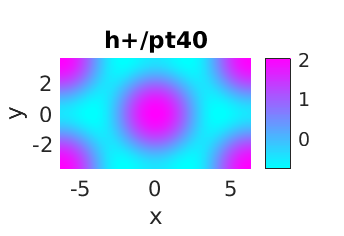}&
\hs{-4mm}\ig[width=0.24\textwidth]{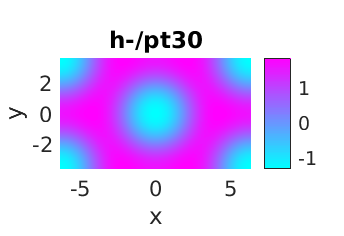}&
\hs{-4mm}\ig[width=0.24\textwidth]{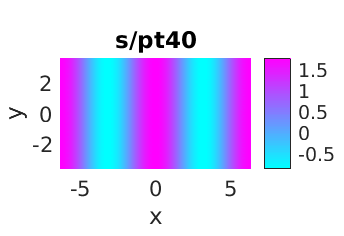}\\
\ig[width=0.24\textwidth]{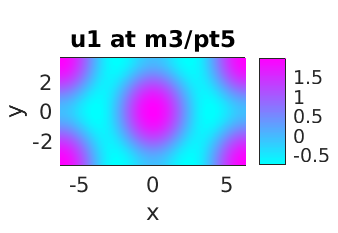}&
\hs{-4mm}\ig[width=0.24\textwidth]{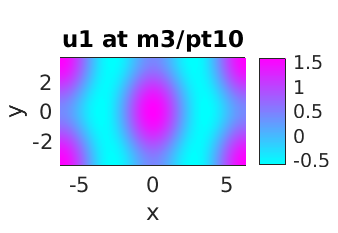}&
\hs{-4mm}\ig[width=0.24\textwidth]{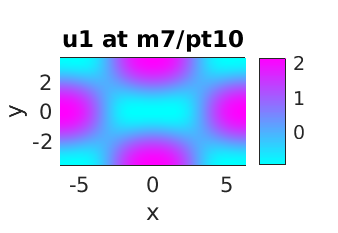}
\end{tabular}}
\end{tabular}
\ece 
\vs{-6mm}
\caption{{\small Example results from {\tt cmds2dhex.m}. Bifurcation diagrams and example plots SH over a small rectangular domain permitting hex solutions. 
For $ \nu=0$ in (a), hex (unstable), str and pq (stable) bifurcate in supercritical pitchforks. The (up and down) hex become stable at $\lam=\lam_1\approx 0.645$, and these points are connected by a mixed 
mode branch m1, which which passes the pq branch near $\lam=0.39$. 
The pq patterns become unstable $\lam=\lam_2\approx 0.37$, where a stable branch 
m2 (light brown) bifurcates. 
For $\nu=1.3$ in (b) the hex are transcritical, and we consider 
secondary bifurcations and mixed mode branches at larger amplitude. See 
text for details.  \label{shf10}}}
\end{figure}

\subsubsection{Hexagonal symmetry} 
The script {\tt cmds2dhex.m} considers \reff{swiho} over  $\Om{=}(-l_x,l_x){\times}(-l_y,l_y)$, $l_x{=}2\pi$, 
$l_y{=}2\pi/\sqrt{3}$. 
This small domain is 4 times the minimal (rectangular) domain $\Om=(0,l_x)\times (0,l_y)$ 
allowing hexagon patterns. 
In Fig.~\ref{shf10}(a) we have $\nu=0$ and hence at $\lam=0$ have pitchforks 
of stripes (stable), pq (stable) and hexagons (unstable). 
We continue these patterns 
to rather large amplitude and find secondary bifurcations. 
The branch m2 (light brown) bifurcates at the loss of stability of the pq branch and gives an example of a (stable) pattern different from those on the 
minimal domain. 

For $\nu=1.3$ in 
(b), the hexagons bifurcate transcritically and the up-hexagons become stable after 
the fold. The stripes are unstable at their (subcritical) pitchfork bifurcation but become stable 
at larger amplitude, $\lam=\lam_1\approx -0.01$, while the up hexagons become 
unstable again at $\lam=\lam_2\approx 1.21$. These two points are connected by 
a mixed mode branch which we call beans. Similarly, the down-hexagons 
become stable at $\lam=\lam_3\approx 0.46$, and the branch bifurcating there 
connects to (shifted) up hexagons at $\lam\approx 1.8$. 
Even on this relatively small domain there are many additional bifurcation 
on the hexagons and stripes branches. At the end of {\tt cmds2dhex} we plot, 
just for illustration, the energies $\CE(u)$ associated 
to some branches from Fig.~\ref{shf10}.
Moreover, we remark that: 
\bci 
\item For $\nu=0$ we obtain all primary branches from {\tt cswibra}. 
For $\nu=1.3$ we naturally use {\tt qswibra} to obtain the hexagons, 
and then, for convenience, {\tt gentau} to generate the stripes. 
\item  It is again crucial to use {\tt pmcont} to continue the 
patterned branches; simply using {\tt cont} results in various 
uncontrolled branch-jumpings. Just one illustrative example is given at the 
end of {\tt cmds2dhex.m}. 
\item As the secondary bifurcations are generically simple, for their  detection we use {\tt p.sw.bifcheck=1}, and a sufficiently small stepsize {\tt ds} to not miss 
bifurcation points via too large steps. 
\eci 
On larger domains, there naturally are more patterns, and both, the 
avoidance of branch jumping and the bifurcation detection become 
more difficult problems. See \S\ref{tntsec}. 




\subsubsection{Planar fronts between hexagons and zero}\label{f1sec}
\begin{figure}[ht]
\bce 
\begin{tabular}{ll}
\ig[width=0.25\textwidth]{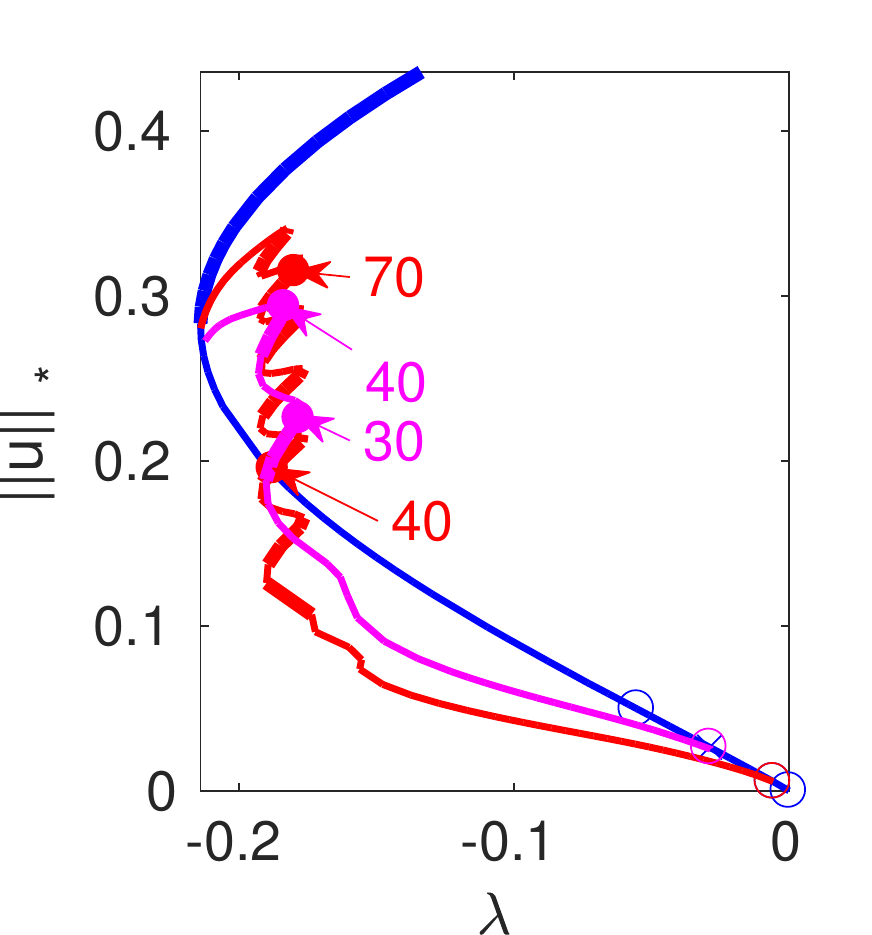}
&\raisebox{22mm}{\begin{tabular}{l}
\ig[width=0.35\textwidth]{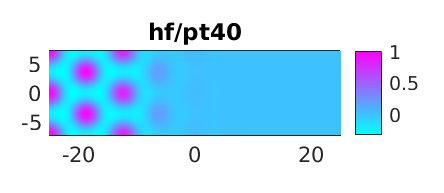}\ig[width=0.35\textwidth]{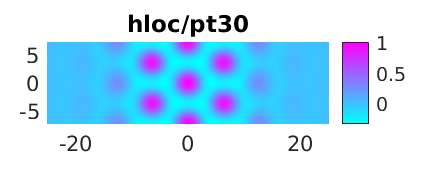}
\\[-2mm]
\ig[width=0.35\textwidth]{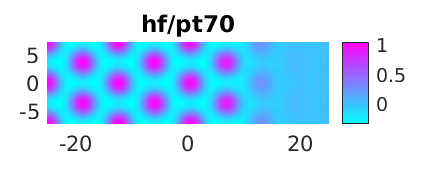}\ig[width=0.35\textwidth]{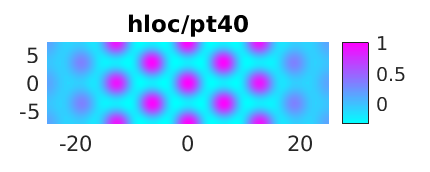}
\end{tabular}}
\end{tabular}
\ece 

\vs{-5mm}
\caption{{\small Results from {\tt cmds2dhexfro}  for the SH over a long 
rectangular domain, $\nu=1.3$. Primary branch of (up) hexagons (blue), and snaking 
branches of fronts (red, {\tt hf}) between hexagons and $u=0$ and of localized hex (magenta, {\tt hloc}). \label{shf11}}}
\end{figure}

Localized patterns similar to Fig.~\ref{shf1} can also occur in 2D and 3D. 
Moreover, while in 1D we basically have localized patterns of 'stripes' connected to $u\equiv 0$, in 2D we can have heteroclinic connections and 
heteroclinic cycles between 
various patterns and $u\equiv 0$, or between different patterns, e.g., between stripes and hexagons. This 
is discussed in more detail in \S\ref{tntsec}, in \S\ref{schnaknum}, and in, e.g., 
\cite{uwsnak14}, and here we restrict to planar fronts between 
hexagons and $u\equiv 0$, and 1D-localized hexagon-patches. Figure \ref{shf11} illustrates the basic idea. Over sufficiently large (long) 
domains, there are secondary bifurcations after the primary subcritical 
bifurcation to (here) hexagons, and these lead to snaking branches 
of localized hexagons, which (over bounded domains) eventually 
reconnect to the primary hexagon branch. 
 Listing \ref{shl4b} gives the main commands. 

\hulst{caption={{\small (Selection from) {\tt \dname/cmds2dhexfro.m}. Main part of the script to 
compute a snaking branch of a front between hexagons and $u=0$.  The remainder of the script deals with the snake of localized hexagons and plotting.}}, 
label=shl4b, language=matlab,stepnumber=0, linerange=2-14}{\dhome/cmds2dhexfro.m}

\begin{figure}[ht]
\bce 
\begin{tabular}{ll}
\hs{-3mm}\ig[width=0.19\textwidth]{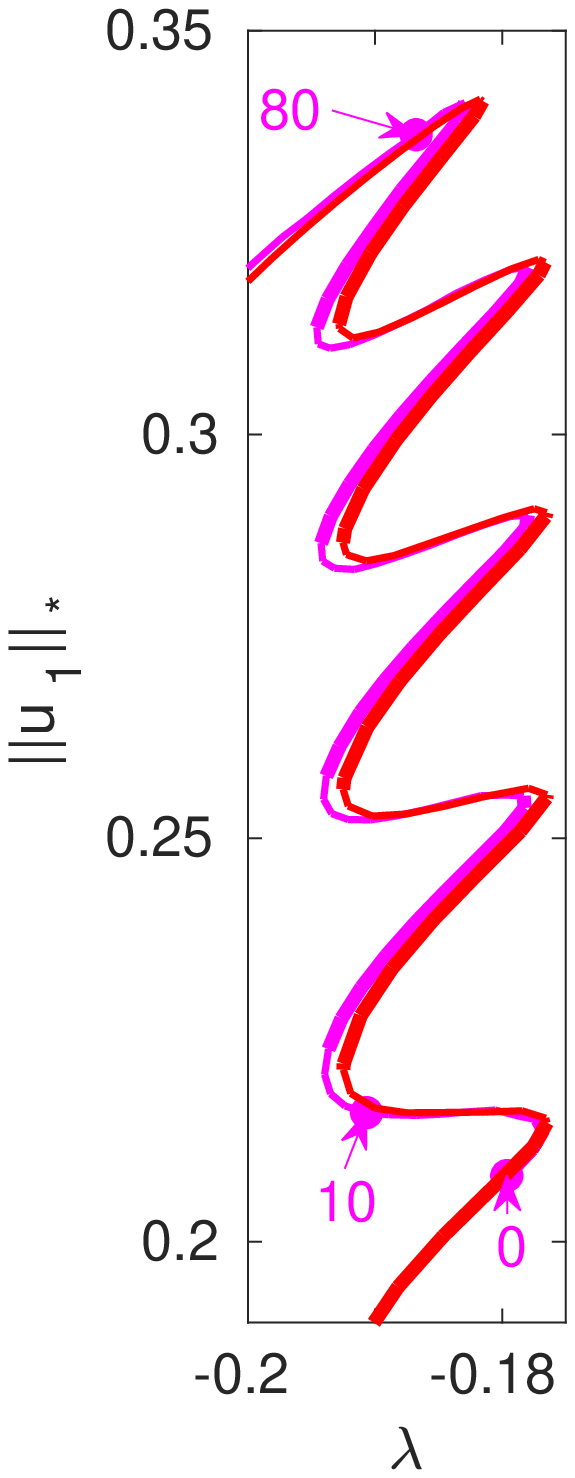}&
\hs{-9mm}\raisebox{44mm}{\begin{tabular}{l}
\ig[width=0.88\textwidth]{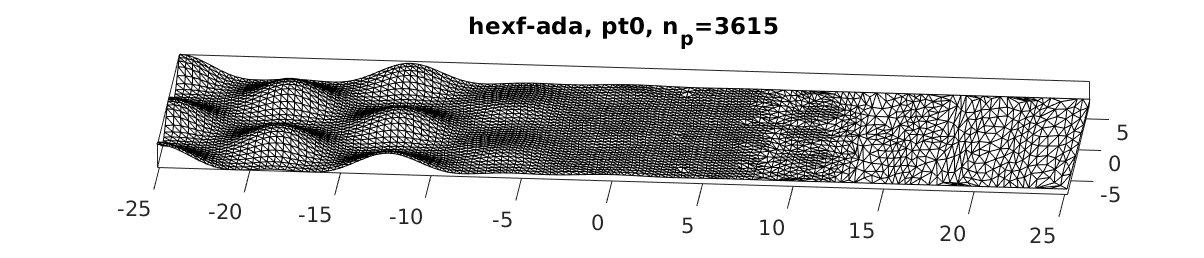}\\[-3mm]
\ig[width=0.88\textwidth]{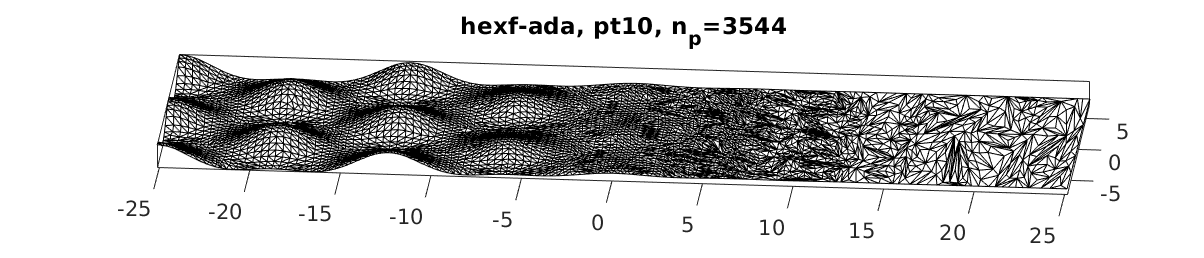}\\[-3mm]
\ig[width=0.88\textwidth]{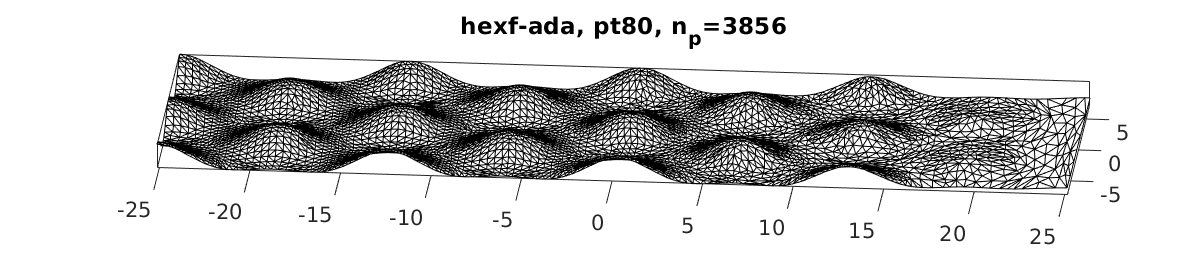}
\end{tabular}}
\end{tabular}
\ece 

\vs{-5mm}
\caption{{\small Results from {\tt cmds2dhexfroada}. We coarsen 
hf/pt40 (with $n_p=4650$) from Fig.~\ref{shf11} to $n_p=3615$,  
then continue with {\tt amod=5} (mesh--adaptation 
each 5th step), and, importantly, ${\tt p.trop.Llow}=0.1$, because 
otherwise the mesh in flat parts of solutions becomes too coarse. 
On the left, the red curve is from the fixed mesh with $n_p=4650$, and 
the magenta curve from adaptation.    \label{shf11a}}}
\end{figure}

In Fig.~\ref{shf11a}, we use \trulle\ \cite{trulletut} 
to first coarsen hf/pt40 (with $n_p=4650$) 
to a coarser mesh, and then continue with \ma\ (see also Remark \ref{marem}) 
each 5th step. Here, a crucial point is to use a rather low 
${\tt p.trop.Llow}=0.1$, because otherwise the flat part of the solution 
(at $x>0$) will become 'too coarse'. This is reasonable for a fixed 
solution, but then the snake fails to grow further hexagons into the flat area, 
and continuation fails via branch jumping or non-convergence. 
As usual, the precise form of the adaptation 
strongly depends on the 
\trulle\ parameter choices \cite{trulletut}, and we refer to {\tt cmds2dhexfroada} for 
details and comments. 

\hulst{caption={{\small (Selection from) {\tt \dname/cmdshexfroada.m}. 
The continuation with \ma\ is set up in lines 12-14; extra coarsening switched 
off via trcop.crmax=0. 
 }}, 
label=shl4c, language=matlab,stepnumber=5, linerange=1-14}{\dhome/cmds2dhexfroada.m} 

\subsection{Two cubes as models for the SC and BCC lattices}\label{3dshnumsec}
As indicated in \S\ref{sh3dsec}, the bifurcations of Turing patterns in 3D 
are in general rather complicated. 
Numerical studies have essentially been restricted to obtaining 
patterns from time integration, 
aka direct numerical simulation (DNS), see, e.g., \cite{SYUO07,EV09}. 
Here we first restrict to 
a simple 3D analog of \S\ref{shsqd}, namely $\Om=(-\pi,\pi)^3$, again 
with homogeneous Neumann BCs $\pa_n u=\pa_n\Delta u=0$ on $\pa\Om$. At the first bifurcation point $\lam=0$ we then have 
a three dimensional kernel $N(G_u)={\rm span}\{\cos(x), \cos(y), \cos(z)\}$, 
i.e., the wave vectors $k^{(1)}=(1,0,0)$, $k^{(2)}=(0,1,0)$, and 
 $k^{(3)}=(0,0,1)$ generate a so called simple cubic lattice. 
By symmetry, i.e., from the amplitude equations \reff{sqae3D}, we also 
know that all bifurcations at $\lam=0$ (and in fact at all subsequent 
bifurcation points) must be pitchforks, and thus we directly use 
{\tt cswibra} to obtain bifurcation directions. 

\begin{figure}[ht]
\bce 
\begin{tabular}{l}
{\small (a) Three kernel vectors}\\
\ig[width=0.2\textwidth]{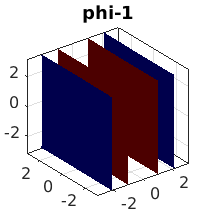}\ig[width=0.2\textwidth]{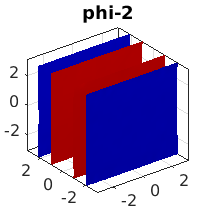}
\ig[width=0.2\textwidth]{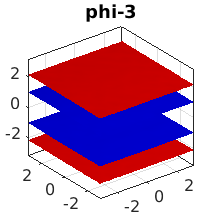}\\
{\small (b) Three bifurcation directions (one from each isotropy class)}\\
\ig[width=0.2\textwidth]{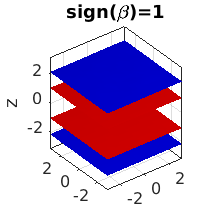}
\ig[width=0.2\textwidth,height=36mm]{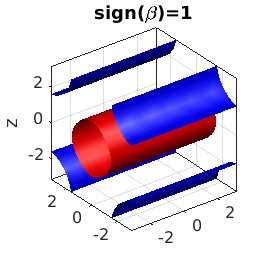}
\ig[width=0.2\textwidth]{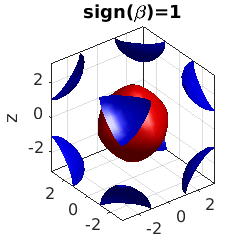}
\end{tabular}
\begin{tabular}{l}
{\small (c) Two bifurcation diagrams}\\
\ig[width=0.23\textwidth,height=40mm]{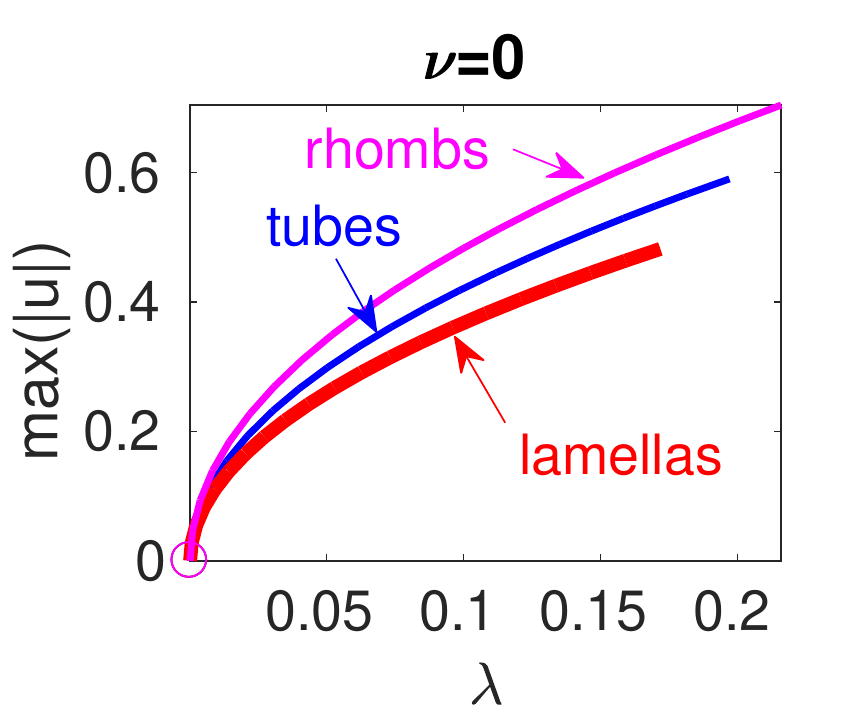}\\
\ig[width=0.23\textwidth,height=40mm]{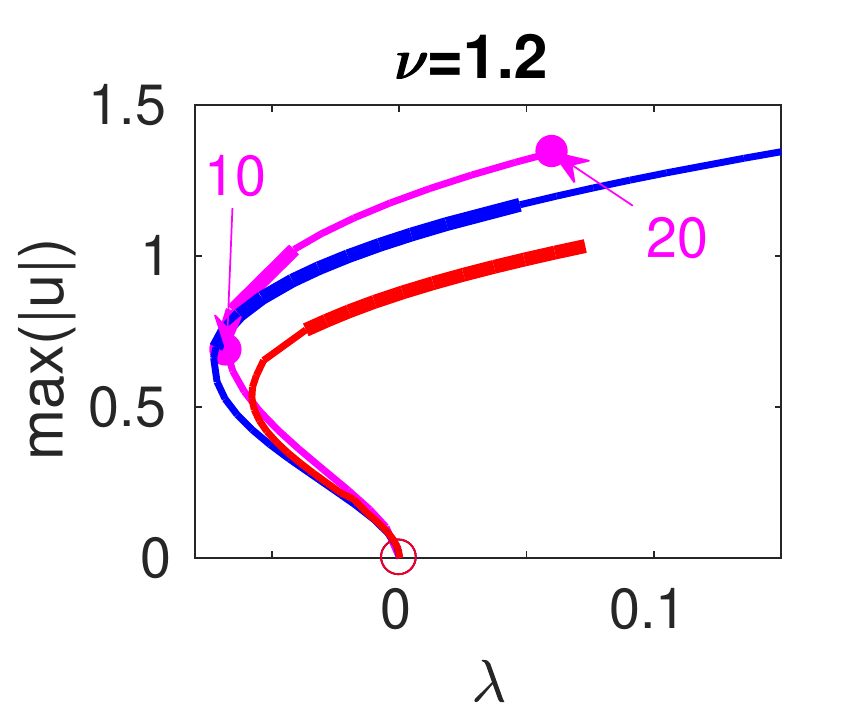}
\end{tabular}
\ece 

\vs{-5mm}
\caption{{\small Selected results from demo {\tt sh/cmds3dcube}. Primary bifurcations at $\lam=0$ in \reff{swiho} over the cube $\Om=(-\pi,\pi)^3$ 
with homogeneous Neumann BCs, pseudo criss-cross mesh of $n_p=6006$ points and $n_t=33000$ tetrahedral elements, see \S\ref{msec} for 
comments on the meshing. 
(a) Isosurface plot of $u$ for the numerical kernel vectors, where blue and red correspond to 
$m_{{\rm blue}}=\frac 3 4 m_0+\frac 1 4 m_1$, $m_{{\rm red}}=\frac 1 4 m_0+\frac 3 4 m_1$, respectively, with $m_0=\min u$, $m_1=\max u$.   (b) Three (of 8) bifurcation directions obtained from {\tt cswibra}, with 
$\al=(0.002, 0.005, 0.914), (-0.03, 0.52, -0.53)$ and $(1.27, 1.29, -1.3)$, respectively.  The other five are 
obtained from symmetry, i.e., rotation and/or translation.  ${\tt sign(\beta)=1}$ refers to the case $\nu=0$. (c) Bifurcation diagrams, $\nu=0$, and $\nu=1.2$, stable branches as thicker lines. The shown branches follow the planforms predicted at bifurcation.
 \label{shf12}}}
\end{figure}

\begin{figure}[h]
\bce 
\begin{tabular}{l}
{\small (a) three kernel vectors on the BCC (sub)lattice, and bifurcation direction for a BCC}\\
\ig[width=0.22\textwidth]{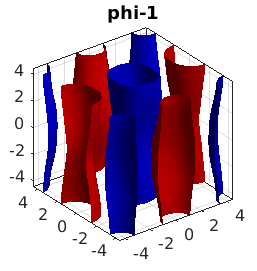}\hs{-0mm}
\ig[width=0.22\textwidth]{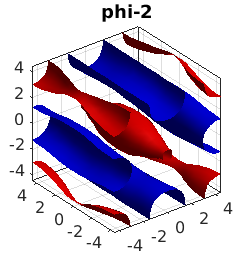}\hs{-0mm}
\ig[width=0.22\textwidth]{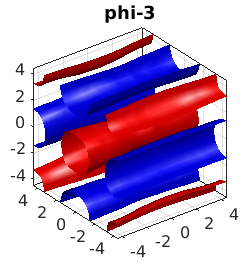}
\ig[width=0.24\textwidth]{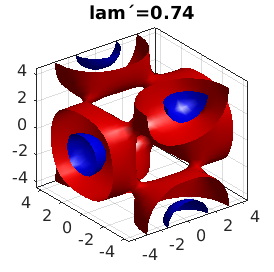}\\
{\small (b) BD of BCCs and square-prisms on BCC lattice, and example solutions}\\[-0mm]
\hs{-5mm}\ig[width=0.22\textwidth,height=36mm]{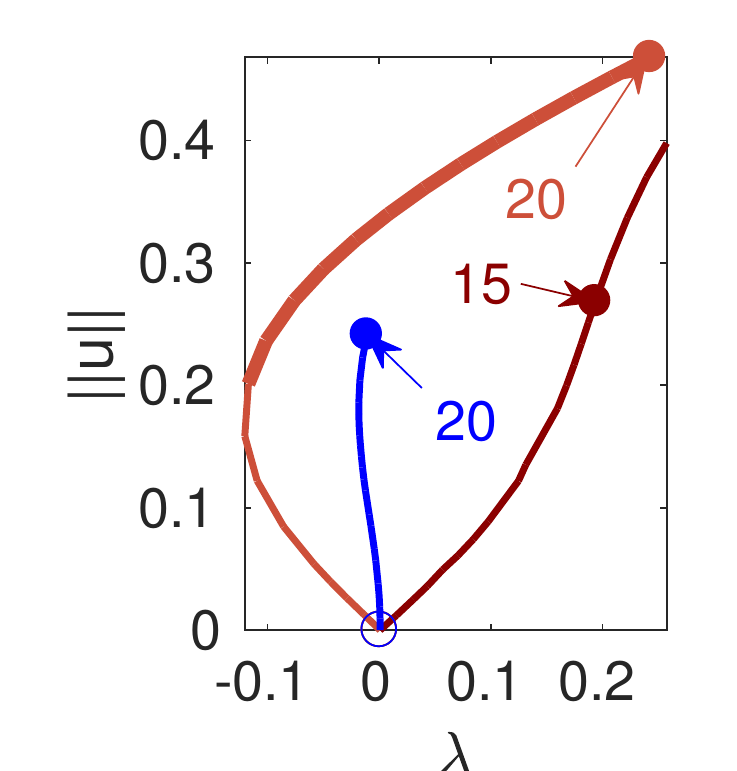}\hs{2mm}
\raisebox{-2mm}{
\ig[width=0.22\textwidth]{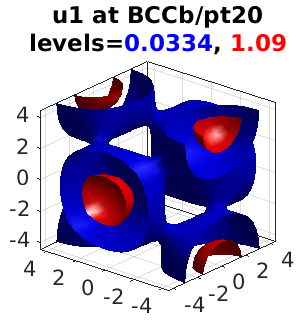}\hs{-0mm}
\ig[width=0.22\textwidth]{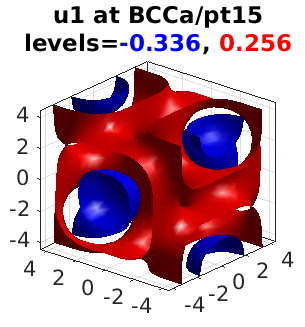}\hs{-0mm}
\ig[width=0.22\textwidth]{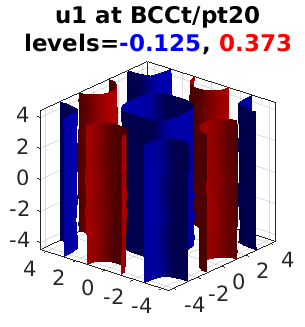}}
\end{tabular}

\ece
\vs{-2mm}
   \caption{{\small \reff{swiho} on a ``BCC lattice cube'' $\Om=(-\sqrt2\pi,\sqrt2\pi)^3$, $\nu=1$. 
(a) kernel vectors (distorted square prisms) and bifurcation direction 
for a BCC obtained from {\tt qswibra}. (b) BD, 'hot' and 'cold' BCCs 
and square prism example solutions. $n_p=10351$, 
$n_t=57024$ tetrahedra, see {\tt cmds3dBCC.m} for details. 
  \label{shf12b}}}
\end{figure}

Figure \ref{shf12} shows some results from {\tt cmds3dSC.m}. In (a) we give 
the three numerically obtained kernel vectors $\phi_1,\phi_2,\phi_3$, 
given by three clean lamellas. This, however, strongly depends on the chosen mesh, 
see \S\ref{msec} for further remarks, and Fig.~\ref{shf12b} for a less clean 
example, and in general it is not obvious how to compose the pertinent three (modulo symmetries) 
bifurcation directions from $\phi_1,\phi_2,\phi_3$. 
Calling {\tt cswibra} yields ten bifurcation directions, of which 
we plot three, one of each isotropy subgroup, i.e., $\tau_1,\tau_2,\tau_3$ in (b). (c) shows the BDs for $\nu=0$ (all branches supercritical, 
with the lamellas stable), and $\nu=1.2$ (all branches subcritical). 
In the latter case, the tubes become stable shortly after their fold, 
but later become unstable again, while the lamellas 
become and stay stable at large amplitude. However, 
even over this 'minimal' domain there are many secondary bifurcations, and 
stable large amplitude solutions without simple symmetries. 

In Fig.~\ref{shf12b} we consider solutions on a cube allowing 
BCC (body-centered cubic) branches. The BCC lattice corresponds to $n{=}6$ wave vectors 
\huga{\label{bcck}\text{
$k_1{=}\frac{1}{\sqrt{2}}\bpm 1\\1\\0\epm$, 
$k_2{=}\frac{1}{\sqrt{2}}\bpm 0\\1\\1\epm$, 
$k_3{=}\frac{1}{\sqrt{2}}\bpm 1\\0\\1\epm$, 
$k_4{=}\frac{1}{\sqrt{2}}\bpm 1\\-1\\0\epm$, 
$k_5{=}\frac{1}{\sqrt{2}}\bpm 0\\1\\-1\epm$, 
$k_6{=}\frac{1}{\sqrt{2}}\bpm -1\\0\\1\epm$,}
} 
leading to a six--dimensional amplitude systems for amplitudes 
$A_1,\ldots, A_6$, including quadratic terms due to resonant triads 
such as $k_1-k_2{=}-k_6$. This amplitude system has a variety of 
solution branches, see \cite{CKnob97, CKnob99}, but on a cube 
with side-length 
$\sqrt{2}l\pi$, $l\in\N$ and homogeneous 
Neumann BCs, only few of the solutions of the amplitude system can 
be realized, namely: 
\bci 
\item Tubes, or more precisely square prisms, corresponding to $0\ne A_1{=}A_4\in\R$, $A_2,A_3,A_5,A_6{=}0$, 
i.e., $u\sim\cos((x+y)/\sqrt{2})+\cos((x-y)/\sqrt{2})$, and of course 
other orientations and spatial shifts. 
\item Balls, or, more precisely, BCCs, $A_1{=}\ldots{=}A_6\in\R$, which correspond to equal amplitude superpositions of tubes.  
\eci 
The three kernel vectors at the primary bifurcation are computed 
as distorted tubes. The BCC branch bifurcates transcritically, and 
bifurcation directions are found by {\tt qswibra}. We call the 
branch to the left (right) hot (cold) balls as they have maxima (minima) in the 
centers. The hot balls become stable at the fold, while cold balls and 
tubes are always unstable.

Figures \ref{shf12} and \ref{shf12b} are just intended as first 
illustrations of 3D pattern formation with \pdep. For instance, 
by extending the domain from Fig.~\ref{shf12b} in one direction we can now produce snaking branches of 'localized hot balls', see \S\ref{tntsec}. 
There, however, we rather focus on 'problems with too 
many solutions', which occur on large (2D and) 3D domains, and generate 
fronts between the hot balls and zero, and other localized solutions, via 
'educated guesses'. 

\subsection{Periodic domains: Discrete and continuous symmetries, demo {\tt shpbc}}\label{sy1sec}
\def\dhome{./pftut/shpbc}\def\dname{shpbc}
If the system $G(u)$ has continuous symmetries described by 
a Lie group $\Ga$, i.e., $G(\ga u)=\ga G(u)$ for all $\ga\in\Ga$, 
then solutions of $G(u)=0$ come in (continuous) group orbits, i.e., 
if $u$ is a solution, then so is $u_\ga:=\ga u$ for all $\ga\in \Ga$. 
In particular, if the group orbit is nontrivial, i.e., $\ga u\ne u$ for 
\def\iid{{\rm Id}}
$\iid \ne \ga\in H$ where $H$ is a subgroup of $\Ga$, then $\pa_\ga u|_{\ga=\iid}$ 
is in the kernel of $G_u(u)$. Thus $G_u$ always has a zero eigenvalue, 
and a robust continuation of $G(u)=0$ requires to remove the symmetry. 
A natural and practical selection of a (locally) unique element in the group orbit $\{\ga u: \ga\in H\}$ goes by adding a constraint that requires the predictor $u$ from a solution $u_{\rm old}$ to lie transverse to the group orbit of $u_{\rm old}$. In a Hilbert space this is naturally an orthogonality relation, the so-called `phase condition' (PC) 
\begin{equation}\label{e:phase}
\langle \partial_g u_{\rm old}, u-u_{\rm old}\rangle=0. 
\end{equation}
See \cite{symtut} for further discussion, and various 
examples with continuous symmetries and suitable PCs. 

Moreover, in case of BPs of higher (discrete) multiplicity with additional 
continuous symmetries $\Ga$, we must also remove 
the symmetries $\Ga$ for branch switching with {\tt q(c)swibra} 
because otherwise the predictors for bifurcating branches cannot be isolated. 
For this, the user can pass the auxiliary list {\tt aux.ali} of 'active' (for 
the branch switching) kernel vectors to {\tt q(c)swibra}, which 
 typically can be identified after a first inspection of the kernel vectors 
using {\tt aux.besw=0}. Here we illustrate this procedure with a simple 
example, namely the SH equation \reff{swiho} on 
$\Om=(-2\pi,2\pi)^2, \text{ with periodic BCs (pBCs) }$ 
\huga{\barr{l} \pa_x^ju(2\pi,y)=\pa_x^ju(-2\pi,y), \quad 
\pa_x^j\Delta u(2\pi,y)=\pa_x^j\Delta u(-2\pi,y), \quad j=0,1,\ \text{ (pBCs in $x$),}\\
\pa_y^j u(x,2\pi)=\pa_y^ju(x, -2\pi), \quad 
\pa_y^j\Delta u(x,2\pi)=\pa_y^j\Delta u(x,-2\pi), \quad j=0,1, \ \text{ (pBCs in $y$),  }
\earr 
}
for all $x,y\in(-2\pi,2\pi)$, where the pBCs for $\Delta u$ naturally arise 
because we have a fourth order problem, or, equivalently, pBCs for the second 
component of the vector $(u,\Delta u)$. 
The SH equation is thus 
translationally invariant in $x$ and $y$, i.e., the (continuous) 
symmetry group is 
$$
\Ga=\{g_{(\rho,\sig)}\in [0,4\pi)^2\}, \text{ with  group action } 
g_{(\rho,\sig)}u(x,y)=u(x-\rho,y-\sig)
$$
and the obvious addition modulo $4\pi$ of the group elements. 
The generators associated to $g_{(\rho,0)}|_{\rho=0}$ and $g_{(0,\sig)}|_{\sig=0}$ are 
$\pa_x$ and $\pa_y$. 

The bifurcation points from the trivial branch are now $\lam=(\lam_1,\lam_2,\lam_3,\ldots)=(0,\frac 1 {16}, 
\frac 1 4,\ldots)$, with kernels spanned by 
\huga{\label{shpker}
\barr{l} \lam_1: \phi_1=\sin(x),\ \phi_2=\cos(x),\ 
\phi_3=\sin(y),\ \phi_4=\cos(y), \text{ (4 dimensional kernel), }\\
\lam_2: \barr{l}\sin(x)\sin(y/2),\ \sin(x)\cos(y/2),\ \cos(x)\sin(y/2),\  \cos(x)\cos(y/2),\\ 
\sin(x/2)\sin(y),\ \sin(x/2)\cos(y),\ \cos(x/2)\sin(y),\ \cos(x/2),\cos(y),\earr \text{ (8 dimensional), }\\
\lam_3: \sin(x)\sin(y),\ \sin(x)\cos(y),\ \cos(x)\sin(y),\  \cos(x)\cos(y), 
\text{ (4 dimensional), }\\
\vdots .
\earr
}
For, e.g., $\lam_1$, clearly $\phi_1$ and $\phi_2$ are related by $\phi_2=
g_{(\pi,0)}\phi_1$, and  span$\{\phi_1,\phi_2\}{=}\{\sin(x{+}\rho): \rho\in [0,4\pi)\}$, 
i.e., the group orbit of all shifts (in $x$) of $\phi_1$, and similar 
relations hold for the kernels at $\lam_2, \lam_3$ and all further $\lam_j$. 
Equivariant bifurcation theory \cite{GoS2002, hoyle} now tells us that 
to find the bifurcations at, e.g., $\lam_1$, it is sufficient to 
restrict to the ansatz $u=(\al_1 \phi_1, \al_2 \phi_3)$ (one representative 
of stripes in $x$ and $y$, respectively), because all other solutions 
(solution branches) are then related to those obtained from $\phi_1,\phi_3$ 
via the symmetries $\Ga$. 

Thus, given a continuous symmetry, to apply {\tt q(c)swibra} at a 
bifurcation point $(u_0,\lam_0)$ we should restrict  to a selection 
of kernel vectors with just one representative from each group orbit. 
Practically we use the following algorithm (see Listing \ref{pshl1} 
for example calls), where w.l.o.g.~we use 
{\tt cswibra} because in our first example we only have 
pitchfork branches; for transcritical branches the algorithm is 
the same, with {\tt qswibra} instead of {\tt cswibra} (see
for instance \S\ref{acs-sec}). 
\bcen
\item Call {\tt cswibra} with {\tt aux.besw=0} (only compute kernel), 
possibly with large {\tt aux.m} (to compute many eigenpairs with $\mu$ 
near $0$), and {\tt aux.ali=[]} (do not select kernel vectors). This 
simply plots the eigenvectors belonging to the $m$ eigenvalues of 
smallest modulus. 
\item Inspect these plots to find the eigenvectors related by $\Ga$, 
and from each class select only one representative by putting its number 
into {\tt aux.ali}. Then call {\tt cswibra} again with {\tt aux.besw=1} 
and {\tt aux.hasker=1}. 
\ecen 
Often, the symmetries are easy to spot after step 1, such 
that {\tt aux.ali} in step 2 can be easily chosen. Additionally, 
a small {\tt aux.ali} can be chosen deliberately to restrict the subspace 
for the search of (predictors for) solutions. 

\begin{figure}[ht]
\bce 
\begin{tabular}{ll}
{\small (a) BD }&{\small (b) Solutions from branches from $\lam_1, \lam_2$, and $\lam_3$. }\\
\hs{-7mm}\ig[width=0.25\textwidth]{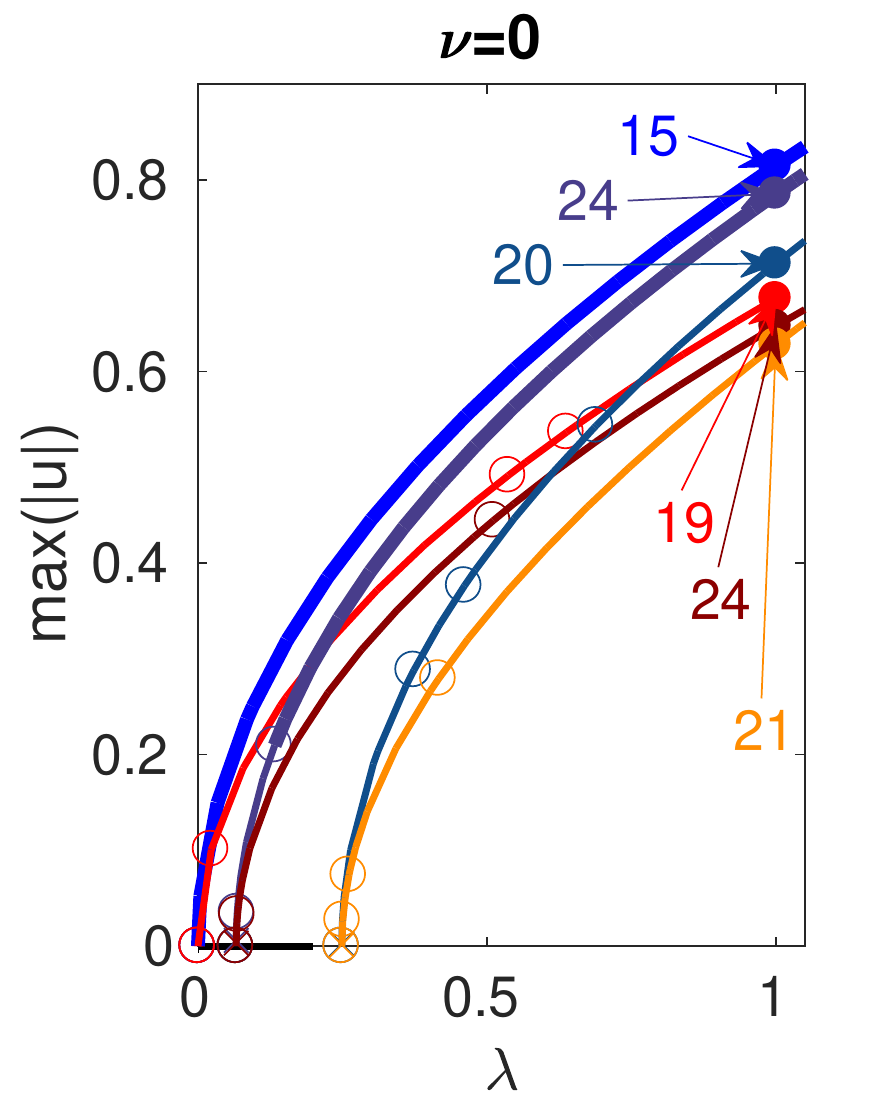}&
\hs{-2mm}\raisebox{25mm}{\begin{tabular}{l}
\ig[width=0.16\textwidth]{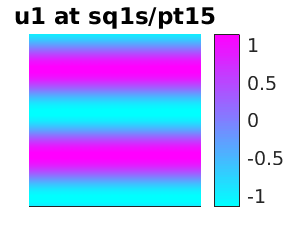}
\ig[width=0.16\textwidth]{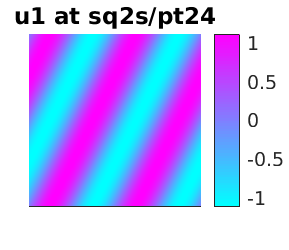}
\ig[width=0.16\textwidth]{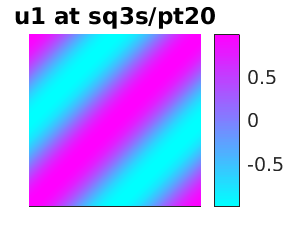}\\
\ig[width=0.16\textwidth]{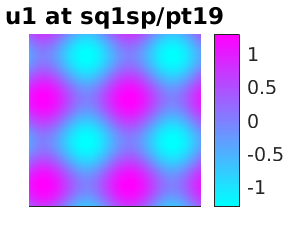}
\ig[width=0.16\textwidth]{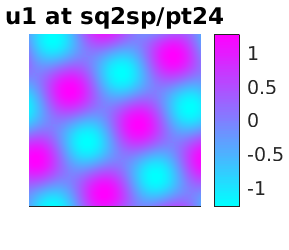}
\ig[width=0.16\textwidth]{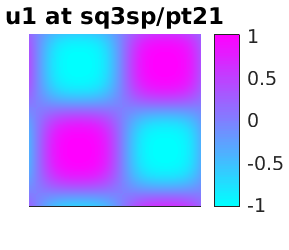}
\end{tabular}}
\end{tabular}
\ece
\vs{-5mm}
   \caption{{\small The SH equation \reff{swiho} over $\Om=(-2\pi,2\pi)^2$ 
with pBCs in $x$ and $y$, $\nu=0$. The kernels at the 1st, 2nd and 3rd BPs 
$u\equiv 0$ and $\lam=0, 1/16, 1/4$ are 4,8, and 4--dimensional, respectively. However, modulo the translational symmetries 
$\Ga$ they are only 2,4 and 2--dimensional, respectively. Modulo $\Ga$, 
at $(0,\lam_1)$ there bifurcate 3 branches (horizontal and vertical stripes, 
and spots), and similarly at $(0,\lam_3)$ (SW to NE and NW to SE  stripes, 
and spots). At $(0,\lam_2)$ we have 6 bifurcating branches, namely 
four types of stripes and two types of spots. 
  \label{shpbcf1}}}
\end{figure}

Figure \ref{shpbcf1} shows results for \reff{swiho} with $\nu=0$ over $\Om=(-2\pi,2\pi)^2$ 
with pBCs and thus kernels \reff{shpker}. At each BP $\lam_1,\lam_2$ and $\lam_3$ we compute just 2 
bifurcating branches, i.e., one stripe branch 
and one spot branch. Regarding the pertinent PCs \reff{e:phase} for the 
continuation of the nontrivial branches we proceed as follows. 
For vertical/horizontal stripes, the PCs read 
\def\uold{u_{{\rm old}}}
\hual{\label{pcx}
&\spr{\pa_x \uold,u}=0 \text{ (to fix translations in $x$), and }\\
&\spr{\pa_y \uold,u}=0 \text{ (to fix translations in $y$), }\label{pcy}
} 
respectively. For diagonal stripes both $\ga_{(\rho,0)}$ and $\ga_{(0,\sig)}$ 
generate the same group orbits such we can use either \reff{pcx} or 
\reff{pcy}, while 
for spots we need both. For the implementation, we therefore 
assemble differentiation matrices {\tt p.mat.Kx} and {\tt p.mat.Ky} at 
startup, and set up routines {\tt qfx, qfy}, and {\tt qf} (both), and their  derivatives. 
After branch-switching from $u\equiv 0$ we first do a few 
(1 to 3) steps without PC. We then switch on the pertinent 
PC by setting, e.g., {\tt p.nc.nq=1} and {\tt p.fuha.qf=@qfx}, 
{\tt p.fuha.qfder=@qfxder} for \reff{pcx}, and 
continue further. The case of translational PC in $y$ or in $x$ and $y$ 
works analogously, and for convenience these commands are collected in 
little functions 
{\tt p=qxon(p), p=qyon(p), p=qxyon(p)}, respectively. See Listing 
\ref{pshl1}, {\tt cmdssq.m} for the main script, and Table \ref{shpbcdir} 
for an overview of the files involved. 
\hulst{caption={{\small  {\tt cmdssq.m} (selection). Using 
{\tt cswibra} with {\tt aux.besw=0} to first inspect the kernel, then 
with {\tt aux.besw=1} and \mbox{${\tt aux.ali}=[1,2]$} to factor out 
the continuous symmetries before deriving and solving the CBE \reff{cbe}. 
Subsequently, we use {\tt seltau} to choose the bifurcation direction, 
and switch on the needed PCs for continuation ({\tt qyon} for horizontal stripes, and {\tt qxyon} for spots). 
}}, label=pshl1, language=matlab,stepnumber=0, linerange=9-16}{\dhome/cmdssq.m}

\begin{table}[ht]\taskip
\caption{Scripts and functions in {\tt pftut/shpbc}. 
\label{shpbcdir}}
\bce\vs{-4mm}
{\small 
\begin{tabular}{l|p{0.77\textwidth}}
script/function&purpose, remarks\\
\hline
cmdssq, cmdshex&scripts, with cmdssq yielding Fig.~\ref{shpbcf1}. \\%
cmdssq\_defl&script to compute further solutions by deflation, see \S\ref{adeflsec}\\
shinit&initialization, including the call {\tt p=box2per(p,[1,2])} to switch 
on pBCs\\
oosetfemops&set FEM matrices, including {\tt filltrafo} to account for the pBCs \\
sG,nodalf,sGjac&rhs and Jacobian as usual\\
qf,qfder&phase condition (PC) function (in $x$ and $y$) and derivative\\
qfx, qfxder&PC and derivative, only in $x$, e.g., for 'vertical stripes', see also qfy, qfyder\\
qxon, qyon, qxyon&convenience functions to switch on PCs in $x$, in $y$, or in 
both, respectively. 
\end{tabular}
}
\ece\vs{-3mm}
\end{table}\teskip

\brem\label{pbcrem1} a) 
The implementation of periodic BCs is explained in 
\cite{pbctut}. Essentially, we only need to call {\tt p=box2per(p,[1 2])} 
during initialization (see {\tt shinit.m}), and {\tt filltrafo} 
after each assembly of a system matrix such as {\tt K} or {\tt M}, see 
{\tt oosetfemops.m}. 

b) A phase condition such as \reff{pcx} is an additional equation and hence requires to free an additional parameter, and to do so we introduce a dummy 
'speed' parameter $s_x$ and add $s_x\pa_x u$ to the rhs of the SH equation. 
Similarly, for \reff{pcy} we add $s_y\pa_y u$, such that 
the augmented problem reads 
\huga{\label{swihosx} 0= -(1+\Delta)^2 u + \lam u  +\nu u^2-u^3+s_x\pa_x u+s_y\pa_y u. 
}
In the parameter vector, the new parameters sit at positions 3 for $s_x$ and 4 for $s_y$, and thus we set {\tt p.nc.ilam=[1 3]} for the $x$--PC, 
{\tt p.nc.ilam=[1 4]} for the  $y$--PC, and {\tt p.nc.ilam=[1 3 4]} if 
both PCs are active. The speeds $s_x,s_y$ are naturally initialized with $(0,0)$, 
and they both stay $\CO(10^{-6})$ or smaller during all continuations. 

c) In {\tt cmdssq\_defl.m} we use deflation to compute further branches of 
solutions for \reff{swiho} over $\Om=(-2\pi,2\pi)^2$ with pBC, see \S\ref{adeflsec}. 
In {\tt shpbc/cmdshex.m} we treat the related case of a periodic domain allowing 
hexagons, where essentially for the hexagons we need to use {\tt qswibra} 
instead of {\tt cswibra}. In  \S\ref{pfsurf} we use 
the ideas explained here to treat related problems for pattern formation 
on spheres, which naturally lead to large kernels. 
\eex\erem 

\def\dhome{./pftut/sh}\def\dname{sh}

\subsection{Remarks on choices of 2D and 3D meshes}\label{msec}
\begin{figure}[ht]
\bce 
\begin{tabular}{lll}
{\small (a) standard}&(b) criss-cross, with refinement&{\small (c) pseudo criss-cross}\\
\hs{-2mm}\ig[width=0.19\textwidth]{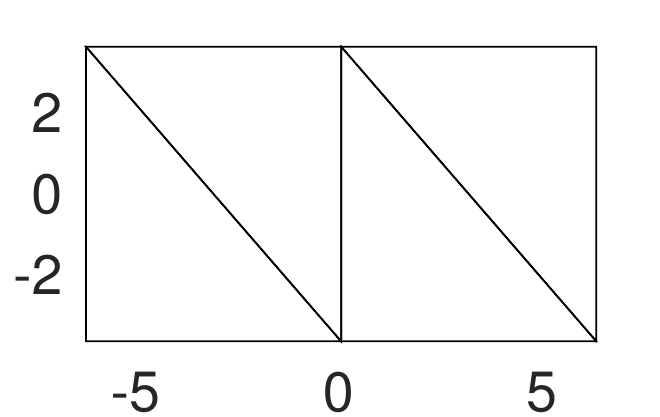}&
\hs{-2mm}\raisebox{-2mm}{\ig[width=0.19\textwidth]{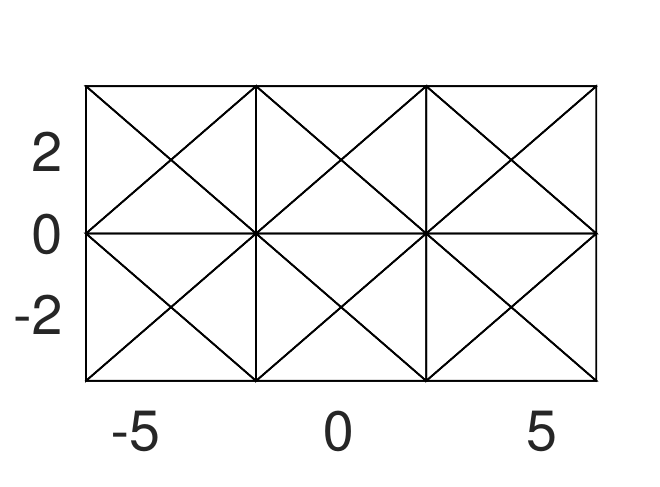}\hs{-2mm}\ig[width=0.19\textwidth]{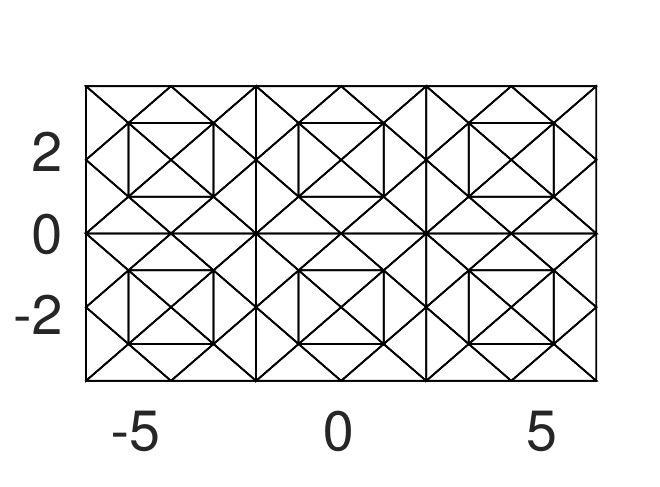}}&
\hs{-3mm}\ig[width=0.19\textwidth]{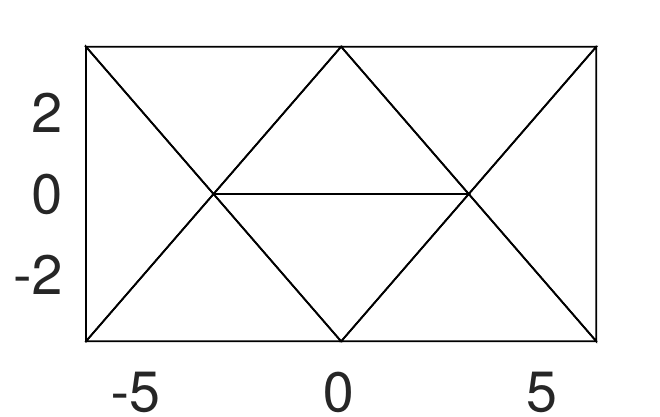}
\hs{-2mm}\ig[width=0.19\textwidth]{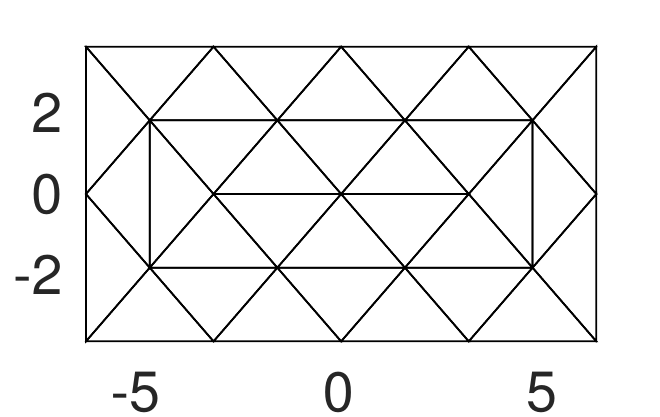}
\end{tabular}
\begin{tabular}{lll}
{\small (d) standard meshing in 3D}&{\small (e) pseudo criss-cross, with 1 refinement (right)}&{\small (f) $\phi_3$ (d-mesh)}\\
\ig[width=0.15\textwidth]{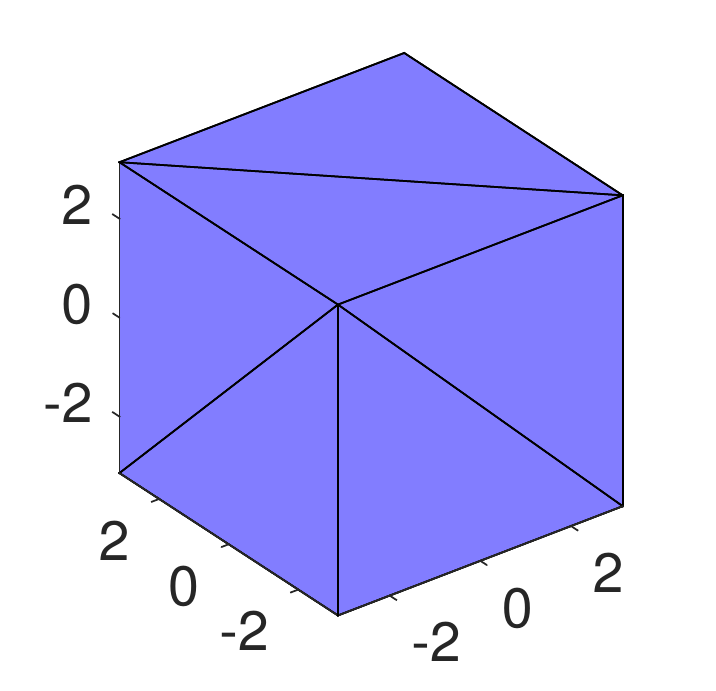}
\ig[width=0.15\textwidth]{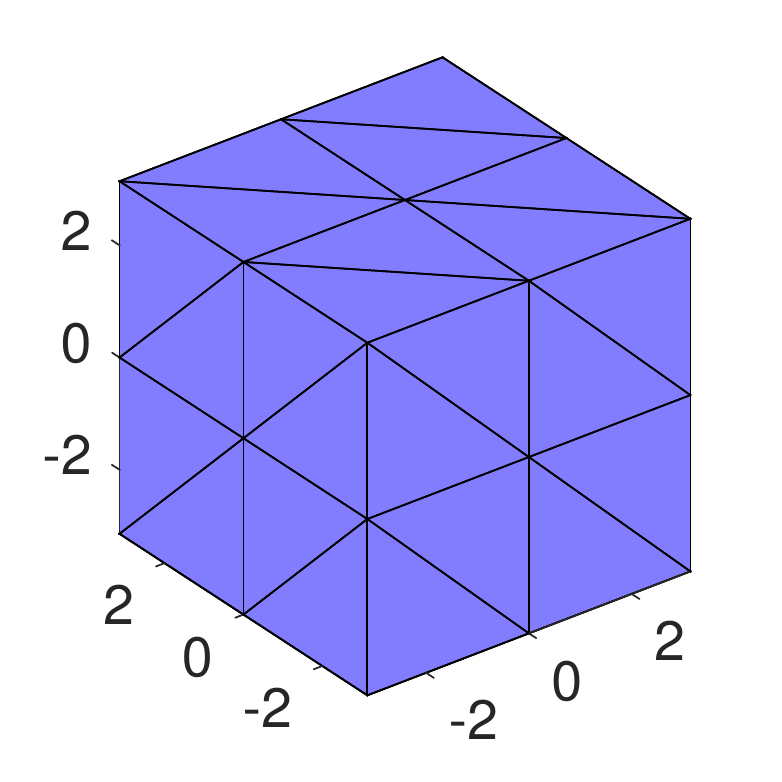}&
\ig[width=0.15\textwidth]{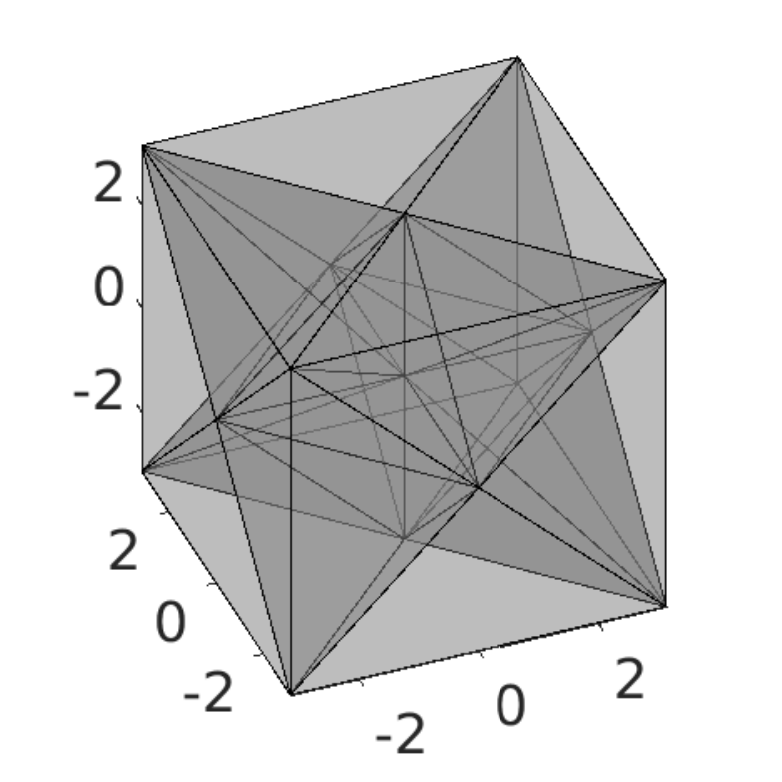}
\ig[width=0.15\textwidth]{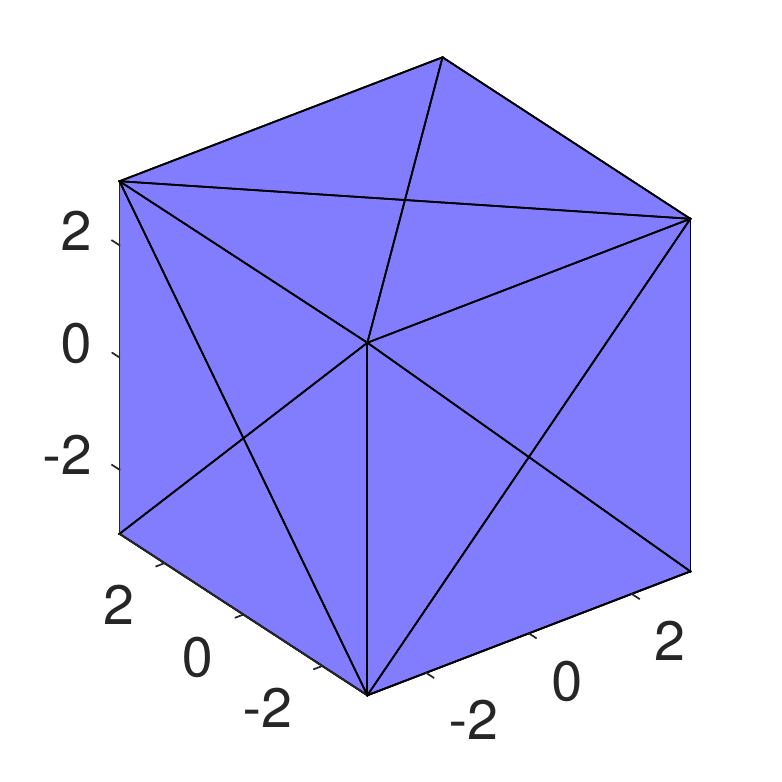}
\ig[width=0.15\textwidth]{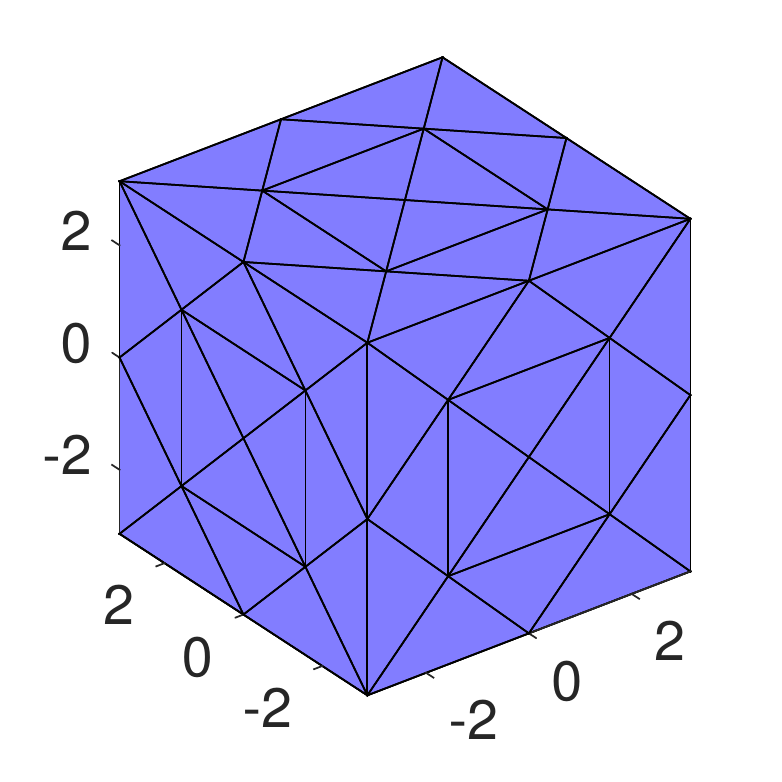}&
\ig[width=0.15\textwidth]{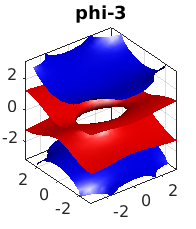}
\end{tabular}
\ece 

\vs{-5mm}
\caption{{\small (a-c) 2D meshes on $\Om=(-l_x,l_x)\times (-l_y,l_y)$, $l_x=2\pi, l_y=2\pi/\sqrt{3}$, starting from a meshgrid with $n_x=3, n_y=2$. 
(a) Standard meshing ($n_p=6$) in 2D destroys reflection and rotation symmetry. 
(b) criss-cross via ``refine-longest'', $n_p=18$, and 
with one additional default refinement, $n_p=59$. This has (locally) the full $D_4$ symmetry (reflections and discrete rotations). 
(c) pseudo criss-cross, where rectangle centers are added to the grid before 
meshing, $n_p=8$, and $n_p=28$ after 1 (red) refinement. This also always 
has the reflection symmetries but in general no discrete rotational symmetry. (d) Standard 3D meshing destroying all symmetries. 
(e) Criss-cross like meshing here keeps all symmetries (discrete rotation and 
reflection), also under (uniform) refinement. (f) $\phi_3$ for SH on a ``standard'' mesh with $n_p=6450$, compare to Fig.~\ref{shf12}(a). 
 \label{mfig}}}
\end{figure}

The default meshing of rectangles in 2D proceeds via Delauney triangulation of a regular rectangular grid. 
As a consequence, these 
meshes have no reflection or rotational (discrete, by $\pi/2$) symmetry, see Fig.~\ref{mfig}(a) for a sketch. However, if we have many solutions (solution branches) 
``close together'', and (multiple) bifurcations distinguished by their 
symmetry, then it is desirable to have meshes as symmetric as possible. 
So called {\em criss-cross meshes}, which using the \ptool\ 
can for instance be generated by calling {\tt refinemesh(..,'longest')} 
on a (default) {\tt poimesh}, 
have a $D_4$ (rotations by $\pi/2$ and reflections) symmetry (locally if the domain does not have $D_4$), 
which also on further uniform refinement stays intact, 
see Fig.~\ref{mfig}(b) for an example. If the \ptool\ is not available, 
a simple but efficient method to obtain similar meshes, which we 
call {\em pseudo criss-cross} is as follows. 
We start with a regular rectangular grid, and then add the rectangle 
midpoints to the grid. A subsequent Delauney triangulation then 
yields meshes of type (c), which are at  least reflection symmetric. 

Similarly, the default meshing in 3D produces asymmetric meshes of type (d). 
Consequently, the continuation of highly symmetric branches such as the tubes 
and rhombs in Fig.~\ref{shf12} may be problematic: The solutions may jump 
(or, near bifurcation points, ``slowly drift'') to a less symmetric 
branch. For instance, over standard grids the tubes often jump to 
lamellas, and the rhombs to tubes (or at least strongly distort at larger amplitude). This can be alleviated 
by choosing meshes of type (e) in Fig.~\ref{mfig}, which we also call 
pseudo criss--cross. Here we 
start with a regular cuboid grid, add all cuboid centers and 
face centers, and then do the Delauney meshing, and afterwards possibly 
some uniform ('red') mesh--refinement. We remark that starting with a coarse 
mesh and refinement vs starting with a fine grid produces similar but in general 
not equivalent results.  

A lack of mesh symmetry is also often reflected in 
distorted eigenvectors at multiple BPs. Figure \ref{mfig}(f) shows 
one of the three distorted lamellas kernel vectors obtained 
for the same settings as in Fig.~\ref{shf12}, but on a standard 
mesh of type (d) with $n_p=6450$ points. This is not a problem for 
{\tt cswibra}, which computes the same $\tau_1,\tau_2,\tau_3$ as in Fig.~\ref{shf12}(b), 
but branch switching more likely fails than on more symmetric meshes in the sense that the initial corrector jumps, e.g., from the rhombs predictor to the 
tubes branch. 

To give the user some easy control over the 
meshing, the calls {\tt pde=stanpdeo2D(lx,ly,nx,ny,sw)} and 
{\tt pde=stanpdeo3D(lx,ly,lz,nx,ny,nz,sw)} have, besides the obvious arguments $l_x, n_x, \ldots$, the struct {\tt sw} as an auxiliary argument. Currently, this 
can have two fields, namely 
\bci 
\item If {\tt sw.sym=1}, then meshes of type (c) (2D) and (e) (3D) from 
Fig.~\ref{mfig} are generated.
\item In 2D, if {\tt sw.sym=2}, then we generate genuine type (b) criss cross meshes.  
\item If ${\tt sw.ref}>0$, then {\tt sw.ref} refinement steps are executed after 
the initial meshing.
\eci 
See {\tt sh/shinit.m} and {\tt cmds2dsq.m, cmds2dhex.m, cmds3dSC.m, cmds3dBCC}, 
and the {\tt cmds*} scripts in our next example {\tt schnakpat} 
for templates and details. 

\brem\label{marem}{\rm 
(a) \oop, like many other FEM packages offers additional elements, for instance bilinear rectangular elements (in 2D) and 
triangular prism elements (in 3D). For some applications, these show some 
advantages, but in this tutorial we restrict to the triangle and tetrahedra elements. 

(b) An important strength of the FEM is the option of adaptive mesh refinement. 
For \pdep, this is discussed in some detail in \cite{actut} (1D and 2D, using error estimators) 
and \cite{trulletut} (2D and 3D, using the anisotropic \ma\ package \trulle). 
However, for the (roughly 
harmonic) periodic patterns considered here, \ma\ is typically 
not very efficient, and, moreover, it may break symmetries and introduce 
anisotropies into the problem. Therefore, \ma\ does not play a big role in this 
tutorial. On the other hand, for fronts between patterns 
and constant solutions, \ma\ may be quite efficient as the 
constant part obviously 
only needs a coarse mesh.  In Fig.~\ref{shf11a} we gave one 
example for this in 2D, and in Fig.~\ref{shtf3} we do the analog in 3D. 
}\eex
\erem

\subsection{Problems with 'many' solutions, 
warnings, tips and tricks}\label{tntsec}
\subsubsection{General remarks}\label{tnt1} 
\begin{figure}[h]
\bce 
\begin{tabular}{lll}
{\small (a) BD}&{\small (b) Zoom of BD}&{\small (c) 'tangents' at {\tt BP16}}\\
\ig[width=0.28\textwidth, height=60mm]{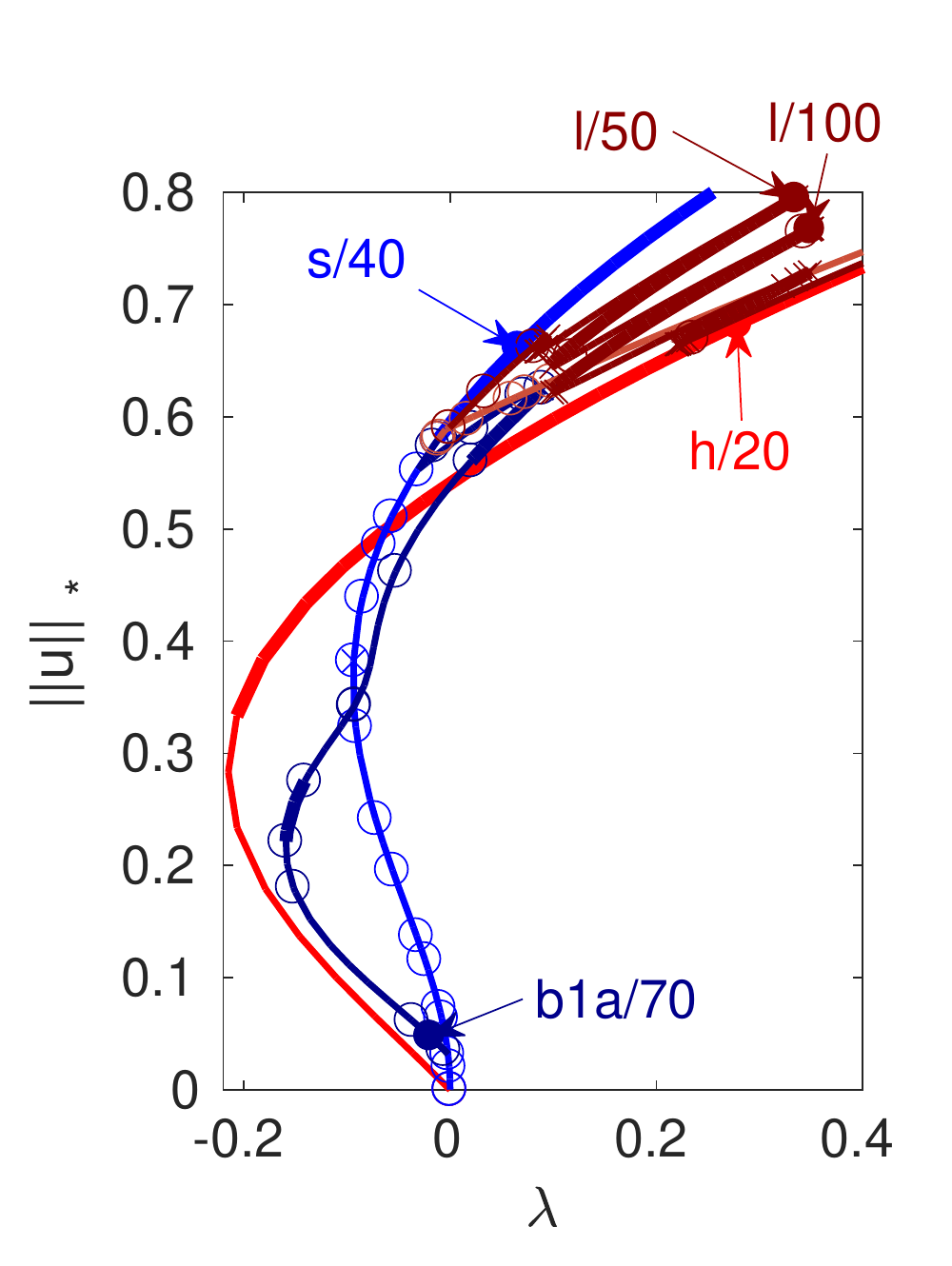}&
\ig[width=0.28\textwidth, height=60mm]{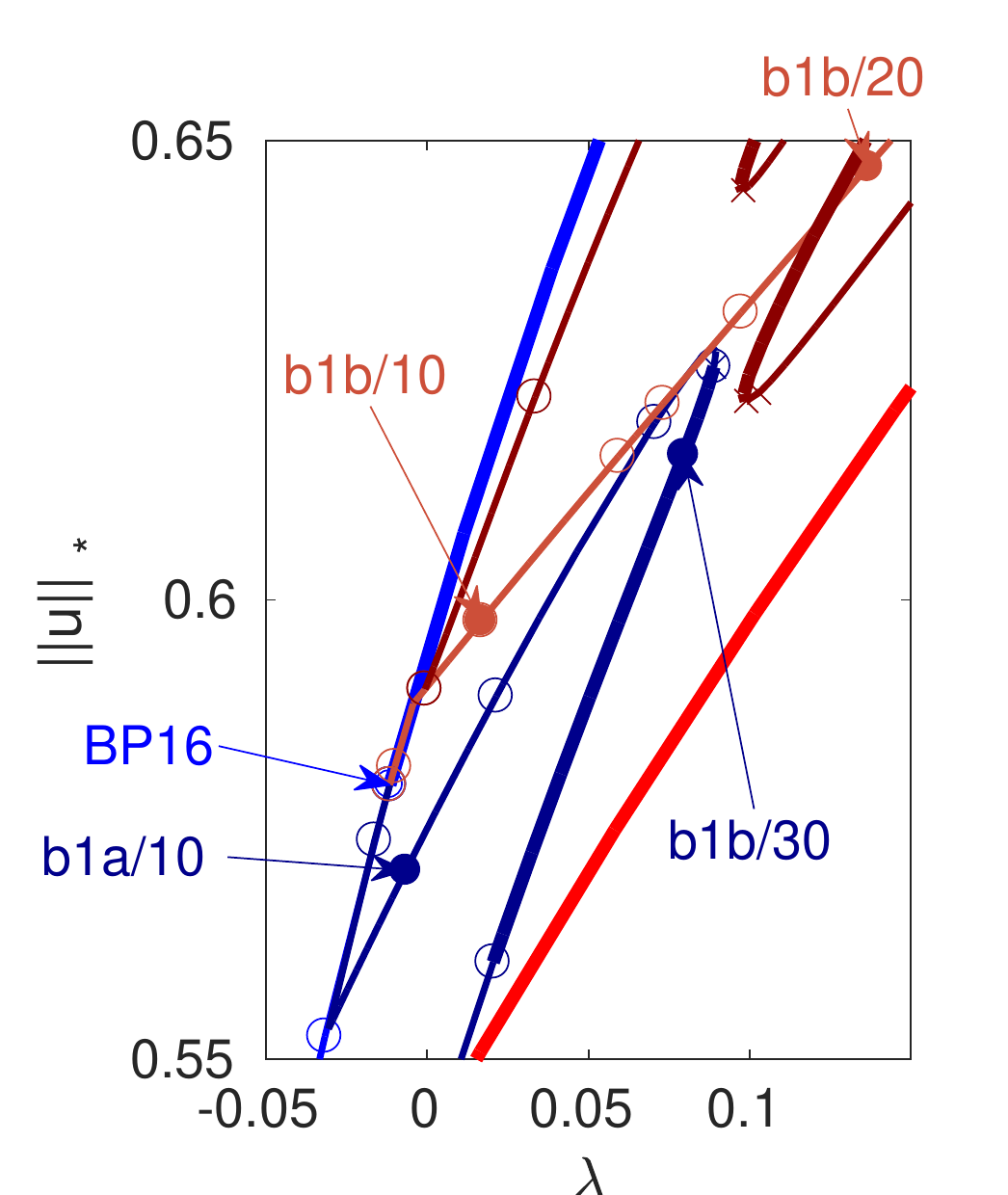}&
\raisebox{28mm}{\begin{tabular}{l}
\ig[width=0.3\textwidth]{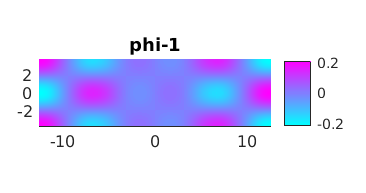}\\[-6mm]
\ig[width=0.3\textwidth]{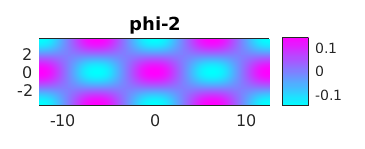}\\[-4mm]
\ig[width=0.3\textwidth]{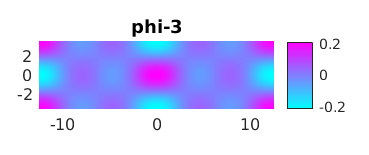}
\end{tabular}}\\
{\small (d) plots of h and s}&{\small (e) plots from b1b (beans)}&{\small (f) plots of h2s front}\\
\raisebox{25mm}{\begin{tabular}{l}
\ig[width=0.28\textwidth]{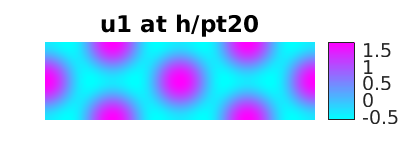}\\[-2mm]
\ig[width=0.28\textwidth]{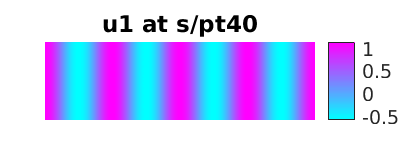}
\end{tabular}}
&\raisebox{25mm}{\begin{tabular}{l}
\ig[width=0.28\textwidth]{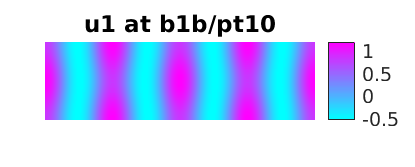}\\[-2mm]
\ig[width=0.28\textwidth]{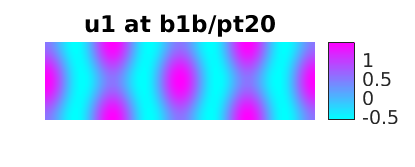}
\end{tabular}}
&\raisebox{25mm}{\begin{tabular}{l}
\ig[width=0.28\textwidth]{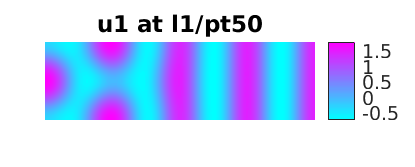}\\[-2mm]
\ig[width=0.28\textwidth]{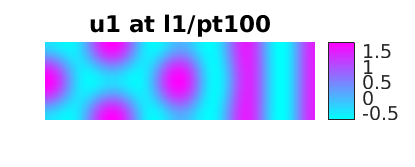}
\end{tabular}}\\[-10mm]
(g) plots from {\tt b1a}&&\\
\ig[width=0.28\textwidth]{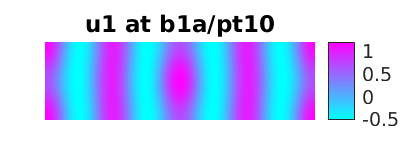}&
\ig[width=0.28\textwidth]{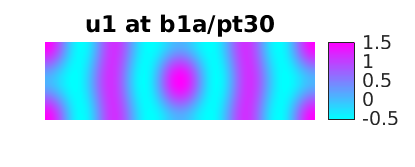}&
\ig[width=0.28\textwidth]{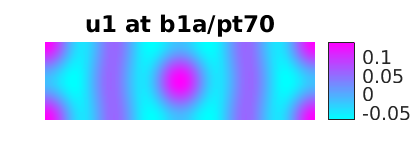}
\end{tabular}
\ece 
\vs{-6mm}
\caption{{\small \reff{swiho} on $\Om{=}(-2l_x,2l_x){\times}(-l_y,l_y)$, $l_x{=}2\pi, l_y{=}2\pi/\sqrt{3}$, $\nu=1.3$, example results from {\tt cmds2dhexb.m}. 
See text and {\tt cmds2dhexb.m} for details.  \label{shf11b}}}
\end{figure}

In \S\ref{sh2dnum} (2D) and \S\ref{3dshnumsec} (3D) we considered 
small (almost minimal) domains. Over larger domains, the number 
of patterns resulting from the Turing instability and secondary bifurcations 
quickly becomes quite large, which can be a serious problem for the numerics. 
For illustration, and for comments on how to deal with these 
problems, here we double the domain from Fig.~\ref{shf10}, see Fig.~\ref{shf11b}, and the script {\tt cmds2dhexb.m}. 
We focus on the stripe branch {\tt s} and its secondary bifurcations, which due 
to the many secondary bifurcation points 
(and in contrast to {\tt cmds2dhex.m}) we now continue using {\tt p.sw.bifcheck=2}. The {\tt s} branch solutions again gain stability near $\lam_1\approx -0.01$, 
but now, with {\tt dsmax=0.1}, the number of unstable eigenvalues jumps from 
2 to 0 in the continuation across $\lam_1$, leading to the 
localization of {\tt BP16} in Fig.~\ref{shf11b}(a,b). The {\em three} smallest eigenvalues then are 
$\mu_{1,2,3}=-0.000024,-0.016013, 0.04903$, and the corresponding 
eigenvectors (1st component) are shown in (c). 
It turns out that: 
\bci 
\item  To each of these eigenvectors there is a branch bifurcating from {\tt s}, 
although only approximately at {\tt BP16}, and these eigenvectors are 
also (approximately) returned by {\tt cswibra} at {\tt BP16}. 
\item 
Thus we can simply call {\tt seltau} after {\tt cswibra}, and 
obtain the bifurcating branches {\tt b1a, b1b}, and {\tt b1c} (not shown). 
\eci 

In particular, the {\tt b1b} branch are the beans (light brown (e)), 
from which a 
snaking branch (dark brown (f)) of fronts between hexagons and stripes bifurcates. 
Example plots from the (dark blue) branch {\tt b1a} associated to {\tt phi-3} are shown in (g). 
This reconnects to the stripes at low amplitude, and has (small) stable segments. 
Similarly, calling {\tt cswibra} at many other BPs on the stripe branch 
yields bifurcations to various branches of patterns which have (small) 
stable segments. In fact, once the bifurcation points become 
'sufficiently dense'  
on a given branch, we can more or less 
\bci 
\item call \mbox{\tt q(c)swibra} at {\em any} point, 
including regular points. Typically, the eigenvectors belonging to  
small eigenvalues are then sufficiently close to the kernel vectors at 
a nearby bifurcation point, and usually branch switching via {\tt seltau} 
works.
\eci   
In this sense, already on this still small domain
\bci 
\item it becomes 
essentially impossible to obtain a 'nearly complete' bifurcation diagram 
that contains at least the stable solutions at small to intermediate 
amplitude, $\lam\in(-0.2,0.4)$, say. 
\eci 

Moreover, we want to stress that in these circumstances, the use of {\tt pmcont} 
instead of {\tt cont} seems crucial to avoid (reduce) uncontrolled branch 
switching. Here we recall that: 
\bci 
\item Branch jumping does not produce 'wrong solutions', but a wrong 
bifurcation picture.
\item For {\tt pmcont} it is important to choose {\tt dsmin} 'sufficiently large', 
as a too small {\tt ds} (obtained via many stepsize reductions possible 
for small {\tt dsmin}) leads to essentially 
the same behavior of {\tt pmcont} as {\tt cont}, e.g., branch jumping.  
\item Rather use {\tt p.pm.resfac} (residual decrease in each Newton step, default $10^{-3}$) and 
{\tt p.pm.mst} (number of different length predictors, default 4) to tune {\tt pmcont}. Smaller {\tt p.pm.resfac} means that more predictors are discarded 
(stricter behavior of {\tt pmcont}). Often, it also helps to (maybe 
only on a 'difficult' segment of the branch) relax {\tt p.nc.tol} 
(residual tolerance, default $10^{-8}$) somewhat, e.g., set ${\tt p.nc.tol}=10^{-6}$. 
\eci 

Another option is adaptive mesh-refinement. This often 
helps if patterns start to 'drift' under continuation, as it introduces 
a (helpful) mesh-inhomogeneity, which may pin patterns at the 'right' 
(desired) positions. It is not needed here, but at the end of {\tt cmds2dhexb.m} we give 
an example of such an adaptive mesh--refinement on the {\tt l1} branch, after which the branch continues qualitatively as before.  
Finally, consider the remarks on mesh--symmetry from \S\ref{msec}. 

In any case, if there are 'too many' solutions close to each other, the 
continuation is more complicated and may fail by, e.g.: 
\bci 
\item Missing (important) bifurcation points due to a too large stepsize 
(which may be necessary to avoid undesired branch switching); 
\item Undesired branch switching even under strict settings for {\tt pmcont}; 
\item Non-convergence of {\tt pmcont} under too strict settings (too small 
{\tt p.pm.resfac} and/or too large {\tt p.nc.dsmin}). 
\eci 
Thus, to study pattern formation in 2D (or 3D, where the above problems 
usually become worse), we recommend to always start with a rather small 
domain. In particular on larger domain, a useful alternative to the 
'continue and bifurcate' strategy used so far may be a direct search for 
patterns of interest, described next. 

\subsubsection{Deflation}\label{adeflsec}
So far we implicitly assumed that 
all 'relevant' solution branches can be 
found by starting from a known (trivial) branch, continuation and 
branch switching. However, there may be (relevant) branches not connected to 
the starting branch, e.g., bifurcations may be imperfect, 
or a trivial branch may not be available and it may be unclear how to 
find a first solution. In this case, we may simply generate, e.g., 
a number of initial guesses for $u$ at fixed $\lam$, run Newton loops 
and hope to converge to a solution.%
\footnote{As Newton loops only converge locally, 
it is sometimes helpful to first run some time-integration to 
decrease the residual, i.e., to get closer to a solution, 
see \S\ref{tnt1} for some concrete examples.} 
However, if we aim to find to several solutions at fixed $\lam$, then using 
several different initial guesses may not work, as we may always converge 
to the same solution, or not converge at all.

There are (at least) two methods aiming to improve this. The first 
are called minimax algorithms, which, essentially, exclude the span of 
previously found solutions from the search directions for new solutions. 
This has been implementated in \mlab\ for (scalar) PDEs 
in \cite{LZ01, LZ02} and linked with \pdep\ in \cite{CK15gluing}. 

A related method which is readily available in \pdep\ is deflation, which 
in the context of continuation has been (re)proposed in \cite{FBF15}, 
see also \cite{FBB17}, and \cite{CKF18}. Assuming that we already have a number of 
solutions $(u^{(1)},u^{(2)},\ldots,u^{(l)})$, $l\ge 1$, of 
$G(u)=0$ (at fixed $\lam$), the idea is to modify the problem to 
\huga{\label{adefl1} 
F(u;u^{(1)},\ldots,u^{(l)}):=\CD(u;u^{(1)},\ldots,u^{(l)})G(u)=0, 
}
with a deflation operator $\CD$ which, essentially, goes to $\infty$ 
fast enough as $u\to u^{(j)}$ for $j\in\{1,\ldots,l\}$ to exclude 
the roots $u^{(j)}$ of $G$ from $F$, i.e., 
\huga{\label{adefl1b}
\liminf_{i\to\infty}\|F(u;u^{(1)},\ldots,u^{(l)})\|>0\text{ for all sequences 
$(u_i)$ such that }
u_i\to u^{(j)}, j=1,\ldots,l. 
} 
At the same time we require 
$F(u;u^{(1)},\ldots,u^{(l)})=0 \aqui G(u)=0$ 
for all further roots of $G$. A simple choice is (the scalar function) 
\huga{\label{adefl2}
\CD(u;u^{(1)},\ldots,u^{(l)})=\left(\al_1+\Pi_{j=1}^l \|u-u^{(j)}\|_q^q\right)^{-1}, 
}
where $\|\cdot\|_q$ is the standard $q$--norm in $\R^n$, 
and $q$ and $\al$ are suitable parameters, and naturally 
the roots $u^{(j)}$ must be considered with their multiplicity 
for \reff{adefl1b} to hold. 
 In \cite{FBF15, FBB17} 
it is shown how this choice 'deflates' the domain of attraction 
of known roots of $G$ for the Newton loops for $F$, and illustrated 
by various examples, including nonlinear PDE problems 
(after FEM discretization). In this context, $\pa_u G$ is sparse, but 
\huga{\label{adefl1}
\pa_u F=\CD\pa_uG+G\pa_u\CD
}
is dense. 
However, since $G\in\R^{n_u\times 1}$ and $\pa_u\CD\in\R^{1\times n_u}$, 
$\pa_u F$ is a rank-one correction of $\CD\pa_u G$, which for 
instance can be handled with Sherman--Morrison--Woodbury (SMW) formulas  
\cite[\S2.7.3]{numrc}, see also \S\ref{gcsec}. 

For the (discretized) PDEs it may be useful to replace $\|u\|_q$ by  
$\|u\|_{L^q(\Om)}$ (discretized), and we choose the 
default $q=2$ and $\al_1=1$. Moreover, we 
found it more robust to modify $\CD$ to 
\huga{\label{adefl2b}
\CD(u;u^{(1)},\ldots,u^{(l)})=1/\left(\al_1+\Pi_{j=1}^l \min(\|u-u^{(j)}\|_q^q,\al_2)\right),
}
with default $\al_2=1$. The $\min$ is used 
because for large $l$ the term $\Pi_{j=1}^l \|u-u^{(j)}\|_q^q$ may 
otherwise become large even if $u$ is close to some $u^{(j)}$ (but far 
from others). This $\CD$ is now longer differentiable on the null sets 
$\|u-u^{(j)}\|_q=\al_2$, but in practice this is not a problem, 
and an advantage is that this way $\CD$ is flat away from the known roots, 
$\CD=1/\al_2$, and hence $G\pa_u\CD=0$ in \reff{adefl1}. 
If $\pa_u\CD\ne 0$, then often lumped 
approximate Jacobians also work as an alternative to SMW formulas 
in the Newton loops for \reff{adefl1}. 

\brem\label{deflrem1} a) The choices of $(q,\al_1,\al_2)=(2,1,1)$ and of the 
norm $\|u\|_2$ vs $\|u\|_{L^2(\Om)}$ are essentially heuristic, and may 
have to be adapted for any given problem, or even depending on the 
initial guess for the deflated Newton loop and the solutions already computed. 
See \cite[\S4.3,\S4.4]{FBF15} for a detailed discussion of the 
hard--to--predict influence of $q$ and $\al_1$ on how many 
solutions will be found. On the other hand, \cite{FBF15} 
reports that for most choices of $(q,\al)$ at least some 
solutions are found, and the same holds for the example below and further 
examples in \pdep. 

b) In \cite{CKF18} the values $(q,\al_2)=(2,1)$, the $\|\cdot\|_{L^2(\Om)}$ 
norm, and $\al_3=\infty$, are used, but with 
a modified $\CD$ that for a given solution deflates the full group 
orbit under rotation. 
\eex\erem 

\begin{figure}[ht]
\bce 
\begin{tabular}{lll}
{\small (a) Four solutions from deflation}&{\small (b) BD}&{\sm sample solutions from branches 2 and 3 }\\
\hs{-2mm}\raisebox{29mm}{\begin{tabular}{l}
\ig[width=0.2\tew]{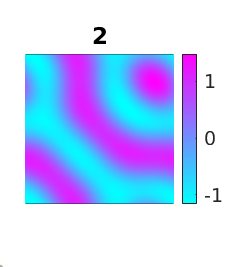}
\ig[width=0.2\tew]{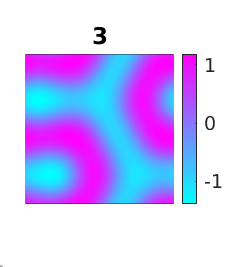}\\[-10mm]
\ig[width=0.2\tew]{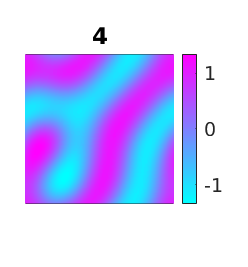}
\ig[width=0.2\tew]{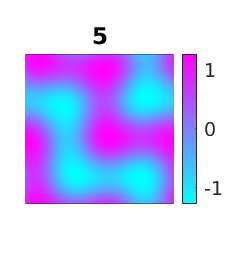}
\end{tabular}}
&\hs{-7mm}\ig[width=0.2\tew]{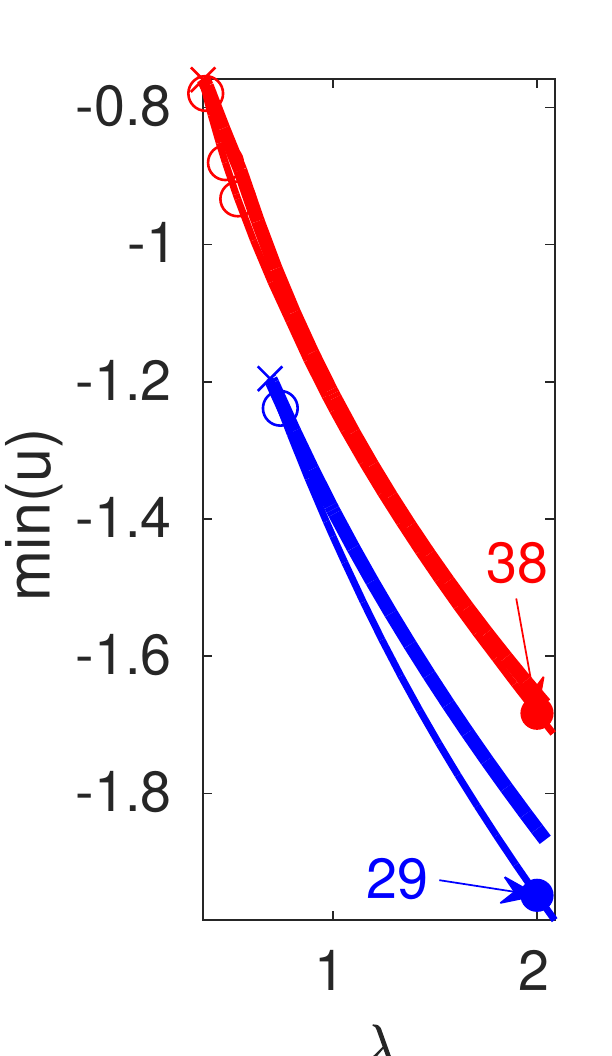}&
\hs{-2mm}\raisebox{29mm}{\begin{tabular}{l}
\ig[width=0.2\tew]{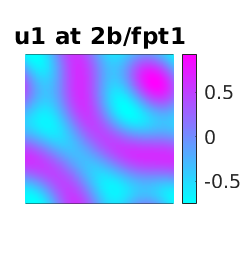}\ig[width=0.2\tew]{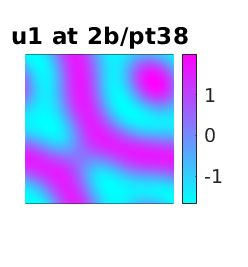}\\[-10mm]
\ig[width=0.2\tew]{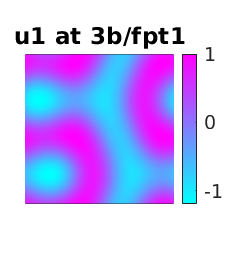}\ig[width=0.2\tew]{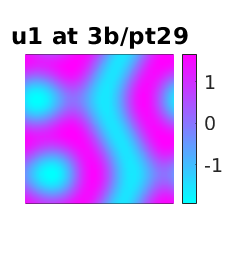}
\end{tabular}}
\end{tabular}
\ece
\vs{-2mm}
   \caption{{\small Deflation for the SH equation \reff{swiho} at $\lam=1$ 
over $\Om=(-2\pi,2\pi)^2$ 
with pBCs in $x$ and $y$, $\nu=0$. Using $u\equiv 0$ as primary known solution, 
ten searches from initial guesses with random ($|k|\approx 1$) Fourier 
coefficients, yield the four solutions from (a). (b,c) Continuation of 
2 and 3 and sample solutions. 
  \label{shpbcf1b}}}
\end{figure}

\paragraph{Deflation for the SH equation on $\Om=(-2\pi,2\pi)^2$ with pBC.} 
 In {\tt shpbc/cmdssq\_defl.m}, as an example we 
use deflation to compute 
solution branches not connected to the trivial branch $u\equiv 0$ for 
the SH equation on $\Om=(-2\pi,2\pi)^2$ with pBC, with sample results given 
in Fig.~\ref{shpbcf1b}. Here we simply init with $u=0$ 
at $\lam=1$, and subsequently run a deflation with standard 
parameters $q=2$, $(\al_1,\al_2,\al_3)=(1,1,1)$ and 
initial guesses obtained from {\em random} Fourier coefficients of 
wave vectors near the unit circle such as $k=(1,0.25), k=(0.25,1)$ and so on, 
combined with random spatial shifts. If 
this yields convergence, then the found solution is added to the 
list {\tt p.defl.u} of known solutions, and the next deflation 
is tried with the same initial guess. After no more solutions are 
found from this initial guess, the next initial guess is tried. 
For 10 initial guesses, this give about 4 solutions on average, 
which, due to the rather unstructured character of the initial guesses 
are typically of the shape in Fig.~\ref{shpbcf1b}(a), rather than 
regular stripes or spots. After the deflation, we continue the 
solutions 2 and 3 in $\lam$, yielding Fig.~\ref{shpbcf1b}(b,c). 
 Continuing in negative direction, the branches 
show folds, and altogether they are not (directly) connected to 
any of the primary bifurcating branches from Fig.~\ref{shpbcf1}. 
Incidentally, at the starting point $\lam=1$, both solutions are stable, 
but generally mostly unstable solutions were found in the deflation. 

\subsubsection{Patterns from (educated) guesses and time--integration, isolas}\label{tnt1} 

In case one is primarily interested in a particular pattern $u^*$, 
which one knows to exist, there is the option to 
use a rough initial guess of $\uti$ for $u^*$ and aim to converge to $u^*$ by 
a Newton loop. If one additionally knows (or expects) the pattern to be stable, 
then it might be helpful or even necessary to first improve the initial guess 
by running some time integration, aka direct numerical simulation (DNS). After the system has then come sufficiently closed to a (the) desired stationary solution $u^*$, again a Newton loop can be tried to compute $u^*$. 

\begin{figure}[h]
\bce 
\begin{tabular}{lll}
{\small (a)}&{\small (b)}&{\small (c)}\\
\raisebox{20mm}{\begin{tabular}{l}
\ig[width=0.3\textwidth]{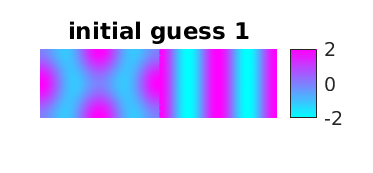}\\[-6mm]
\ig[width=0.3\textwidth]{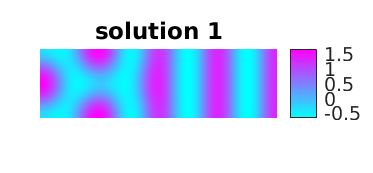}
\end{tabular}}&
\raisebox{20mm}{\begin{tabular}{l}
\ig[width=0.3\textwidth]{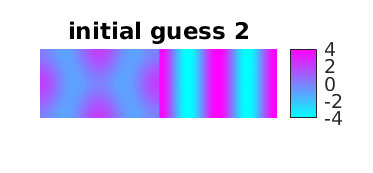}\\[-6mm]
\ig[width=0.3\textwidth]{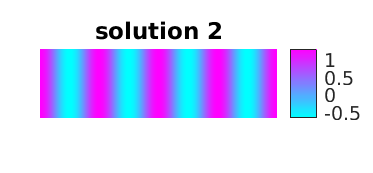}
\end{tabular}}&
\ig[width=0.2\textwidth]{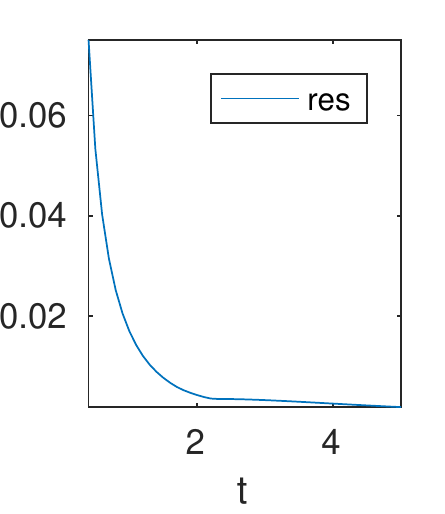}
\end{tabular}
\ece 
\vs{-6mm}
\caption{{\small Obtaining solutions from guesses, if necessary including time-integration. (a) A 'reasonable' initial guess for a hex-to-stripes front, yielding the desired solution directly from a Newton loop, $\nu=1.3$, $\lam=0.2$. (b) A 'bad' initial guess for a hex-to-stripes front; here 
the Newton loop gives the stripe solution. However, if we run {\tt tintxs} 
on initial guess 2, then at $t=5$ the solution is sufficiently close to 
the front for a Newton loop to converge to this desired solution. \label{shtf}}}
\end{figure}

\paragraph{2D.} A simple example is given in {\tt sh/cmds2dtint.m}, with 
some results plotted in Fig.~\ref{shtf}. Here we use same domain as in 
Fig.~\ref{shf11b} and aim to directly obtain a 'hex2stripes' front as 
in Fig.~\ref{shf11b}(f). For this we use initial guesses of the form 
\huga{\label{h2sig} u_1(x,y)=\left\{\barr{ll} 
\cos(x)+\cos(x/2)\cos(\sqrt{3}y/2)&x\le 0\\
\al\cos(x)&x\ge 0\earr\right.,\quad u_2(x,y)\equiv 0, 
}
$\al{=}2$. Then, even though we only seed $u_1$, and only with a rough guess,  
a Newton loop on this $\uti{=}(u_1,u_2)$ takes us directly to our 'desired 
pattern' {\tt solution 1} in Fig.~\ref{shf11b}(a). On the other hand, 
if we take a guess $\uti$ too far off, then a direct Newton loop may not converge, or 
may converge to an 'undesired pattern'. For instance \reff{h2sig} with $\al{=}4$ 
in Fig.~\ref{shf11b}(b), leads to the stripe pattern \mbox{{\tt solution 2}.} 

Often, it helps to use the guess $\uti$ as an initial condition and 
run some time--steppers. Time integration is not a core feature of 
\pdep, but we do provide a number of simple semi-implicit time steppers, 
described in \cite{p2p2man}, which essentially only need the struct {\tt p} 
as main input. The main time-steppers are 
{\tt tint, tintx} (general purpose, where the {\tt x} stands 
for more comprehensive output such as time-series of the residuals res$(t)=\|G(u(t)\|_\infty$), 
and {\tt tints, tintxs} (for semilinear systems, with pre-factoring 
of the stiffness matrix). If we use this on {\tt guess 2} until $t=5$, 
then we obtain the residuals shown in Fig.~\ref{shf11b}(c), and a 
subsequent Newton loop takes us to the desired {\tt solution 1}.

\hulst{caption={{\small {\tt \dname/cmds2dtint.m}. Obtaining 
solutions from initial guesses, possibly combined with some time-integration 
(lines 17-21).  
}}, 
label=sl10, language=matlab,stepnumber=5}{\dhome/cmds2dtint.m}

\begin{figure}[h]
\bce 
\begin{tabular}{ll}
{\small (a) BD and zooms}&{\small (b) BCC, tube, guess for b2z front, 
and 1st solution}\\
\ig[width=0.25\textwidth,height=60mm]{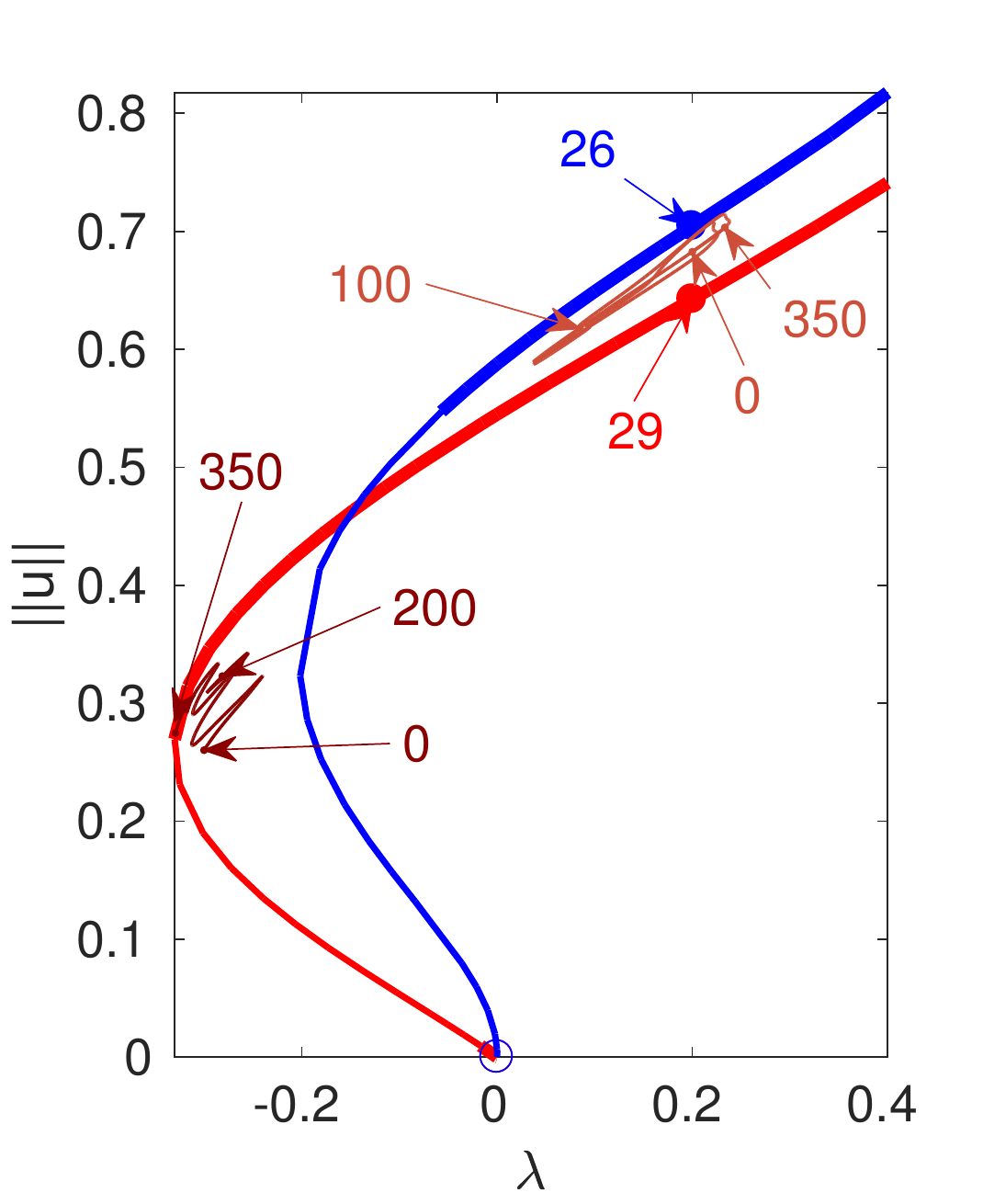}
\hs{-4mm}
\raisebox{25mm}{\begin{tabular}{l}
\ig[width=0.23\textwidth]{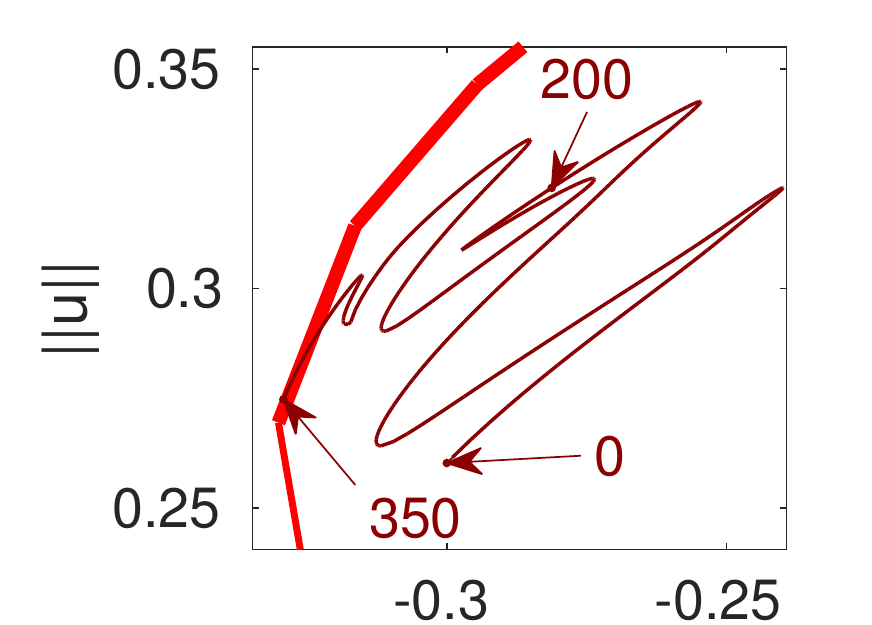}\\
\ig[width=0.23\textwidth]{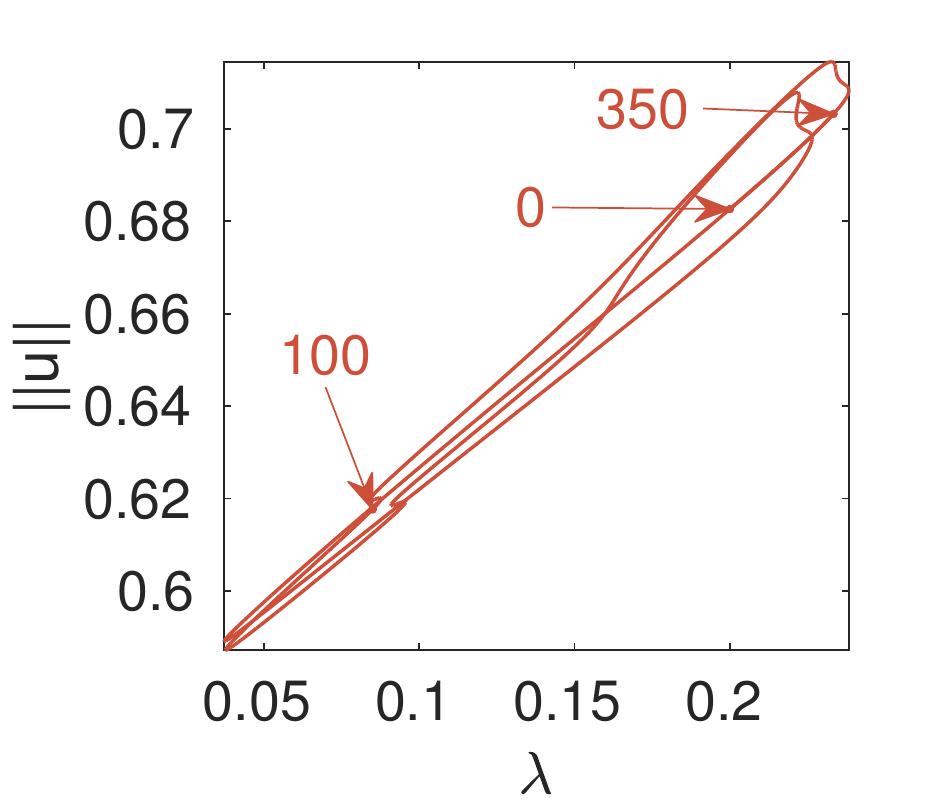}
\end{tabular}}
&\raisebox{25mm}{\begin{tabular}{l}
\ig[width=0.12\textwidth]{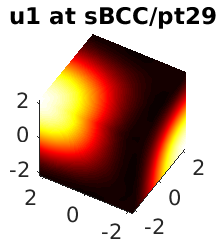}\\[-2mm]
\ig[width=0.12\textwidth]{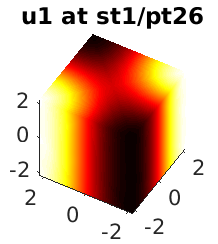}
\end{tabular}}
\ig[width=0.15\textwidth]{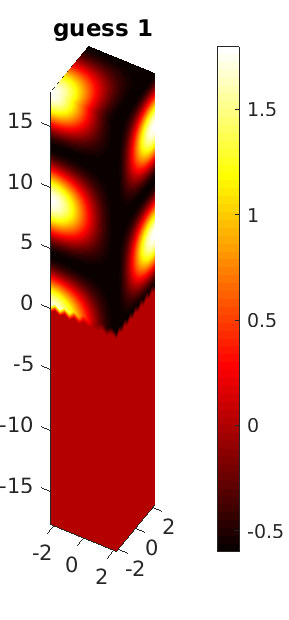}\ig[width=0.15\textwidth]{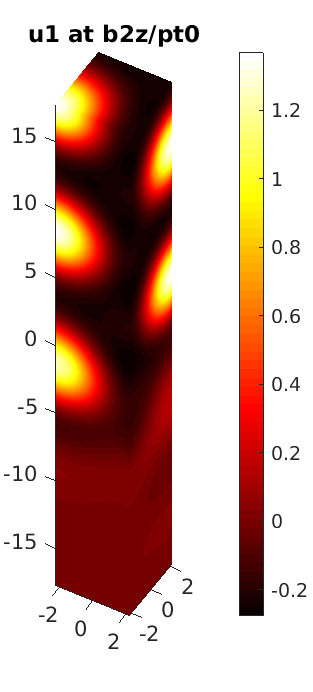}
\end{tabular}\\
\begin{tabular}{ll}
{\small (c) two more solutions on b2z branch}&
{\small (d) guess for bcc-to-tubes (b2t) solution, and three sol.~on b2t branch.}\\
\ig[width=0.16\textwidth]{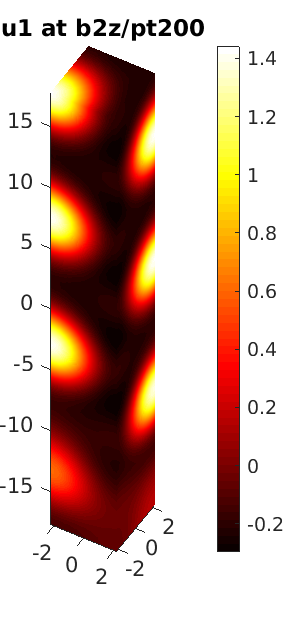}
\ig[width=0.16\textwidth]{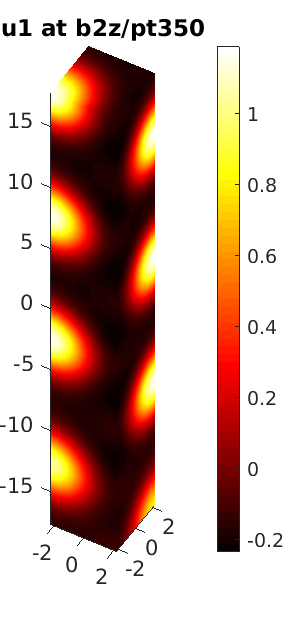}&
\ig[width=0.15\textwidth]{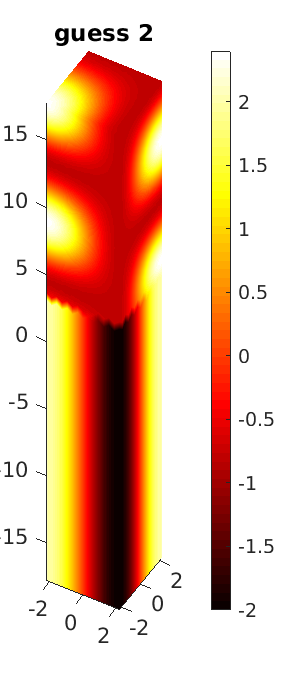}
\ig[width=0.15\textwidth]{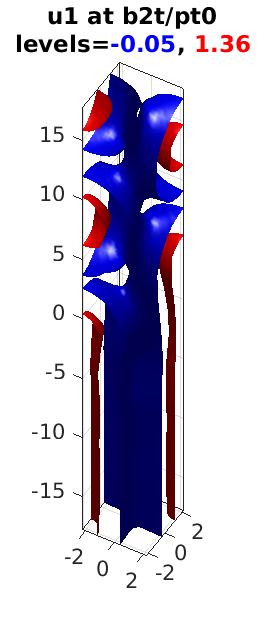}
\ig[width=0.15\textwidth]{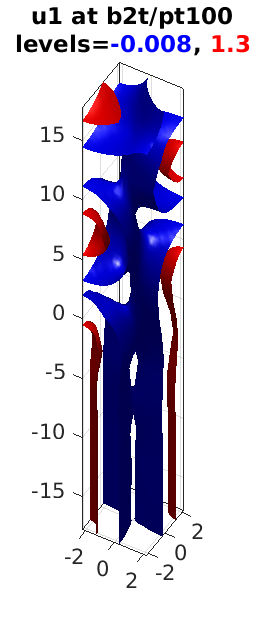}
\ig[width=0.15\textwidth]{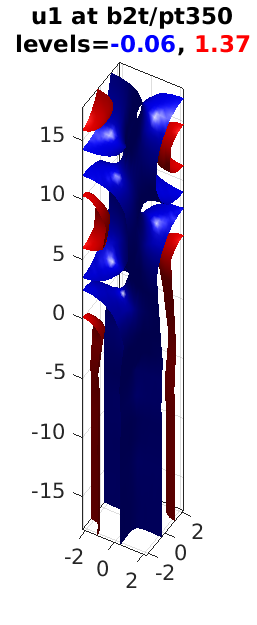}
\end{tabular}
\ece 
\vs{-6mm}
\caption{{\small Results from {\tt  \dname/cmdsBCClong.m} for \reff{swiho} on $\Om=(-l_x,l_x)^2\times(-l_z,l_z)$, $l_x=\pi/\sqrt{2}$, $l_z=8l_x$, $\nu=1.5$. 
(a) BD of BCCs (red), tubes (blue), and {\tt b2z} front  branch (dark brown) 
and {\tt b2t} isola (light brown). (b) BCC and tube plotted over small domain; 
guess for {\tt b2z}  front, and solution obtained from Newton loop. 
(c) two more solutions on the {\tt b2z} branch. 
(d) {\tt b2t} guess, and solution plots. $n_p=17375$ grid points and 
$n_t=101577$ tetrahedral elements. Computation of {\tt b2z} and {\tt b2t}  branches takes about 20Min on an i7 laptop computer.\label{shtf2}}}
\end{figure}

\paragraph{3D.} 
In Fig.~\ref{shtf2} we proceed similarly in 3D. Here,  as in Fig.~\ref{shf1} (1D) and in Fig.~\ref{shf11} (2D), for 
'sufficiently long' 3D cuboids we expect snaking branches of localized BCCs in the bistable range of BCCs and the trivial solution. In {\tt cmdsBCClong.m} 
we first compute a BCC branch (red) and a tubes branch (blue) on the minimal 
domain $\Om=(-l_x,l_x)^3$, $l_x=\pi/\sqrt{2}$, see the first two plots in 
 Fig.~\ref{shtf2}(b).%
\footnote{For speed and convenience, the red and blue BCC and tubes branches in the left panel of (a) are from the minimal domain  $\Om=(-l_x,l_x)^3$, including the stability, 
but we can obtain the same branches on $\Om=(-l_x,l_x)^2\times 
(-8l_x,8l_x)$, with the same stability for the BCCs and almost the same 
stability for the tubes.}
Then we let $\Om=(-l_x,l_x)^2\times (-8l_x,8l_x)$ and $\lam=-0.3$, and try 
the guess 
\huga{\label{b2zig} u_1(x,y,z)=\left\{\barr{ll} 
0.4\re\biggl[\sum_{j=1}^6 \exp(\ri k_j\cdot (x,y,z))\biggr], &z\ge 0\\
0&z\ge 0\earr\right.,\quad u_2(x,y,z)\equiv 0, 
}
with $k_j$ from \reff{bcck} to obtain a BCC-to-zero front {\tt b2z}. 
A Newton loop takes us to {\tt b2z/pt0}. Continuing this branch 
we find it connected to the BCC branch near zero (not shown) and near the 
BCC fold (zoom in (a), top right panel). For speed, in this continuation we switch off 
bifurcation detection and spectral computations, and instead here remark 
that this snaking branch consists of alternating stable and unstable 
segments, as expected.

As an additional example for the multitude of patterns on this long domain,  similar to Fig.~\ref{shtf} we seek a 
BCC-to-tubes front branch {\tt b2t} in the bistable range of BCCs and tubes. We let $\lam=0.2$, and as a guess use 
\huga{\label{b2tig} u_1(x,y,z)=\left\{\barr{ll} 
0.4\re\biggl[\sum_{j=1}^6 \exp(\ri k_j\cdot (x,y,z))\biggr],&z\ge \pi\\
\cos((x+y)/\sqrt{2})+\cos((x-y)/\sqrt{2}),&z\le \pi\earr\right.,\quad u_2(x,y,z)\equiv 0, 
}
see the first plot in Fig.~\ref{shtf2}(d). A Newton loop takes us to the solution {\tt b2t/pt0}. 
Continuing this branch we find that it forms an isola (bottom right panel 
in (a)): After going back and forth twice, near the 340th continuation point it returns to {\tt b2t/pt0}. 
For speed we again switch off the stability and bifurcation detection 
computations, and remark that by checking stability a posteriori we find that significant segments of this branch consist 
of stable solutions. 

\brem\label{brrem} In \cite{UW18ab}, similar results can be found 
for the 3D Brusselator RD-system, i.e., snaking branches of fronts between 
BCCs and the trivial (spatially homogeneous) solution ({\tt b2z}--branch), 
and between BCCs and tubes ({\tt b2t}--branch). 
There, the {\tt b2z}--branches were obtained via bifurcation from 
(subcritical) BCC--branches, and a {\tt b2t}--branch via ``educated'' initial guesses. 
\eex\erem 

\begin{figure}[h]
\bce 
\begin{tabular}{ll}
{\small (a)}&{\small (b)}\\
\raisebox{5mm}{\ig[width=0.23\textwidth]{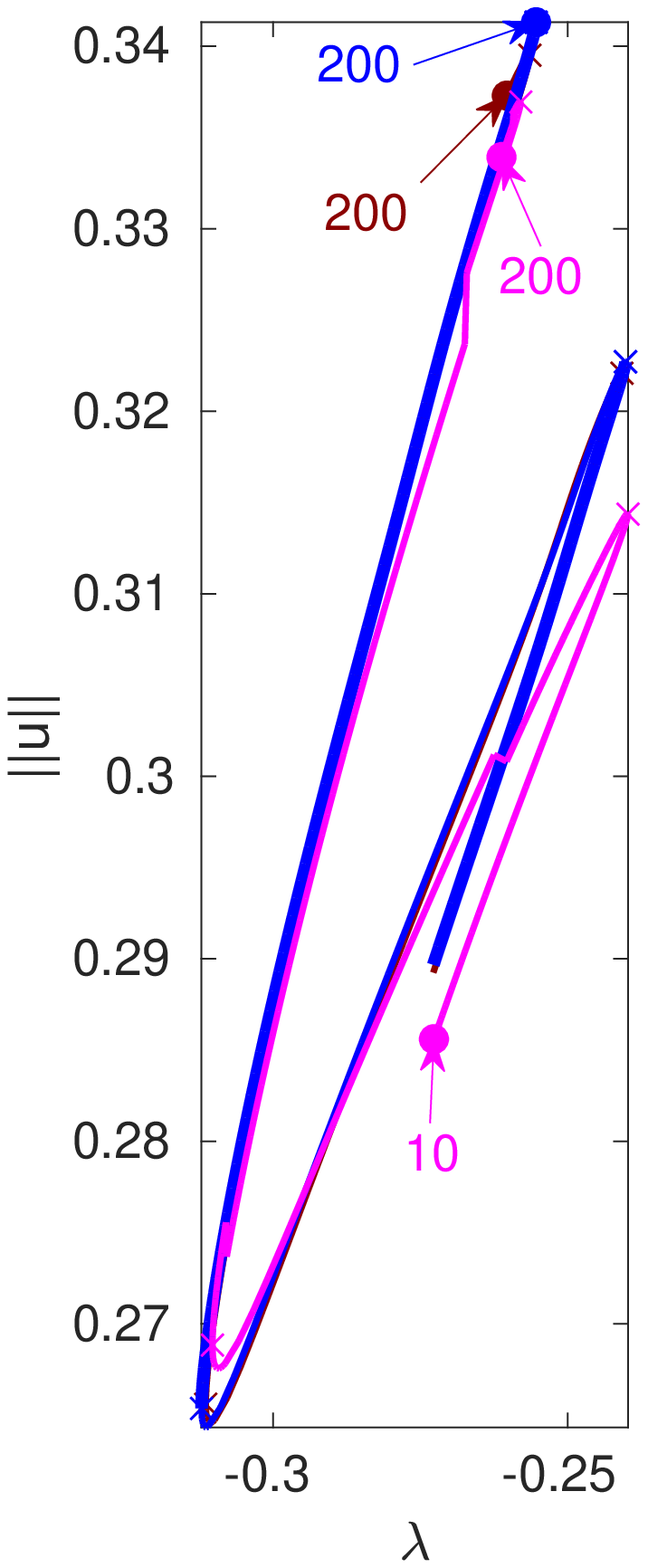}}&
\ig[width=0.19\textwidth]{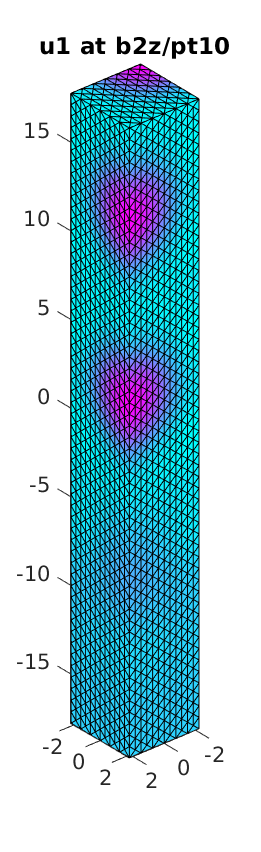}
\ig[width=0.19\textwidth]{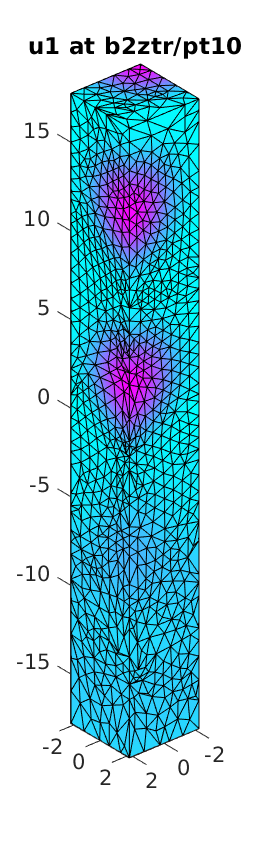}
\ig[width=0.19\textwidth]{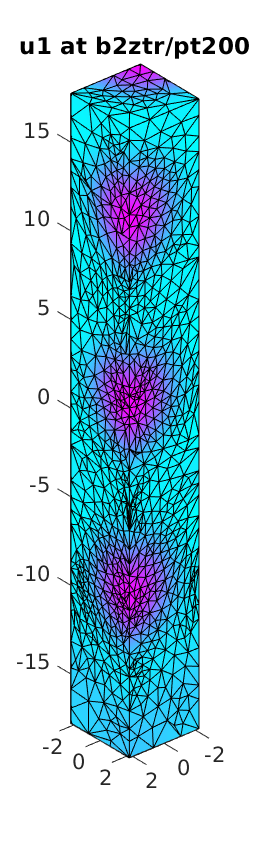}
\end{tabular}
\ece 
\vs{-6mm}
\caption{{\small Results from {\tt  \dname/cmdsBCClongref.m} for \reff{swiho} on $\Om=(-l_x,l_x)^2\times(-l_z,l_z)$, $l_x=\pi/\sqrt{2}$, $l_z=8l_x$, $\nu=1.5$. 
(a) comparison of the same branch on different meshes. Brown (b2z): 
uniform mesh, $n_p=11360$; blue: original uniform mesh, $n_p=17375$, on top 
of brown branch; 
magenta (b2ztr): adaptation by \trulle, first reducing {\tt b2z/pt10} 
to $n_p=5982$, 
then gradually increasing to $n_p=8484$ (at pt200, due to the growth of the 
next spot). The initial adaptation 
reduces the $L^2$ norm form the brown branch, and 
each subsequent adaptation also yields a small 'jump' on the magenta branch. 
(b) sample plots.\label{shtf3}}}
\end{figure}

In Fig.~\ref{shtf3} we compare results from \ma\ by \trulle\  (magenta branch) 
with the original {\tt b2z} branch (brown) and the same branch on a finer 
uniform mesh (blue, on top of the brown branch). The result is 
that using \trulle\ we can compute the branch about 50\% faster than 
on the original mesh, while the blue branch takes about three times as long 
as the brown one. We refer to {\tt cmdsBCClongada.m} 
for details and the 
\trulle\ parameter choices \cite{trulletut}. See also Fig.~\ref{shf11a} and 
Listing \ref{shl4c} for the analogs in 2D. 

\section{Demo {\tt schnakpat}}\label{schnaknum}
\def\dhome{./pftut/schnakpat}
\def\dname{schnakpat}
In the demo {\tt schnakpat} we consider the (modified) Schnakenberg reaction diffusion 
system  
\huga{\label{tur1}
\pa_t U=D\Delta U + F(U),\quad U=\bpm u\\v\epm, 
\quad F(U)= \bpm-u{+}u^2v \\\lambda{-}u^2v\epm
+\sigma \left(u{-}\frac 1 v\right)^2\bpm 1\\-1\epm, 
}
with  diffusion matrix $D=\ssmatrix{1 & 0\\ 0 & d}$ and parameters $\lam>0$ and $\sig\in\R$. In suitable parameter regimes, \reff{tur1} shows Turing 
bifurcations from the homogeneous branch $(u,v)=(\lam,1/\lam)$, and, in a 
nutshell, we may more or less expect all phenomena explained for the 
SH equation in the previous sections also in \reff{tur1}. 
The term involving $\sig$ does not change the 
homogeneous branch or the linearization around 
it, but has been introduced to tune the primary bifurcation from super-- to subcritical. 
The system \reff{tur1} has already been considered as a \pdep\ and pattern formation model problem in \cite[\S4.2]{p2p} (with $\sig=0$), and in \cite{uwsnak14} and \cite{sftut}. Here we want to give a concise and updated demo which besides illustrating the use of {\tt qswibra} and {\tt seltau}  also recovers the main 
results from \cite{uwsnak14}.  Table \ref{schnakdirtab} gives an overview of the involved files.  We fix 
$d=60$ throughout, and focus on five tasks, namely: 
\bci
\item explain a trick to let \pdep\ display the dispersion relation 
for homogeneous states; 
\item compute a  basic bifurcation diagram of 1D patterns, including snaking branches of localized patterns; 
\item generate a  basic bifurcation diagram of 2D patterns 
over small 2D domains; 
\item compute snaking branches of localized 2D patterns over long 2D domains;  
\item compute the primary bifurcations in 3D for the SC and BCC lattices. 
\eci 

\begin{table}[ht]\taskip
\caption{Scripts and functions in {\tt demos/schnakpat}. 
\label{schnakdirtab}}
\bce\vs{-4mm}
{\small 
\begin{tabular}{p{0.22\textwidth}|p{0.72\textwidth}} 
script/function&purpose, remarks\\
\hline
cmds1d&display the dispersion relation, compute a basic 1D bifurcation diagram, and fold continuation, yielding Fig.~\ref{schnakf00}--\ref{schnakf0b}. \\%
cmds2dsq&compute a  basic bifurcation over a square, primary bifurcations are pitchforks. \\
cmds2da&compute a  basic bifurcation over a small rectangle, Fig.~\ref{schnakf1}. \\%
cmds2db&compute a branch of spots embedded in stripes over a long 2D domain, Fig.~\ref{schnakf2}. \\
cmds3dSC/BCC&compute basic bifurcation diagrams over SC and BCC cubes.\\
\hline
schnakinit&initialization, 1D and 2D\\
sG, nodalf, sGjac&as usual\\
spjac&Jacobian for fold continuation\\
spufu&auxiliary function for plotting the dispersion relation\\
schnakbra&modification of {\tt stanbra} (to include the $L^8$ norm on the branch)
\end{tabular}
}
\ece
\end{table}\teskip

\subsection{1D: computing the dispersion relation, basic branches, and snaking}\label{schnak1dsec}
\begin{figure}[ht]
\bce 
\begin{tabular}{l}
\ig[width=0.3\tew,height=30mm]{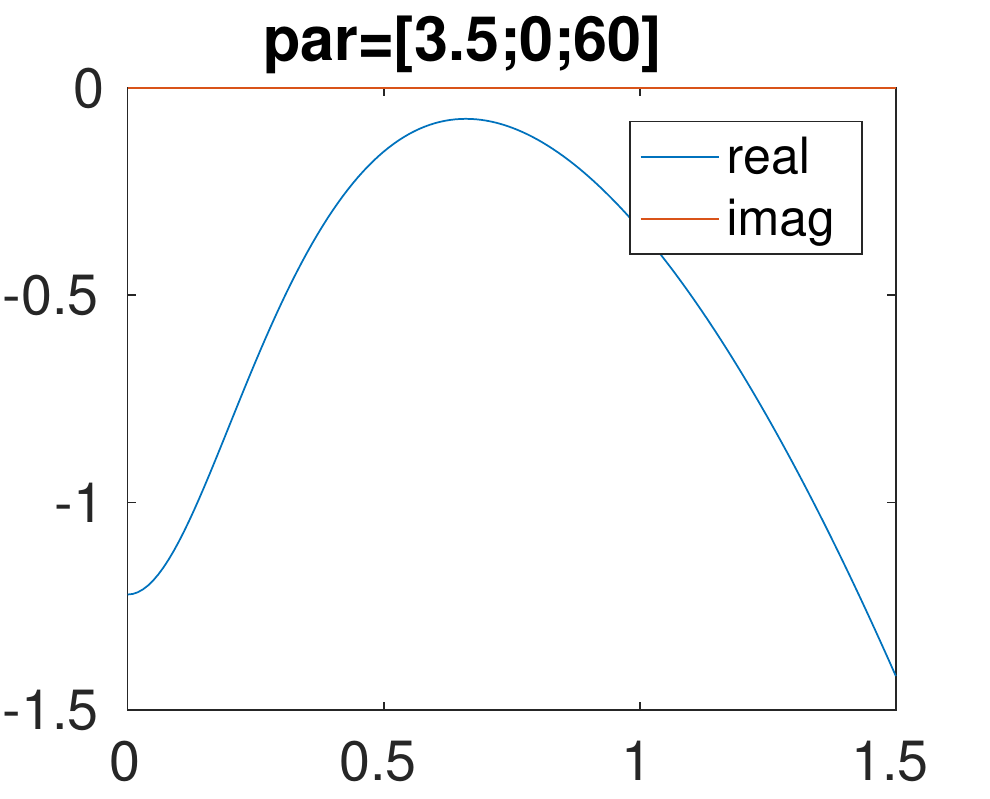}
\ig[width=0.3\tew,height=30mm]{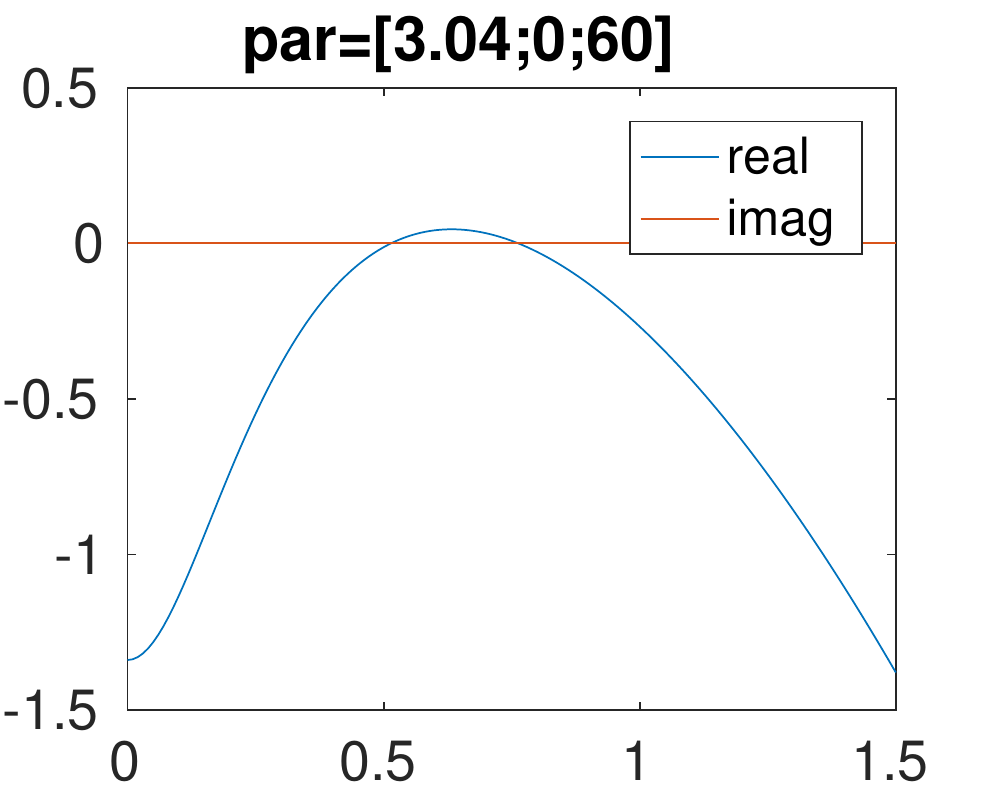}
\ig[width=0.3\tew,height=30mm]{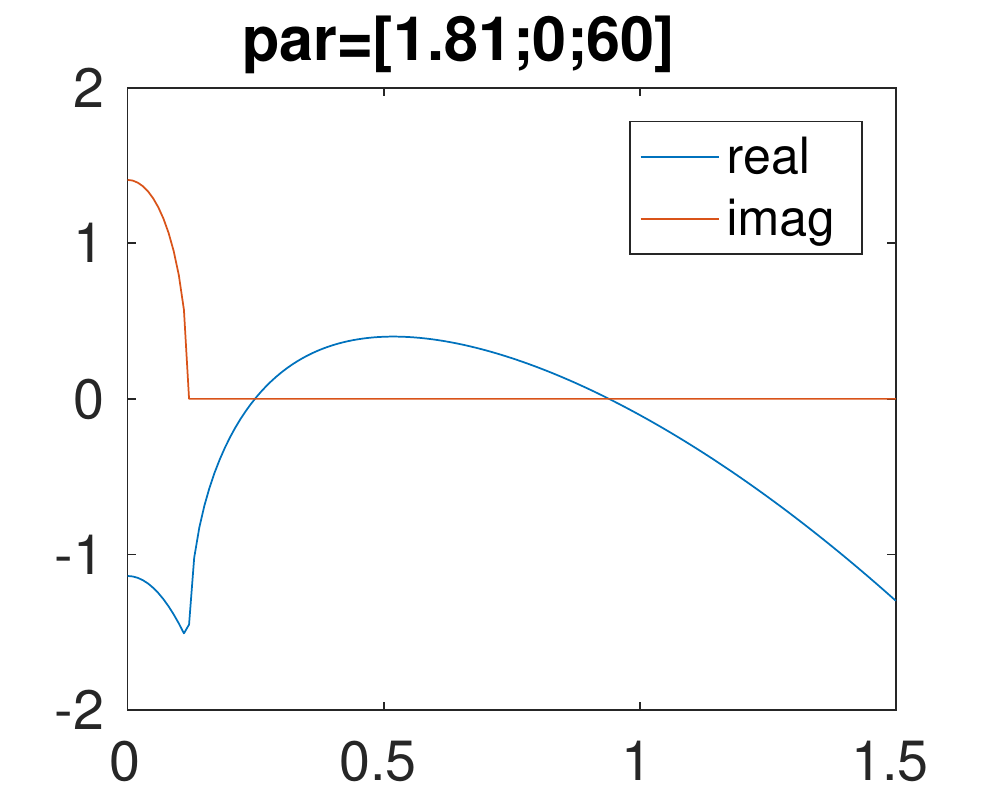}
\end{tabular}
\ece 

\vs{-5mm}
\caption{{\small Preparatory step (not strictly necessary) for \reff{tur1}: plotting the dispersion for different $\lam$. \label{schnakf00}}}
\end{figure}

For \reff{tur1} (with $d=60$ fixed) we know the parameter value for first 
Turing bifurcation from the homogeneous branch $(u,v)=(\lam,1/\lam)$ 
and the critical wave number, namely $\lam_c=\sqrt{60}\sqrt{3-\sqrt{8}}\approx 
3.21$ and $k_c=\sqrt{\sqrt{2}-1}$. 
Nevertheless, in {\tt cmds1d} we start the 1D computations on a small domain to 
illustrate the usage of {\tt spufu} (see Listing \ref{sl10b}) to plot the dispersion relation, see Fig.~\ref{schnakf00}. 
{\tt spufu.m} is a modification of the \pdep\ library function {\tt stanufu}, 
 and should be easily adaptable to any RD system. See also demo 
{\tt hopfdemos/extbru} explained in \cite{hotut} for a 3 component case.

\hulst{caption={{\small (Selection from) {\tt \dname/spufu.m}. Modification of (addition to) {\tt stanufu} to plot the dispersion relation.  In 
line 22 we extract $(u,v)$ at just one point, shorten the vector of unknowns accordingly, and compute the local Jacobian $\pa_{\vec{u}} f$ of the 'nonlinearity' $f$ given in {\tt nodalf}, where we use that this 
is already encoded in {\tt njac} (and called accordingly in sGjac). 
To compute $\mu(k)$ we then loop over $k$ and numerically solve the pertinent 
$2\times 2$ eigenvalue problem. This can be modified to 
other two--component or general $N$--component systems in a straightforward way, 
where essentially $N$, the pertinent wave-number range {\tt kv}, and 
the diffusion constant(s) are the problem dependent points in {\tt spufu}. 
}}, 
label=sl10b, language=matlab,stepnumber=5, linerange=21-31, 
firstnumber=21}{\dhome/spufu.m}

\hulst{
label=schnl4, language=matlab,stepnumber=5, linerange=2-12}{\dhome/cmds1d.m}
\hulst{caption={{\small (Selection from) {\tt \dname/cmds1d.m}. In C1 we use 
{\tt spufu} (on a small domain) to display the dispersion relation. 
In C2 we then start the computations on a large domain, $l_x=5\pi/k_c$, 
which means that the primary Turing branch T1 ($k=k_1:=k_c\approx 0.64$) 
has 10 periods in $\Om$. 
Then we follow the Turing branches T2 ($k=k_2\approx 0.61$), T3 ($k=k_3\approx 0.7$) and T6 
($k=k_6\approx 0.58$) since in particular 
the branch T6 with only $7.5$ periods in $\Om$ moves furthest to the right. 
Additionally, we follow a front bifurcating on T1. 
 The remainder of the script deals with plotting, and with fold 
continuation.}}, 
label=schnl4b, language=matlab,stepnumber=0, linerange=16-16}{\dhome/cmds1d.m}

Listing \ref{schnl4b} shows the start of {\tt schnakpat/cmds1d.m}. After finding $k_c$ in C1, in C2 we restart the computations on a domain 
tuned to the critical mode $\cos(k_c x)$, i.e, of length 
$10\pi/k_c$, 
see Fig.~\ref{schnakf0}. We follow the Turing branches T1, T2, T3 and T6, associated to the first three and the sixths branch point (counting from the right), and 
a snaking branch S1 bifurcating from T1. The rather large value of $\sig$ has the disadvantage that the periodic patterns 
are somewhat nonphysical because $u$ does not stay positive. However, 
an interesting feature of $\sig=-0.6$  is that the 'most subcritical' 
branch is not the primary Turing branch T1, but (here) T6 with $k=k_6\approx 0.58$. 
Moreover, on S1 the periodic patterns have wave-number $k$ near $k_6$, and in particular the snake reconnects not to T1 but to T6.

\begin{figure}[ht]
\bce 
\begin{tabular}{ll}
\ig[width=0.38\tew]{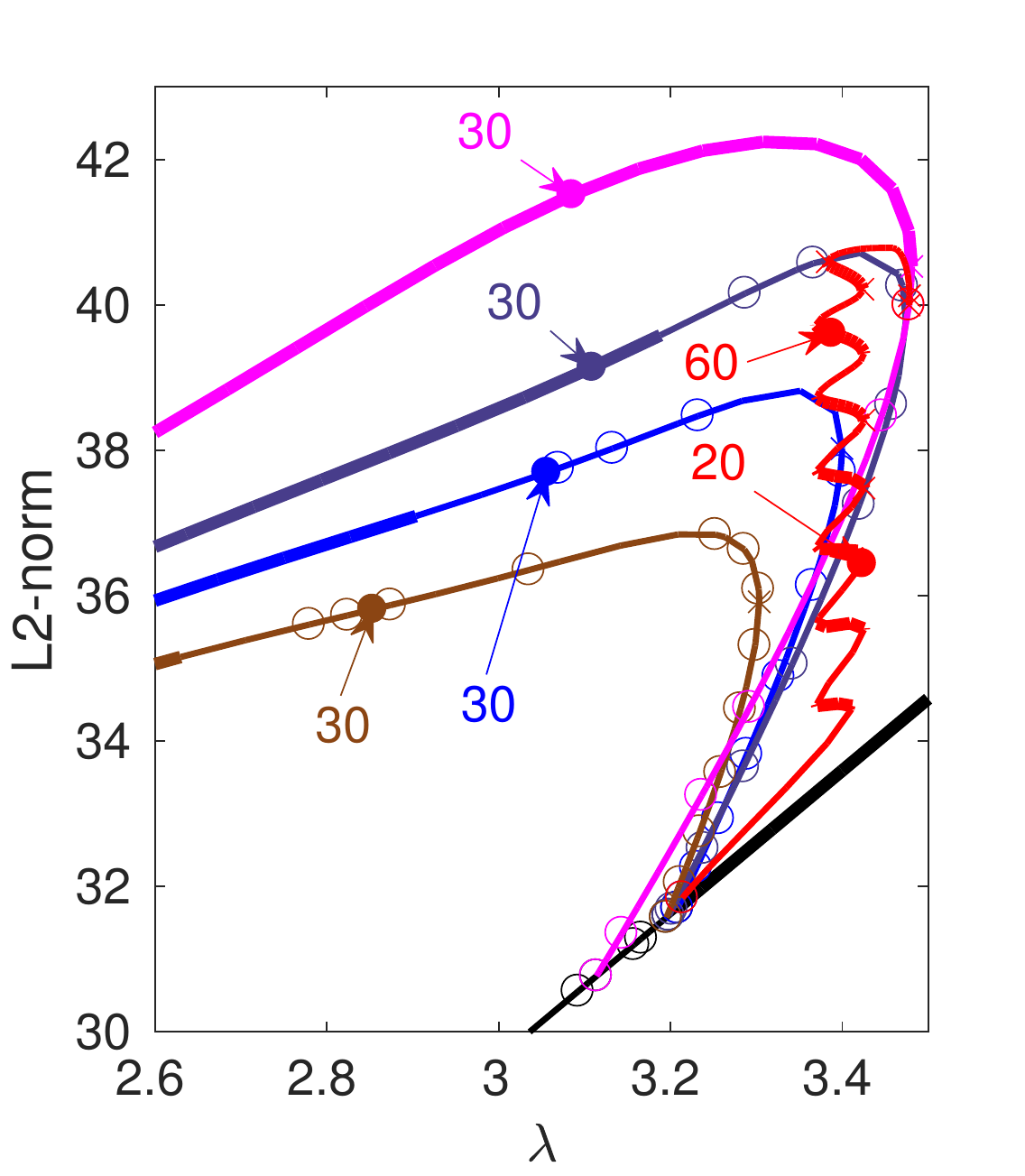}
&\raisebox{35mm}{\begin{tabular}{l}
\ig[width=0.25\tew]{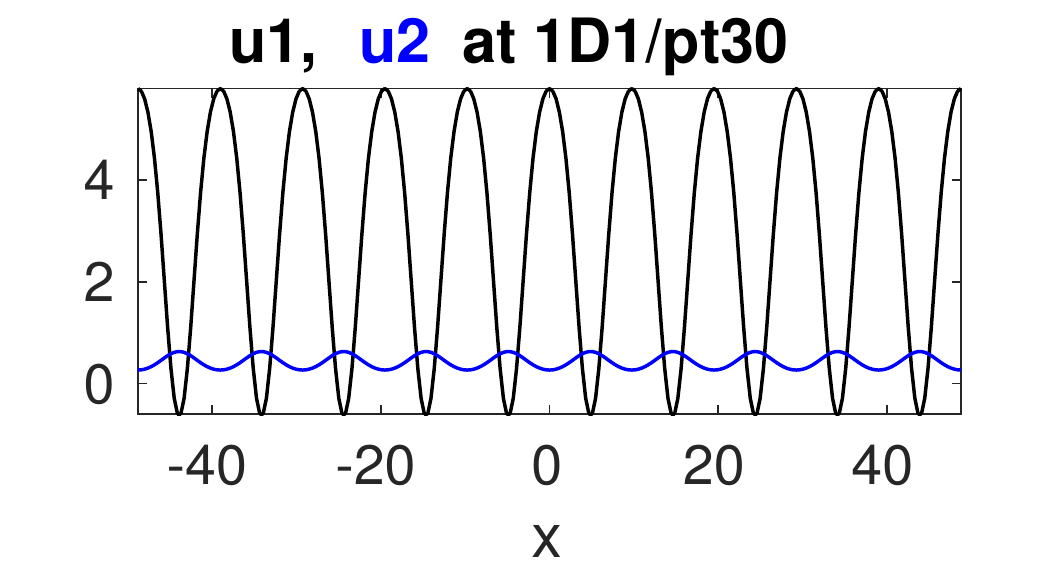}
\ig[width=0.25\tew]{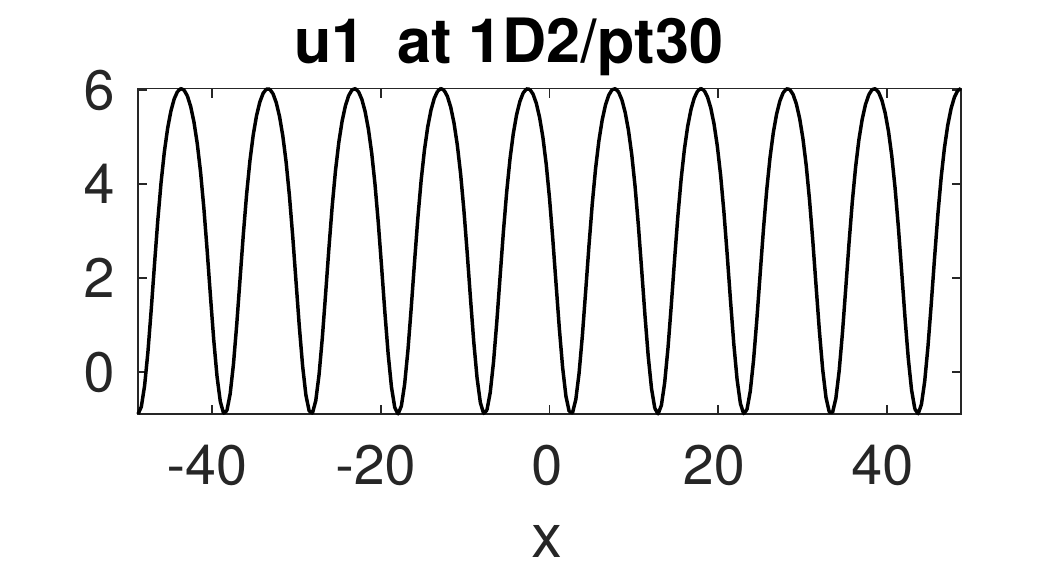}\\
\ig[width=0.25\tew]{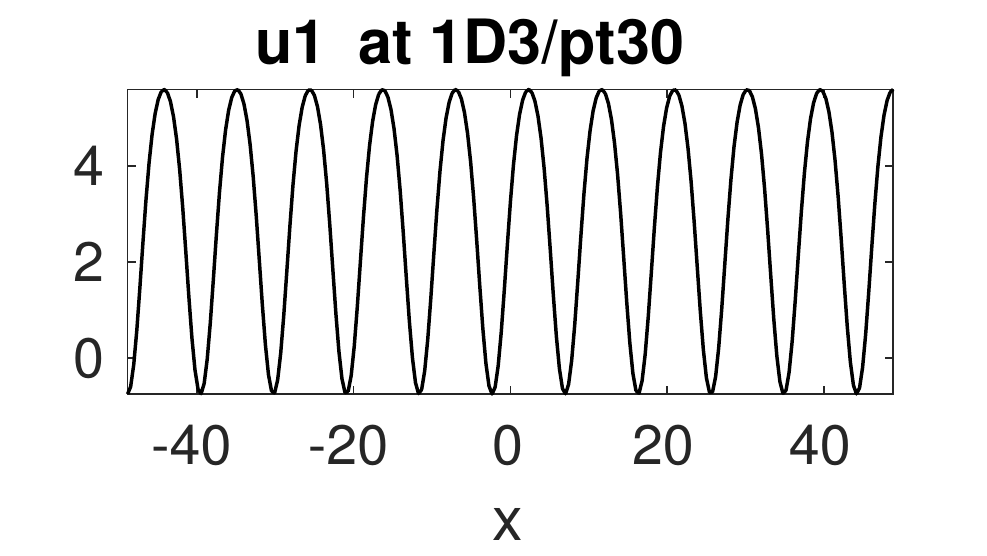}
\ig[width=0.25\tew]{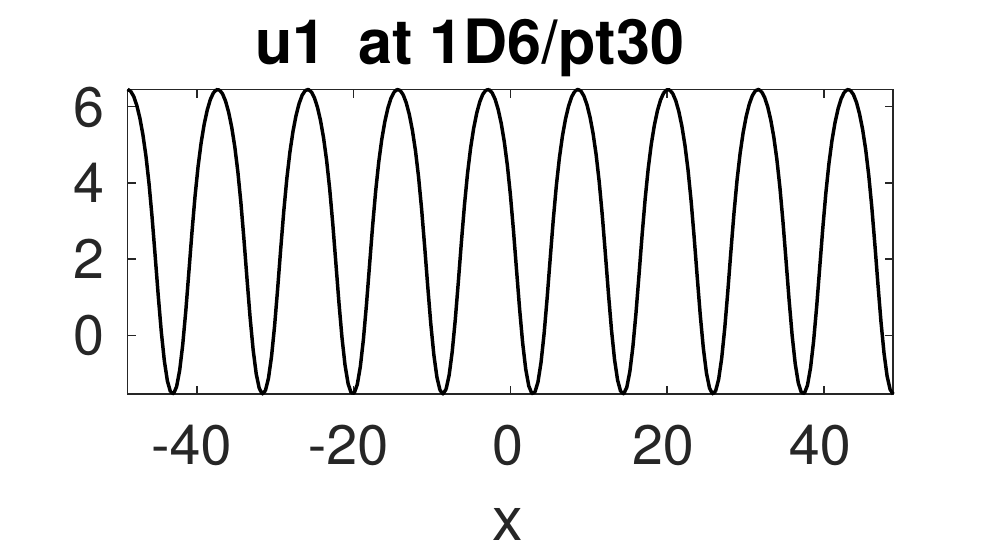}\\
\ig[width=0.25\tew]{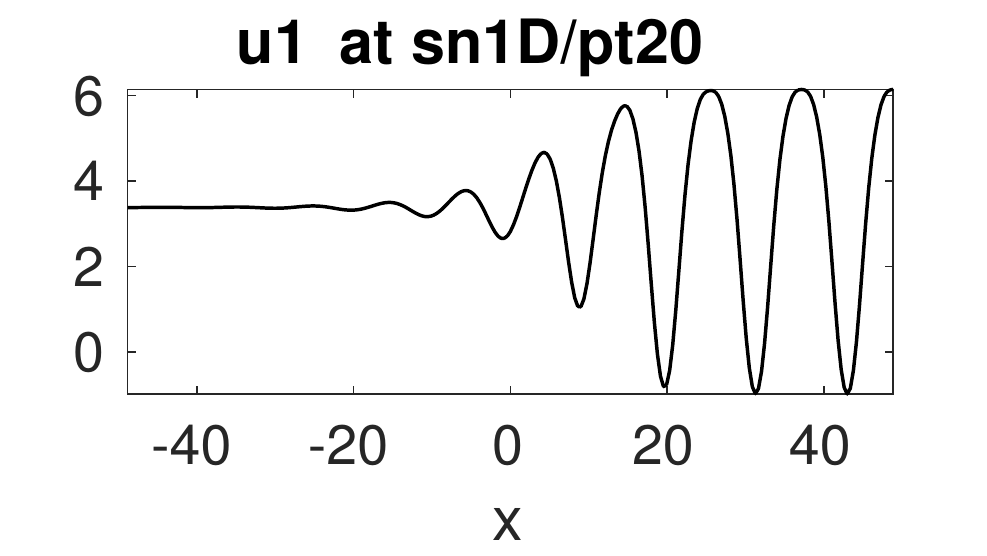}
\ig[width=0.25\tew]{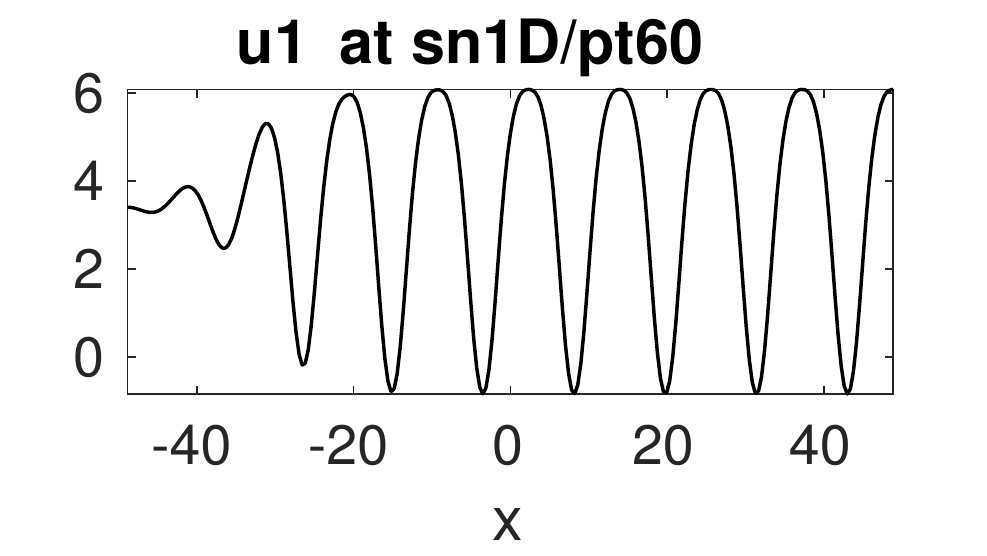}
\end{tabular}}
\end{tabular}
\ece 
\vs{-4mm}
\caption{{\small \reff{tur1}, $(\sig,d)=(-0.6,60)$, $l_x=10\pi/kc\approx 48.8132$. Turing branches T1 (blue), T2 (dark blue), T3 (brown) and T6 (magenta), and a snaking branch of a front bifurcating from T1 but reconnecting to T6. In the solution plot of 1D1/pt30 we use the setting
{\tt plotsol('1D1','pt30',1,[1 2],'cl',{'k','b'});} to plot both components. 
 \label{schnakf0}}}
\end{figure}

In the remainder of 
{\tt cmds1d} we follow the folds on T1, T3 and T6 as functions of $\sig$, see Fig.~\ref{schnakf0b}. 
This illustrates the role played by $\sig$ for 
the structure of the bifurcation diagram: The primary branch T1 bifurcates 
subcritically only for $\sig{<}\sig_0{\approx}-0.3$. Moreover T1 
extends furthest to the right for all $\sig{>}\sig_1{\approx}-0.5$ and 
becomes stable at its fold, respectively is stable directly after 
bifurcation for $\sig>\sig_0$.  

\begin{figure}[ht]
\bce 
\begin{tabular}{ll}
\raisebox{3mm}{\ig[width=0.254\tew]{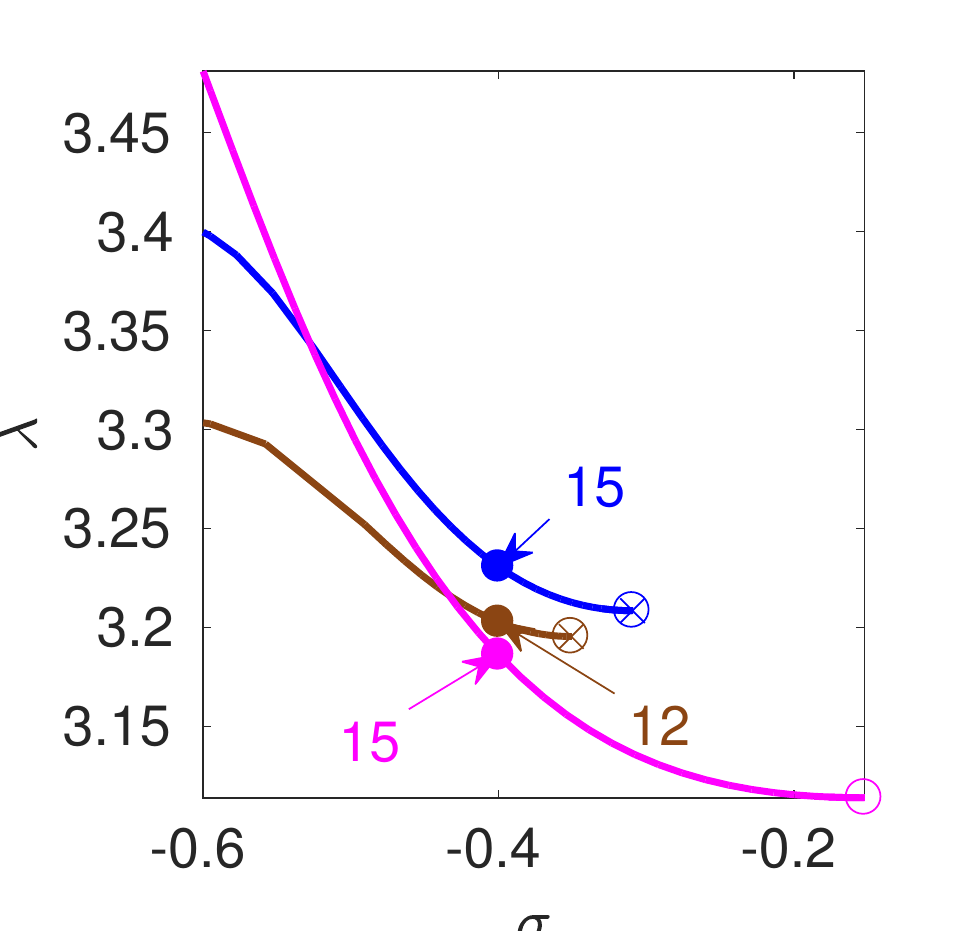}}&
\ig[width=0.24\tew]{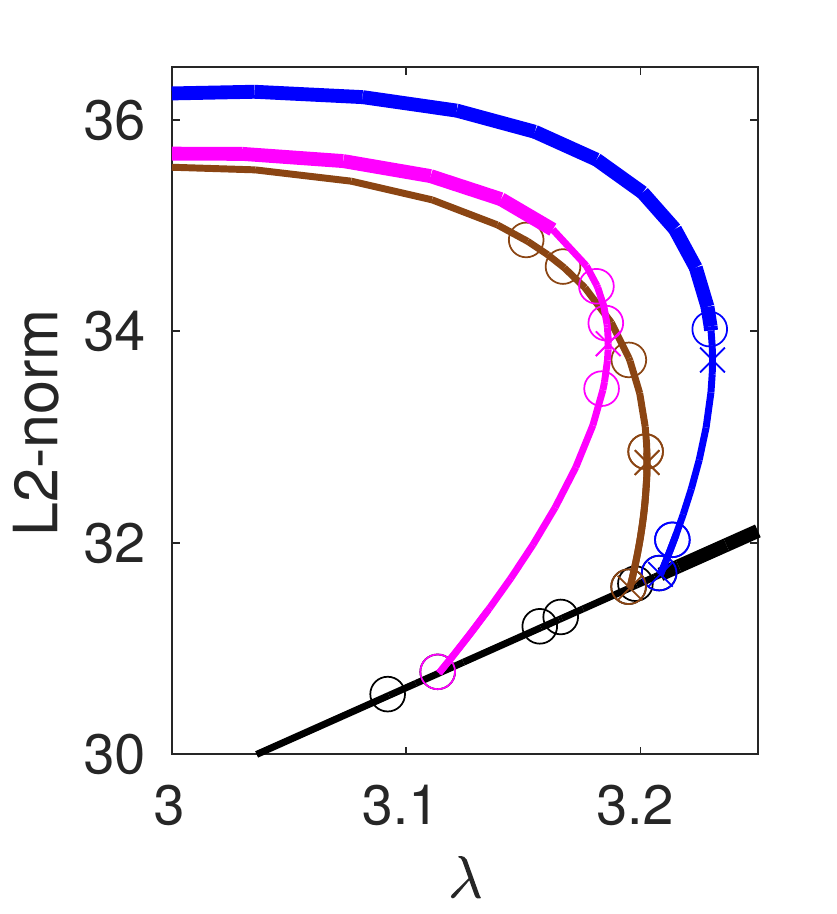}
\end{tabular}
\ece 
\vs{-3mm}
\caption{{\small Continuation of folds on T1,T3 and T6 in $\sig$, and behavior 
of branches at $\sig=-0.4$.  \label{schnakf0b}}}
\end{figure}

\subsection{2D: basic bifurcation diagram, and branches of localized patterns}\label{ssec1}
\begin{figure}[ht]
\bce 
\begin{tabular}{l}
{\small (a) Two numerical kernel vectors at the first BP, 
and a tangent $\tau$ from {\tt qswibra}}\\
\ig[width=0.25\tew]{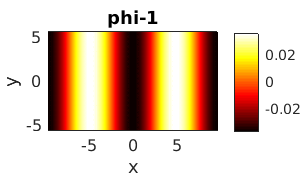}
\ig[width=0.25\tew]{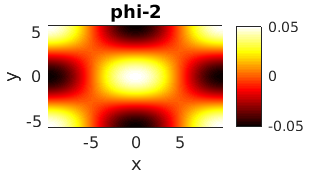}\hs{5mm}
\ig[width=0.25\tew]{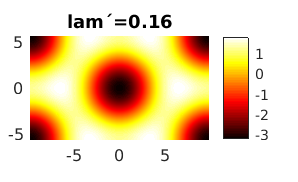}\\
{\small (b) Basic bifurcation diagram, including 'bean' (mixed mode) branches}\\
\ig[width=0.3\tew,height=60mm]{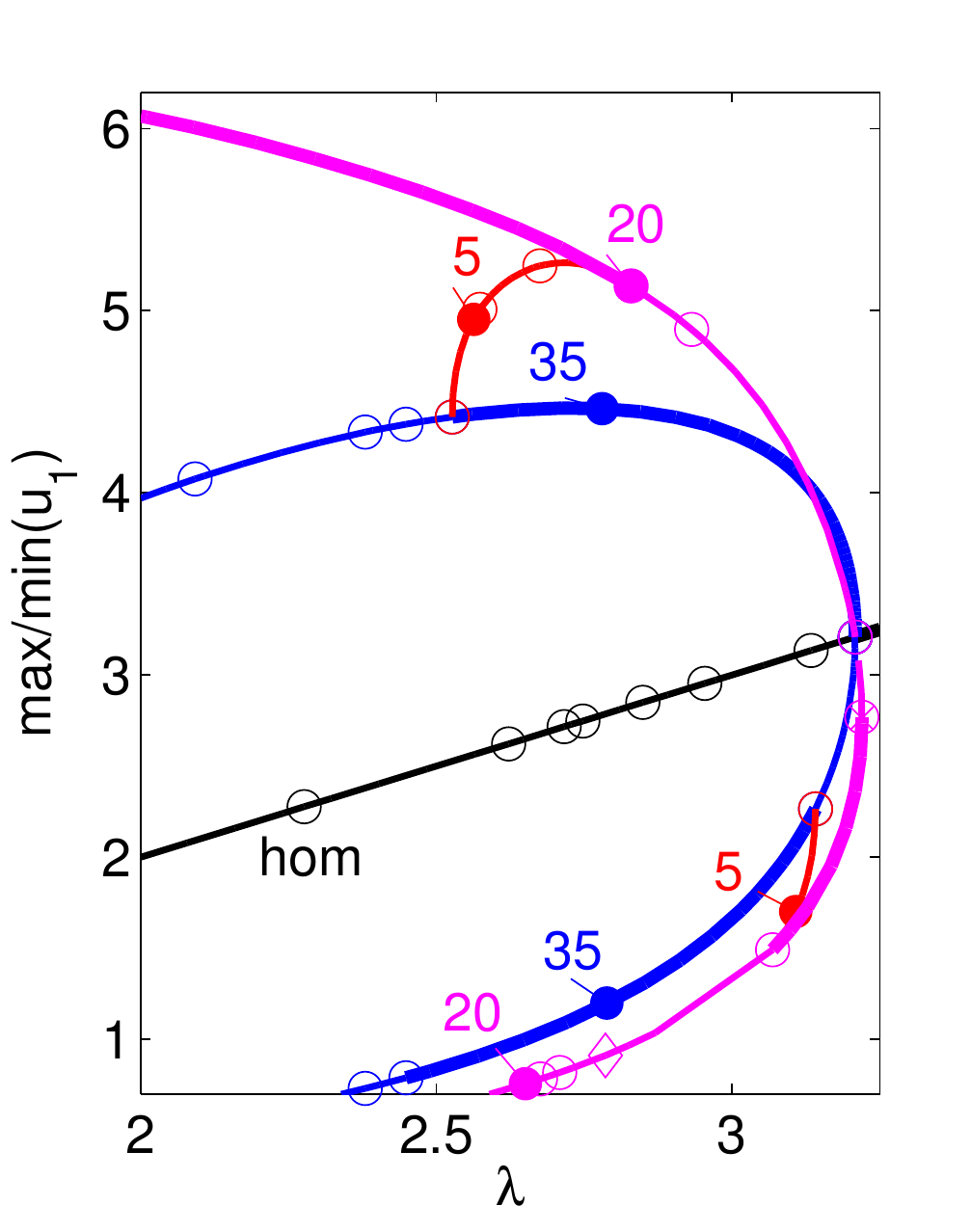}
\raisebox{32mm}{\begin{tabular}{ll}\hs{-4mm}
\ig[width=0.25\tew]{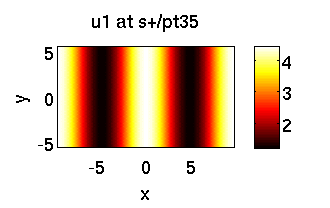}&
\hs{-6mm}\ig[width=0.25\tew]{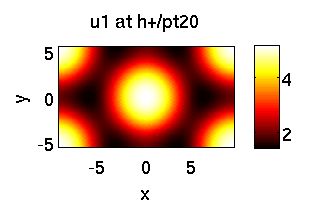}\\
\hs{-4mm}\ig[width=0.25\tew]{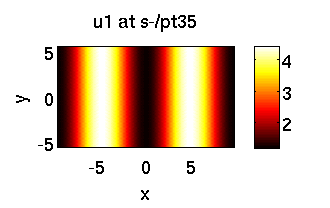}&
\hs{-6mm}\ig[width=0.25\tew]{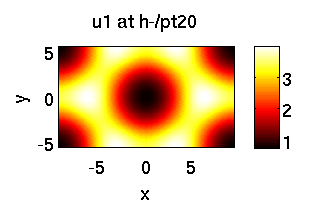}\\
\end{tabular}}\\ 
\ig[width=0.25\tew]{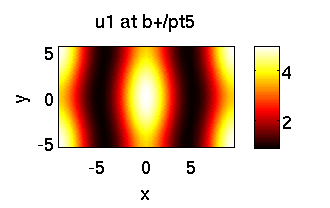}\hs{-4mm}
\raisebox{6mm}{\ig[width=0.17\tew]{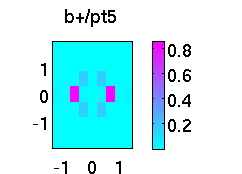}}\hs{-2mm}
\ig[width=0.25\tew]{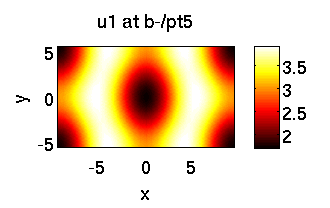}\hs{-2mm}
\raisebox{6mm}{\ig[width=0.17\tew]{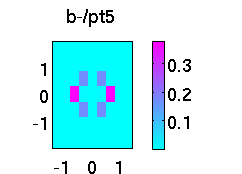}}
\end{tabular}
\ece
\vs{-6mm}
   \caption{{\small Results from {\tt \dname/cmds2da.m} for \reff{tur1} on a small rectangular domain $\Om=(-l_x,l_x)\times (-l_y,l_y)$, $l_x=\pi/k_c$, $l_y=l_x/\sqrt{3}$, 
corresponding to a hexagonal dual lattice. (a) kernel at the first BP, 
 and 'hex' bifurcation direction obtained from qswibra. (b) Bifurcation diagram 
and example solution plots; stripes (blue),   
hexagons (magenta), and mixed modes or beans (red), including Fourier plots.  
  \label{schnakf1}}}
\end{figure}

To compute the 'standard' bifurcation diagram of stripes and hexagons in 
Fig.~\ref{schnakf1} we first let $\sig=0$ and proceed similarly as for the SH stripes and hexagons 
in Fig.~\ref{shf10}: Following the homogeneous branch over a domain 
$\Om=(-l_x,l_x)\times (-l_y,l_y)$ with $l_x=\pi/k_c$ and $l_y=l_x/\sqrt{3}$, 
we find a double branch point at $\lam=\lam_c=\sqrt{60}\sqrt{3-\sqrt{8}}\approx 
3.21$. We then use {\tt qswibra} to switch to the hexagon branch, which  
we follow in ``both directions'' (positive and negative {\tt ds}) to subsequently 
discuss secondary connections between the spots and ``+'' stripes, and 
the gaps and ``-'' stripes.
Since here the kernel vectors $\phi_{1,2}$ are clean stripes, we 
skip a call to {\tt cswibra} and use {\tt gentau} to follow the 
stripe branches. 
The mixed mode connections between stripes and spots are obtained from 
branch--switching where the stripes lose/gain stability. See {\tt cmds2da.m}  and Fig.~\ref{schnakf1}. Additionally we remark that 
on the gap branch there is a Hopf bifurcation point  near $\lam=2.75$. 
We do not discuss this here, but the bifurcating branch of oscillating 
gaps can be obtained from the commands at the end of {\tt \dname/cmds2da.m} 

\begin{figure}[ht]
\bce 
\ig[width=0.28\tew,height=60mm]{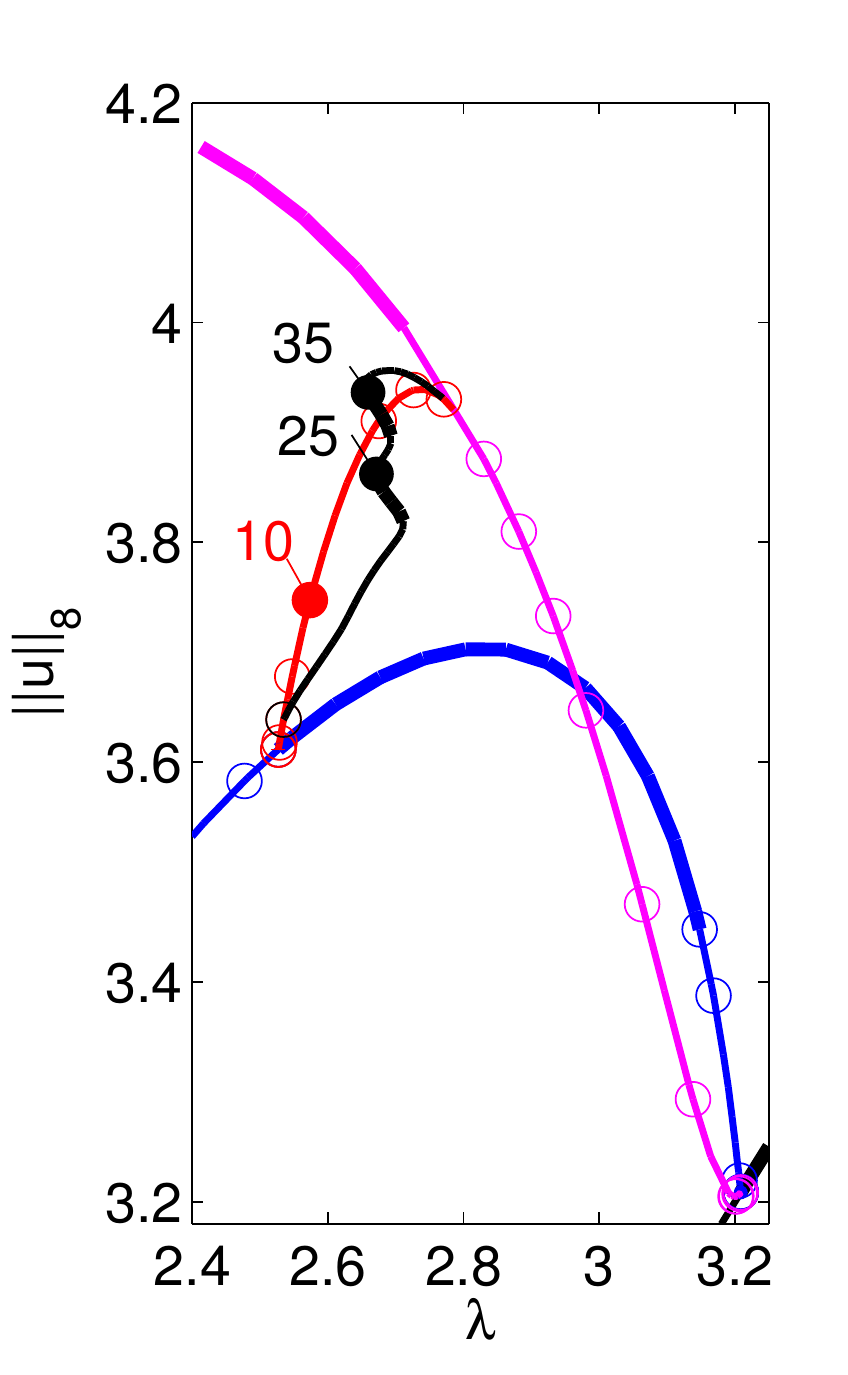}\hs{-7mm}
\raisebox{28mm}{\begin{tabular}{l}
\ig[width=0.5\tew]{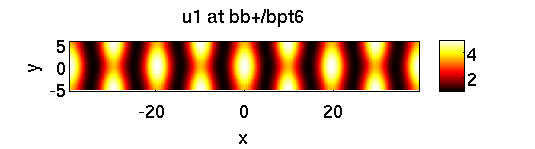}\hs{-9mm}\raisebox{5mm}{\ig[width=0.27\tew]{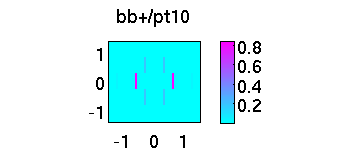}}\\[-6mm]
\ig[width=0.5\tew]{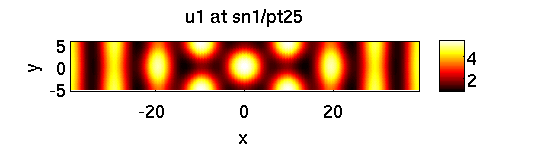}\hs{-9mm}\raisebox{5mm}{\ig[width=0.27\tew]{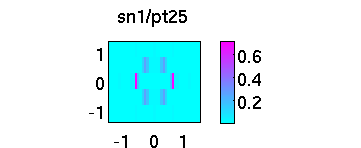}}\\[-6mm]
\ig[width=0.5\tew]{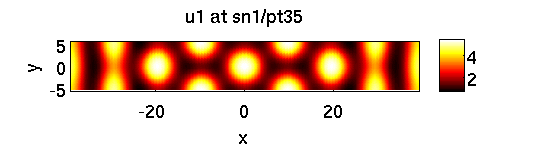}\hs{-9mm}\raisebox{5mm}{\ig[width=0.27\tew]{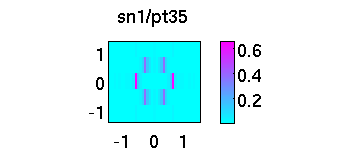}}
\end{tabular}}

\ece
\vs{-2mm}
   \caption{{\small \reff{tur1} on a long rectangular domain $\Om{=}(-l_x,l_x){\times}(-l_y,l_y)$, $l_x=8\pi/k_c$, $l_y=\pi/(\sqrt{3}k_c)$. 
On the bean branch (red) between stripes (blue) and spots (magenta), there are bifurcation points 
leading to snaking branches of localized patterns. Right column: Fourier 
spectrum of $u_1-\spr{u_1}$. 
  \label{schnakf2}}}
\end{figure}

The mixed mode branches and the associated bistability ranges, for instance 
between ``+'' stripes and spots, by analogy with the SH equation suggest 
the existence of localized patterns over patterns, e.g., of spots embedded 
in stripes. A multitude of such solutions has been discussed in \cite{uwsnak14}, 
and in Fig.~\ref{schnakf2} we only illustrate one example, computed 
in {\tt cmds2db.m}. Here we essentially increase the domain length in $x$, 
and then find bifurcation points on the mixed mode branches where 
branches of localized patterns bifurcate, which return to the mixed 
mode branch at the other end. The only non-standard setup in the software 
is that we modify {\tt stanbra} to {\tt schnakbra} and set 
{\tt p.fuha.outfu=@schnakbra}. Here we append the (normalized) $L^8$ norm 
$\ds \|u\|_8=\left(\frac 1 {|\Om|}\int_\om u^8\dd x\right)^{1/8}$ to the branch output, because this yields a bigger difference between spots and stripes than 
the $L^2$ norm, and is therefore more suitable for plotting. 

\hulst{caption={{\small (Selection from) {\tt \dname/cmds2da.m}. Here the kernel 
vectors are $\phi_1$=stripes and $\phi_2$=patchwork quilt, and {\tt qswibra} 
computes the pertinent linear combination and $\lam'(0)$ for the hex branch. 
From the inspection of $\phi_1, \phi_2$, for the stripe branch we then 
directly use {\tt gentau}. 
}}, 
label=sl10c, language=matlab,stepnumber=5, linerange=2-11, 
firstnumber=1}{\dhome/cmds2da.m}

\brem\label{s3Drem}
The analogs of Fig.~\ref{shf12} and Fig.~\ref{shf12b} for the 
Schnakenberg model in 3D are 
computed in {\tt cmds3DSC.m} and {\tt cmds3DBCC.m}, and we only remark that: 
\bci 
\item For  \reff{tur1} in 3D, the choice of a pseudo criss-cross meshes seems even 
more vital than for \reff{swiho}; over standard meshes, solutions quickly 
loose symmetry. 
\item The SC lamellas and rhombs 
can be continued very robustly via {\tt cont}, while the tubes need {\tt pmcont}, and at larger amplitude still tend to drift to lamellas. 
\item On the BCC domain, the tubes continue very robustly and become stable 
at large amplitude, while the continuation of the BCCs becomes more difficult 
because they tend to loose symmetry also over pseudo criss-cross meshes. 
\eci  
\erem 

\section{Five intermezzi} 
\label{intersec}
We give five short intermezzi dealing with more or less 
classical problems in pattern formation. 
The main purpose is to explain in a concise way 
how additional features of \pdep\ can be exploited, for instance branch-point continuation in the demo {\tt shEck} (\S\ref{shecksec}) to approximate 
the Eckhaus instability of rolls in the 1D SH equation, and  
the coupling with additional equations in 
the Cahn--Hilliard demo {\tt CH} (\S\ref{chsec}). The demo {\tt hexex} (\S\ref{hexsec}) illustrates the 
use use of {\tt gentau} to deal with multiple BPs of higher indeterminacy,  
while {\tt chemtax} (\S\ref{chemsec}) revisits a chemotaxis system from \cite[\S4.1]{p2p} 
to explain how to deal with quasilinear terms in the \oop\ setting. 
Finally, in the demo {\tt shgc} (\S\ref{gcsec}) we consider a SH equation 
with a global coupling, which thus requires customized linear system solvers 
and eigenvalue solvers. 

\subsection{Approximation of the Eckhaus curve by 
BP continuation: 
{\tt shEck}}
\label{shecksec}
\def\dhome{./pftut/shEck}
\def\dname{shEck}
\def\wt{\tilde}
An important result for the classical 1D SH equation (with a scaling 
parameter $\ell$, which we first fix to $\ell=1$) 
\huga{\label{sh3} \pa_t u = -(1+\ell^2\pa_x^2)^2 u + \lam u -u^3, 
}
is the so--called Eckhaus instability of stripes \cite{eck65}. 
In detail, for $x\in\R$ such that we have the continuum $k\in\R$ of 
admissible wave numbers, we have the supercritical bifurcation 
of stripes with wave number $k$ at $\lam=(1-k^2)^2$. However, 
except for $k=1$ the bifurcating stripes are not stable directly 
at bifurcation, but only for 
\def\Eck{\lam_{{\rm Eck}}}
\huga{\label{kEck}
\lam>\Eck(k):=3\kap^2-\kap^3+\CO(\kap^4), \quad \kap=k^2-1. 
}
Here we illustrate how this Eckhaus curve can be approximated on a 
finite domain via BP continuation. 
In \pdep, this is done similar to fold continuation, i.e., via the extended system \cite[\S3.3.2]{mei2000}
\huga{\label{bpe}
H(U)=\bpm G(u,\lam)+\mu M\psi\\
G_u^T(u,w)\psi\\
\|\psi\|_2^2-1\\
\spr{\psi,G_{\lam}(u,w)}\epm 
=\bpm 0\\0\\0\\0 
\epm, \quad U=(u,\psi,w),  
}
where $(u,\lam)$ is a (simple) BP (for the continuation in $\lam$), 
$\psi$ is an adjoint kernel vector, 
$w=(\lam,\mu)$ with $w_1=\lam$ the primary active parameter and 
$w_2=\mu$ as additional active parameter.  

\brem\label{BPlocrem} (a) The extended system \reff{bpe} is regular at simple 
BPs, see \cite{M80, Mei89, mei2000}, and thus can be used 
for localization of (simple) BPs if a sufficiently good initial guess $(u,\psi,\lam,0)$ 
is available. However, we (currently) hardly use this option, mainly 
because the implementation of $\pa_u(G_u^T\psi)$ (see below) requires some 
additional effort, and the localization by bisection is usually 
fast and accurate enough. See \cite{lsstut} for an example for BP 
localization via extended systems. 

(b) Freeing a second parameter, i.e., setting $\wt w=(\wt\lam,w)$ 
with {\em a new} $\wt\lam$, dropping the\ $\wt\ $ and augmenting \reff{bpe} with the usual arclength condition $p(U,\rds)=0$, $U=(u,\psi,\lam,w)\in 
\R^{2n_u+3}$, \reff{bpe} can also be used for BP continuation, in which we are interested here. 
\eex \erem 

To prepare the use of \reff{bpe} for BP continuation, \pdep\ provides the 
call {\tt p=bpcontini(p,newpar,...)}, which (internally) doubles the number $n_u$,  
of unknowns, stores $\psi$ at $u_{n_u+1,\ldots,2n_u}$, and shifts the 
parameters to the pertinent new positions $2n_u+1$:end (with $n_u=$old $n_u$). 
The main task then is to set up  $\pa_u(G_u^T\psi)$ for  the Jacobian 
\huga{\label{JH}
J_H=\bpm G_u&\mu M&G_{\lam}&M\psi\\
\pa_u(G_u^T\psi)&G_u^T&\pa_{\lam}(G_u^T\psi)&0\\
0&2\psi^T&0&0\\
\psi^T\pa_\lam G_u^T&G_\lam^T&\psi^T\pa_{\lam} G_\lam&0
\epm, 
}
while $G_u$ is already available, and all other derivatives can efficiently 
be done numerically and hence automatically. However, for semilinear problems 
$G(u,\lam)=K(\lam)u-Mf(u,\lam)$, where the stiffness matrix $K(\lam)$ does 
not depend on $u$, $\pa_u(G_u^T \psi)$ has a simple form. For instance, 
for a 2-component system with 
$\dst f=\bpm f(u_1,u_2)\\f_2(u_1,u_2)\epm$ we have 
\huga{
\pa_u(G_u^T\psi)=\left(\pa_u\bpm f_{1,u_1}\psi_1{+}f_{2,u_1}\psi_2\\
f_{2,u_1}\psi_1{+}f_{2,u_2}\psi_2\epm\right)M^T
=\bpm f_{1,u_1u_1}\psi_1{+}f_{2,u_1u_1}\psi_2&
f_{1,u_1u_2}\psi_1{+}f_{2,u_1u_2}\psi_2\\
f_{1,u_1u_2}\psi_1{+}f_{2,u_2u_1}\psi_2&
f_{1,u_2u_2}\psi_1{+}f_{2,u_2u_2}\psi_2
\epm M^T, 
}
where expressions such as $f_{i,u_j,u_k}\psi_m$ are to be understood as 
pointwise multiplication. Thus, {\tt bpjac} from Listing \ref{shEckl1} 
yields the desired $\pa_u(G_u^T\psi)$, 
while Listing \ref{shEckl2} shows pertinent cells from the script {\tt cmdsEck.m}, and Listing \ref{shEckl3} shows how to put the 
effective wave number $k$ onto the branch for plotting.

\begin{figure}[h]
\bce 
\begin{tabular}{llll}
{\small (a)}&{\small (b)}&{\small (c)}&{\small (d)}\\
\ig[width=0.25\tew,height=5cm]{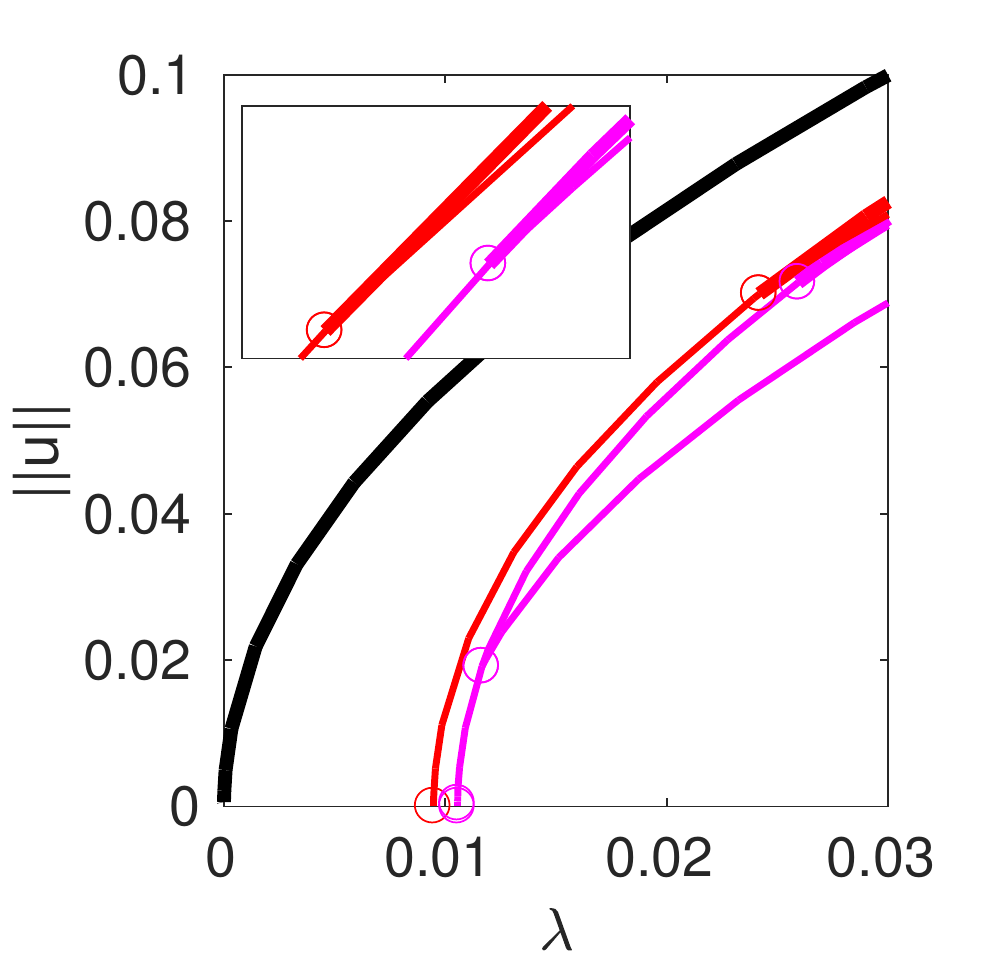}&
\raisebox{25mm}{\begin{tabular}{l}
\ig[width=0.2\tew]{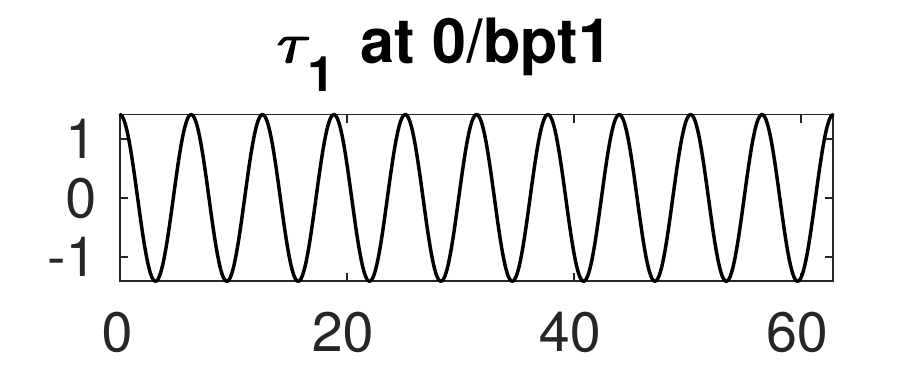}\\
\ig[width=0.2\tew]{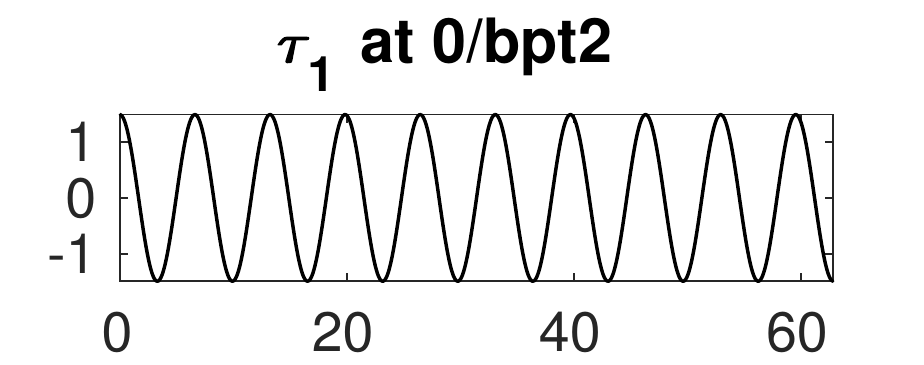}\\
\ig[width=0.2\tew]{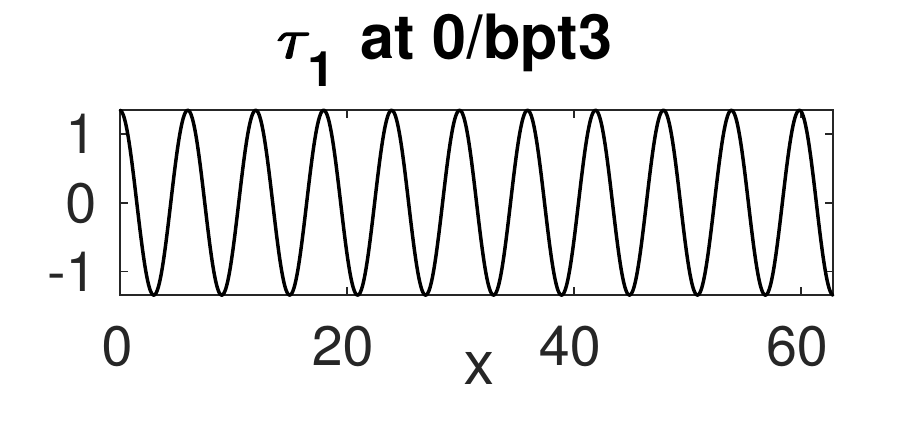}
\end{tabular}}&
\hs{-8mm}\raisebox{25mm}{\begin{tabular}{l}
\ig[width=0.2\tew]{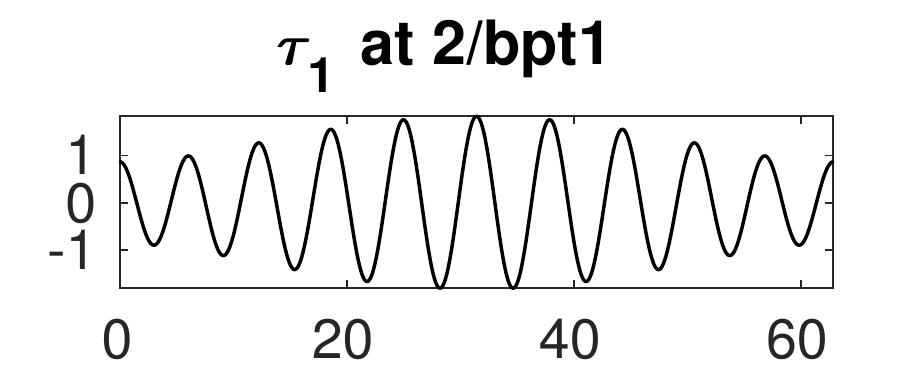}\\
\ig[width=0.2\tew]{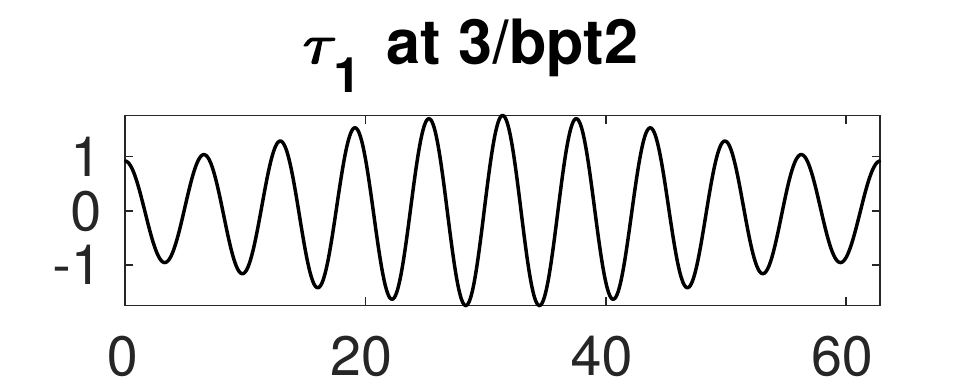}\\
\ig[width=0.2\tew]{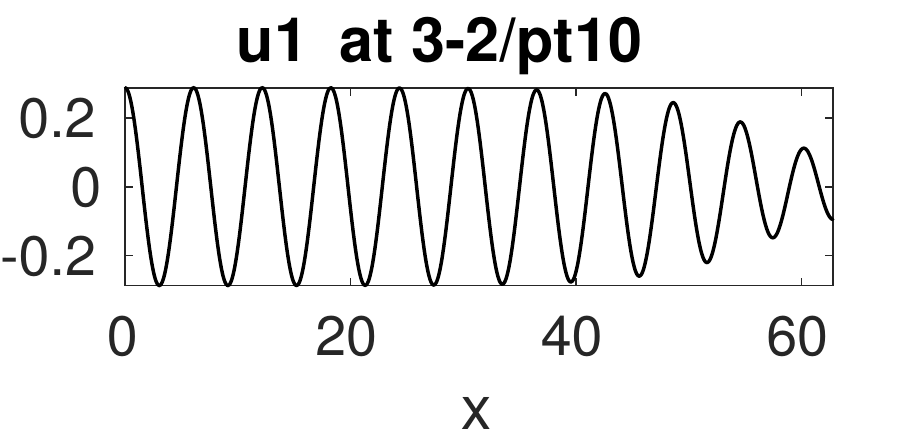}
\end{tabular}}
&\ig[width=0.25\tew,height=5cm]{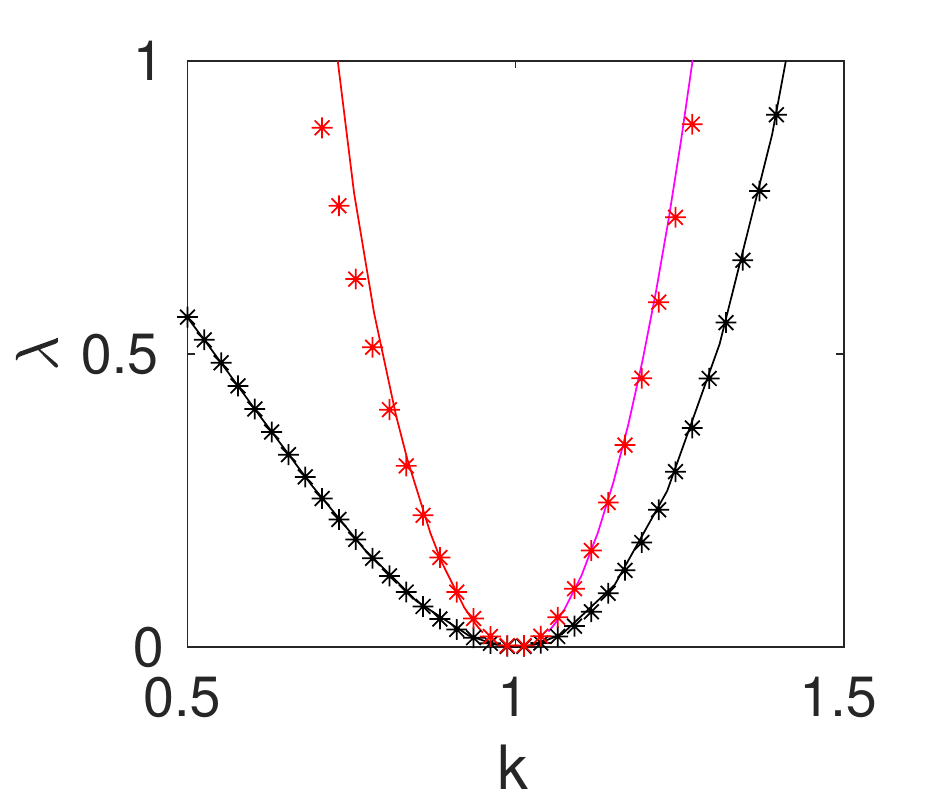}
\end{tabular}
\ece 
\vs{-6mm}
\caption{{\small (a) First 3 bifurcating branches on $\Om=(0,20\pi)$, with 
wave numbers $k=1,19/20$ and $21/20$, and secondary bifurcations. Inset zoom near $\lam=0.025$. (b) tangent plots at the BPs 
on $u\equiv 0$. (c) tangent plots at the 'Eckhaus points' {\tt b2/bpt1} 
and {\tt b3/bpt2} (first two panels), and continuation of the branch 
bifurcating at {\tt b3/bpt2}. (d) BP continuation, yielding the 
black existence curve, and two approximations of $\Eck(k)$. These three 
curves are also compared with the formulas $\lam=(1-k^2)^2$ 
and $\lam=\Eck(k)$ from \reff{kEck} (dots). \label{shBBf1}}}
\end{figure}

To compute (approximate) $\Eck$ we consider \reff{sh3} on $\Om=(0,l_x)$ 
with $l_x=20\pi$ and homogeneous Neumann BCs. The first three BPs from $u\equiv 0$ then have 
$k=1$, $k=19/20$ (dilated pattern) and $k=21/20$ (compressed pattern), 
respectively, see Fig.~\ref{shBBf1}(a,b).  The 2nd and 3rd branches {\tt b2} and {\tt b3} start 
with 1 and 2 unstable eigenvalues, respectively, and they both 
gain stability at $\lam\approx 0.025$. The bifurcating branches 
at these secondary bifurcation correspond to long wave modulations 
of the patterns, see  Fig.~\ref{shBBf1}(c). In (d) we then continue 
the primary bifurcation point {\tt 0/bpt1} ($(u,\lam)=(0,0)$) and the 
secondary BPs {\tt b2/bpt1} and {\tt b3/bpt3} in the scaling parameter $\ell$, i.e., set $w=(\ell,\lam)$ in \reff{bpe}. For {\tt 0/bpt1} we naturally obtain 
the (continuous) existence-of-patterns curve $\lam_{{\rm ex}}=(1-k^2)^2$ 
(black), where 
$k=\ell k_0$ with $k_0=1$, i.e., periodic solutions with minimal period $k$ 
exists for $k>\lam_{{\rm ex}}(k)$. Similarly, for 
{\tt b2/bpt1} (with $k_0=19/20$) and {\tt b3/bpt2} (with $k_0=21/20$) 
we obtain the red/magenta approximations of $\Eck$. 

\hulst{caption={{\small  {\tt \dname/bpjac.m}. 
$\psi$ is in {\tt u(p.nu+1:2*p.nu)}, and the parameters 
are at {\tt u(2*p.nu+1:end)}. 
}}, 
label=shEckl1, language=matlab, stepnumber=100}{\dhome/bpjac.m}

\hulst{language=matlab, linerange=34-36,stepnumber=100}{\dhome/cmdsEck.m}
\hulst{caption={{\small  Two cells from {\tt \dname/cmdsEck.m}. 
In line 1 of the top cell we initialize BP continuation in the scaling 
parameter $\ell$ by calling {\tt bpcontini}, and save the original 
wave number $k_0=19/20$ of {\tt b2} in p.k. This is used to put the 
correct $k=k_0\ell$ onto the branch. 
The bottom cell contains the 
plot commands to obtain Fig.~\ref{shBBf1}(d). 
}}, 
label=shEckl2, language=matlab,linerange=45-53,stepnumber=100}{\dhome/cmdsEck.m}

\hulst{caption={{\small  {\tt \dname/shbra1d.m}, 
to put $k=k_0\ell$ onto the output branch. 
}}, 
label=shEckl3, language=matlab, stepnumber=100}{\dhome/shbra1d.m}

\brem\label{BBrem}
In 1D, the Eckhaus curve $\lam{=}\Eck(k)$ is the (lower) boundary of the Busse 
balloon $\lam{>}\Eck(k)$, which is the parameter set of stable (1D) patterns. 
In 2D, the Busse balloon for the SH equation 
is further delimited by the so called zig-zag instability curve $k>\sqrt{1-\lam^2/512}+\CO(\lam^4)$, for small $\lam$. 
In general, the boundary of the Busse balloon for patterns consists of various 
instability curves, such as Eckhaus--, zig-zag--, cross-roll-- and 
other instabilities, and the Busse balloon is typically a bounded set, 
even if, as in the SH equation, the existence region of the pattern is 
unbounded. The (asymptotic) computation of Busse 
balloons is an important but 
complicated problem, see, e.g., \cite[\S8]{hoyle}. The above 
example and further tests suggest that at least in simple cases the continuation 
of BPs (and of Hopf BPs) may be a simple but efficient method 
to approximate Busse balloons. However, further details will 
appear elsewhere. 
\eex\erem 

\subsection{Cahn-Hilliard: Demo {\tt CH}}\label{chsec}
\def\dhome{./pftut/CHa}\def\dname{CHa}
The Cahn--Hilliard problem models 'spinodal 
decomposition' of an alloy, and consists in finding stationary 
points (in particular minimizers)  of the energy 
\huga{\label{ch1} 
E_\eps(u)=\int_\Om \frac 1 2 \eps \|\nabla u\|^2+W(u)\dd x,
\text{ 
under the mass constraint }
\frac 1 {|\Om|}\int_\Om u\dd x=m
} 
and zero flux--BCs. 
Here $\Om\subset\R^d$ is a bounded domain,  $\eps>0$ 
is a parameter for the so--called interface energy, and $W$ is a double well 
potential, e.g., $W(u)=-\frac 1 2 u^2+\frac 1 4 u^4$. A detailed 
bifurcation {\em analysis} of the problem on the unit square can 
for instance be found in \cite[\S III.2.6]{kiel2012}. 
Here we are mainly interested in the implementation of the constraint 
$\frac 1 {|\Om|}\int_\Om u\dd x=m$, and give a few numerical results for 
illustration. 

We let $\Om=(-1/2,1/2)^d$ such that $|\Om|=1$. The first variations 
wrt $u$ and $\lam$ of the Lagrangian $\dst 
L(u,\lam)=E_\eps(u)+\lam(\int_\Om u\dd x-m)$ 
then yield the Euler--Lagrange equations 
\huga{\label{ch2}
{\rm (a)}\quad G(u):=-\eps\Del u+W'(u)-\lam\stackrel!=0, \qquad {\rm (b)}\quad q(u):=\int_\Om u\dd x-m\stackrel!=0. 
} 
(\ref{ch2}a) is a version of the Allen--Cahn equation, see, e.g., 
\cite{actut}, with the additional parameter $\lam$. 
There are (at least) two different ways to add the constraint 
(\ref{ch2}b) to the PDE  (\ref{ch2}a): We can either (a) use the designated 
function handle {\tt p.fuha.qf} for (\ref{ch2}b), or we can 
directly append (\ref{ch2}b) to {\tt p.fuha.sG}. We generally 
recommend (a), but in some cases (b) is more convenient, 
for instance here to also do BP continuation for \reff{ch2}. 
Before we start, we note that $\int_\Om u(x)\dd x$ is on the FEM level most 
conveniently implemented as {\tt vM*u}, where 
{\tt vM=sum(M,1)} and $M$ is the mass matrix. 

\paragraph{Implementation of additional equations via {\tt p.fuha.qf} (demo CHa).}
To implement additional 
equations such as (\ref{ch2}b) we provide an interface in the 
form of function handles ${\tt p.fuha.qf=}q$, where $q:\R^{n_u}\times\R^p\ra 
\R^{n_q}$, $p=$number of parameters, $n_q=$number of additional eqns, 
here $n_q=1$. The additional $n_q$ equations then require $n_q$ additional 
parameters $w$, such that the extended system takes the form 
\huga{\label{genq}
H(u,\lam,w))=\bpm G(u,\lam,w)\\q(u,\lam,w)\\p(u,\lam,w,{\rm ds})\epm=0\in\R^{n_u+n_q+1}, 
}
where we take $\lam$ as the name of the primary parameter, $w$ is  
the vector of non-primary active parameters, and the last equation 
$p(u,\lam,w,{\rm ds})=0$ is the usual arclength condition. 
Additional equations $q(U)=0$, $U=(u,\lam,w)$ also often occur as 
phase conditions in case of continuous symmetries, see \cite{symtut}, 
and thus are an important feature of \pdep. For efficiency, it is recommend 
to also provide a function handle ${\tt p.fuha.qder}=\pa_u q(u,\lam,w)$. 

\begin{figure}[h]
\bce 
\begin{tabular}{lll}
{\small (a)}&{\small (b)}&{\small (c)}\\
\ig[width=0.26\tew,height=5cm]{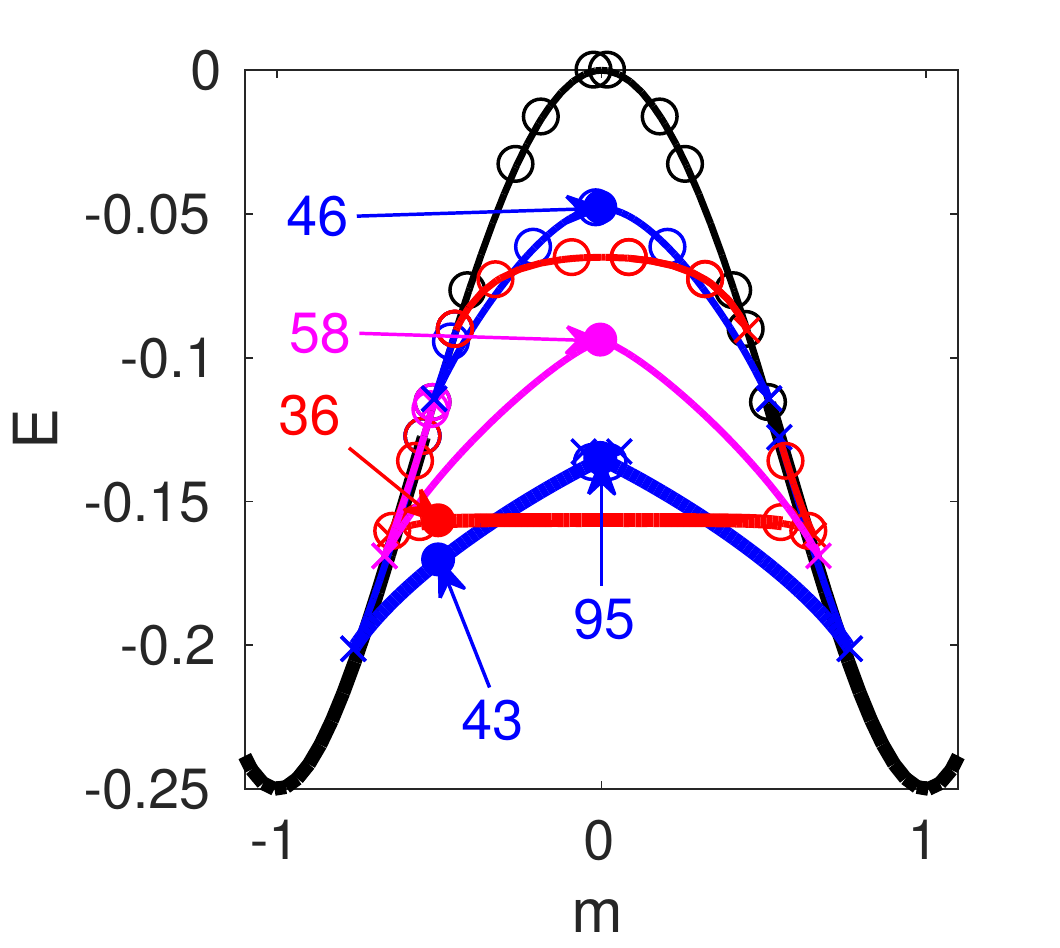}&
\hs{-5mm}\ig[width=0.26\tew,height=5cm]{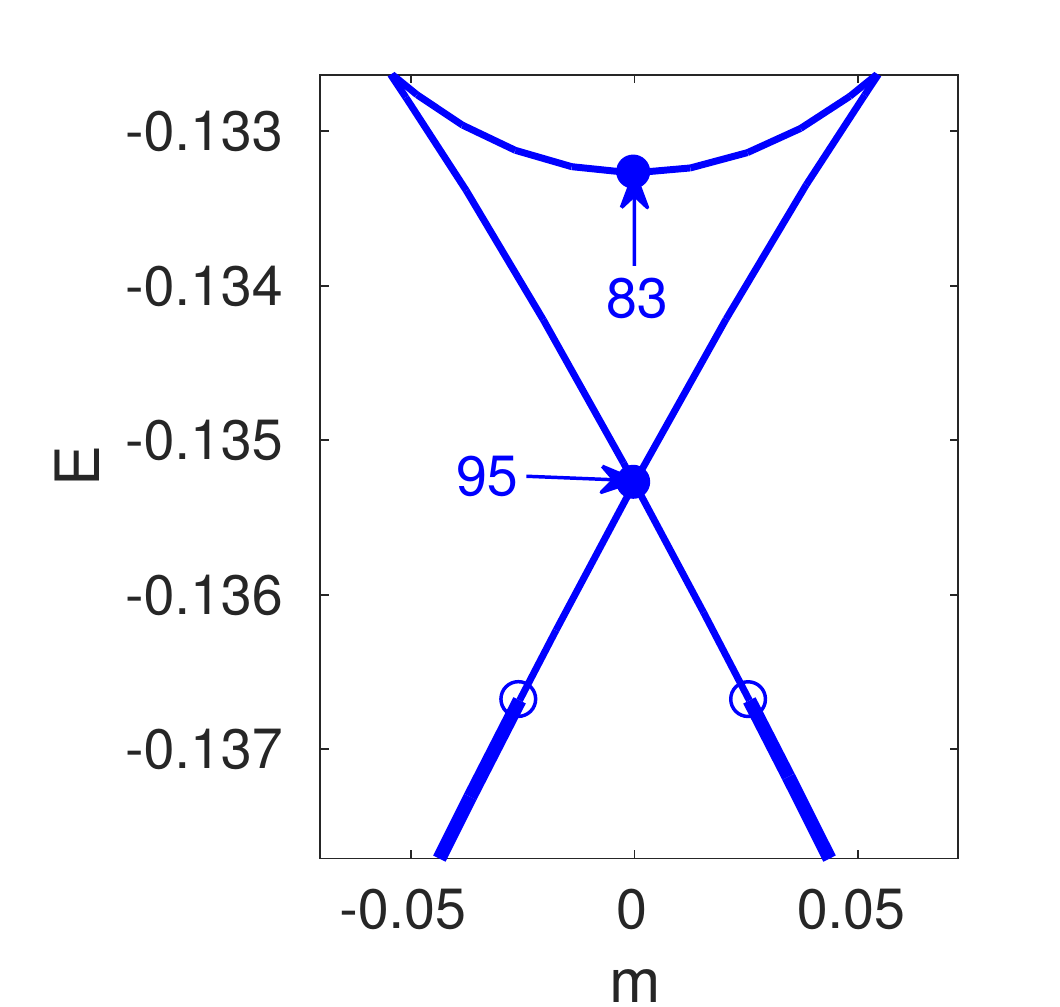}&
\raisebox{25mm}{\begin{tabular}{l}
\ig[width=0.15\tew]{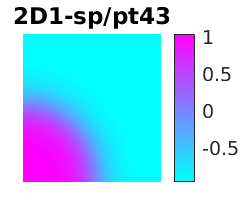}\ig[width=0.15\tew]{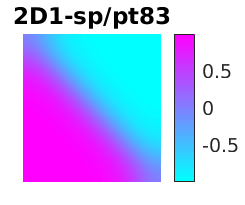}
\ig[width=0.15\tew]{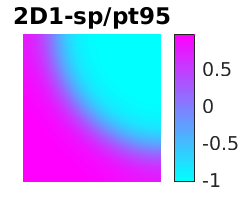}\\
\ig[width=0.15\tew]{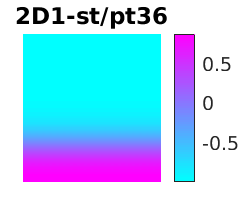}\ig[width=0.15\tew]{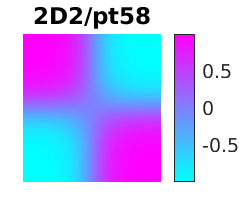}\ig[width=0.15\tew]{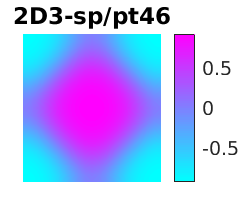}\\
\end{tabular}}
\end{tabular}
\ece 
\vs{-6mm}
\caption{{\small The Cahn--Hilliard problem \reff{ch1} with $W(u)=-\frac 1 2 u^2+\frac 1 4 u^4$ on the unit square, $\eps=1/100$. (a) BD $E$ over $m$. The spinodal 
region (where the alloy can lower its energy by decomposing) extends from 
$m=-m_0$ to $m=m_0$, where $-m_0\approx -0.55$ is the first BP on 
the homogeneous branch (black) (a). This BP is double, with spots (blue) 
and stripes (red) bifurcating subcritically, and both becoming stable in folds. 
The further BPs may be simple (BP2, magenta) or double (BP3, 
again with spots and stripes). Example plots in (c). However 
the branches bifurcating at higher BPs never become stable, and 
thus we concentrate at the first BP. For $m$ near $-m_0$, the spots 
have the lowest energy, but they become unstable near $m=0$, see the 
zoom in (b).  \label{chf1}}}
\end{figure}

We proceed by example, first in 2D (see the next \S\ for 1D results) and 
use {\tt par=[m,eps,lam]} as the parameter vector for \reff{ch2}. 
We fix $\eps=1/100$, and continue the trivial branch $u\equiv m$ in $m$ 
with the Lagrange multiplier $\lam$ as second active 
parameter, such that {\tt p.nc.ilam=[1 3]}. Listing \ref{chl2}  shows the implementation of $q$ and $\pa_u q$, and of the 
energy $E$, respectively, where $E$ is put on the branch for plotting 
as for the SH equation. Listing \ref{chl4} 
show some pertinent parts of the script {\tt cmds2D}, and 
some results from {\tt cmds2D.m} are shown in Fig.~\ref{chf1}. 

\hulst{language=matlab, stepnumber=100}{\dhome/qf.m}
\hulst{language=matlab, stepnumber=100}{\dhome/qfder.m}
\hulst{caption={{\small Functions {\tt qf.m}, {\tt qfder.m} 
and {\tt chE.m} from {\tt \dname}, computing the constraint (\ref{ch2}b), its derivative 
$\pa_u q$, and the energy $E$ from \reff{ch1}, respectively. The 
1st order differentiation matrices {\tt Dx} and {\tt Dy}, and 
the 'integration vector' {\tt p.mat.vM=sum(p.mat.M,1)}, are generated 
in {\tt oosetfemops}}.}, label=chl2, language=matlab, stepnumber=100}{\dhome/chE.m}

\hulst{caption={{\small {\tt \dname/cmds2D.m}, first two cells. 
{\tt chinit} as usual, and in lines 4,5 we set the function handles 
for the auxiliary equation $q=0$ and the derivative $\pa_u q$, and 
{\tt ilam=[1,3]} 
as we now have two active parameters. The first BP (on a square) is double, 
and thus in the second cell we use cswibra to find 2 (altogether 4) tangent 
directions (spots and stripes). The remainder of {\tt cmmds2D.m} deals 
with the other BPs and plotting, and works as usual.}}, label=chl4, language=matlab, stepnumber=100,linerange=2-10}{\dhome/cmds2D.m}

\paragraph{Direct implementation of the mass constraint in $G$, and BP 
continuation.} 
\def\dhome{./pftut/CHb}\def\dname{CHb}
Alternatively to using {\tt p.fuha.qf}, we may directly append the equation $q=0$ to the PDE 
$G$, and one parameter, e.g., $\lam$, to the {\em vector of PDE unknowns 
$u$}. Thus, we let, on the discrete level, 
\huga{
u=(u_{\rm PDE},u_\lam)\in\R^{n_u}\text{ with }n_u=n_p+1,
} write \reff{ch2} in the form 
\huga{\label{ch3}
G(u)=\bpm \tilde{G}(u)\\ \frac 1 {|\Om|} v_Mu_{\rm PDE}-m 
\epm=
\bpm \eps K u_{\rm PDE}+MW'(u_{\rm PDE})-u_\lam v_M^T\\
\frac 1 {|\Om|} v_Mu_{\rm PDE}-m\epm=0\in \R^{n_u}, 
}
where $\tilde{G}$ means the PDE (\ref{ch2}a). 
The remaining parameters are $(m,\eps)$, and we need to: 
\bci 
\item modify ${\tt p.nu}\mapsto {\tt p.nu}+1$ at init, append $\lam$ 
to $u$, and also modify the mass--matrix {\tt p.mat.M} in {\tt oosetfemops}. 
\item modify ${\tt sG}$ and ${\tt sGjac}$ (and, for BP continuation,
also {\tt bpjac}) accordingly. 
\eci 
See Listing \ref{chl5} for examples of these (easy) changes. 
The respective scripts in {\tt \dname} (naturally) give the same results as in 
{\tt CHa}, and we only 
give a few comments pertaining to BP continuation. We find 
\huga{
\pa_u G=\bpm \pa_{u_{\rm PDE}}\tilde{G}&-v_m\\
\frac 1 {|\Om|}v_m^T&0\epm,
} 
In particular, since the constraint 
is linear in $u_{\rm PDE}$ and $\lam$ we obtain 
\huga{
\pa_u(G_u^T\psi)=\bpm {\rm diag}(f_{uu}\psi_{\rm PDE})&0\\
0&0\epm\in\R^{n_u\times n_u}, 
}
which is implemented in {\tt \dhome/bpjac}. 

\hulst{language=matlab, stepnumber=100}{\dhome/sG.m}
\hulst{caption={{\small {\tt \dname/sG.m} and {\tt sGjac}}. 
Most importantly, the proper PDE $u$ is in {\tt u(1:p.nu-1)}.}, label=chl5, language=matlab, stepnumber=100}{\dhome/sGjac.m}

\begin{figure}[h]
\bce 
\begin{tabular}{ll}
{\small (a)}&{\small (b)}\\
\ig[width=0.24\tew, height=4.5cm]{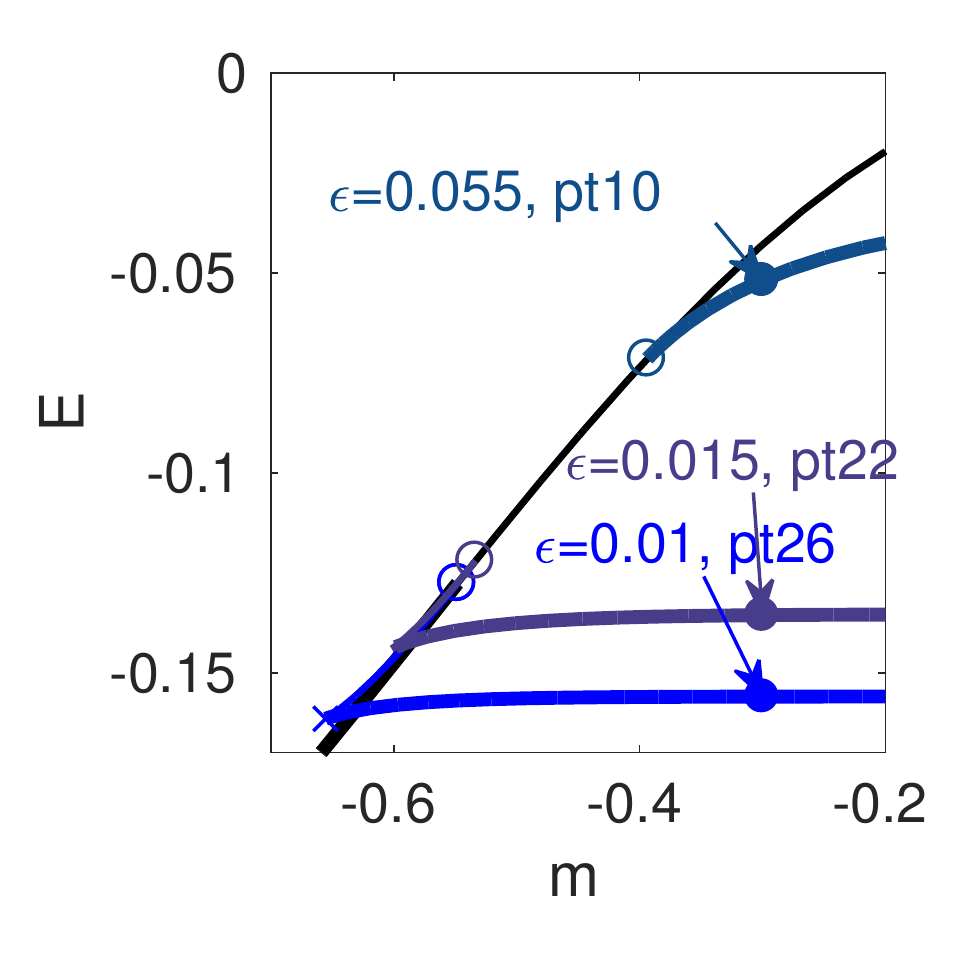}&
\raisebox{3mm}{
\ig[width=0.15\tew]{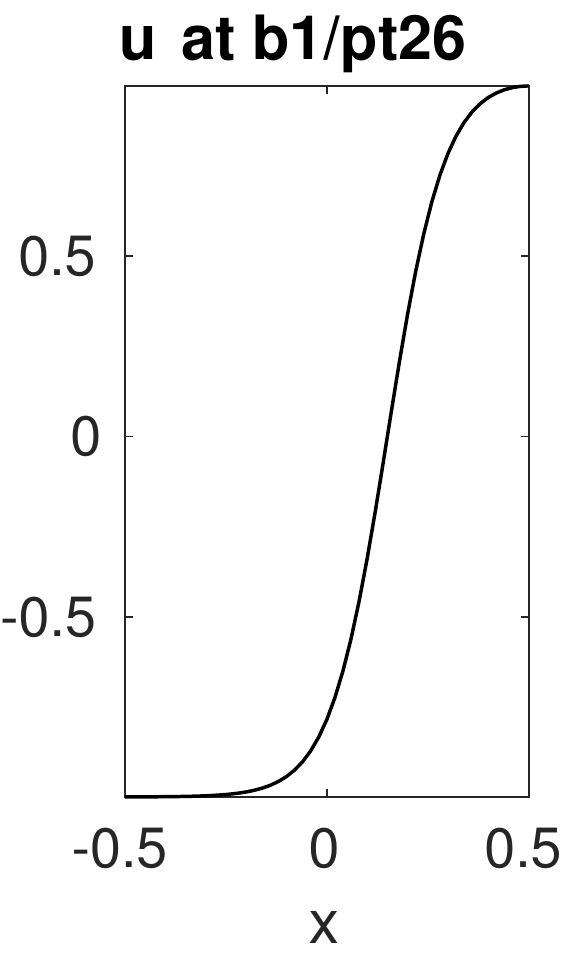}\ig[width=0.15\tew]{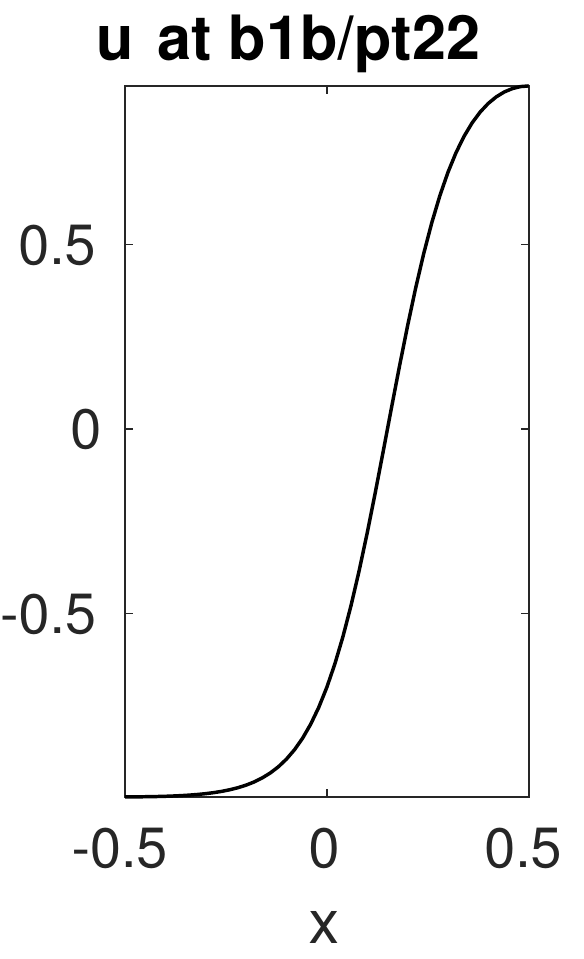}
\ig[width=0.15\tew]{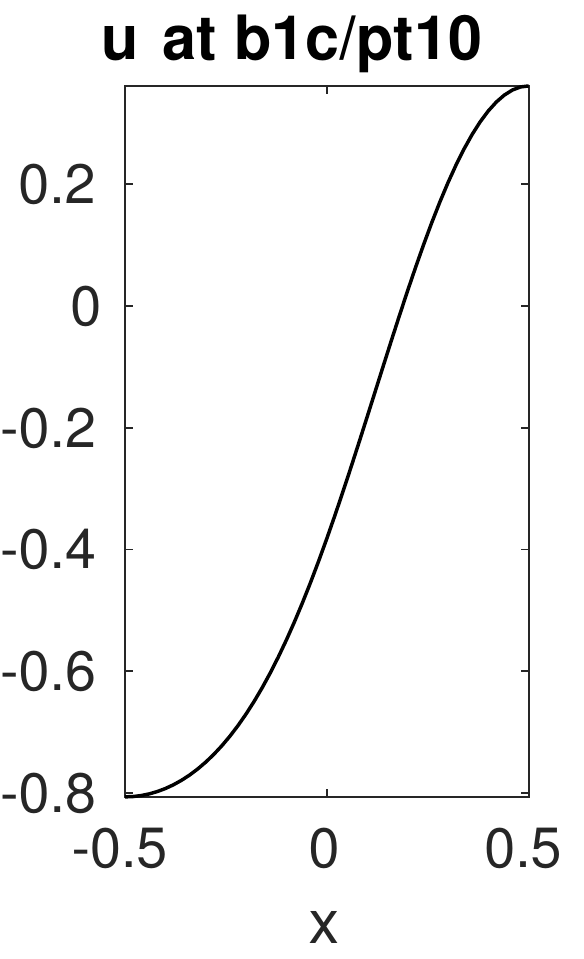}}
\end{tabular}
\ece 
\vs{-6mm}
\caption{{\small Branch point continuation (and subsequent return to 
normal continuation) from {\tt \dname/cmds1D.m}. Cahn--Hilliard problem \reff{ch1} with $W(u)=-\frac 1 2 u^2+\frac 1 4 u^4$ on $(-1/2,1/2)$. 
(a) The primary nontrivial branch(es) (blue) for different $\eps$. (b) 
example plots at $m=-0.3$ for increasing $\eps$. 
 \label{chf2}}}
\end{figure}

Fig.~\ref{chf2} then shows some results on BP continuation in $\eps$ 
of the first BP on the homogeneous branch (from continuation in $m$), 
i.e., the black branch in Fig.~\ref{chf2}(a). We focus on 1D since 
the influence of $\eps$ is easier to see. The BP continuation 
is switched on via 
{\tt p=bpcontini('tr','bpt1',2,'bp1c')}.  Upon varying $\eps$, 
with $m$ as a secondary parameter, we again run along the black branch in 
the $E$ over $m$ plot, but with increasing $\eps$ the BPs appear later 
on the branch. Using, e.g., {\tt p=bpcontexit('bp1c','pt5','trb')};  
{\tt p=swibra('trb','bpt1','b1b',0.5); cont(p)} we obtain the branch in darker 
blue, corresponding to $\eps\approx 0.015$, and similarly for a 
third branch at $\eps\approx 0.055$. The solution plots in (b) 
show that these are indeed 'the same branches at different $\eps$', i.e., 
the larger $\eps$ gives smoother interfaces. 

\brem\label{qrem}(a) In summary, the implementation of \reff{ch2} via  {\tt p.fuha.q} in {\tt CHa} 
is more flexible, and thus generally recommended, and we mainly set up the 
alternative {\tt CHb} for illustration, and to do BP continuation in a 
straightforward way. 

(b) The dynamic CH equation is obtained from taking the conserved flux 
\huga{\label{ch4}
\pa_t u=-G(u):=\nabla\cdot[\nabla\del_\phi E_\eps(u)]=-\Delta[\eps\Delta u-W'(u)], 
}
see \cite{EGU19} for discussion. 
For zero-flux BCs, i.e., $\pa_n u=\pa_n\Delta u=0$ on $\pa\Om$, 
this conserves the 
mass $\int_\Om u\dd x$, but for the steady problem 
$0=G(u)$ we again need to explicitly enforce the mass conservation. 
\eex
\erem 

\subsection{Pearling in the functionalized Cahn--Hilliard equation: Demo {\tt fCH}}\label{fchsec}
A binary mixture of a solvent (e.g.~water) and (hydrophobic) molecules 
(polymers or lipids) can be 'functionalized' by adding hydrophilic 
side chains to the molecules. Intuitely, this may favor the formation 
of bilayer interfaces (``channels'', see Fig.~\ref{fchf1}(a) for examples 
in the model given below), where the 
hydrophilic (hydrophobic) ends of the molecules 
point to the solvent (to the inside of the channel). In Fig.~\ref{fchf1},  
black corresponds to the low and yellow to a high concentration 
of molecules. 
If the mixture is dilute, the channels show a pearling instability, and if 
there is 'excess mass', i.e., 
a relatively high volume fraction of molecules, then channels tend 
to bend, which is also called meandering. 

The model we consider is from 
\cite{GJPY11, DHPW12}, to which we also refer for further background. 
Letting $u:\Om\to \R$ be the volume fraction of solvent and molecules, 
with $u=-1$ corresponding to pure solvent and 
some $u_+>0$ corresponding to saturation, 
the free energy of the mixture reads 
\huga{\label{fchE} 
\CF(u)=\int_\Om \frac 1 2 (\eps^2 \Delta u-W'(u))^2
-\eps^\beta\left(\frac{\eta_1} 2
\eps^2|\nabla u|^2+\eta_2W(u)\right)^2\dd x, 
}
where $0<\eps\ll 1$ is related to interface thickness, and $W:\R\to\R$ is 
a double--well potential with typically unequal strict 
minima at $-1$ and $u_+>0$, and a strict maximum at $u=0$. The first summand 
is the squared variational derivative of the Cahn--Hilliard energy $E_\eps(u)$ 
(with $\eps^2$ instead of $\eps$), 
cf.~\reff{ch1}, and can thus be seen as a bending energy, also called 
Willmore functional. The second 
summand represents the functionalization, where $\eta_1>0$ and $\eta_2\in\R$ 
model the strenght of the hydrophilicity in terms of interfaces and volumes, 
respectively. Thus, $\eta_1>0$ means that interfaces lower the 
free energy. 

For $\beta=1$ ($\beta=2$) we have so called strong (weak) 
functionalization, and as in \cite{DHPW12} we focus on the strong case. 
The evolution of the system is assumed to be given by a 
gradient flow $\pa_t u=-\CG\del_u\CF(u)$, 
where 
the choice of the gradient operator $\CG$ together with the BCs must ensure 
mass conservation, i.e., $\ddt \int_\Om u(t,x)\dd x=0$. 
The simplest choice is the projection 
\huga{\label{fchg1}
\CG f=\Pi f:=f-\frac 1 {|\Om|}\int_\Om f(x)\dd x, 
}
leading to the evolution equation 
\huga{\label{tfch} 
\pa_t u=-\CG[(\eps^2\Del-W''(u)+\eps\eta_1)(\eps^2\Del u-W'(u))+\eps\eta_dW'(u)],
} 
where $\eta_d=\eta_2-\eta_1$. 
In suitable parameter regimes, \reff{tfch} 
is very rich in pattern formation. The basic building blocks (in 2D) 
are straight and curved  bilayer 
interfaces between $u\equiv -1$ and $u$ near $u_+$, which 
show ``pearling'' and ``meander'' instabilities as illustrated in 
Fig.~\ref{fchf1}(a). 

\taskip
\begin{table}[ht]\caption{Selected scripts and functions in {\tt pftut/fCH}. 
Other files, fchinits, oosetfemops, sG (with nodalf), and sGjac, 
more or less as usual. 
\label{fchtab}}
{\small 
\begin{tabular}{l|p{0.81\textwidth}}
script/function&purpose,remarks\\
\hline
cmds1&script to continue a channel, and meandering and pearling 
bifurcations from it, Fig.~\ref{fchf1} \\
cmds2&script for continuation in $\eta_1$ and DNS, Fig.~\ref{fchf2}\\
e2rs\_ad\_hoc&trivial elements-to-refine-selector based on $|u+2|$\\
qf2, qf2jac&the phase conditions \reff{sfchs3} and \reff{fchpc}, and the derivatives\\
nodalft&modification of {\tt nodalf} (the 'nonlinear terms' in \reff{sfchs}), 
setting $\ga=0$ and including \reff{fchg1} for mass conservation. 
\end{tabular}
}
\end{table}
\teskip

A \pdep --setup for \reff{tfch} and some results were already presented 
in \cite{p2p2}, but without stability considerations. Moreover, 
here we present a simpler setup, and want to explain further tricks to 
compute patterns in \reff{tfch} more robustly. See Table \ref{fchtab} 
for an overview of the used files. 
Setting $v=\eps^2\Del u-W'(u)$, the problem can be written 
as the two component system 
\begin{subequations}\label{sfchs}
\hual{
\pa_t u=&-\eps^2 \Delta v+W''(u)v-\eps\eta_1 v-\eps\eta_dW'(u)+\eps\ga,\\
0=&-\eps^2 \Delta u+W'(u)+v,
}
where $\ga$ is the (scaled) Lagrange multiplier for mass conservation 
in \reff{tfch}, over domains $\Om$ with Neumann BCs for $u$ and $v$. 
We take $\ga$ as an additional unknown,  and add the equation 
\huga{\label{sfchs3}
q(u):=\frac 1 {|\Om|}\int_\Om u\dd x-m=0, 
}
\end{subequations}
where $m$ is a reference mass, also taken as a parameter, 
which here we shall use as the main continuation parameter 
as in the CH equation in \S\ref{chsec}.  
Moreover, as we shall be interested in rather dilute mixtures 
$m$ near $-1$, it turns out that often phase conditions are needed 
or at least very useful for robust continuation of patterns. 
For instance, for the channels in Fig.~\ref{fchf1} we use the 
phase condition 
\huga{\label{fchpc}
\spr{\sin(x),u}=\int_\Om \sin(x)u(x,t)\dd x=0, 
}
fixing the channel in the middle of the domain, and add $s\pa_x u$ 
to the rhs of (\ref{sfchs}a). 
Thus, we now have four parameters $(\eta_1,\eta_2,\eps,m)$, two 
additional unknowns $(\gamma,s)$, and two additional 
equations \reff{sfchs3} and \reff{fchpc}, i.e., $n_q=2$.

For $W$ we follow \cite[\S5]{DHPW12} and let 
$$
W(u)=W_p(u+1)+20(u-m_p+1)^{p+1}H(u-m_p+1),\text{ where }
W_p(u)=\frac 1 {p-2}(pu^2-2u^p)
$$
with $p=3$, $m_p=(p/2)^{1/(p-2)}=3/2$, 
and $H$ being the Heaviside function. In \cite[\S5]{DHPW12}, additional 
to a strong analysis 
which establishes the existence of straight and curved channels and 
gives certain instability criteria for them, numerical 
time integrations are presented with $\eps=0.1$, 
$\eta_2=2$, and $\eta_1$ between 1 and 2. 
We use similar parameter 
regimes, but a somewhat larger $\eps=0.25$ to avoid very steep interfaces. 

\begin{figure}[ht]
\bce
\btab{l}{
\btab{ll}{
{\sm (a) straight channel, pearling and meandering (two examples)}&{\sm (b) BD}\\
\hs{-2mm}\ig[height=47mm]{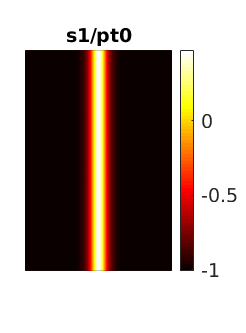}\hs{-2mm}
\ig[height=47mm]{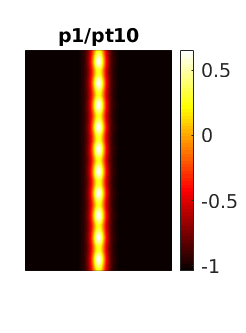}\hs{-2mm}
\ig[height=47mm]{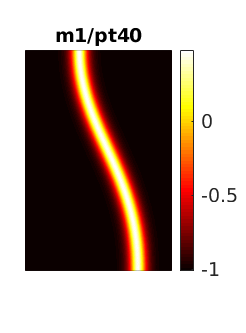}\ig[height=47mm]{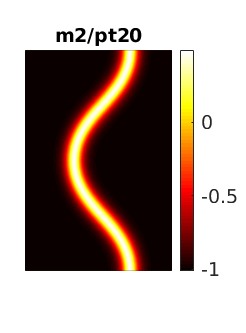}&
\hs{-4mm}\ig[height=49mm]{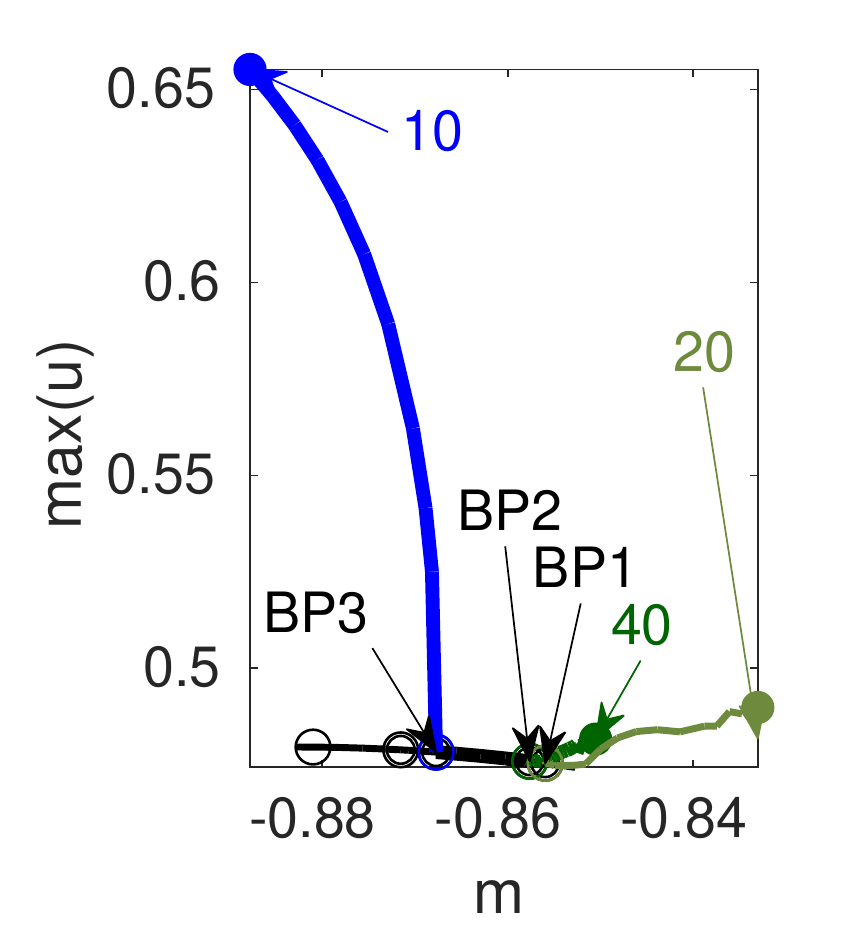}
}\\
\btab{ll}{
{\sm (c) bifurcation directions at primary bifurcations}
&{\sm (d) mesh}\\
\raisebox{6mm}{\hs{-3mm}\ig[height=33mm]{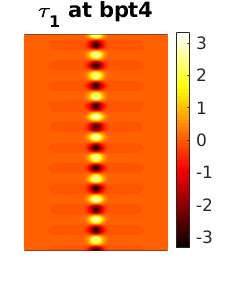}
\hs{-3mm}\ig[height=33mm]{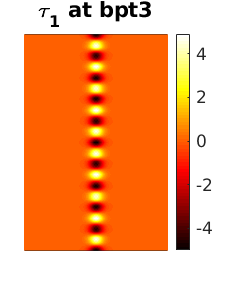}
\hs{-3mm}\ig[height=33mm]{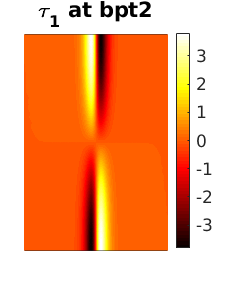}\hs{-3mm}\ig[height=33mm]{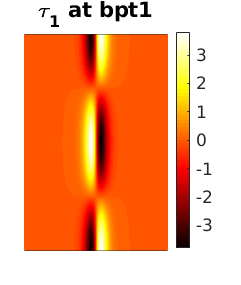}
}
&
\hs{-6mm}\ig[height=36mm]{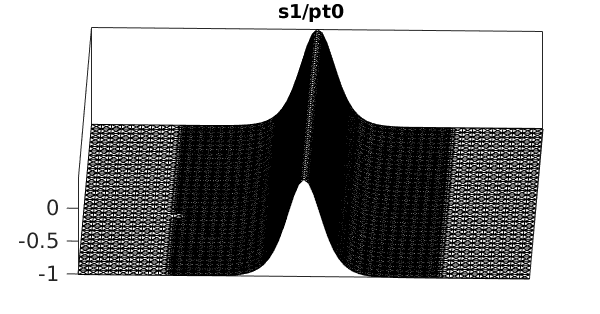}
}}
\ece 

\vs{-5mm}
\caption{{\small Example results from {\tt cmds1}, $\Om=(-2,2)\times (-3,3)$, 
$\eps=0.25, \eta_1=1, \eta_2=2$, continuation in $m$ starting with initial  
in the form of straight chanel with $\approx=-0.853$, which is unstable. 
Continuation to smaller $m$ first yields meandering BPs (green branches), 
then a 
stable segment for the straight channel, then pearling 
(blue branch {\tt p1}), as also illustrated in the  
bifurcation directions in (c). (d) is meant to illustrate the straight 
channel shape and the used mesh. 
 } \label{fchf1}}
\end{figure}


The starting straight channel in Fig.~\ref{fchf1} is obtained from the guess 
\huga{\label{fchiguess}
\text{$u_{ig}(x)=-1+a/\cosh(bx), \quad 
v_{ig}=-W'(u_{ig})$}}
with $a=1.25$, $b=20/3$, which yields the initial mass $m=m_0\approx -0.853$, 
followed by a Newton loop for \reff{sfchs}--\reff{fchpc}. 
It turns out that due to the large variety of patterns 
possible for \reff{sfchs}, a careful choice and refinement of meshes 
is important, 
in particular mesh symmetry. Thus, here we use a 
{\em very} ad--hoc refinement strategy, i.e., we replace 
the error estimator {\tt e2rs} by {\tt e2rs\_ad\_hoc}, 
which simply selects the triangles according $|u+2|$. 
For the initial guess \reff{fchiguess}, a 1--step mesh refinement 
from an initial criss--cross mesh of about 3500 points  
then yields the first solution $u_0$ in Fig.~\ref{fchf1}(a), on a mesh 
of about 9000 points, refined around the $x=0$ line, and we keep 
this mesh for all the subsequent computations.%
\footnote{At the end of {\tt cmds2.m} we use \trulle\ 
to generate meshes which are clearly much better adapted to specific 
chosen solutions. However, using these meshes for continuation, 
with or without further adaptation during continuation, makes 
the continuation less robust: Often, genuine bifurcations on 
the symmetric mesh (which are also expected from theory) become 
imperfect bifurcations on the (anisotropic) \trulle\ meshes.} 

The straight channel solution $u_0$ at $m=m_0$ 
is unstable with $2$ unstable eigenvalues of $\pa_u G$ 
for \reff{sfchs} (which do not take into account the constraints 
\reff{sfchs3} and \reff{fchpc}), and for DNS of \reff{tfch}. 
These DNSs are again based on the formulation as the DAE system 
\reff{sfchs}, with both \reff{sfchs3} and the phase condition \reff{fchpc} 
switched off, and with $\ga=0$. We 
implement this by a small modification of 
the library function {\tt tintxs} and of the 
nonlinearity function {\tt nodalf} used for \reff{sfchs}  
to account for $\Pi$ from \reff{fchg1}. 
Continuing $u_0$ to smaller $m$, we first find  two BPs BP1 and BP2 
at $m\approx -0.856$  and $m=-0.858$, respectively, 
after which the straight channel 
is stable until BP3 at $m\approx -0.868$, followed by 
BP4 at $m\approx -0.872$ and further BPs at lower $m$. 
BP1 and BP2 are associated to meandering, while BP3 and BP4 yield 
pearling, see (c) for the bifurcation directions, 
which illustrate the dichotomy between pearling 
at low $m$ (dilute mixture) and bending at 'high' $m$ (excess mass of 
the molecules).
The primary bifurcating meandering dark green branch {\tt m1} is stable after a 
fold. The primary blue pearling branch {\tt p1} bifurcates supercritically 
and is stable until $m\approx -0.9$ (secondary bifurcation not shown).  
Similar 
pearling branches bifurcate from the further BPs on the straight 
channel branch at lower $m$.   
Note that all these (pearling and meandering) 
bifurcations respect the (physical) 
constraint \reff{sfchs3}, and the numerical constraint \reff{fchpc}, 
which is needed for the continuation of the steady 
branches: Omitting it, we get an eigenvalue very close to $0$ 
from shifting a solution in $x$, 
and relatedly we observe that some solutions drift during the 
continuation, even on the rather narrow domain. 
The pearling instabilities at low $m$ only depend 
weakly on the transverse length $l_y$, as they are of relatively 
high wave number in $y$, but of course the bending may set in earlier on 
longer domains. 

\begin{figure}[ht]
\bce
\btab{l}{
\btab{ll}{
{\sm (a)}&{\sm (b)}\\
\ig[height=42mm]{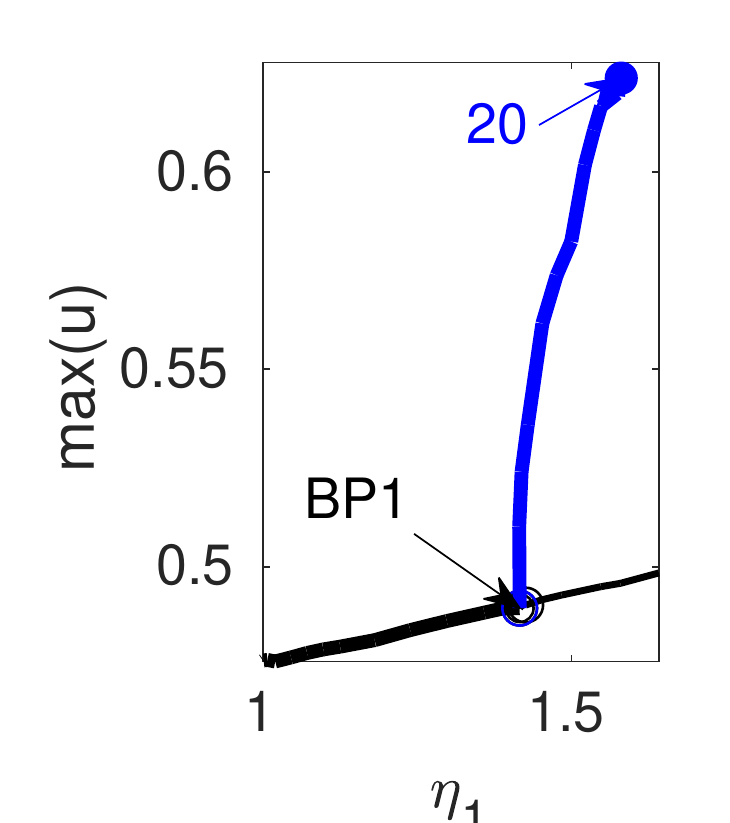}&\ig[height=42mm]{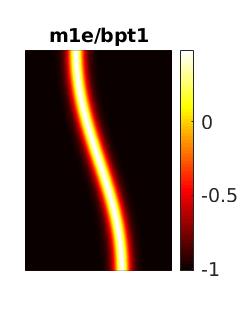}
\ig[height=42mm]{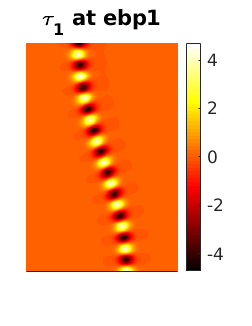}\ig[height=42mm]{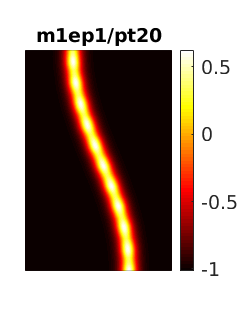}
}\\
{\sm (c)}\\
\ig[height=42mm]{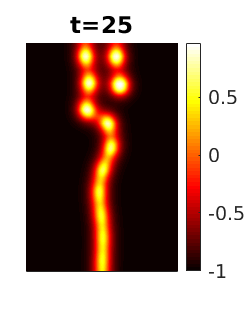}\ig[height=42mm]{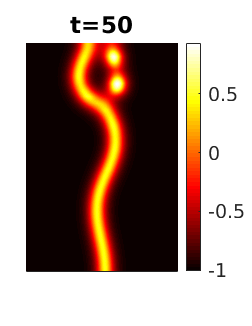}
\ig[height=42mm]{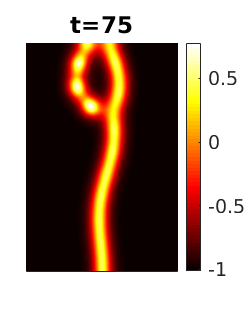}\ig[height=42mm]{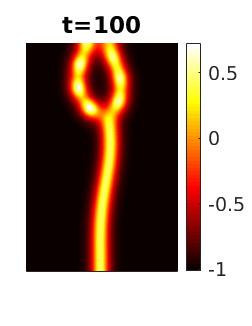}
\ig[height=42mm]{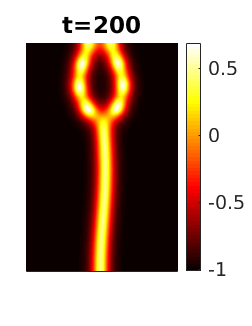}
}
\ece 

\vs{-5mm}
\caption{{\small  (a) Continuation of {\tt m1/pt10} in $\eta_1$, giving 
pearling, with sample solutions and tangent in (b). Snapshots from the 
evolution starting from near {\tt s1} at $m=-0.8$. At $t=200$ the solution 
is quasi steady, and a Newton loop for the steady problem yields a 
(stable) steady solution with no visible change. The pearled loop at 
the top is characteristic for such DNS, i.e., occurs (located somewhere 
along a channel) in the vast majority of simulations.  
} \label{fchf2}}
\end{figure}


In \cite[Fig.5.1, Fig.5.2]{DHPW12}, an example is given where 
increasing $\eta_1$  yields pearling of a curved (circular) channel. 
To see this, in {\tt cmds2} and Fig.~\ref{fchf2}(a,b) we continue 
the meandering solution {\tt m1/pt10} in $\eta_1$. In 
Fig.~\ref{fchf2}(c) we show snapshots from a 
DNS starting near the unstable straight channel 
at $m=-0.8$ (with zero mass perturbation) 
which illustrates the typical behaviour at 'high' $m$ ($m>-0.82$, say;  
of course, the $m$ here also depends on the domain size in $x$, i.e., 
the same channels on domains twice as large in $x$ would yield $m$ values 
much closer to $-1$). 
Initially, some (non--small wave number) pearling occurs, 
but most of the spots recombine to a curved channel, with a 
 pearled loop at the top. 
At $t=200$ the solution is already quite close to a steady state, i.e., 
the residual is below $10^{-8}$, and 
further DNS does not yield visible changes. Thus, we use  
this solution as initial guess for a Newton loop and obtain 
convergence to a steady state without visible change. Conversely, 
similar experiments at low $m$ (close to $-1$) 
 typically yield convergence to straight pearled 
channels, i.e., no meandering at all.

\brem\label{fchrem}
The results in Fig.~\ref{fchf1} and Fig.~\ref{fchf2} only 
scratch the surface of a numerical exploration of the fCH equation 
\reff{tfch}, for which continuation is difficult due to the different 
length scales involved, and the many solutions supported. 
We further refer to \cite{CKP19} for remarkable 
DNS in 2D and 3D, which combined with analysis yield ``qualitative 
bifurcation diagrams'', which partition the parameter space into 
regions where pearling or meandering dominates. See also 
\cite{PW17} for modeling of biological systems (consisting of 
different lipids) by multi--component fCH systems. 
Nevertheless, we believe that numerical continuation and bifurcation 
as presented here may help to sort out the different parameter regimes 
more quantitatively. 
\eex\erem 

\subsection{Higher indeterminacy: Demo {\tt hexex}}\label{hexsec}
\def\dhome{./pftut/hexex}\def\dname{hexex}
As already said, while {\tt qswibra} and {\tt cswibra} work robustly 
for most of the example problems we considered, they are not fail safe: 
the underlying QBE \reff{qeqa} and CBE \reff{cbe} are only solved numerically via {\tt fsolve}, 
and whether a solution is found may depend on the initial guesses for 
the Newton loops, and thus may require some trial and error. 
Similarly, whether a solution $\al$ is correctly or incorrectly identified 
as isolated or non-isolated may depend on the tolerance for the 
Jacobian determinant. 
Therefore we provide the auxiliary arguments in {\tt aux} from Table \ref{swauxtab}, and the fallback routine {\tt gentau} from Algorithm \ref{mswiba}.

\begin{figure}[ht]
\bce 
\begin{tabular}{lll}
{\small (a) Two kernel vectors}&{\small (b) Bif.~ diagram}&{\small (c) solution plots}\\
\raisebox{29mm}{\begin{tabular}{l}
\ig[width=0.24\tew]{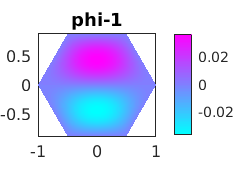}\\[-4mm]
\ig[width=0.24\tew]{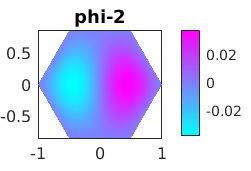}
\end{tabular}}&
\hs{-5mm}\ig[width=0.22\tew]{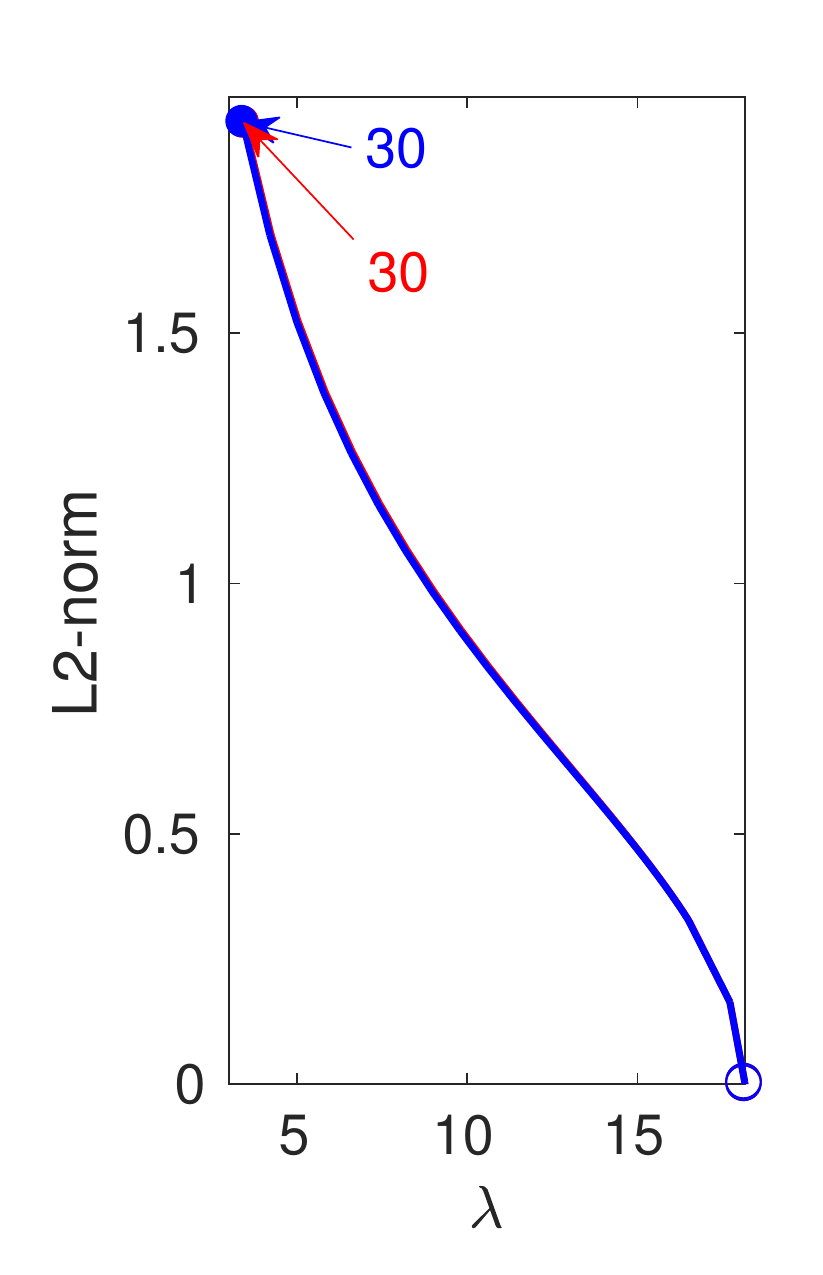}&
\raisebox{29mm}{\begin{tabular}{l}
\ig[width=0.22\tew]{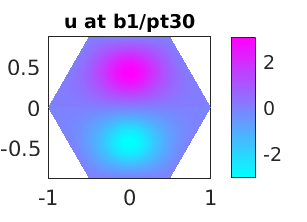}\\
\ig[width=0.22\tew]{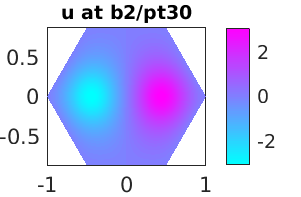}
\end{tabular}}
\end{tabular}
\ece
\vs{-5mm}
   \caption{{\small Results from the demo {\tt hexex} for bifurcations at the 
second BP for \reff{shex}. 
  \label{hf1}}}
\end{figure}

As an example, in the demo {\tt hexex} we consider 
a problem from \cite[\S6.8.2]{mei2000}, namely 
\huga{\label{shex}
G(u,\lam):=\Delta u+\lam(u+u^3)=0
}
on a hexagon with unit side-lengths and Dirichlet BCs. On the trivial branch 
$u\equiv 0$, there is a simple bifurcation point at $\lam=\lam_1\approx7.14$, a double bifurcation point at $\lam=\lam_2\approx 18$, and further bifurcations 
at $\lam=\lam_3\approx 32.5$ (double), $\lam=\lam_4\approx 37.6$ (simple), $\ldots$.  
See Fig.~\ref{hf1}(a) for the kernel vectors at $\lam_2$.  
At the simple BPs we can use {\tt swibra}. However, the problem is $D_6\times Z_2$ equivariant, and thus we expect pitchfork bifurcations at the multiple bifurcation points which are at best $5$-determined, cf.~\cite[Remark 3.1]{mbiftut}, and thus the bifurcation 
directions cannot be computed with {\tt cswibra},  which correctly reports 
that only non--isolated solutions $\al$ are found (with the default setting 
of ${\tt isotol}=10^{-10}$). 

Therefore we try {\tt gentau}, for instance with the natural choice 
$\ga=(1,0)$ and $\ga=(0,1)$. This turns out to immediately yield two 
bifurcating branches, i.e., the tangents to these branches coincide with the numerical kernel vectors. Moreover, for mixed choices 
of $\ga$, i.e., $\ga=(\ga_1,\ga_2)$ with $\ga_1\ga_2\ne 0$, if 
the first Newton loop converges, then the convergence is to one (isotropy class) of these two branches. In fact, this convergence occurs for a large majority 
of $\ga$ values, and only selected large vectors $\ga$ give non-convergence. 
In summary we conclude that 
exactly the two (classes) of distinct branches bifurcate, which fully agrees 
with the high-order determinacy analysis in \cite[\S6.8.2]{mei2000}. 
Thus,  {\tt gentau}, possibly 
with some trial and error, can be an efficient method to find all pertinent 
bifurcating branches of determinacy $k\ge 4$. 

The implementation of {\tt pftut/hexex} is fairly standard, and thus we 
refrain from detailed comments. The only non-obvious issue is how to 
generate the hexagonal domain in the OOPDE setting. For this we use 
the class definition {\tt hexpdeo}, for which we modify {\tt stanpdeo2D} 
and use the method {\tt grid.freeGeometry([x;y])}, see Listing \ref{hexl1}. 
Alternatively we could just pass {\tt x,y} from line 5 to {\tt freegeompdeo.m}. 

\hulst{caption={{\small {\tt \dname/hexpdeo.m}. Using {\tt grid.freeGeometry} to generate a hexagonal domain and coarse mesh, then 
do some uniform refinement. 
}}, 
label=hexl1, language=matlab,stepnumber=5, linerange=1-10, 
firstnumber=1}{\dhome/hexpdeo.m}

\subsection{A quasilinear system: Demo {\tt chemtax}}
\label{chemsec}
\def\dhome{./pftut/chemtax}\def\dname{chemtax}
In \cite[\S4.1]{p2p} we considered a reaction diffusion 
system with cross--diffusion as a quasilinear model problem, with the 
implementation based on the \mlab\ \ptool.  
The (stationary) problem reads 
\huga{\label{chem1} 
0=G(u):=-\bpm d\Delta u_1-\lam \nabla\cdot(u_1\nabla u_2)
\\\Delta u_2\epm 
-\bpm r u_1(1-u_1)\\ \frac{u_1}{1+u_1}-u_2\epm, 
}
with homogeneous Neumann BCs for $u_1, u_2$. The trivial branch 
for \reff{chem1} is $(u_1,u_2)=(1,1/2)$. We take the chemotaxis coefficient
$\lam\in\R$ as bifurcation parameter, fix 
$d=1/4$ and $r=1.52$, and now implement 
\reff{chem1} in the \oop\ setting, and, moreover, explain a general 
method for the efficient setup of Jacobians of nonlinear diffusion via {\tt numjac}. 

We split \reff{chem1} as 
\huga{\label{chem2}
G(u)=-D\Delta u-f(u)+\bpm\lam C(u)\\0\epm,
}
where $D=$diag$(d,1)$, and hence the first two terms are standard 
semilinear, as in, e.g., the Schnakenberg model \reff{tur1}. However, 
\huga{\label{cdt1} 
C(u)=\nabla\cdot(c(u_1)\nabla u_2), \quad c(u_1)=u_1, 
}
is a quasilinear cross--diffusion term, where we introduced $c(u)$ 
for generality. On the FEM level, \reff{chem2} becomes 
\huga{\label{chem3}
G(u)=\bpm dK u_1-\lam K_{12}(u_1)u_2\\Ku_2\epm -Mf(u), 
}
where $K$ is the standard 1-component Neumann-Laplacian, $M$ is 
mass matrix, and $K_{12}(u_1)u_2$ implements 
$\nabla\cdot(c(u_1)\nabla u_2)$. Thus, $K_{12}$ has to be assembled 
in each call to {\tt sG},  see 
Listing \ref{ctl1} below. 

The Jacobian reads 
\hual{
G_u(u)\bpm v_1\\ v_2\epm{=}&
\left[{-}\bpm d\Delta&{-}\lam\nabla\cdot (u_1\nabla \cdot)\\
0&\Delta\epm
{+}\bpm r(2u_1{-}1)&0\\
{-}(1+u_1)^{{-}2}&1\epm\right]
\bpm v_1\\ v_2\epm
{+}\lam\bpm \pa_{u_1}\nab\cdot(v_1 \nabla u_2)\\0 \epm, \label{chem1j}
}
and the last term is problematic from the FEM point of view as it contains 
$\Delta u_1$ and the FEM is based on the weak formulation. 
In \cite[\S5]{actut} (demo {\tt acsuite/acql}), 
we treat the related problem of a quasilinear Allen--Cahn equation, where 
we approximate a term 
$$
\nab\cdot((c_u(u)\nab u) v)=\pa_x((c_u(u)u_x) v)+\pa_y((c_u(u)u_y) v)
$$ 
as $K_x(c_u(u)u_x v)+K_y(c_u(u)u_y v)$. 
Thus, in {\tt acql} we use the first order differentiation FEM matrices $K_x, K_y$, generated 
via {\tt p.mat.Kx=fem.convection(grid,[1;0])} 
and {\tt p.mat.Ky=fem.convection(grid, [0;1])}, respectively, 
and approximations of $\pa_x u$ and $\pa_y u$ via 
differentiation matrices {\tt Dx} and {\tt Dy}. This gives a rather large 
relative error between the Jacobians thus obtained and the numerical 
Jacobians, but the approximation is fast, and the continuation (the 
Newton loops) with these rough approximations of Jacobians works. 

On the other hand, this trick (cf.~Listing \ref{ctl1}) does not work 
in general for \reff{chem1}, i.e., the continuation fails for 
some of the branches. Thus, we use {\tt numjac} to obtain the problematic 
term $\pa_{u_1}\nab\cdot(c(u_1)\nabla u_2)$ in an extra function {\tt getKuvd}. 
For this we put the (here very simple) 
function $c(u_1)=u_1$ into the function handle {\tt p.fuha.cfu=@cfu}. 
The resulting ``hybrid'' {\tt sGjac} gives a relative error of order 
$10^{-8}$ to the full {\tt numjac} (called for {\tt p.sw.jac=0}), 
but is about one magnitude faster, and continuation (and BP detection/localization and subsequent branch switching) work without problems. Clearly, this setup 
can be generalized to other quasilinear problems, and in this sense 
the library functions {\tt Kuv.m} and {\tt getKuvd.m} (which also work for the quasilinear Allen-Cahn model, see {\tt acsuite/acql}) should be seen as templates for adaption to 
a given problem. 


\hulst{language=matlab,stepnumber=0, linerange=1-8, 
}{\dhome/sG.m}
\hulst{caption={{\small {\tt \dname/sG.m} and {\tt sGjac}, using {\tt getKuvd} 
to numerically get $\pa_u\nab\cdot(c(u)\nabla v)$. 
}}, 
label=ctl1, language=matlab,stepnumber=0, linerange=1-16}{\dhome/sGjac.m}

\begin{figure}[ht]
\bce 
\begin{tabular}{ll}
{\small (a) BD}&{\small (b) sample solutions}\\
\raisebox{-4mm}{\hs{-2mm}\ig[width=0.29\tew]{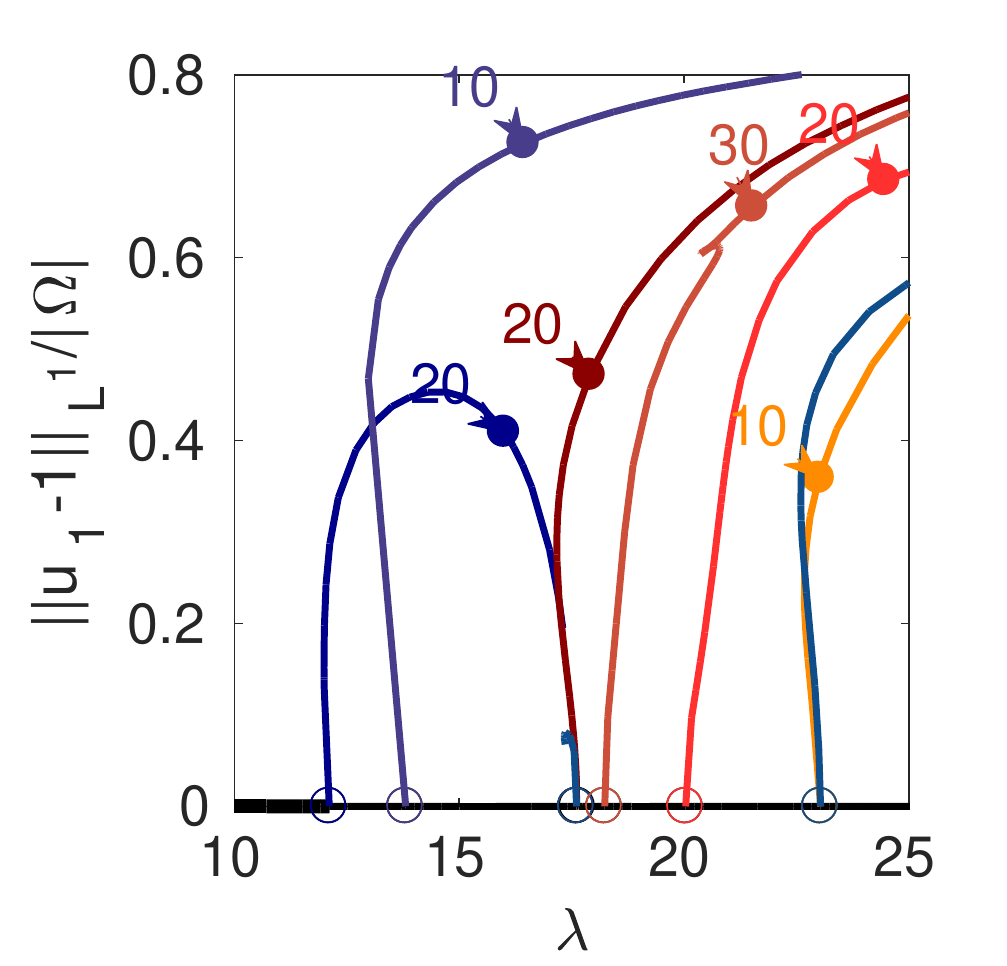}}&
\hs{-2mm}\ig[width=0.12\tew]{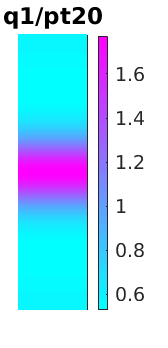}
\hs{-1mm}\ig[width=0.12\tew]{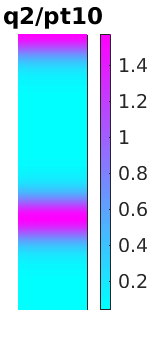}
\hs{-1mm}\ig[width=0.12\tew]{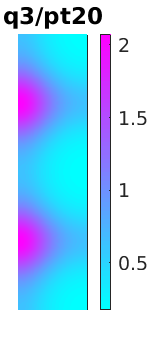}\hs{-1mm}\ig[width=0.12\tew]{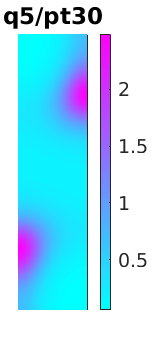}
\hs{-1mm}\ig[width=0.12\tew]{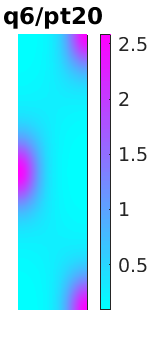}\hs{-1mm}\ig[width=0.12\tew]{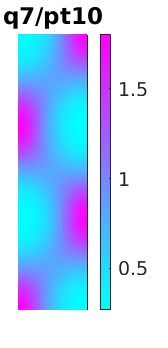}
\end{tabular}
\ece
\vs{-5mm}
   \caption{{\small Basic BD and example solutions for the chemotaxis model 
\reff{chem1} over $\Om=(-0.5,0.5)\times (-2,2)$. Stripe branches {\tt q*} 
in shades of blue, and spots {\tt q*} in shades of red, with * increasing 
left to right.   \label{cf1}}}
\end{figure}

Finally, in the script file {\tt cmds2D.m} we also use a dirty little 
trick wrt to the localization of BPs and branch-switching: for BP detection and localization we use the {\tt p.sw.bifcheck=2} setting with rather large {\tt ds}. However, some  
of the BPs on the trivial branch are rather close together, and thus in some 
steps multiple (distinct) eigenvalues cross the imaginary axis. In this case, 
after localization we use {\tt cswibra} with {\tt aux.besw=0} to only 
compute (approximate) kernel vector. Subsequently calling {\tt gentau} 
with the different respective kernel vectors succeeds, even though the 
branches in general do not ``start'' at the correct $\lam$ values. 
See Fig.~\ref{cf1} for the basic BD. 

\subsection{Global coupling, and customized linear system 
solvers: Demo {\tt shgc}}\label{gcsec}
\def\dhome{./pftut/shgc}\def\dname{shgc}
Interesting phenomena in pattern formation for RD systems (or SH type of 
equations) can occur under additional nonlocal or global coupling \cite{FCS07, MORGAN2014, KT17, SIERO2018}. Here 
we explain a setup such that equations of type $M\pa_t u=-G(u)$, 
respectively the steady version $G(u,\lam)=0$, $u:\Om\ra\R^N$,  can augmented 
by {\em global} coupling in the fairly general form 
\def\fnl{f_{\rm nl}}\def\ddu{\frac{{\rm d}}{{\rm d}u}}
\huga{\label{gcgen}
0=G(u)+\fnl(u,a), \quad a=\spr{h(u)}, 
}
where $\fnl:\R^{N+1}\ra\R^N$ and $h:\R^N\ra\R$ are general functions, 
and $\spr{v}=\frac 1 {|\Om|}\int v(x)\dd x$ denotes a global average.  
See Remark \ref{nlrem} for comments on {\em non--local} (but also non--global) 
coupling. Naturally, $\fnl$ and $h$ may also depend on parameters, and 
on $x$, such the averaging in $\spr{h}$ can be weighted. 

A naive implementation of \reff{gcgen} yields full Jacobians, i.e., 
\huga{\label{gcgj}
\ddu\fnl(u,a)v=\pa_u\fnl(u,a)v+\pa_a\fnl(u,a)\spr{h_u(u) v}.
}
In the FEM discretization, the first term is sparse, 
and the second 
is a full matrix, but of rank 1. Thus, the 
purpose of this section is to explain a setup where this rank-1-correction 
can be treated efficiently by using Sherman--Morrison--Woodbury (SMW) formulas  \cite[\S2.7.3]{numrc}. 
This extends \cite[\S4.3]{p2p}, where the idea was already used for a 
simple scalar problem with a simple linear global coupling%
\footnote{this has also been generalized in the demo {\tt acsuite/acgc}}. 
Moreover, we also provide customized interfaces to the \mlab\ function 
{\tt eigs} for computing eigenvalues, 
such that also spectral computations and hence bifurcations can be treated 
without ever forming the full Jacobian. 

\begin{figure}[ht]
\bce 
\begin{tabular}{ll}
(a)&(b)\\
\hs{-3mm}\ig[width=0.27\tew]{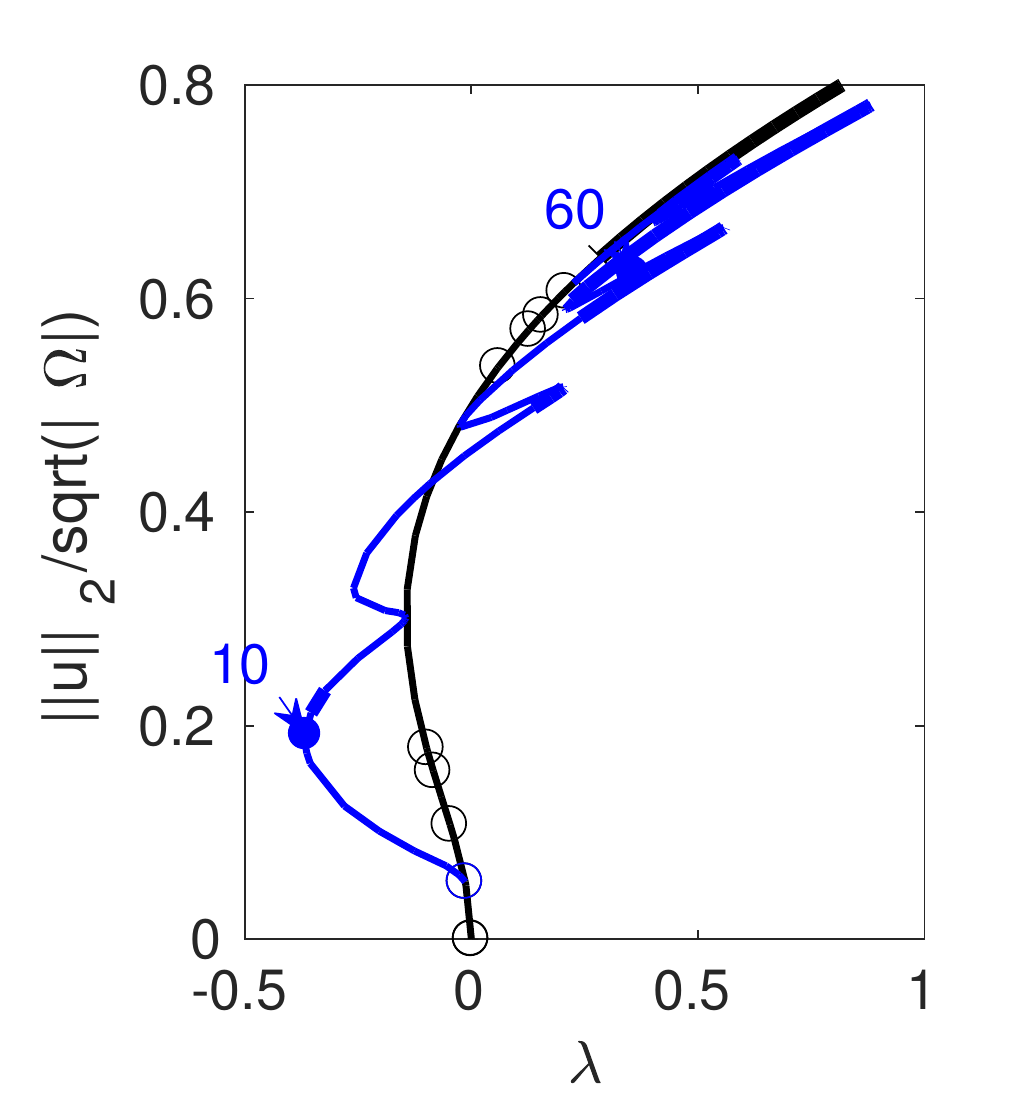}
\hs{-5mm}\raisebox{24mm}{\begin{tabular}{l}
\ig[width=0.26\tew]{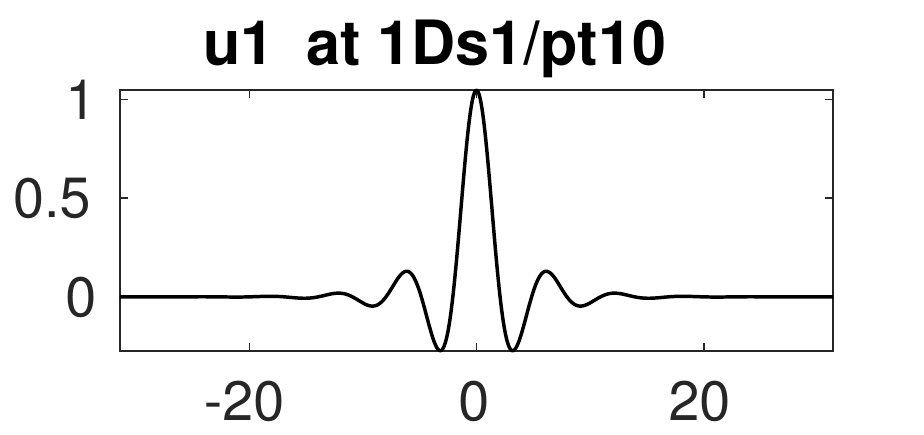}\\
\ig[width=0.26\tew]{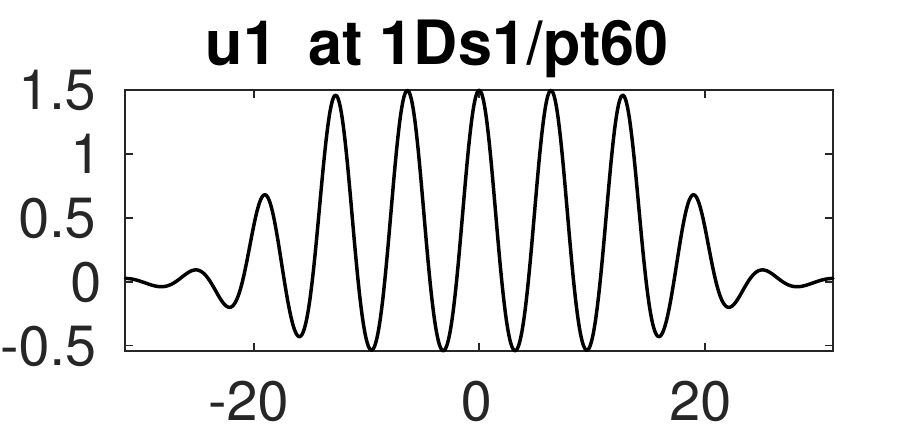}
\end{tabular}}
&\hs{-9mm}\ig[width=0.26\tew]{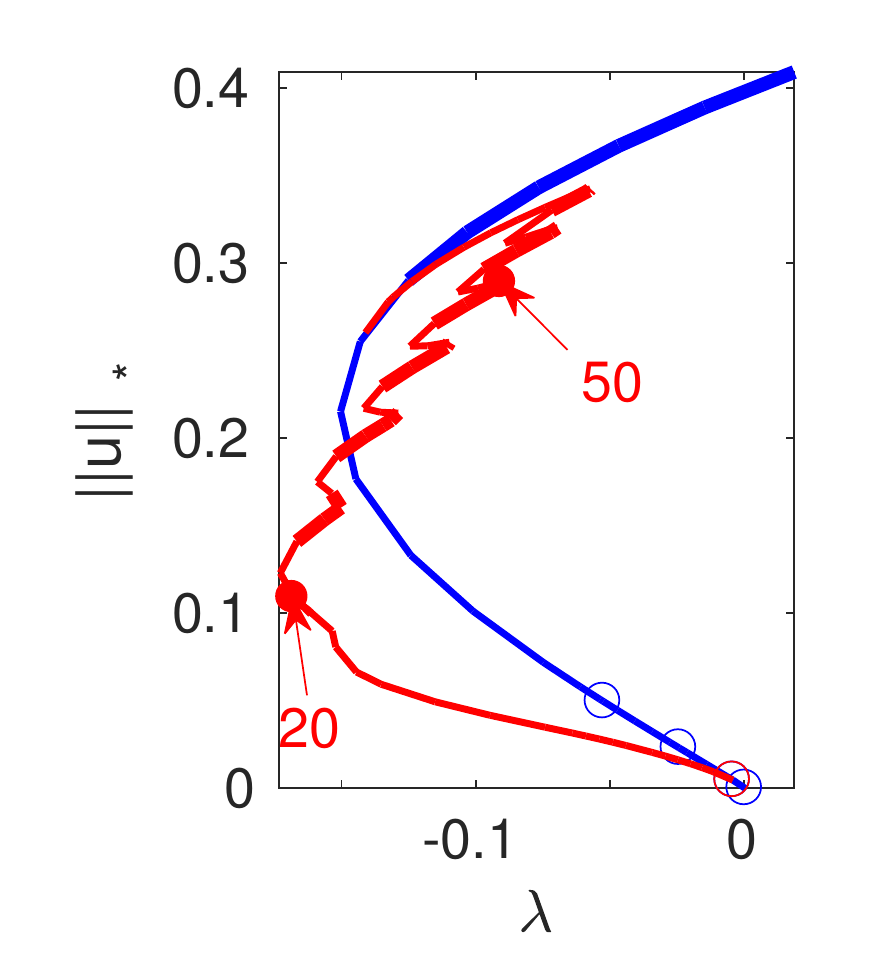}
\hs{-5mm}\raisebox{25mm}{\begin{tabular}{l}
\ig[width=0.27\tew]{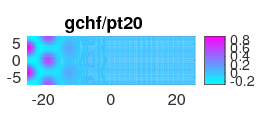}\\
\ig[width=0.27\tew]{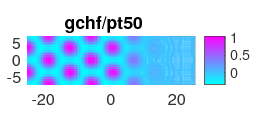}
\end{tabular}}
\end{tabular}
\ece 

\vs{-5mm}
\caption{{\small Slanted snaking for the globally coupled 2-3 SH equation 
\reff{gcsh}. (a) $\Om=(-10\pi,10\pi)$, $\nu=2, \ga=2$, compare to 
Fig.~\ref{shf1}. (b) $\Om=(-l_x,l_x)\times(-l_y,l_y)$, $l_x=8\pi, l_y=4\pi/\sqrt{3}$, $\nu=1.3$, $\ga=1$, compare to 
Fig.~\ref{shf11}. \label{shf13}}}
\end{figure}

A prototype problem is the globally coupled (quadratic--cubic) SH equation 
\huga{\label{gcsh} \pa_t u = -(1+\Delta)^2 u + \lam u  +\nu u^2-u^3-\ga\|u\|^2u, 
}
with parameters $\nu,\ga\in\R$, $\|u\|^2:=\frac 1 {|\Om|}\int u^2(x)\dd x$, 
and (again) Neumann BCs $\pa_n u=\pa_n\Delta u=0$, extending \reff{swiho}, and for instance considered in \cite{FCS07}. 
For \reff{gcsh} with $\ga>0$, the nonlocal term simply acts 
as a reduction of the instability parameter $\lam$: steady solutions  
$u(x;\lam,\ga)$ of \reff{gcsh} correspond to steady solutions 
$u(x;\lam-\ga\|u\|^2,0)$ of \reff{gcsh} with $\ga=0$. In particular, for 
instance the branches of periodic and localized solutions from Figs.~\ref{shf1} and \ref{shf11} get slanted to the right for $\ga>0$. Thus we obtain 
a slanted snaking, and in particular the snakes can move out of the 
bistable range of $0$ and the periodic patterns, which in \cite{FCS07} 
is proposed as a mechanism for the prevalence of localized states in certain 
systems. See Fig.~\ref{shf13} for some exemplary results obtained from numerical continuation of \reff{gcsh}.

Here we are mainly interested in the efficient implementation of \reff{gcgen} in \pdep, and use \reff{gcsh} only as an example, essentially ignoring the scaling relation to \reff{swiho}.  
Again setting $(u_1,u_2)=(u,\Delta u)$, 
\reff{gcsh} becomes, 
\huga{\label{gcshsys}
\bpm 1&0\\0&0\epm \pa_t \bpm u_1\\ u_2\epm=
\bpm -\Delta u_2-2u_2-(1-\lam)u_1+f(u_1)-\ga\|u_1\|^2u_1\\-\Delta u_1+u_2\epm, 
}
$f(u)=\nu u^2-u^3$, which is of the form \reff{gcgen} with 
$\dst \fnl(u,a)=\bpm -\ga a u_1\\ 0\epm\text{ and } h(u)=u_1^2$. 
On the FEM level we obtain 
\huga{\label{gcshfem}
\CM\dot u=-(\CK u-F(u)-\CM \fnl(u,a)),} 
$\CM, \CK, F$ as in \reff{shfem}, where $\fnl(u,a)$ now means the 
vector $\fnl(u,a)=-\ga a(u_{1,1},\ldots,u_{1,n_p},0,\ldots,0)^T$, 
and where $a=\spr{h(u)}$ is evaluated as%
\huga{\label{huform}
\spr{h(u)}=a_{{\rm av}}h(u),\quad a_{{\rm av}}=\frac 1 {|\Om|}
{\tt sum}(\CM),\quad 
h(u)=(h(u_{1,1}),\ldots,h(u_{1,n_p}),0,\ldots,0). 
}
Since here $\CM=\bpm M&0\\0&0\epm$, the last $n_p$ entries of $a_{{\rm av}}$ are 
zero anyway, but in a general setup we would rather recommend 
$a_{{\rm av}}=\frac 1 {|\Om|}({\tt sum}(M),\ldots,{\tt sum}(M))$ ($N$ copies 
of the column sums of the 1-component mass matrix $M$), which of course 
could be used here as well. 
Similarly, the nonlocal part $A_{{\rm nl}}=\pa_a\fnl(u,a)\spr{h_u(u) \cdot}$ of the Jacobian 
from \reff{gcgj} becomes 
\huga{\notag 
A_{{\rm nl}}=(\CM\pa_a\fnl(u,a)) a_{{\rm jac}}\in\R^{n_u\times n_u}, \text{ where }
\pa_a\fnl\in\R^{n_u\times 1}, \quad a_{{\rm jac}}={\tt sum}(\CM{\rm diag}(h_u))\in \R^{1\times n_u},}
which again illustrates that $A_{{\rm nl}}$ is a full matrix but has but rank 1. 
The $\CM$ in $(\CM\pa_a\fnl)$ comes from the $\CM$ multiplying $\fnl$ in \reff{gcshfem}, while the $\CM$ in $a_{{\rm jac}}$ is as in \reff{huform}. 

The SMW formula for solving linear systems 
$(K-c a^T)z=b$ reads 
\huga{\label{smf} 
z=K^{-1}b+\al(K^{-1}c)(a^TK^{-1})b, \quad \al=\frac 1{1-a^TK^{-1}c}.  
}
Thus, the idea is as follows: in {\tt sGjac} we separately 
assemble the sparse part $G_u(u)+\pa_u\fnl$ 
of $G_u$ corresponding to $K$ in \reff{smf}, 
and the vectors $c=\pa_a\fnl$ and $a_{{\rm jac}}$. We then pass these 
to SMW linear system solvers, 
also in the inverse vector iteration underlying {\tt eigs} for 
spectral computations. Since the default interfaces 
for the linear system solvers do not account for the vectors 
$c$ and $a=a_{{\rm jac}}$, these are passed via the 
global \pdep\ struct {\tt p2pglob}. 
Table \ref{gcshtab} lists the pertinent 
files from demo {\tt gcsh}, and some functions from {\tt libs} which 
have not yet been documented otherwise, and Listings \ref{gcshl1}--\ref{gcshl3} 
show the essential modifications compared to the demo {\tt sh}.  

\begin{table}[ht]\taskip
\caption{Scripts and functions in {\tt /demos/gcsh}, with comments on the changes 
compared to {\tt /demos/sh}, and functions from {\tt /libs/linalg} and 
{\tt libs/p2p} pertinent to global coupling problems. 
\label{gcshtab}}
\bce\vs{-4mm}
{\small 
\begin{tabular}{l|p{0.72\textwidth}}
script/function&purpose, remarks\\
\hline
cmds1d, cmds2dhexfro&scripts for 1D and 2D examples, see Fig.~\ref{shf12} \\%
shinit&init, sets {\tt p.fuha.lss=@gclss;p.fuha.blss=@gcblss} 
as linear system solvers \\
oosetfemops&set FEM matrices, also stores the 'averaging vector' 
{\tt p.avvec}\\
sG,nodalf,sGjac&encodes $G$ with 'nonlinearity' in nodalf, and Jacobian; sG also 
 sets the {\tt global cvec}, and {\tt sGjac} the {\tt global avjvec} needed 
by the linear system solvers. \\
shbra1d&modification of {\tt stanbra} for putting the normalized $L^2$ on the branch \\
fnl, hfu&functions $\fnl$ and $h$ for global coupling from \reff{gcsh}\\
fnljac, hjac, fnl\_a&Jacobians of $\fnl$ and $h$, and $\pa_a\fnl$.\\
\hline
gclss&implements \reff{smf}, with vectors $c={\tt cvec}$ and $a={\tt avec}$ 
passed in {\tt p2pglob} \\
gcblss&implements the version of \reff{smf} for the bordered systems of 
arclength continuation; here $a$ and $c$ are augmented by a single $0$\\
gclsseigs&version of \reff{smf} for {\tt eigs} (inverse vector iteration); 
uses {\tt lsslueigs} to solve $Az{=}b$, where $A$ is $LU$-prefactored due for repeated solves with the same $A$.\\
gcafun&interface routine for {\tt eigs} which contains the actual call to {\tt gclsseigs}\\
lsslueigs&version of {\tt lsslu} which stores $LU$ factorizations as globals. 
\end{tabular}
}
\ece
\end{table}\teskip

\hulst{caption={{\small {\tt \dname/shinit.m} (lines 1-6). The 
diffferences to {\tt sh/shinit.m} are in lines 3-5. }}, 
label=gcshl1, language=matlab,stepnumber=10, linerange=1-5, 
firstnumber=1}{\dhome/shinit.m}

\hulst{
label=gcshl2, language=matlab,stepnumber=10, linerange=1-4, 
firstnumber=1}{\dhome/sG.m}

\hulst{stepnumber=0, language=matlab}{\dhome/fnl.m}
\hulst{caption={{\small {\tt \dname/sG.m}, {\tt \dname/fnl.m} and 
{\tt \dname/hfu.m}, which are straightforward, as are the Jacobians and fnl\_a. {\tt p.avvec} is precomputed in {\tt oosetfemops}.}}, 
label=gcshl4, language=matlab, stepnumber=10}{\dhome/hfu.m}

\hulst{caption={{\small {\tt \dname/sGjac.m}. 
In {\tt sG}, the term {\tt fnl} is first added to the 'nonlinearity' $f$, and then  multiplied by $\CM{=}{\tt p.mat.M}$. In other words, {\tt fnl} and 
hence also {\tt fnljac} and {\tt fnl\_a} 
contain no $\CM$, and for the derivatives $\CM$ is taken into account here. 
}}, 
label=gcshl3, language=matlab,stepnumber=10, linerange=1-8, 
firstnumber=1}{\dhome/sGjac.m}

With the modification of the lss setup, the script files for \reff{gcsh} are as for \reff{swiho}, see {\tt gcsh/cmds1d.m} and {\tt gcsh/cmds2dhexfro.m}, 
and the computations for \reff{gcsh} run almost as fast as for \reff{swiho}. In 3D (or, more generally, for large $n_u$) it turns out that combining {\tt gclsseigs} 
with iterative linear system solvers yields further speed advantages, but 
this will be described elsewhere. 


\brem\label{nlrem}
More general nonlocal couplings are often given as $\dst f_{{\rm nl}}(u)(x)=
\int_\Om \kappa(x,\xi)h(u(\xi))\dd \xi$, with a kernel $\kappa:\Om\times\Om\ra \R$, 
often of the form $\kappa(x,\xi)=k(x-\xi)$, specifically with Gaussians  
$k_{{\rm Gauss}}(y)=\al_1\er^{-\|y\|^2/\al_2}$. Our global coupling corresponds to $k\equiv 1$. 
For general nonlocal couplings, Jacobians naturally are again nonlocal, and importantly 
no longer of the form ``local + rank--1--correction''. However, 
preliminary results indicate that at least for fast-decaying kernels, 
e.g., small $\al_2$ in $k_{{\rm Gauss}}$, the full Jacobians can be well 
approximated by reasonably sparse Jacobians by dropping entries below a certain 
threshold. This way, nonlocal nonlinearities can still be treated 
efficiently in \pdep, including a bifurcation analysis. 
Details will be given elsewhere. \eex 
\erem

\section{Pattern formation on curved surfaces}\label{pfsurf}
In \cite[Chapter 15]{Mur}, see also,  
e.g., \cite{VAB99, KEG18, KBF18} and the references therein,  
and \cite{NLCS17} for a review, 
the influence of the geometry and specifically curvature 
on biological pattern formation on surfaces is discussed. 
Similarly, in \cite{Stoop15}, a Swift-Hohenberg equation 
on spheres and tori is used as a model for elastic surface 
patterns, and one interesting result, obtained via time--integration, 
is that a higher curvature favors hexagons, and lower 
curvature favors stripes, possibly with defects, or, 
more generally, labyrinths. Additionally, there are a number of results 
on pattern formation on curved surfaces in certain singular 
limits where spots become strongly localized (i.e., far from onset), 
and where the 
interaction of spots is described by DAEs for the spot locations. 
See, e.g., \cite{JTW16, TT19} and the references therein. 
For this, one major step in the analysis are accurate approximations 
of certain Green's functions on the curved surface, and an interesting 
(semi-analytical) result is that on tori spots slowly drift towards the 
inner radius. 

Already in 1953  \cite{tur53}, see also \cite{tura}, Turing aimed at explaining the patterns formed by the silica skeletons of radiolaria (see Fig.~\ref{ssf2}(a) on page \pageref{ssf2}) via two component RD models, 
and computed the diffusion driven instability of RD systems 
based on the spherical harmonics as eigenfunctions of the 
spherical Laplace--Beltrami operator (LBO) $\DelS$. Since then, 
a lot of 'model-independent' theory on 
bifurcations with $O(3)$ symmetry has been developed, 
see, e.g., \cite{IG84,CLM90,Matt03, Matt04,Call04}, 
and one of our aims here 
is to use \pdep\ to recover some of these theoretical results, 
and to follow some pertinent branches, focusing on the Schnakenberg 
model on spheres, and on tori, which is simpler in the symmetry sense. 

Additionally, in \S\ref{accylsec} we patch together two surfaces via a 
common boundary, and in \S\ref{cpolsec} we study a simple 
model for cell polarization from \cite{CEM19}, consisting of 
linear diffusion in 
the 3D bulk $\Om$ of the cell (the cytosol), coupled a reaction--diffusion 
equation on the surface $\Ga=\pa\Om$ (the cell membrane). 

However, we start with (quadratic--cubic) Allen--Cahn (AC) equations on tori  
and spheres as simpler toy problems, namely 
\huga{\label{ACS} 
\pa_t u=-G(u)\stackrel !=0, \quad G(u)=-\DLB u-\lam u-u^2+\ga u^3,\quad \lam\in\R, 
\ga>0,}
where $\DLB$ denotes the LBO over 
the given surface. Equations similar to \reff{ACS} have 
also been used as model problems (over 1D, 2D and 3D 'flat' domains, i.e., 
with the standard Laplacian and various BCs) in \cite{actut}, and of course 
\reff{ch2} and \reff{shex} are also quite similar. 

We use a standard parametrization 
\huga{\label{top}
\bpm \xti\\\yti\\\zti\epm=\bpm (R+\rho\cos y)\cos x\\
(R+\rho\cos y)\sin x\\\rho\sin y\epm\in\R^3,\quad (x,y)\in\Omega=[-\pi,\pi)^2, 
}
of the (surface of the) torus $\CT_{R,\rho}$ 
with major radius $R>0$ and minor radius $0<\rho<R$. The associated 
LBO is denoted by $\Del_{\CT_{R,\rho}}$ and given by 
\huga{\label{LBt}
\Del_{\CT_{R,\rho}} u(x,y)=\frac 1 {\rho^2(R+\rho\cos y)}\pa_y((R-\rho\cos y)\pa_y u)
+\frac 1 {(R+\rho\cos y)^2}\pa_x^2 u, 
}
with periodic BCs in $x$ (think azimuth $\phi$) and $y$ (think elevation $\theta$). Thus, $\Del_{\CT_{R,\rho}}$ is translationally invariant in $x$, 
i.e., 
\huga{\label{Tx}
\text{$\Del_{\CT_{R,\rho}}S_\xi u(x,y)=S_\xi\Del_{\CT_{R,\rho}}u(x,y)$, 
where $S_\xi u(x,y)=u(x-\xi,y)$,}
}
and invariant under the two discrete mirror symmetries
\huga{\label{mxy} \text{$m_x u(x,y)=u(-x,y)$\quad and \quad
$m_y u(x,y)=u(x,-y)$},}
which will be important to determine whether bifurcations 
are pitchforks or transcritical. 

We parameterize spheres $S_R=\{\tilde x\in\R^3: \|\tilde x\|_2=R\}$ in a slightly 
non--standard way by 
\huga{\label{sphp}
\bpm \xti\\\yti\\\zti\epm=\phi(x,y):=R\bpm \cos y\cos x\\
\cos y\sin x\\\sin y\epm\in\R^3,\quad (x,y)\in\Omega=[-\pi,\pi)\times 
(-\pi/2,\pi/2),  
}
such that $S_R=\phi(\Om)\cup\{N,S\}$, where $N,S=(0,0,\pm R)$, and the 
equator is at $y=0$. 
Here the associated LBO is 
\huga{\label{LBS}
\Del_{S_R} u(x,y)=\frac 1 {R^2}\left(\frac 1 {\cos^2 y}\pa_x^2 u+\frac 1 {\cos y}\pa_y(\cos y\pa_y u)\right), 
}
with periodic BCs in $x$, and one reason why we first consider the torus 
is that the coordinate singularity of $\Del_S$ at $y=\pm\pi/2$ warrants 
some tricks. Moreover, the symmetry group of PDEs of type \reff{ACS} 
(and of system of such type) over $S_R$ is $O(3)$, 
which  makes the bifurcation behavior richer 
than over the torus. 

\subsection{An Allen--Cahn equation on tori: Demo {\tt actor}}\label{act-sec}
The (quadratic--cubic) Allen--Cahn equation \reff{ACS} over $\CT_{R,\rho}$ 
reads 
\huga{\label{tAC} 
G(u)=-\Del_{\CT_{R,\rho}} u-\lam u-u^2+\ga u^3\stackrel!=0,\quad \lam\in\R, 
\ga>0,}
with trivial solution  $u\equiv 0$. 
We choose 
$(R,\rho,\ga)=(2,1,1) \text{ as base parameters}$, 
and first continue in $\lam$, 
but subsequently also continue in $R$. The eigenvalues $\lam$ 
of $-\Del_{\CT_{R,\rho}}$ have for instance been computed (and continued in the 
ratio $R/\rho$) numerically in \cite{glo08}, 
and one crucial observation is that they are generically of multiplicity 
$1$ (with eigenfunctions independent of 'angle' $x$) or $2$ (with eigenfunctions 
dependent on $x$, which thus gives a two dimension kernel via shift 
by half a period). At certain ratios $R/\rho$ there are crossings of eigenvalues 
and hence eigenvalues of higher multiplicity, but we do not discuss these here. 

\begin{figure}[ht]
\bce 
\begin{tabular}{lll}
{\small (a) Bifurcation directions (kernel vectors) at {\tt b2,\ldots,b7} }&{\small (b) Basic BD}&{\small (c) Example soln}\\
\raisebox{28mm}{\begin{tabular}{l}
\ig[width=0.16\tew,height=24mm]{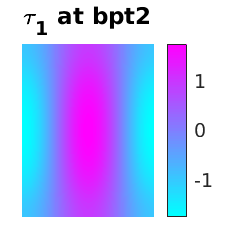}
\ig[width=0.16\tew,height=24mm]{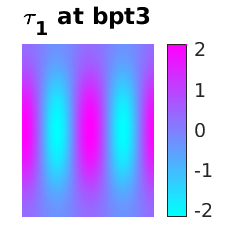}
\ig[width=0.16\tew,height=24mm]{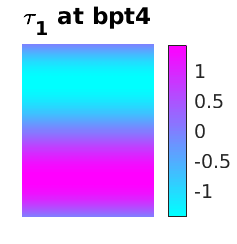}\\
\ig[width=0.16\tew,height=24mm]{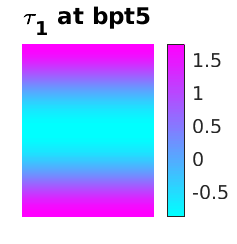}
\ig[width=0.16\tew,height=24mm]{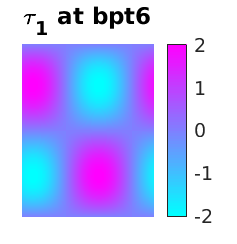}
\ig[width=0.16\tew,height=24mm]{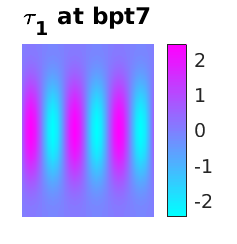}
\end{tabular}}&
\hs{-7mm}\ig[width=0.3\tew]{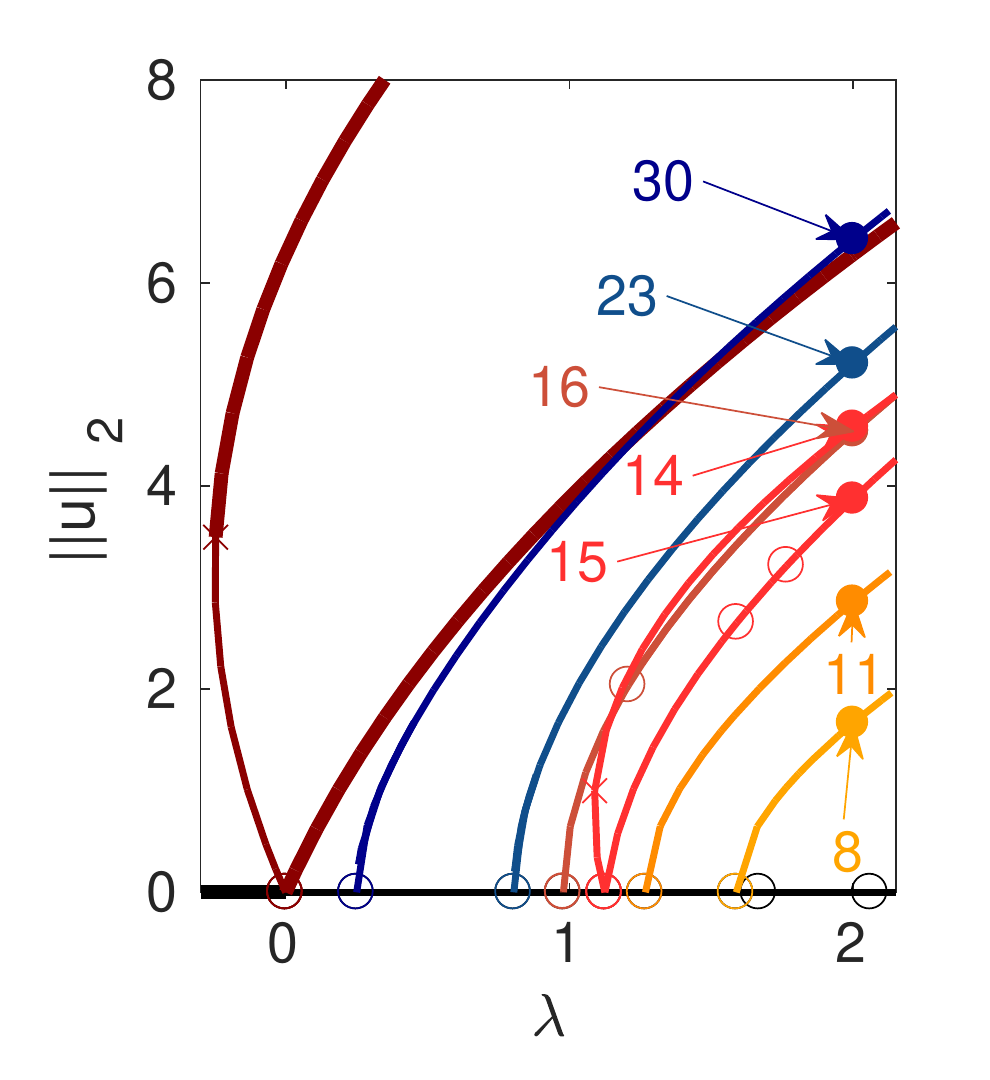}&
\hs{-2mm}\raisebox{26mm}{\begin{tabular}{l}
\ig[width=0.16\tew,height=24mm]{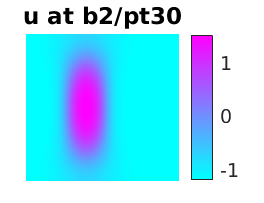}\\
\hs{-2mm}\ig[width=0.17\tew,height=24mm]{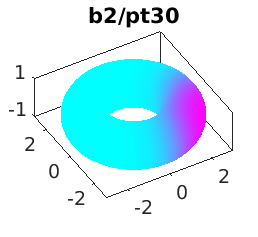}
\end{tabular}}
\end{tabular}\\
\begin{tabular}{l}
{\small (d) Further example solutions}\\
\ig[width=0.16\tew,height=24mm]{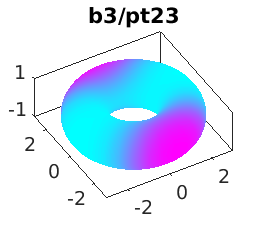}
\ig[width=0.16\tew,height=24mm]{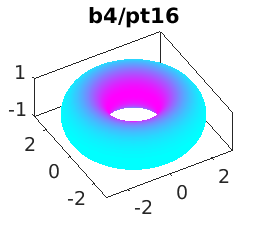}
\ig[width=0.16\tew,height=24mm]{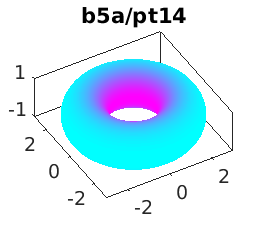}
\ig[width=0.16\tew,height=24mm]{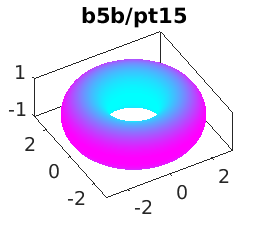}
\ig[width=0.16\tew,height=24mm]{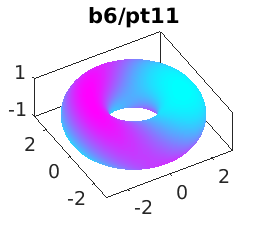}
\ig[width=0.16\tew,height=24mm]{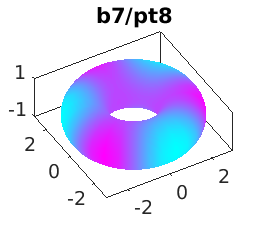}
\end{tabular}\\
\begin{tabular}{l}
{\small (e) Continuation of {\tt b2/pt30} and {\tt b7/pt8} in $R$}\\
\ig[width=0.2\tew]{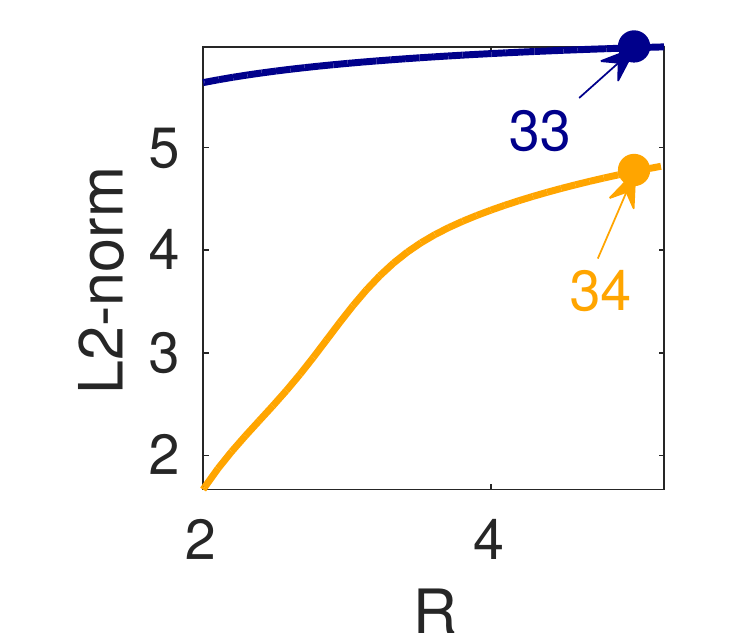}
\raisebox{5mm}{\ig[width=0.18\tew,height=26mm]{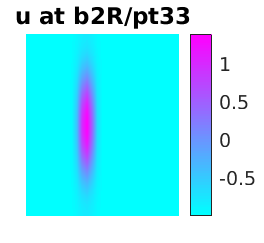}
\raisebox{-5mm}{\ig[width=0.21\tew,height=35mm]{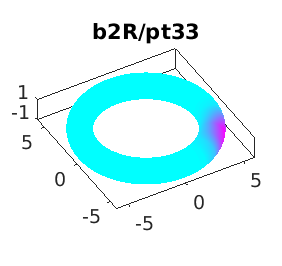}}
\ig[width=0.18\tew,height=26mm]{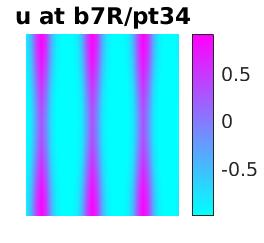}
\raisebox{-5mm}{\ig[width=0.2\tew,height=35mm]{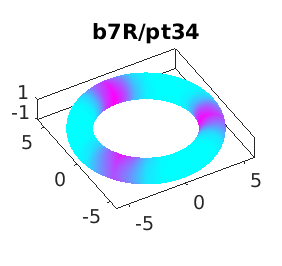}}
}
\end{tabular}
\ece
\vs{-5mm}
   \caption{{\small The Allen--Cahn equation \reff{tAC} on tori, {\tt nt=3600} 
triangular elements. (a) Kernel vectors at BPs 2 to 7 (only one shown 
at the double BPs 2,3,6 and 7). (b-d)  
Continuation in $\lambda$ and example solutions. 
(e) Continuation in $R$ of two solutions from (b). 
  \label{actf2}}}
\end{figure}

Figure \ref{actf2}(a) shows kernel vectors at the branch points 2 to 7, 
from the $u\equiv 0$ branch%
\footnote{all named $\tau_1$ because it's the first (and only) component 
of the function $u$; however in the following we call them $\phi_2,\ldots \phi_7$ for convenience. The eigenfunction at the first branch point $\lam=0$ is $\tau_1=\phi_1\equiv 1$.}. 
The bifurcation points have multiplicities 
$1,2,2,1,1,2,2,\ldots$, with $\lam$--values agreeing with those in \cite{glo08}. 
We label the associated branches {\tt b1,\ldots,b7}, and 
(b) shows a basic bifurcation diagram of \reff{tAC}. 
The symmetries \reff{Tx} and \reff{mxy} explain why only the first and fifths bifurcations are transcritical, 
and the others are pitchforks, even though we have quadratic terms in \reff{tAC} 
and thus no up-down symmetry of the equation. For instance, $S_\pi\phi_2=-\phi_2$,
and hence the two direction $\pm\phi_2$ for {\tt b2} at bifurcation 
are related by symmetry, and this is inherited by {\tt b2}. Similarly,  $S_{\pi/2} \phi_3=-\phi_3$, 
$m_y\phi_4=-\phi_4$, $m_y\phi_6=-\phi_6$ (or again $S_\pi \phi_6=-\phi_6$), 
and $S_{\pi/4} \phi_7=-\phi_7$, with $m_y$ from \reff{mxy}. However, no such hidden up-down symmetry holds 
for $\phi_1$ or $\phi_5$, and hence the bifurcations with these tangents 
are transcritical. In the (spatially homogeneous) 
bifurcation at $\lam=0$, the subcritical leg corresponds to $u>0$ and becomes 
stable after the fold, and the leg with $u<0$ is stable throughout. 
In (c,d) we show a selection of solutions at $\lam=2$, and 
in (e,f) we continue two of these solutions in $R$, just for illustration. 

Regarding the implementation, we comment that:  
\bcen 
\item To generate the LBO \reff{LBt} we use the function {\tt K=LBtor(p,R,rho)}; see its source for the straightforward implementation. As long $R$ and $\rho$ 
are kept fixed, {\tt K=LBtor(p,R,rho)} can be preassembled as usual, but for continuation 
in $R$ or $\rho$ we naturally need to call {\tt LBtor} in each step; 
see {\tt sG.m} and {\tt sGjac.m}, which otherwise are completely standard. 
\item The implementation of periodic BCs is as in \S\ref{sy1sec}. 
Again we call {\tt p=box2per(p,[1 2])} 
during initialization (see {\tt acinit.m}), and {\tt filltrafo} 
after each assembly of a system matrix such as {\tt K} (associated to \reff{LBt}) or {\tt M}. 
\item The translational invariance of \reff{tAC} in $x$ 
requires a phase condition (PC) for the continuation of $x$--dependent 
solutions as in \reff{e:phase}, see also Remark 
\ref{pbcrem1}. However, in contrast to \S\ref{sy1sec} we do not 
need to use {\tt q(c)swibra} for these branches, because modulo 
translations the multiplicity is 1. 
\item To produce the surface plots in Fig.~\ref{actf2} we use the convenience 
function {\tt torplot}. 
\ecen

\subsection{Spheres: Demo {\tt acS}}\label{acs-sec}
\def\dhome{./pftut/acS}
The symmetry group $\Ga$ of a PDE with space ($\xti$) independent coefficients on a 
sphere $S_R$ of radius $R$ is $O(3)$, generated by two rotations (around the $\tilde{x}_1$ axis and 
the  $\tilde{x}_3$ axis, say) and one reflection ($\tilde{x}_1\mapsto -\tilde{x}_1$, say). 
The consequences of these symmetries for bifurcations from a branch 
of spatially homogeneous solutions have been studied in detail in, 
e.g., \cite{IG84,CLM90,Matt03, Matt04,Call04}. Here we want to recover and illustrate 
some of these results numerically, again first using the AC equation 
as a toy model, i.e.,
\huga{\label{ACS2} 
0=G(u)=-\DelS u-\lam u-u^2+\ga u^3,\quad \lam\in\R, 
\ga>0.}
We parameterize $S_R$ by \reff{sphp} with $\DelS$ given in \reff{LBS}, i.e., 
$\Del_{S_R} u(x,y)=\frac 1 {R^2}\left(\frac 1 {\cos^2 y}\pa_x^2 u+\frac 1 {\cos y}\pa_y(\cos y\pa_y u)\right)$. 
The eigenfunctions of $\DelS$  
are the spherical harmonics 
\huga{\label{sharm}
Y^m_\ell(x,y)=\er^{\ri m\phi}P^m_\ell(\sin y), \quad m\in\N,\ \ell=m,m+1,\ldots, 
}
where $P^m_\ell(s)$ are the associated Legendre functions, and the associated 
eigenvalues are 
\huga{\label{spev}
\mu_\ell=\frac 1 {R^2}\ell(\ell+1), \text{ of multiplicity $2\ell+1$}. 
}
For given $\ell\in\N$ we have one eigenfunction with $m=0$ (no $x$--dependence) and two eigenfunctions $\cos(mx)P^m_\ell(\sin y)$ and 
$\sin(mx)P^m_\ell(\sin y)$ for each $1\le m\le \ell$. Thus, for \reff{ACS2} 
we have bifurcation points at 
\huga{
\barr{llllll}
\ell&0&1&2&3&\ldots\\
R^2\lam&0&2&6&12&\ldots\\
\text{multiplicity}&1&3&5&7&\ldots
\earr
}
Moreover, from symmetry and properties of the $Y^m_\ell$ it is known that  bifurcations at even $\ell$ are generically transcritical while those at odd $\ell$ are all pitchforks.

To study these bifurcations numerically, using \reff{sphp} and the ideas from \S\ref{sy1sec} 
for branch switching at BPs of higher (discrete) multiplicity and with additional continuous symmetries, we need to deal with the coordinate singularity of $\Del_S$ at 
$y=\pm\pi/2$, and we should also choose meshes that are coarser 
for $y$ near $\pm \pi/2$ than near $y=0$. For the first problem we use a standard trick and 
slightly cut off the sphere near the poles, and instead of 
$\Om=(-\pi,\pi)\times (-\pi/2,\pi/2)$ choose 
\huga{
\Om=(-\pi,\pi)\times (-\pi/2+\del,\pi/2-\del) 
}
with a small $\del>0$ (typically $\del=10^{-3}$), and  pBCs in $x$ 
and homogeneous Neumann BCs in $y$. 
The pertinent branches for \reff{ACS2} are still like on the full sphere, 
and the choice of $\del$ (in the range $10^{-4}$ to $10^{-2}$) plays 
no visible role. 
Concerning suitable meshes, we use a modification {\tt sppdeo} of {\tt stanpdeo2D}, where we provide a set of points in $\Om$ which becomes coarser 
as $|y|\ra \pi/2-\del$, and then use \mlab's {\tt delaunayTriangulation} 
method to set up the mesh. 

\begin{figure}[ht]
\bce 
\begin{tabular}{lll}
{\small (a) mesh (coarse) }&{\small (b) BD }&{\small (c) example solns on 2nd branch}\\
\raisebox{25mm}{\begin{tabular}{l}
\ig[width=0.23\tew,height=26mm]{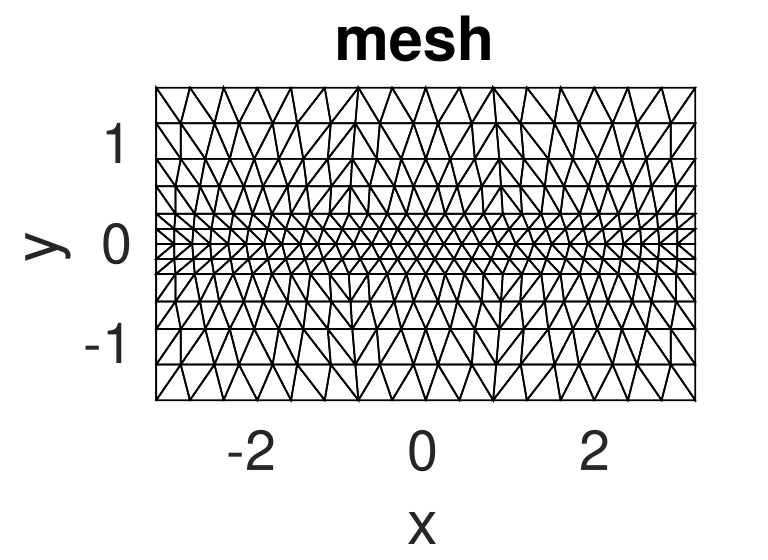}
\\
\ig[width=0.25\tew]{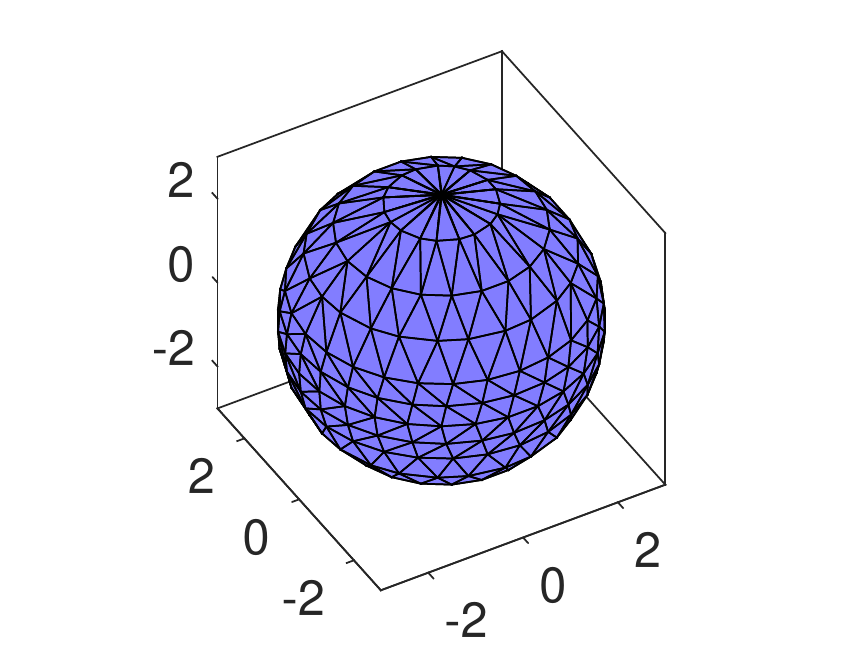}
\end{tabular}}&
\hs{-7mm}\ig[width=0.27\tew]{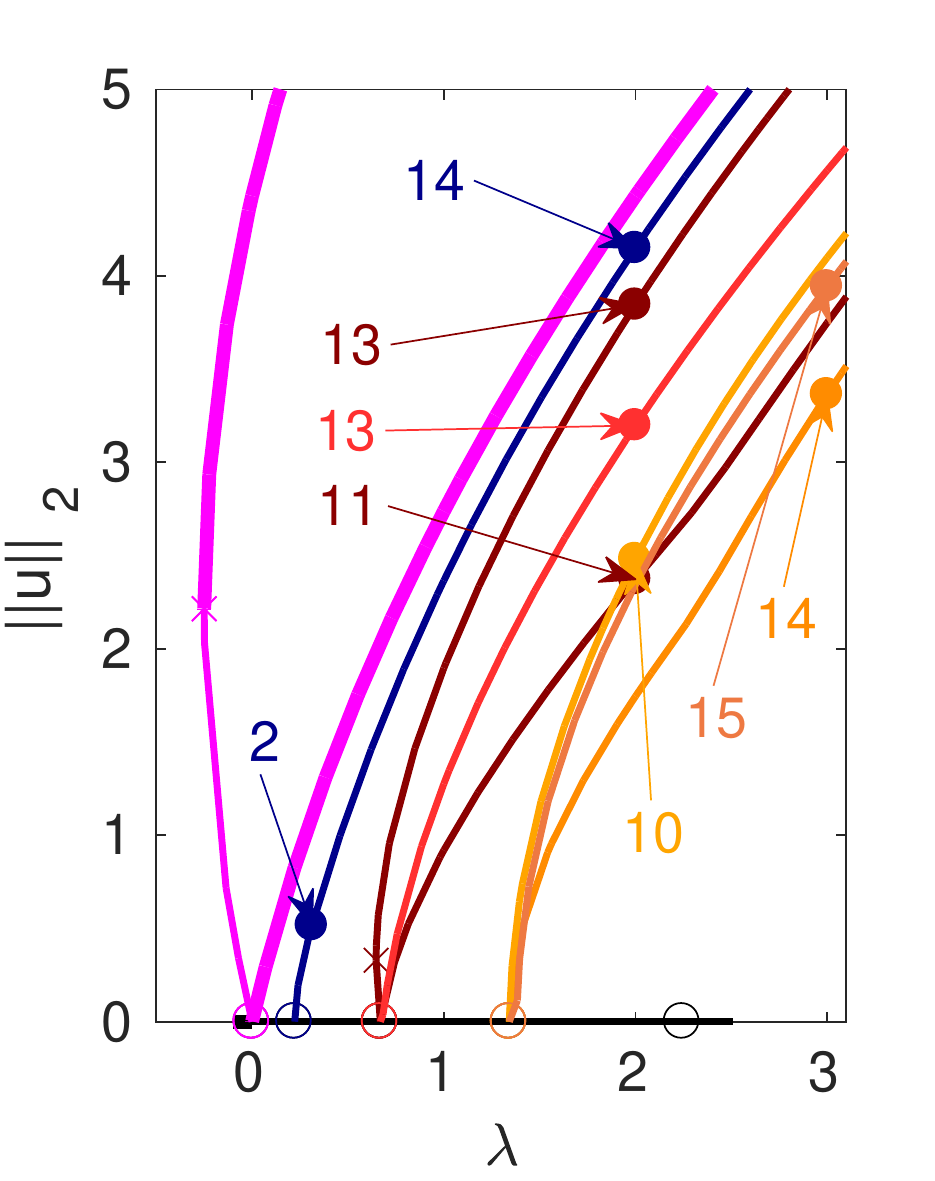}&
\hs{-2mm}\raisebox{25mm}{\begin{tabular}{l}
\ig[width=0.18\tew,height=22mm]{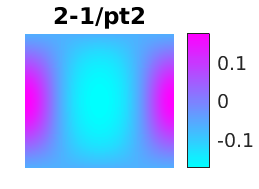}
\ig[width=0.18\tew,height=22mm]{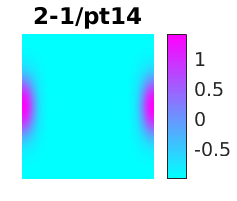}
\\
\ig[width=0.18\tew]{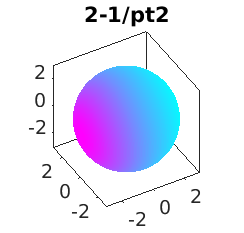}
\ig[width=0.18\tew]{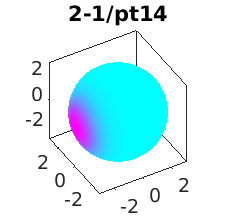}
\end{tabular}}
\end{tabular}\\
\begin{tabular}{l}
(d) further example solutions\\
\begin{tabular}{l}
\ig[width=0.16\tew,height=22mm]{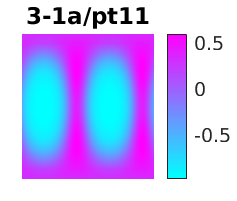}
\ig[width=0.16\tew,height=22mm]{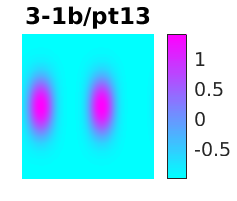}
\ig[width=0.16\tew,height=22mm]{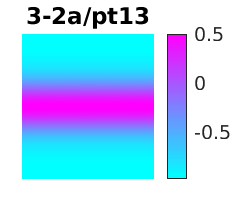}
\ig[width=0.16\tew,height=22mm]{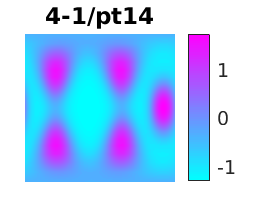}
\ig[width=0.16\tew,height=22mm]{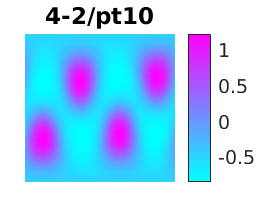}
\ig[width=0.16\tew,height=22mm]{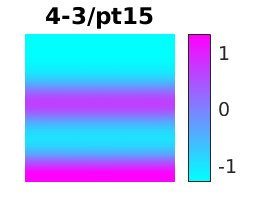}
\\
\ig[width=0.16\tew]{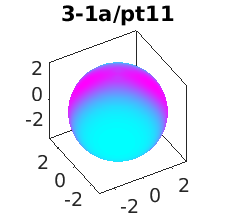}
\ig[width=0.16\tew]{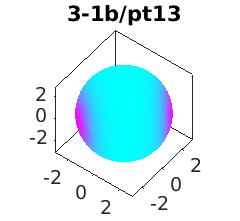}
\ig[width=0.16\tew]{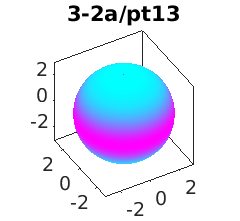}
\ig[width=0.16\tew]{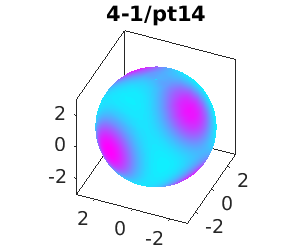}
\ig[width=0.16\tew]{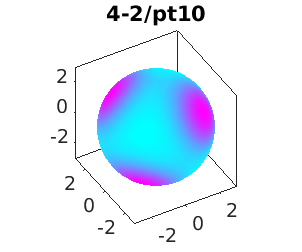}
\ig[width=0.16\tew]{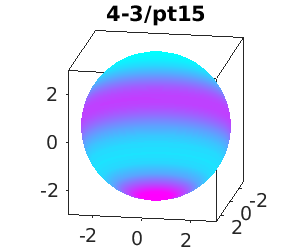}
\end{tabular}
\end{tabular}\\
\ece
\vs{-5mm}
   \caption{{\small The Allen--Cahn equation \reff{tAC} on a sphere of 
radius 3 (with cutoff $\del=10^{-3}$). (a) Idea of the meshing, with coarser meshes for $|y|\ra \pi/2-\del$. 
(b) Basic bifurcation diagram of branches from the first 4 BPs. (c,d) example 
solutions, where {\tt 2-1, 3-1a, 3-1b, 4-1, 4-2, 4-3} correspond to 
the blue (twice), brown (twice), red, and orange (triple) branches, respectively. 
{\tt nt=4470} triangular elements used for the computations. 
  \label{acSf1}}}
\end{figure}

Figure \ref{acSf1} shows results for the continuation of steady solution branches of the AC equation on a sphere of radius $3$. In (a) we just 
sketch the idea of the meshes, (b) shows a basic bifurcation diagram, 
and (c,d) show example solutions. The results completely agree with theoretical predictions. 
In the script {\tt cmds.m} (see Listing \ref{acSl1}) we use {\tt qswibra(\ldots,aux)} (for $l=2$) and {\tt cswibra(\ldots,aux)} (for $l=1,3$) to remove the spurious multiplicity from the 
translational invariance in $x$ by choosing appropriate active kernel lists 
{\tt aux.ali}, cf.~\S\ref{sy1sec}.  We then obtain 1,2, and 3 distinct 
branches, at the 2nd, 3rd and 4th BP, 
respectively. To follow the branches with $x$--dependence we 
use the convenience function {\tt conpc} which switches 
on the $x$ phase-condition after a few initial steps. 
In {\tt cmds.m} we also 
compute a number of branches at further BPs, yielding interesting patterns 
as predicted from the $Y^m_\ell$, but naturally these are all unstable 
in the AC equation.

\hulst{language=matlab,stepnumber=0, 
caption={{\small {\tt acS/conpc.m}; convenience function to continue 
with phase condition in $x$}}}{\dhome/conpc.m}

\hulst{caption={{\small {\tt acS/cmds.m}, using {\tt aux.ali=[1 3]} in {\tt qswibra(\ldots,aux)}} 
for $\ell=1$ to select 
the 'active' kernel vectors. The function {\tt spplot} is a convenience function for the surface plot.},
label=acSl1,language=matlab,stepnumber=0, linerange=5-21}{\dhome/cmds.m}

\subsection{Schnakenberg on spheres: Demo {\tt schnakS}}\label{ss-sec}
The Schnakenberg model on spheres $\CS_{R}$ is obtained by replacing 
$\Delta$ in \reff{tur1}  by $\Delta_{\CS_{R}}$ 
from \reff{LBS}. Thus we consider 
\huga{\label{tur1b}
\pa_t U=D\Delta_{\CS_{R}} U{+}F(U),\ U=\bpm u\\v\epm,\
D=\bpm 1 & 0\\ 0 & d\epm, \
F(U)= \bpm-u{+}u^2v \\\lambda{-}u^2v\epm
{+}\sigma \left(u{-}\frac 1 v\right)^2\bpm 1\\-1\epm. 
}
In Fig.~\ref{ssf2}(b)-(d) we fix $(\sig,d)=(0,60)$, and use $\lam$ as our 
primary bifurcation parameter, on a sphere of radius $R=10$. At the first 
BP1 {\tt bpt1} from the homogeneous branch $U=(\lam,1/\lam)$ at $\lam_c\approx 3.208$ we then have $\ell=6$. Hence $m=\dim N(G_u)=13$ 
and the bifurcations are generically transcritical. At the second BP {\tt bpt2} we have $\ell=5$, and hence $m=11$ and all bifurcations are pitchforks. To find (a selection of) the 
bifurcating branches we use similar ideas as in \S\ref{acs-sec}, 
i.e., we first call {\tt qswibra} (at {\tt bpt1}) 
with {\tt aux.besw=0} and inspect the kernel vectors. 
Then selecting appropriate {\tt aux.ali} we call  {\tt qswibra} with 
{\tt aux.besw=1} to find bifurcation directions. At {\tt bpt2} 
we proceed analogously with {\tt cswibra}. However, both algorithms 
do run into problems for the still high--dimensional kernels (of dimension $\ell+1$), and to find solutions of the branching equation 
we need to increase {\tt aux.soltol}, and to identify solutions as isolated we 
need to decrease {\tt aux.isotol}. Still, depending on the choices of these 
parameters, typically only a subset of the expected branches is found, 
and thus we also use {\tt gentau} to, e.g., start the horizontal stripes 
branches {\tt a13*} (with *=a,b for the two 'legs')  and {\tt b1}. 

\begin{figure}[ht]
\bce 
\begin{tabular}{lll}
{\small (a) Real life}&{\small (b) Some primary bifurcations}
&{\small (c) Three example solutions, the first one stable}\\
\raisebox{25mm}{
\begin{tabular}{l}
\ig[width=0.11\textwidth]{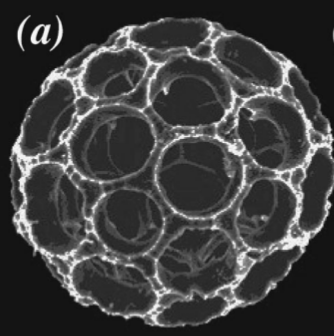}\\\ig[width=0.11\textwidth]{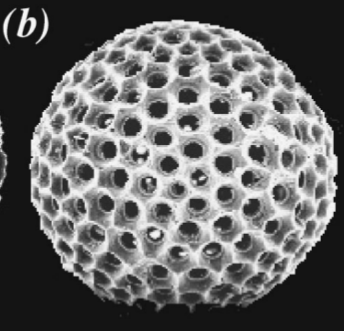}
\end{tabular}}&
\hs{-1mm}\ig[width=0.27\textwidth,height=60mm]{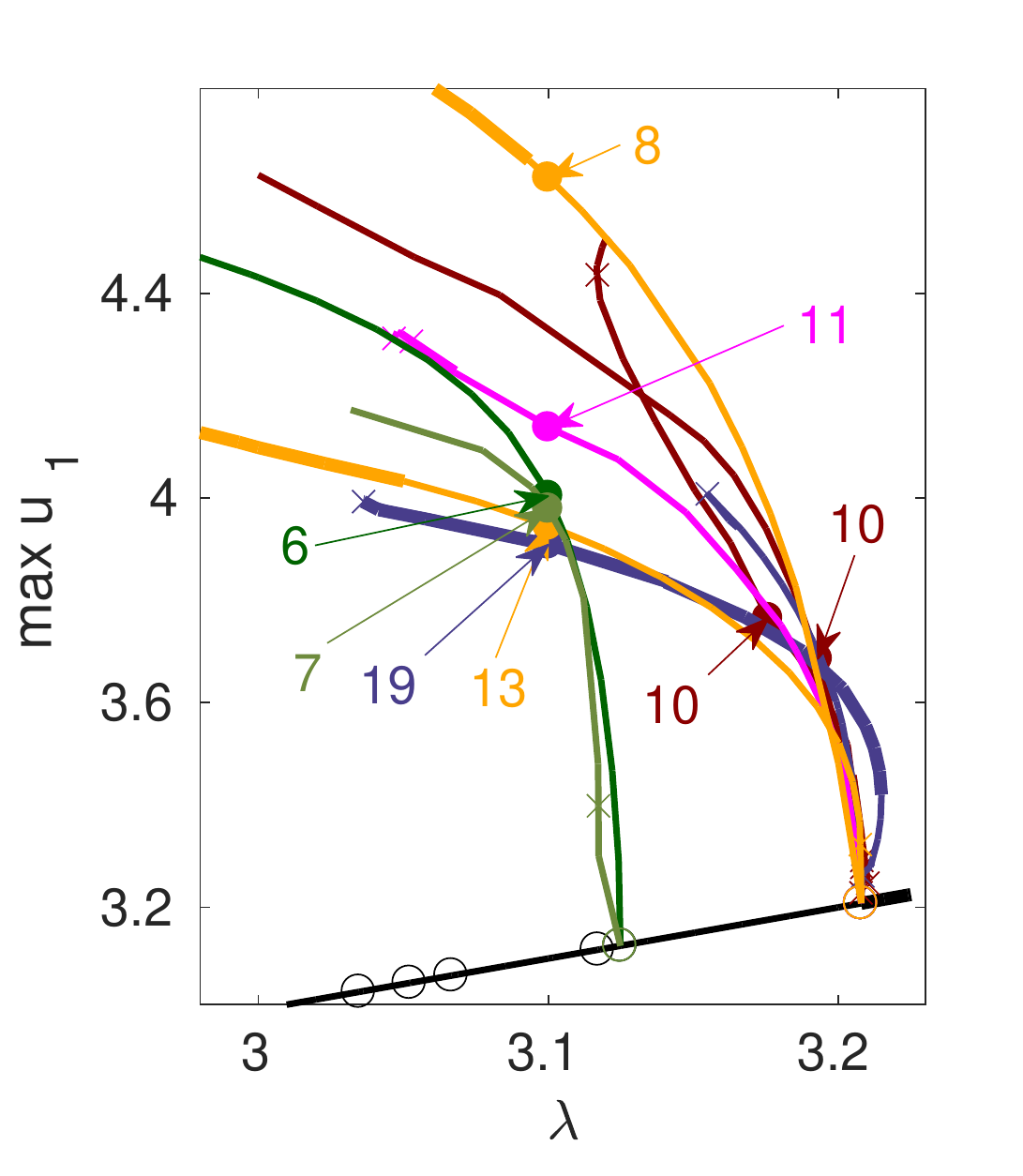}
&
\hs{-6mm}\raisebox{25mm}{\begin{tabular}{ll}
\ig[width=0.18\textwidth]{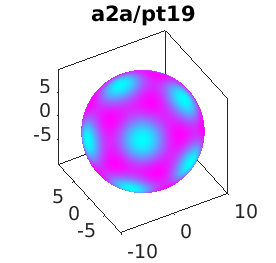}
\ig[width=0.18\textwidth]{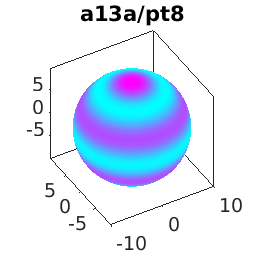}
\ig[width=0.18\textwidth]{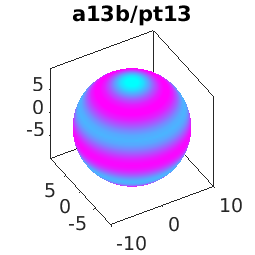}
\\
\hs{-2mm}
\raisebox{-2mm}{\ig[width=0.18\textwidth]{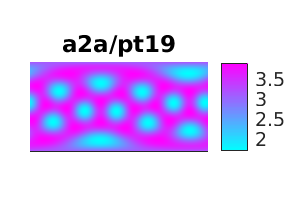}}
\ig[width=0.18\textwidth]{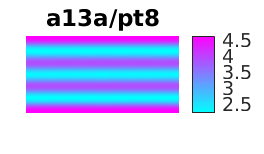}
\ig[width=0.18\textwidth]{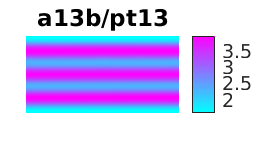}
\end{tabular}}
\end{tabular}
\\
\begin{tabular}{l}
{\small (d) Further example solutions}\\
\hs{-4mm}\ig[width=0.2\textwidth]{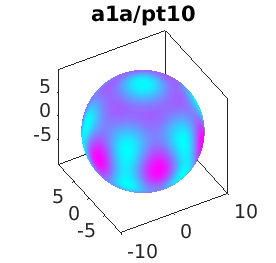}
\ig[width=0.2\textwidth]{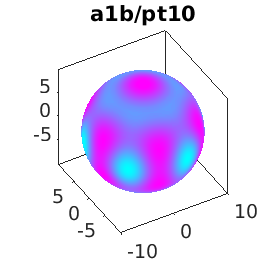}
\ig[width=0.2\textwidth]{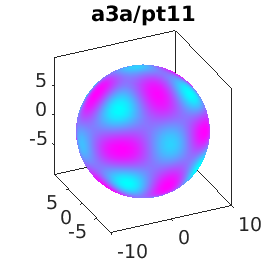}
\ig[width=0.2\textwidth]{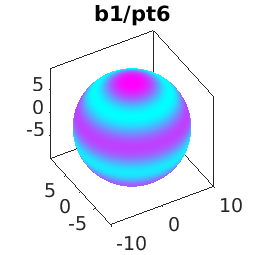}
\ig[width=0.2\textwidth]{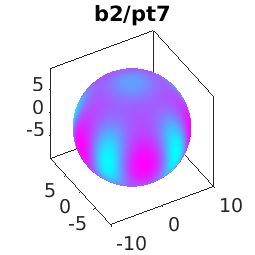}\\
\hs{-4mm}\ig[width=0.2\textwidth]{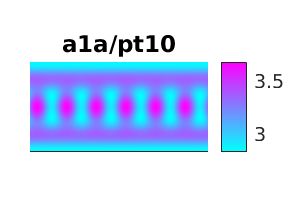}
\ig[width=0.2\textwidth]{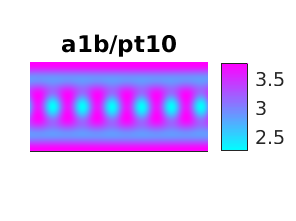}
\raisebox{-2mm}{\ig[width=0.2\textwidth]{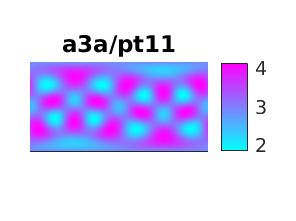}}
\ig[width=0.2\textwidth]{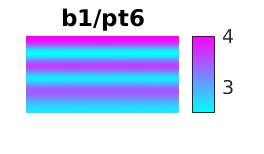}
\ig[width=0.2\textwidth]{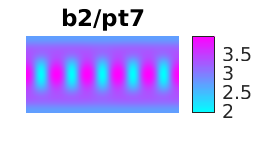}
\end{tabular}

\ece
\vs{-5mm}
   \caption{{\small Pattern formation on spheres. (a) 
Silica skeletons of two radiolaria, scale about 1mm, 
 from \cite{VAB99}. (b)-(d) \reff{tur1b} on a sphere of radius $R=10$, {\tt nt=10000} 
triangular elements. (b) {\em Some} branches bifurcating 
at the first BP ($l=6$, hence transcritical) and second BP ($l=5$, hence 
pitchfork)), colors: {\tt a1} red, {\tt a2} blue, {\tt a13} orange, 
{\tt a3} magenta. 
{\tt b1,b2} green.  (c,d) example solutions. 
  \label{ssf2}}}
\end{figure}

\brem\label{Srem1}
a) For $R=10$, we have $|S_R|=4\pi R^2\approx 1257$, which roughly 
corresponds to a 'not--too small' flat square of side-lengths $7\pi/k_c$, where 
$k_c{=}\sqrt{\sqrt{2}{-}1}$ is the critical wave number from the flat case, 
cf.~\S\ref{schnaknum}. Moreover, $\lam_c(R)|_{R=10}{\approx}3.208$ is already quite 
close to the (flat) infinite domain value $\lam_c{=}\sqrt{60}\sqrt{3{-}\sqrt{8}}\approx 3.21$, 
and $\lam$ only changes slightly for larger $R$. 
The wave number $\ell$ changes in 
the size parameter $R$, but the wavelength (on $S_R$) essentially stays the same. Thus, we conclude that for $R=10$ (and larger) we are in the regime 
where the patterns from the first (and second) BP are intrinsic, and not an artifact of 
the domain size. See also the discussion of intrinsic versus extrinsic patterns 
in \cite[\S6]{LBFS17}. 

b) Besides the difficulties due to larger $\ell$, the main 
consequence of a larger $R$ is that the BPs come closer together, and 
that the spectral gaps between the $2\ell+1$ (approximate) zero eigenvalues and the rest 
of the spectrum decrease, exactly as in the flat case. For $R=10$, the 
gap size at {\tt bp1} is about $0.025$ (the value of the $14$th eigenvalue), 
and at {\tt bpt2} the gap size is about $0.013$. 

c) The case $\ell=6$ has for instance been discussed in a detailed model--independent way in \cite[\S5.1]{Matt03}. The 1(real)+6(complex)--dimensional 
amplitude equations of quadratic order yield four branches (modulo 
isotropy) of patterned solutions, namely: 

\begin{tabular}{lll}
Isotropy&\#neg.~Eigenval.~close to bif.&Example in Fig.~\ref{ssf2}\\
$O(2)$ (e.g., rotation around $\tilde{x_3}$ axis)&3&{\tt a13}\\
$I$ (icosaeder group)&1&{\tt a2}\\
$O$ (octaeder group)&5&{\tt a3}\\
$D_6$ (symmetry group of reg.~hexagon)&4&{\tt a1}
\end{tabular}

In particular, while all bifurcating branches are unstable at bifurcation, 
the $I$ branches are preferred in the sense of a certain 
variational principle \cite{Bu75}, 
and in the sense that they have only one unstable 
direction and hence the potential to stabilize in a fold. This is what 
happens to the {\tt a2a} branch in Fig.~\ref{ssf2}. We also recover 
the other numbers of unstable eigenvalues near bifurcations, and note that the stripes 
{\tt a13a} and {\tt a13b} both become stable at rather large amplitude. 

d) Additionally, and in contrast 
to the Schnakenberg model over flat domains in \S\ref{ssec1}, we 
note that patterned branches 
typically do not extend to small $\lam$, but show a number of folds for 
$\lam$ still 'not far' from bifurcation, i.e., near $\lam>3$, say. See, e.g., 
branches {\tt a1b} and {\tt a2a/b}. This may partly be due to 'branch jumping' 
in the continuation, which as usual we try to mitigate by using 
{\tt pmcont}, cf.~\S\ref{tnt1}, but it seems that the spherical geometry gives more 
restrictions on the $\lam$ range of solutions than 'flat boxes' with Neumann BCs, cf., e.g., Fig.~\ref{schnakf1}. A similar effect also 
occurs if we try to continue the branches from $R=10$ to larger $R$ and smaller $R$, see Fig.~\ref{ssf3}.  \\
(e) In any case, if one wants 
to study patterns and bifurcations on (significantly) larger 
spheres ($R>20$, say),  then time 
integration (cf.~\S\ref{tnt1}) to identify stable branches should be used first.  
\eex\erem

\begin{figure}[ht]
\bce 
\begin{tabular}{l}
\raisebox{0mm}{
\hs{-5mm}\ig[width=0.25\textwidth]{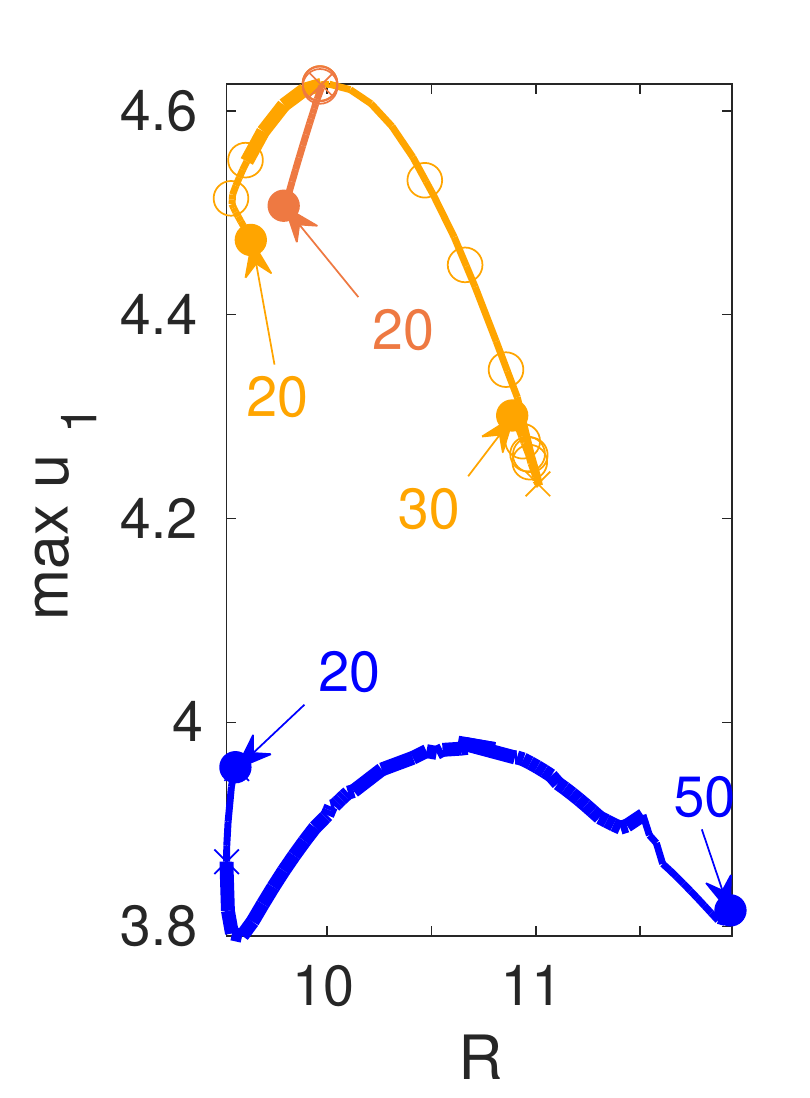}}
\hs{-0mm}\raisebox{30mm}{
\begin{tabular}{l}
\ig[width=0.22\textwidth]{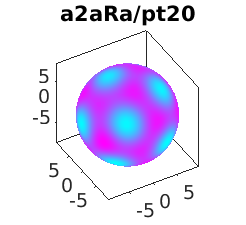}
\ig[width=0.22\textwidth]{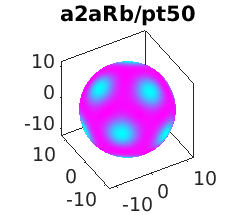}\\
\ig[width=0.22\textwidth]{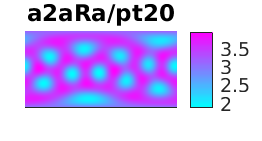}
\ig[width=0.22\textwidth]{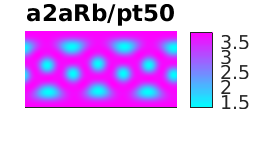}
\end{tabular}}\\
\begin{tabular}{l}
\ig[width=0.22\textwidth]{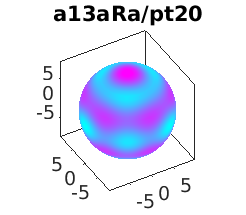}
\ig[width=0.22\textwidth]{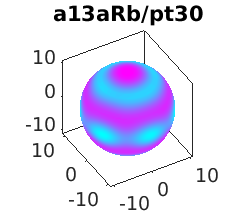}
\ig[width=0.22\textwidth]{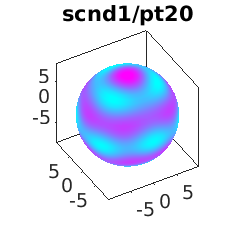}\\
\ig[width=0.22\textwidth]{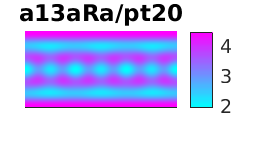}
\ig[width=0.22\textwidth]{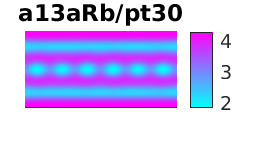}
\ig[width=0.22\textwidth]{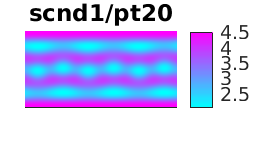}
\end{tabular}
\end{tabular}
\ece
\vs{-5mm}
   \caption{{\small Continuation in $R$ of a pentagonal (blue branch) and a stripe (orange branch) solution from Fig.~\ref{ssf2}}, and 
one secondary branch (brown). Both types of primary branches solutions seem to  exist only in a rather small range in $R$, i.e., both show folds at $R$ not much 
smaller/larger than the starting value $R=10$, and there is a strong tendency to 
'branch jumping' during the continuation. See text for further comments. 
\label{ssf3}. 
}
\end{figure}

\subsection{Schnakenberg on tori: Demo {\tt schnaktor}}
\label{st-sec}
Following \S\ref{act-sec}, in Figures \ref{stf2} and \ref{stf3} we 
consider \reff{tur1} on tori, now replacing $\Delta$ by $\DelT$, i.e., 
\huga{\label{tur1c}
\pa_t U=D\DelT U{+}F(U),
}
with $D$ and $F(U)$ as in \reff{tur1b}, and $U=U(x,y)$, $(x,y)\in [-\pi,\pi)^2$ with pBCs in $x$ and $y$. We use 
the base parameters $(\sig,d)=(-0.1,60)$, and again $\lam$ 
as the primary bifurcation parameter. The torus 
is simpler than the sphere in the sense that the bifurcations from the trivial branch are 
generically simple (modulo the double multiplicity due to translational 
invariance in azimuth), and naturally \reff{tur1b} inherits the symmetries \reff{Tx}, \reff{mxy} from $\Delta_{\CT_{R,\rho}}$, and these can be used to 
decide which bifurcations are transcritical and which are pitchforks. 
 However, even over the relatively 
small torus in Fig.~\ref{stf2} with $(R,\rho)=(12,4)$ and hence $|\CT_{R,\rho}|=4\pi \rho R\approx 600$, 
already more than 25 eigenvalues cross $0$ between $\lam=3.21$ and $\lam=3.15$, 
and thus in the implementation (see {\tt cmds1.m}) we now use a trick 
as for \reff{chem1} 
to obtain  a first bifurcation diagram close to the first bifurcation from $U\equiv (\lam,1/\lam)$ in a simple way: 

\begin{figure}[ht]
\bce 
\begin{tabular}{ll}
{\small (a) Basic bif.~diagram}&{\small (b) Example plots from first 4 branches}\\
\hs{-3mm}\ig[width=0.24\tew,height=60mm]{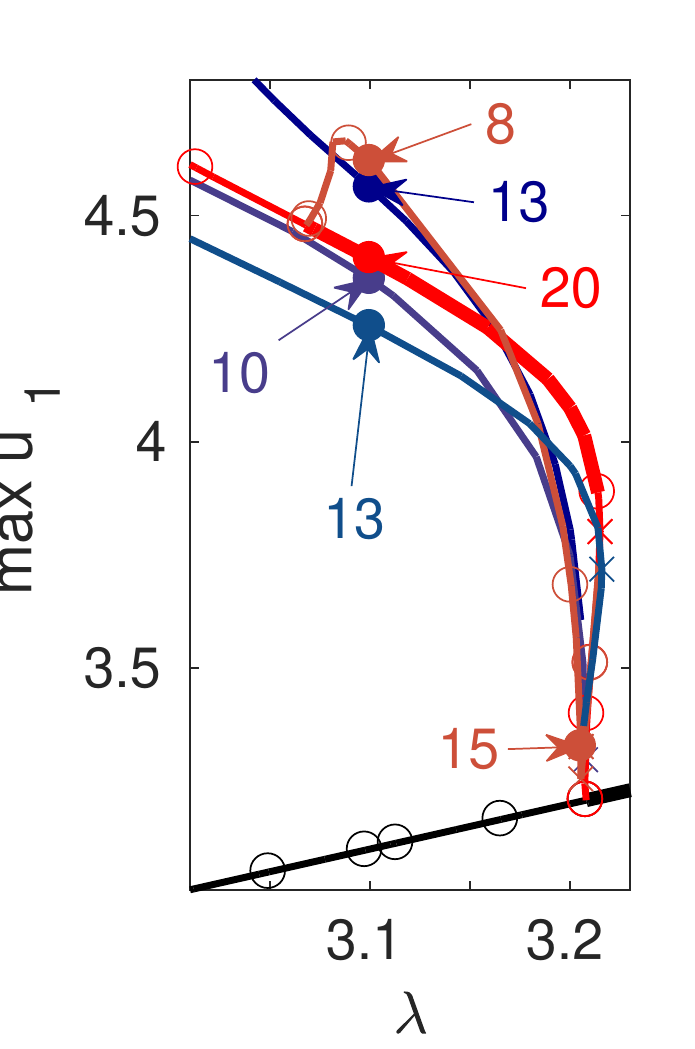}&
\hs{-4mm}\raisebox{25mm}{\begin{tabular}{l}
\ig[width=0.185\tew]{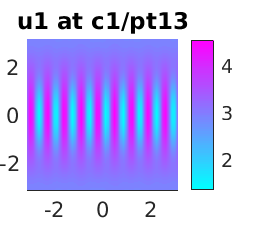}
\ig[width=0.185\tew]{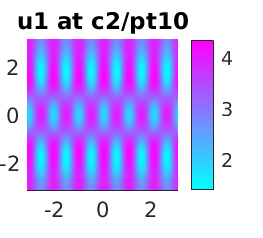}
\ig[width=0.185\tew]{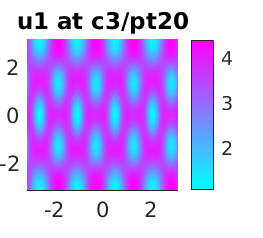}
\ig[width=0.185\tew]{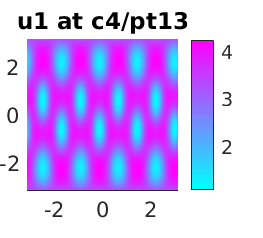}\\
\ig[width=0.185\tew]{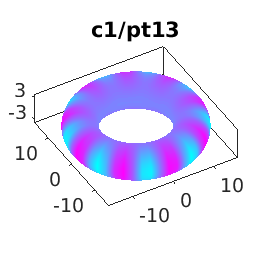}
\ig[width=0.185\tew]{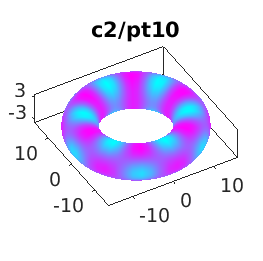}
\ig[width=0.185\tew]{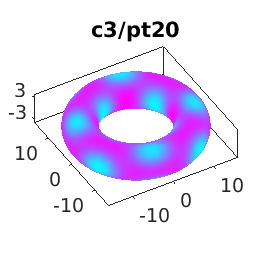}
\ig[width=0.185\tew]{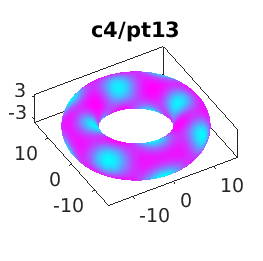}
\end{tabular}}
\end{tabular}\\[-4mm]
\begin{tabular}{p{33mm}l}
{\small (c) Solutions from the secondary branch {\tt c3-5}}&
\raisebox{-20mm}{
\ig[width=0.185\tew]{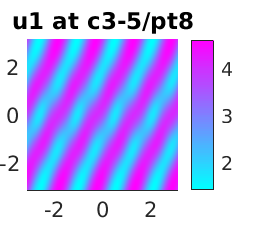}
\hs{-1mm}\ig[width=0.2\tew]{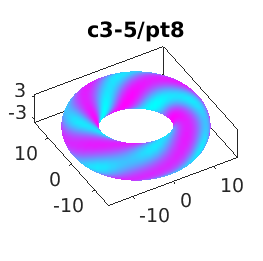}\quad 
\ig[width=0.185\tew]{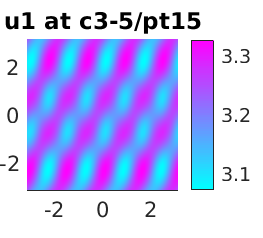}
\hs{-1mm}\ig[width=0.2\tew]{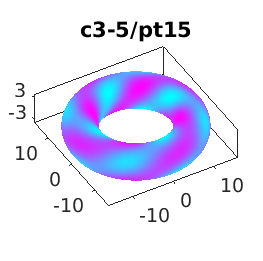}}
\end{tabular}
\ece
\vs{-5mm}
   \caption{{\small \reff{tur1c} on a rather small torus, where however 
already many primary bifurcation occur close together near $\lam\approx 3.2$. 
Branches {\tt b1} (dark blue), {\tt b2} (light blue), {\tt b3} (red), {\tt b4} 
(blue again), and 2 secondary branches {\tt c3-4} (orange) and {\tt c3-5} (brown).  This BD is only a small selection, of which only {\tt c3} 
is stable in a certain $\lam$ range, see (b) for example plots at $\lam=3.1$. (c) shows solutions from secondary branches bifurcating from {\tt c3}, 
where {\tt c3-5} reconnects to {\tt c3} at small amplitude.  {\tt nt=10000} 
triangular elements.
  \label{stf2}}}
\end{figure}

\bci 
\item  
We only localize ``the first'' (where we don't care if it is really the first) BP using the {\tt bifcheck=2} setting. We then call {\tt cswibra} 
but only to compute (here {\tt m=8}) 'nearby' eigenvectors, 
i.e., with the argument {\tt aux.besw=0}. Then we use 
{\tt gentau} to select different approximate numerical kernel vectors as predictors for the 
branch switching.
\eci 
 Figure \ref{stf2}(a,b) shows results for the first four branches thus obtained. 
The bifurcations are pitchforks for symmetry reasons as explained 
for \reff{tAC}. 
In {\tt cswibra} we use $m=8$ because the first four BPs are all $x$--dependent 
and hence double, such that, e.g., to obtain {\tt c3} we use 
{\tt p=gentau(p0,[0 0 0 0 1],'c3')}. Due to the $x$--dependence we 
also again switch on a 
PC similar to \reff{pcx} after one step. 
The 4 branches are all unstable throughout, except for {\tt c3} which becomes stable after the fold, before 
again loosing stability in a bifurcation to a branch of solutions with patterns 
wrapping around the torus, see Fig.~\ref{stf2}(c). In summary:
\bci 
\item Already on $\CT_{12,4}$, \reff{tur1c} is very rich in pattern formation,  
but the only stable pattern we found consist of a rather regular 
arrangement of hexagons, see {\tt c3/pt13} in Fig.~\ref{stf2}(b). 
There are also patterns which consist of stripes, and patterns which combine 
spots and stripes, but these are illustrated for a larger torus below. 
\item The implementation re--uses {\tt LBtor} and other ideas from \S\ref{act-sec} 
in a straightforward way, and additionally we use {\tt cswibra} with {\tt aux.besw=0}  as explained above. 
\eci 

In Fig.~\ref{stf3} we consider \reff{tur1c} on a larger half torus with 
$(R,\rho)=(25,10)$, and hence $|\CT_{R,\rho}|=1000\pi$. Here 'half' means that $(x,y)\in (-\pi/2,\pi/2)\times [-\pi,\pi)$, and we use pBCs only in $y$ but homogeneous Neumann BCs in $x$, which is 
switched on via {\tt p=box2per(p,2)}. The idea is 
to reduce computational costs using 
that solutions that are even in $x$ can be mirrored at $x=\pm \pi/2$ to 
obtain solutions on the full torus. Otherwise the implementation with 
script {\tt cmds2.m} uses the same ideas as {\tt cmds1.m}. In (b) we 
show two exemplary solutions, where the branch {\tt d10} consists of 
pure stripes, obtained from a primary bifurcation at the 4th BP from 
the homogeneous branch (in the sense of the trick explained after \reff{tur1b}). 
In (c) we show solutions from the red branch in (a), which becomes stable 
roughly at $\lam=3.1$. At the outer radius there are somewhat regular hexagons, which get closer to stripes for $|\vt|\ra \pi$, i.e., near the inner radius. 
For smaller $\lambda$ the branch shows a fold 
where the 'inner spots' merge to stripes and the branch looses stability. 

This last result at least phenomenologically agrees with the results from \cite{Stoop15}, 
where, in the bistable regime between spots and stripes for a SH equation, 
the spots tend to sit on the 'outside' ($-\pi/2<\vt<\pi/2$), and the stripes on the 'inside' ($|\vt|>\pi/2$). The mean curvature $\dst H=-\frac {1} 2 
\left(\frac {\cos\vt}{R+\rho\cos\vt}+\frac 1 \rho\right)$ is smaller on the inside, 
and thus these results also agree with similar results on spheres, 
for which high curvature favors hexagons, and low curvature favors 
(labyrinthine) stripes.

\begin{figure}[ht]
\bce 
\begin{tabular}{l}
{\small (a) Partial BD}\hs{25mm}{\small (b) Example solutions on {\tt d1} and {\tt d10}, all unstable }\\
\hs{-1mm}\ig[width=0.24\tew, height=50mm]{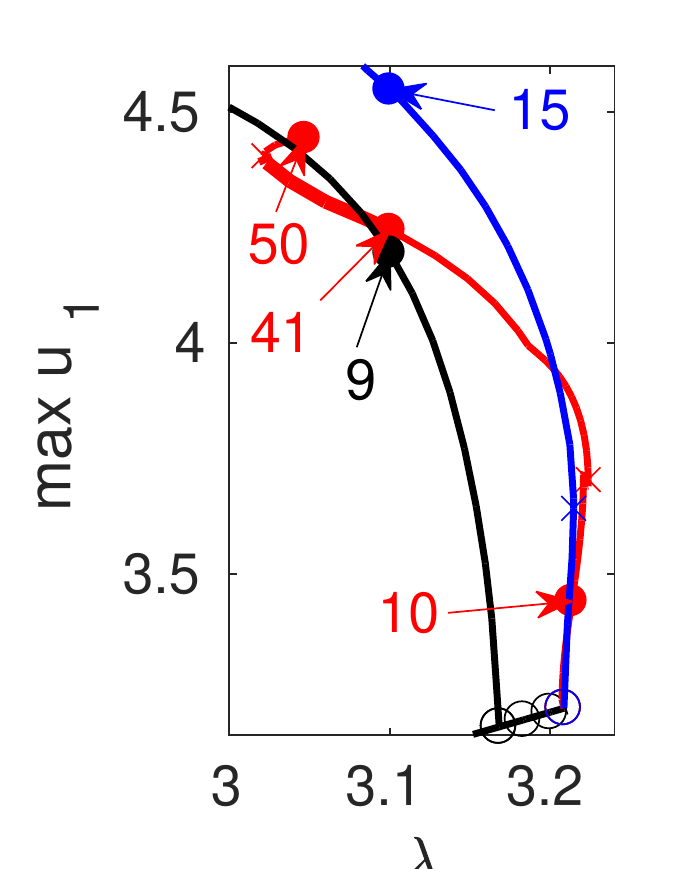}
\hs{4mm}\raisebox{5mm}{
\ig[width=0.2\tew]{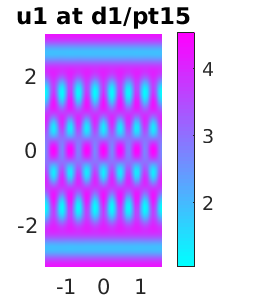}
\hs{-6mm}
\raisebox{-3mm}{\ig[width=0.26\tew]{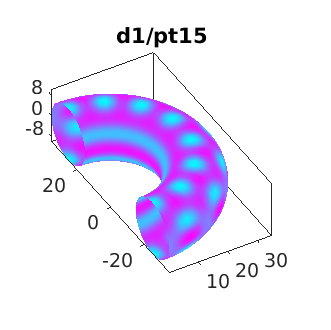}\hs{-2mm}
\ig[width=0.3\tew]{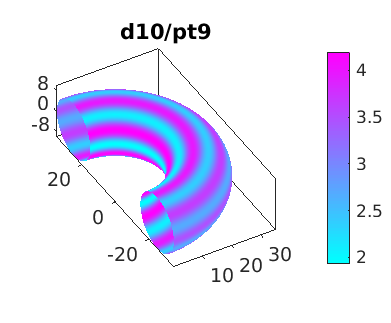}}}\\[-0mm]
{\small (c) Example solutions on {\tt d7}, where, e.g., {\tt pt41} 
corresponds to a stable solution}\\
\ig[width=0.21\tew]{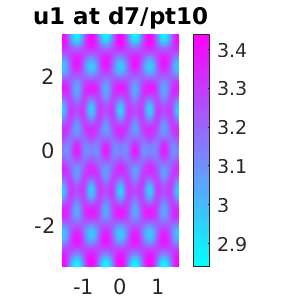}
\hs{-6mm}
\raisebox{-5mm}{\ig[width=0.23\tew]{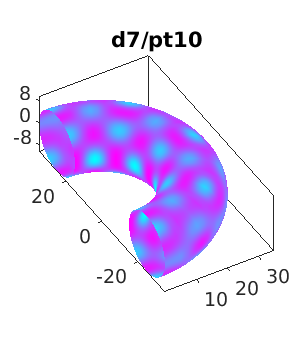}}\quad 
\ig[width=0.27\tew]{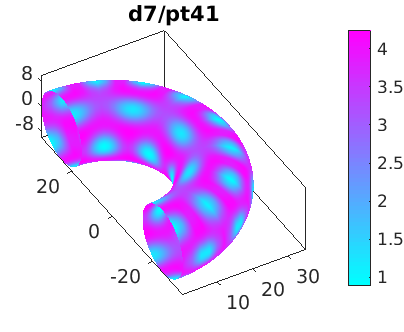}\quad
\ig[width=0.27\tew]{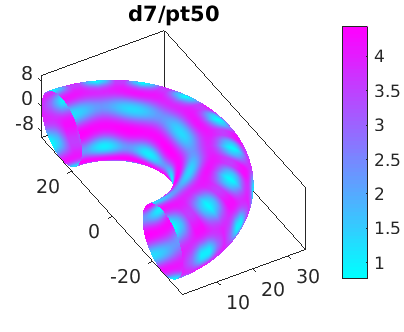}
\end{tabular}
\ece
\vs{-5mm}
   \caption{{\small \reff{tur1c} on a 'half torus', with Neumann BCs in $x$ direction. (a) BD of three primary bifurcations from the trivial branch, 
with two example solutions from the unstable branches {\tt d1} (blue) and 
{\tt d10} (black) in (b). (c) shows example solutions from the red branch {\tt d7} with for instance {\tt pt41} stable. {\tt nt=12800} 
triangular elements.
  \label{stf3}}}
\end{figure}

\subsection{A coupled problem on a cylinder with a lid: Demo {\tt accyl}}
\label{accylsec}
\def\dhome{./pftut/accyl}
As an example how to patch together two surfaces, or, more generally, two problems 
defined on two domains, we consider 
the coupled problem 
\begin{subequations}\label{cpp}
\hual{
0&=G_1(u_1):=-c\Del_{C_{R}} u_1-\lam u_1-u_1^3+\ga u_1^5 
&\text{\quad in\quad $\Om_1=\{\xt\in\R^3: \xt_1^2+\xt_2^2=R^2, |\xt_3|\le l_y\},$}
\label{e1}\\
0&=G_2(u_2):=-c\Del_{R} u_2 
&\text{\quad in\quad $\Om_2=\{\xt\in\R^3: \xt_1^2+\xt_2^2\le 
R^2, \xt_3=l_y\}$}, \label{e2}
}
with BCs given by 
\hual{
 u_2=u_1 \text{ on $\Ga_1=\pa\Om_2\cap \pa\Om_1$},\quad \pa_n u_1=0 \text{ on $\pa\Om_1$, \label{cppbc}}
}
\end{subequations} 
where $n$ is the outer normal to the surface, i.e., here $n=(0,0,1)$. 
As $\Om_1$ is a cylinder 
with radius $R$ and height $2l_y$, we use the parametrization 
$$
\Om_1=\phi(Q_1),\ Q_1=(-\pi,\pi)\times (-l_y,l_y), \quad 
\phi(x,y)=(R\cos x,R\sin x,y).$$
Then $\Del_{C_{R}}$ in \reff{e1} is given by $\Del_{C_{R}}u_1(x,y)=\frac 1 {R^2}\pa_x^2 u(x,y)+\pa_y^2 u(x,y)$, and $\pa_n u_1=\pa_y u_1$. 
For the disk $\Om_2$ as the top lid of the cylinder we choose 
\huga{\text{
$\Om_2=\psi(D)$, $D=\{x^2+y^2<1\}$, 
$\psi(x,y)=(R x,Ry,l_y)$, such that $\Del_{R} u_2=\frac 1{R^2}(\pa_x^2+\pa_y^2)u_2$.}
}
The BCs \reff{cppbc} on $\Ga_1$ 
can be classified as 'Kirchhoff BCs', i.e., 
of type 
\huga{
\text{$A\bpm u_1\\\pa_n u_1\epm=B\bpm u_2\\ \pa_n u_2\epm$ with 
$A=\bpm 1&0\\0&1\epm$ and $B=\bpm1&0\\0&0\epm$.}
} 

Essentially, $u_1|_{\Ga_1}$ 
gives the boundary values for the Poisson problem \reff{e2} for $u_2$. 
However, we do not claim any physical significance for \reff{cpp}, 
but rather use it as an example to illustrate how to set up a coupling 
via BCs of two otherwise unrelated PDEs, which live over 
different domains. This is different from the systems considered 
so far, where $u_1,u_2$ were components of a vector valued function $u=(u_1,u_2)$ over a common domain. One motivation is that problems 
of type \reff{cpp} may occur when one wants to patch together charts for a 
surface. Also note that in \reff{e1} we now have a cubic-quintic AC equation 
instead of the quadratic-cubic case in \S\ref{act-sec}. 

Figure \ref{accf1} shows some results for \reff{cpp}, 
and Table \ref{accyldtab} gives an overview of the files used, 
including some local modifications of standard \pdep\ library functions. 
The main ideas  
of the implementation are: 
\begin{figure}[ht]
\bce 
\begin{tabular}{lll}
{\small (a) }&{\small (b) }&{\small (c) }\\
\raisebox{5mm}{\ig[width=0.23\tew]{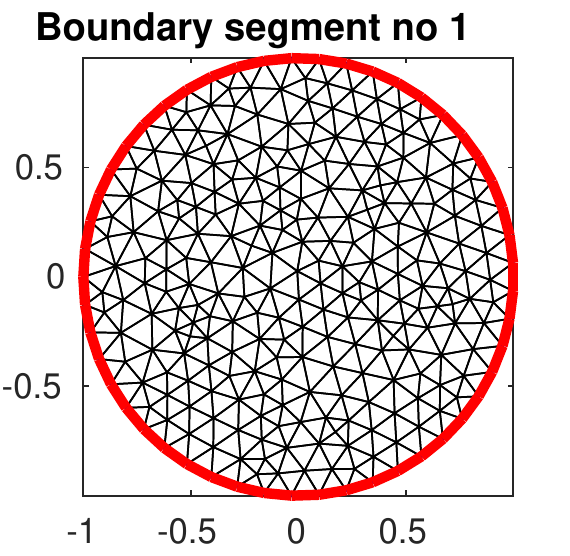}}&
\hs{-1mm}\ig[width=0.23\tew,height=50mm]{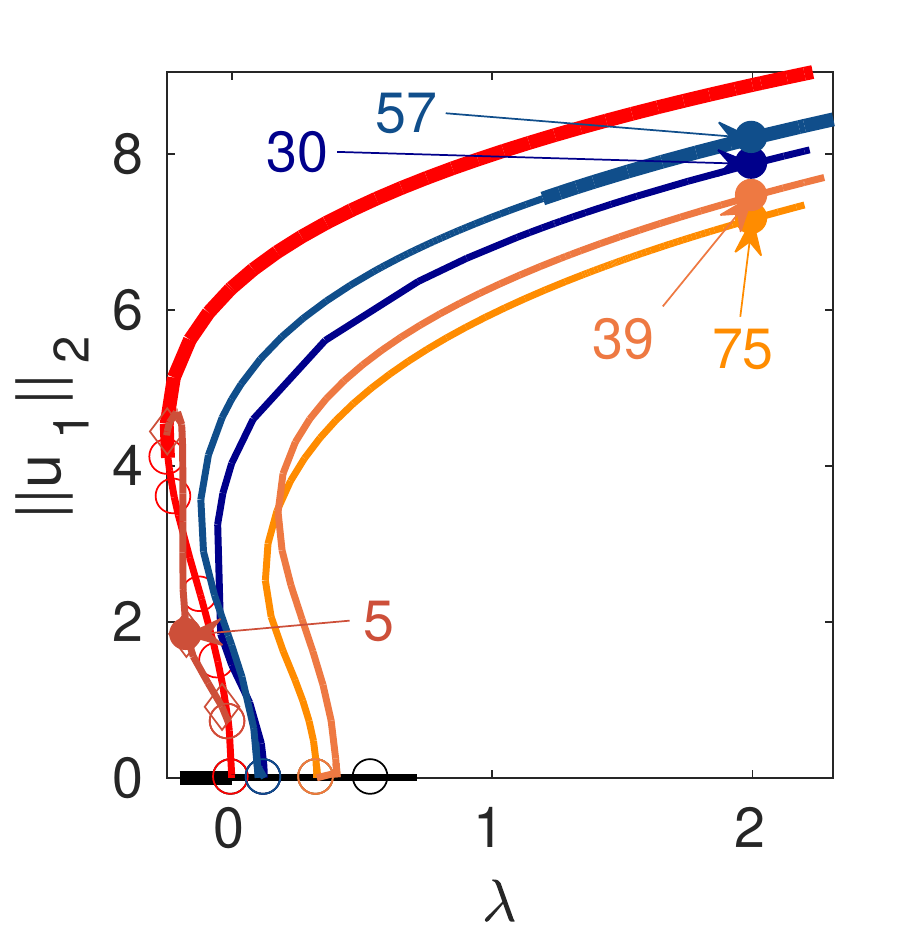}
&\hs{0mm}\raisebox{0mm}{
\ig[width=0.17\tew]{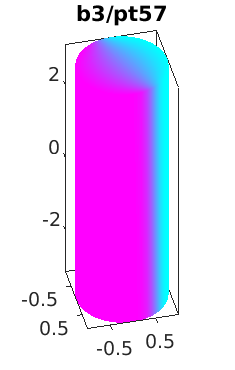}
\ig[width=0.17\tew]{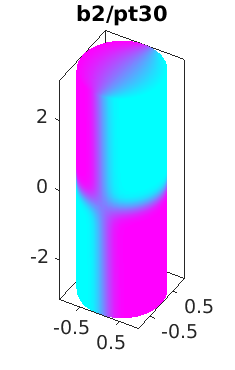}}
\end{tabular}
\begin{tabular}{lll}
(d)&(e)&(f)\\
\ig[width=0.16\tew]{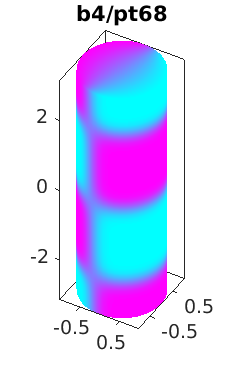}
\ig[width=0.16\tew]{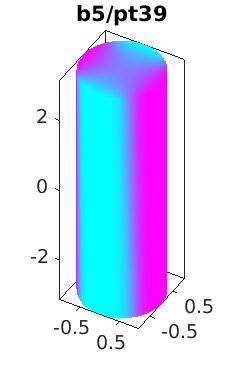}&
\hs{-3mm}\ig[width=0.28\tew,height=48mm]{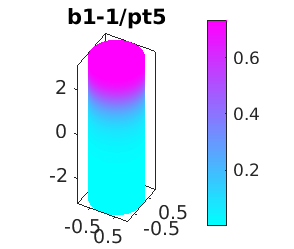}&
\hs{-3mm}\ig[width=0.2\tew]{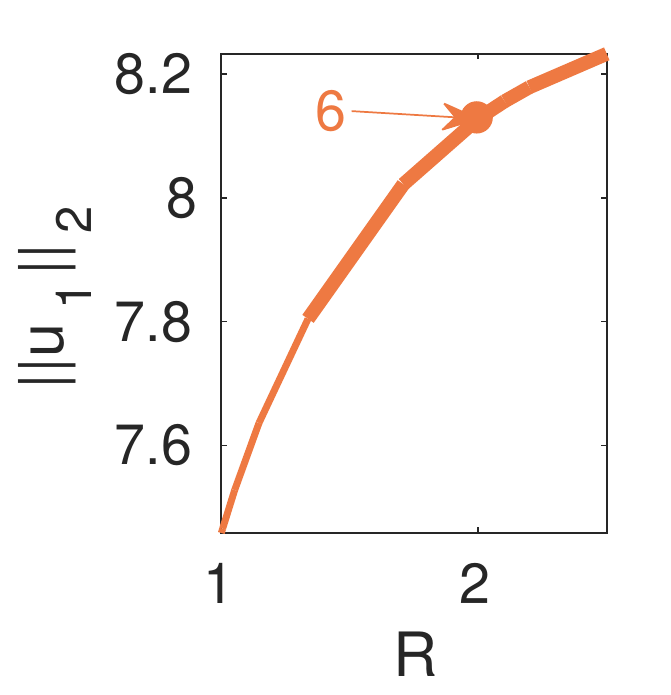}
\ig[width=0.18\tew]{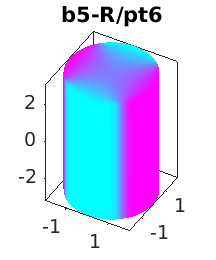}
\end{tabular}
\ece
\vs{-5mm}
   \caption{{\small Results for \reff{cpp}, $(c,\ga,l_y)=(0.1,1,\pi)$ 
throughout, discretization with 4800 (cylinder $\Om_1$) + 500 (top lid $\Om_2$) triangles. (a) shows the mesh of the top disk, and its boundary, 
plotted via 
{\tt [x2,i2]=p.p2.grid.bdseg(1)}. (b) shows a basic bifurcation 
diagram, with five primary branches ({\tt b1} (red), {\tt b2} (light blue), 
{\tt b3} (dark blue), {\tt b4} (light yellow), {\tt b5} (dark yellow)) 
and one secondary branch {\tt b1-1} (brown). (c--e) show associated 
solution plots, where in (c,d) the color scale is roughly $-1.3$ to $1.3$. 
(f) shows an example continuation in $R$. 
  \label{accf1}}}
\end{figure}
\bcen 
\item We first define the cylinder (mantle) as a standard PDE object 
{\tt p.pdeo=stanpdeo2D(lx,ly,nx,ny)}, $l_x=l_y=\pi$, {\tt nx,ny}=numbers  
of discretization points in $x$ and $y$, and switch on the periodic BCs in $x$ 
via {\tt p=box2per(p,1)}; see {\tt acinit.m}. 
\item We then identify the top boundary $\Ga_1$ (its $(x,y)$ values {\tt x1} and its 
point indices {\tt i1} in the grid for $\Om_1$) via the method 
{\tt [x1,i1]=pde.grid.bdseg(3)} from {\tt grid2D}, and 
use {\tt t=x1(1,1:end)} and {\tt 
geo=polygong(cos(t),sin(t))} to generate (the geometry for) a discretization for the top 
disk, i.e., a {\em second} PDE object {\tt p.p2}. 
\item To conveniently encode the PDEs \reff{cpp} and Jacobians (see {\tt sG.m} and {\tt sGjac}), and specifically the coupling conditions 
$u_1|_{\Ga_1}=u_2|_{\Ga_2}$, we store the indices {\tt p.i1} and {\tt p.i2} 
of the points on $\Ga_1$ for both meshes, such that the associated 
solution components are {\tt u1(i1(:))} and {\tt u2(i2(:))}, respectively. 
More specifically, let $n_i, n_{p}$ and $n_{p2}$ be the number 
of points on the interface $\Ga_1$, and in the meshes for $\Om_1$ and $\Om_2$, 
respectively. We then 
generate matrices ${\tt p.Q2}\in \{0,1\}^{n_{p2}\times n_i}$, 
${\tt p.S1}\in\{0,1\}^{n_i\times n_p}$, and 
${\tt p.S2}\in\{0,1\}^{n_i\times n_{p2}}$  such 
that $\uti_j=S_j u_j=u_j|_{\Ga_1}$ 
and such that $Q_2\uti_2(x)=\uti_2(x)$ if $x\in\Ga_1$, $0$ else. 

\item In {\tt acinit}, it remains to initialize {\tt u=[u1;u2;pars]} in 
a standard way (with $u_1\equiv 0$ and $u_2\equiv 0$). Additionally, 
since $\Om_1$ is periodic (in $x$) but $\Om_2$ is not, it is 
useful to remember {\tt p.nu1} (compared to {\tt p.np}) as the number 
of actual unknowns in $\Om_1$. 
\item With these preparations we can set up {\tt oosetfemops}, see Listing 
\ref{accl1}. Essentially, here we first set up the matrices {\tt Kphi} and {\tt Kz} 
corresponding to $-\pa_\phi^2$ and $-\pa_z^2$ in $\Om_1$ to later build 
the LBO $-\Del_{C_R}={\tt Kphi/R^2+Kz^2}$, and {\tt Dphi}$=\pa_\phi$ to 
implement a PC similar to \reff{pcx} on $\Om_1$. Afterwards, we assemble the stiffness and 
mass matrices {\tt p.mat.K2} and {\tt p.mat.M2} on $\Om_2$, and finally 
put together the total mass matrix {\tt p.mat.M}. 
\item The main issue in {\tt sG} is the coupling $u_1-u_2=0$ on $\Ga_1$, 
for which we use a stiff spring approximation \cite{actut}. We identify 
the boundary nodes of $u_1$ and $u_2$ via {\tt p.S1} and {\tt p.S2} and 
add {\tt p.sf*p.Q2*(p.S1*p.mat.fill*u1-p.S2*u2)} to the rhs of \reff{e2}. 
This can easily be differentiated with respect to {\tt u1} and {\tt u2} 
and accordingly be put into {\tt sGjac.m}. 
\item In the script file {\tt cmds.m} we 
start with continuation of the trivial branch $u=(u_1,u_2)=(0,0)$ in $\lam$. 
The first bifurcation at $\lam=0$ is simple, and yields 
a spatially constant branch  $(u_1,u_2)=(\al,\al)$, which becomes stable at the fold. There are 
a number of BPs up to the fold, and we follow one of these, see {\tt b1-1} 
in Fig.~\ref{accf1}(e). 
For $(R,l_y)=(\pi,\pi)$, after the first bifurcation 
there are many 
close together bifurcation points $(\lam_j,0)$ 
with $\lam_j>0$ on the trivial branch, and we use the same trick 
as in {\tt schnaktor}. I.e., we simple localize some of the BPs using {\tt bifcheck=2}, then use {\tt cswibra} with {\tt aux.besw=0} to only compute 
nearby kernel vectors, which we then use as (approximate) predictors. 
If the solutions depend on $x$ on $\Om_1$, then we need to set a phase 
condition as in \reff{pcx}, see {\tt qf}, {\tt qfder}. In these continuations 
we switch off bifurcation detection, but note that a large number of further 
bifurcations occur. In particular, {\tt b3} (with angular wave number 1) stabilizes 
at a certain amplitude, i.e., near $\lam=1$. 
\item At the end of {\tt cmds.m} we also do one continuation in $R$, which 
exemplarily shows that many patterned solutions stabilize at large $R$, 
as expected from the standard AC equation \cite{actut}. 
\ecen 

\begin{table}[ht]\taskip
\caption{Main scripts and functions in {\tt accyl/}; at the bottom we list 
local modifications of library functions to use {\tt p.nu1} instead of the standard {\tt p.nu}.
\label{accyldtab}}
\bce\vs{-4mm}
{\small 
\begin{tabular}{l|p{0.73\tew}}
script/function&purpose, remarks\\
\hline
cmds&main script, essentially yielding Fig.~\ref{accf1}. \\
acinit&initialization, here setting up {\em two} PDE objects {\tt pde} and 
{\tt p2}, containing meshes etc for the two domains.\\
oosetfemops&set two sets of FEM matrices {\tt M,K} etc, and coupling matrices.  \\
sG,sGjac&rhs and Jacobian\\
qf,qfder&phase condition (and Jac) for rotational invariance\\
userplot&plotting cylinders as in Fig.~\ref{accf1}, called by {\tt plotsol} 
due to {\tt p.plot.pstyle=-1}\\
\hline
box2per, stanbra&mods of respective library functions {\tt box2per} \\
\end{tabular}
}
\ece
\end{table}\teskip

\hulst{caption={{\small {\tt oosetfemops.m} from {\tt pftut/accyl}  with some non standard steps to deal with the different dimensions (and properties) 
of the two fields {\tt u1} and {\tt u2}. See 5.~in the above comments.}},
label=accl1,language=matlab,stepnumber=0, linerange=1-13}{\dhome/oosetfemops.m}

\subsection{Bulk--surface coupling in a model for cell polarization: demo {\tt cpol}}\label{cpolsec}
In \cite{CEM19}, a model for so-called cell polarization is set up 
and studied analytically and numerically by time--integration using 
a bulk--surface FEM. The model consists of (fast) linear diffusion 
of inactive GTPase (variable $w$) in 
the 3D bulk $\Om$ of the cell (the cytosol), coupled a reaction--diffusion 
equation for the active GTPase (variable $u$) 
on the surface $\Ga=\pa\Om$ (the cell membrane). 
After nondimensionalzation it reads 
\allowdisplaybreaks
\begin{subequations}\label{cpol}
\begin{alignat}{2}
\pa_t u&=\eps\Del_\Ga u+\frac 1 \eps f(u,w),\quad &x\in\Ga,\\
\pa_t w&=\frac 1 \eps \Del w,&x\in\Om,\\
\pa_n w&=-f(u,w),&x\in\Ga, 
\end{alignat}
\end{subequations}
where $f(u,w):=(k_0+\frac{\ga u^2}{1+u^2})w-u$, with $k_0,\ga>0$ and 
$0<\eps=\sqrt{D_u/D_w}\ll 1$ a small parameter such that $\eps^2$ is 
the ratio of the diffusion constants of $u$ and $w$. 
In the limit $\eps\to 0$, a scalar equation for $u$ can be formally 
derived, but here we shall study the full coupled problem for 
small but finite $\eps$, choosing $\Om=B_1(0)$. Cell polarization 
in this and similar models (see Remark \ref{cpolrem1}) roughly 
means that $u$ concentrates in some part of $\Gamma$, and $w$ concentrates 
in some part of $\Om$.

Importantly, (\ref{cpol}) conserves the mass 
\huga{\label{cpolm}
m=\int_\Ga u\dd\Ga+\int_\Om w\dd x,
}
and $m$ turns out to be a convenient bifurcation parameter for 
steady states of \reff{cpol}, like in 
the Cahn--Hilliard equation in \S\ref{chsec}. We thus write the steady state problem 
for \reff{cpol} as 
\begin{subequations}\label{cpol2}
\begin{alignat}{2}
0&=\eps\Del_\Ga u+\frac 1 \eps f(u,w)+\lam_m+s\pa_\phi u,\qquad&x\in\Ga,\\
0&=\frac 1 \eps \Del w+\lam_m,&x\in\Om,\\
\pa_n w&=-f(u,w),&x\in\Ga, 
\end{alignat}
\end{subequations}
where $\lam_m$ acts as a Lagrange multiplier for the mass--conservation, 
and we also introduced a rotational wave--speed $s$ to deal 
with the rotational invariance (in $x_1=$azimuth) of the sphere. 
In contrast to the CH problem in \S\ref{chsec},  
\reff{cpol} is not variational, and thus the $\lam_m$ in 
\reff{cpol2} are somewhat formal Lagrange multipliers corresponding 
to the functional $\lam_m(\int_\Ga u\dd\Ga+\int_\Om w\dd x-m)$. 
Nevertheless, like in the KS equation they work efficiently to fulfill \reff{cpolm}, 
and $|\lam_m|<10^{-7}$ for all our continuations below. 

\begin{figure}[ht]
\bce 
\begin{tabular}{ll}
{\small (a) }&{\small (b) }\\
\ig[width=0.3\tew]{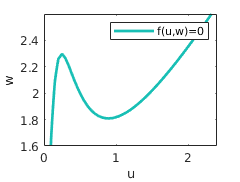}&
\hs{-1mm}\ig[width=0.21\tew]{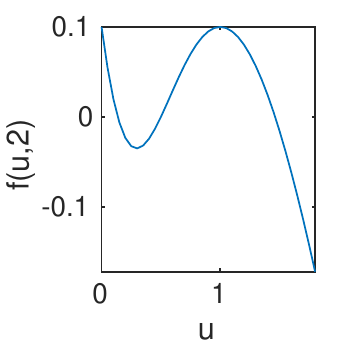}
\end{tabular}
\ece
\vs{-4mm}
   \caption{{\small Bistability for $f$ with $k_0=0.05$ and $\ga=1$, for which 
$w_1\approx 1.8$ and $w_2\approx 2.3$. 
  \label{cpolf0}}}
\end{figure}

Provided that $8k_0<\ga$ 
the reaction $f(u,\ov{w})$ shows bistability for $\ov{w}\in (w_1,w_2)$, 
see Fig.~\ref{cpolf0}, i.e., $f(u,\ov{w})$ has three zeros $u$, which 
together with a sufficiently small $\eps$ is the crucial ingredient 
for pattern formation (i.e., cell polarization) for \reff{cpol}. 

\brem\label{cpolrem1}
Problems similar to \reff{cpol} are also studied in, e.g., 
\cite{RR14,GKRR16,MC16,NV19}. In \cite{RR14}, (fast) linear diffusion 
for $w$ in the bulk $\Om$ is coupled to a RD {\em system} for species 
$u,v$ on 
the surface $\Ga=\pa\Om$, and a number of interesting results are 
derived analytically and numerically, indicating that also 
there cell polarization is triggered by $D_w$ being much larger 
than $D_u\approx D_v$, i.e., $0<\eps^2=D_w/D_u\ll 1$. In particular, \cite{NV19} 
contains a detailed asymptotic analysis of this problem, reducing 
the system to a scalar obstacle problem in the limit $\eps\to 0$, and 
obtaining conditions for cell polarization to occur. 
In \cite{GKRR16} a 
Cahn--Hilliard like phase--field model is studied, and 
\cite{MC16} coupleds a RD system for $w=(w_1,w_2)$ in the bulk 
to a RD system for $(u,v)$ on the surface, focussing on patterns 
generated by a Turing instability in $\Om$ or on $\Ga$. 

The numerical analysis in \cite{RR14, GKRR16, MC16} (and also in 
\cite{CEM19}) then proceeds by time--integration, where 
\cite{RR14, GKRR16} use a diffusive interface approach to 
approximate the surface diffusion, while \cite{MC16, CEM19} 
constructs compatible meshes for $\Om$ and $\Ga$. However, 
no (numerical or analytical) bifurcation analysis seems to exist for 
any of these systems. 
\eex
\erem

\begin{figure}[ht]
\bce 
\begin{tabular}{lll}
{\small (a) }&{\small (b) }&{\small (c) }\\
\ig[width=0.25\tew]{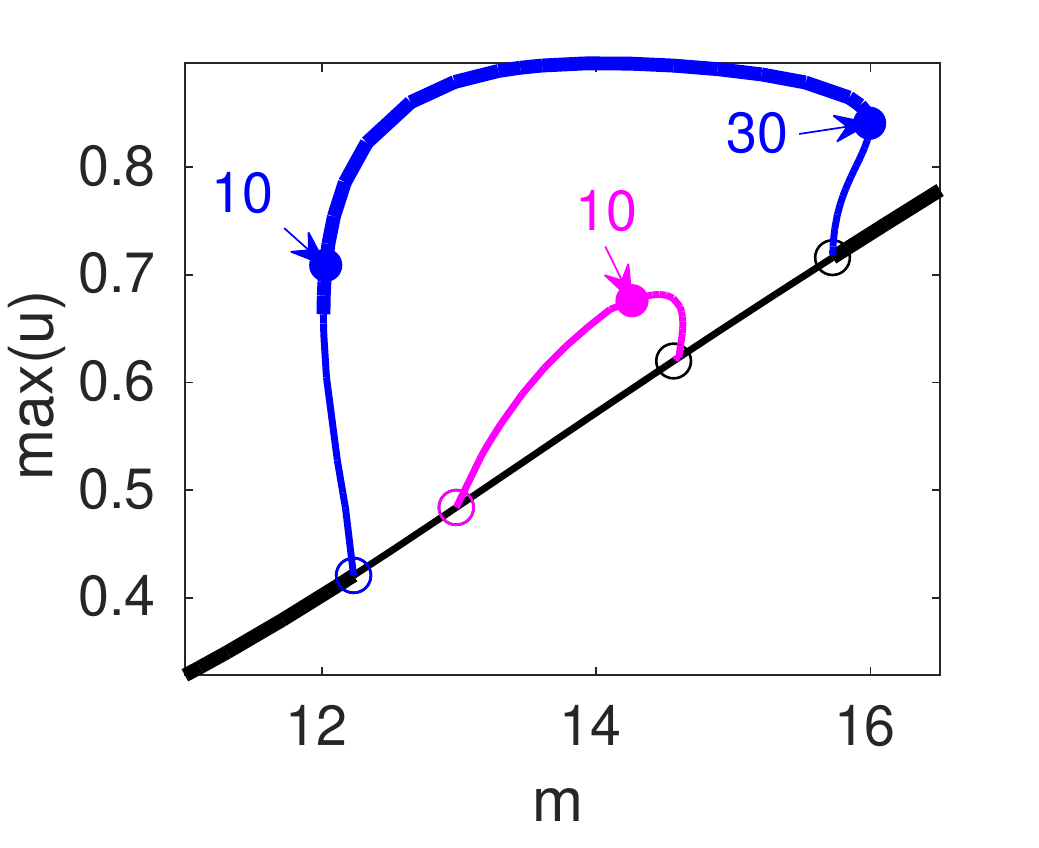}&\ig[width=0.25\tew]{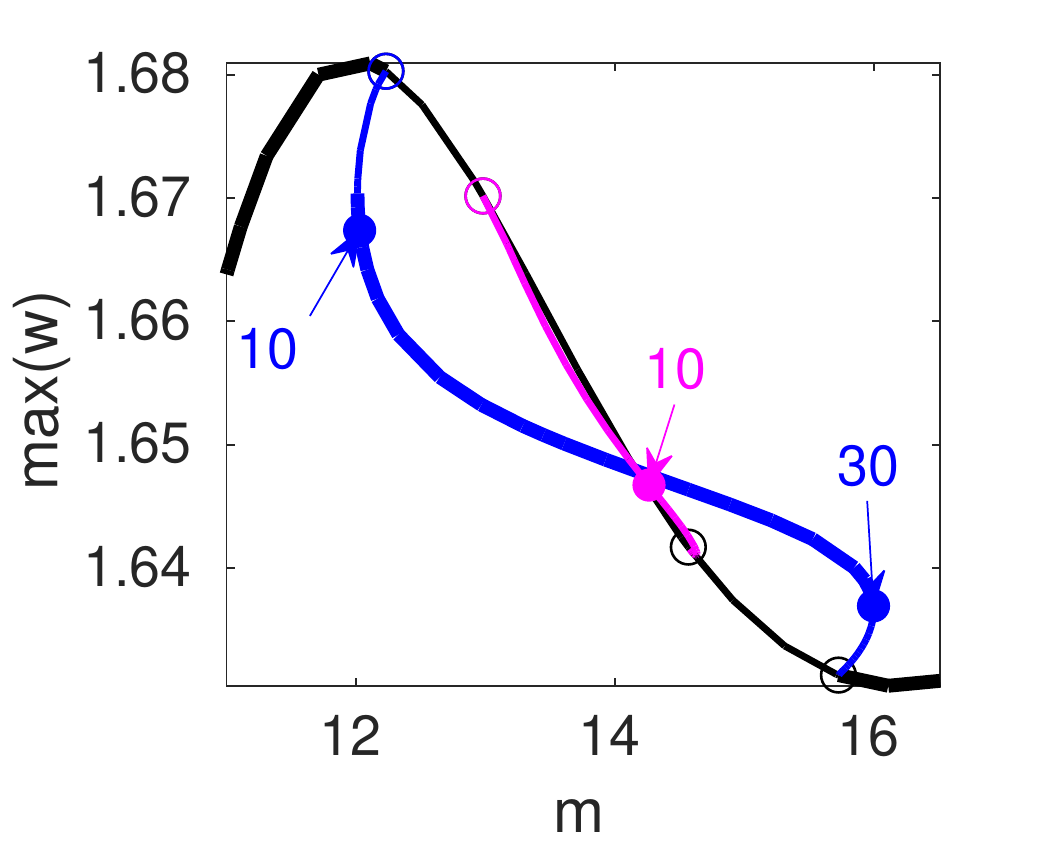}&
\ig[width=0.25\tew]{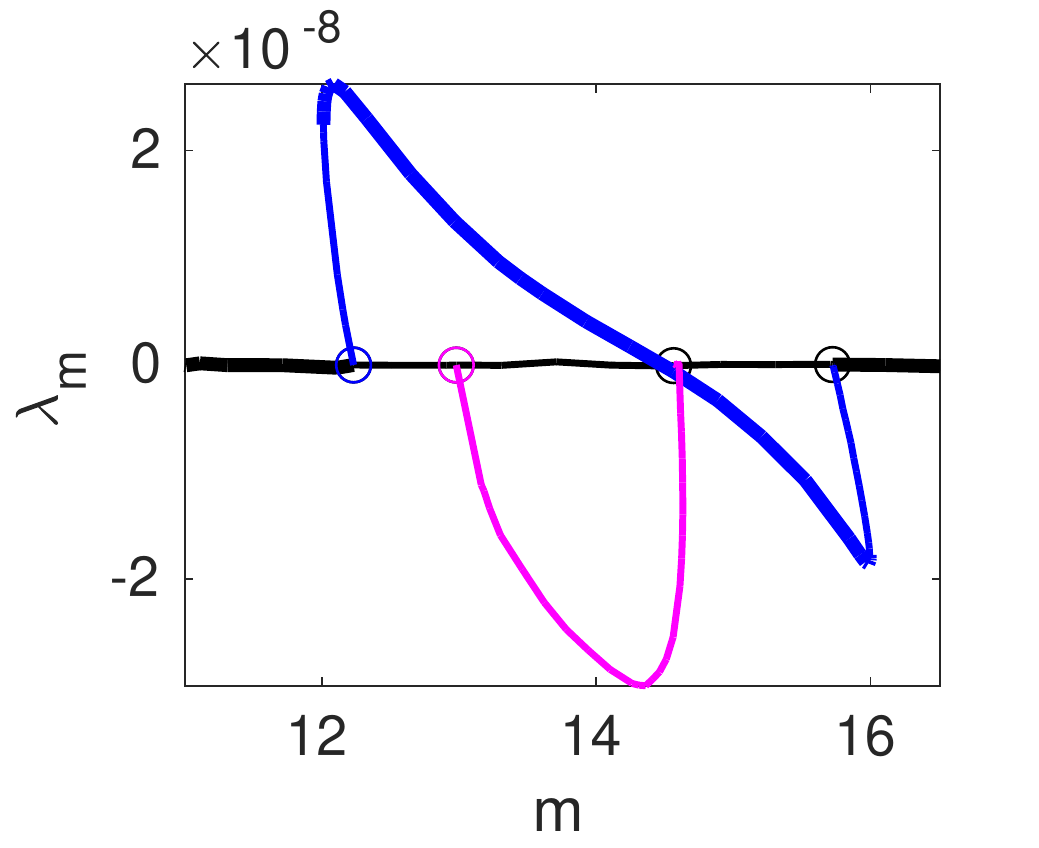}\\
(d)\\
\ig[width=0.28\tew]{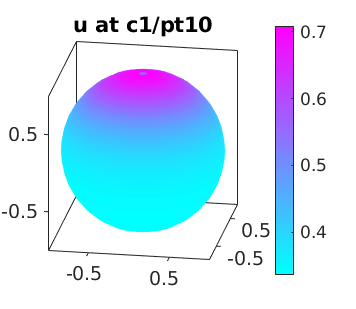}&
\ig[width=0.28\tew]{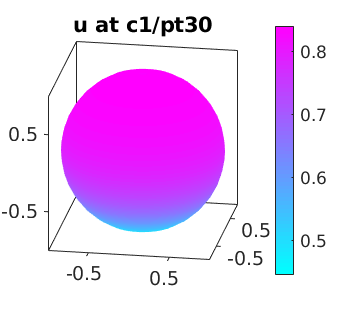}&
\hs{-1mm}\ig[width=0.28\tew]{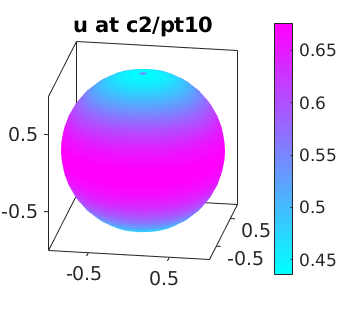}\\
\hs{-9mm}\ig[width=0.35\tew]{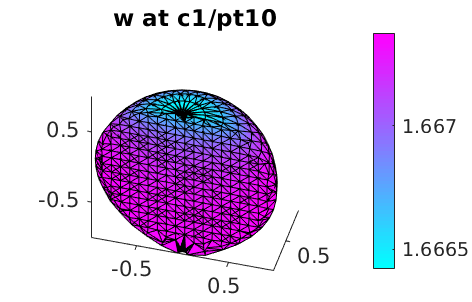}&
\hs{-4mm}\ig[width=0.35\tew]{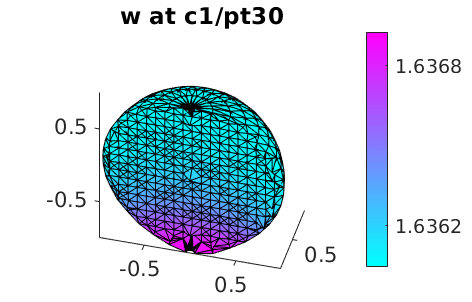}&
\hs{-4mm}\ig[width=0.35\tew]{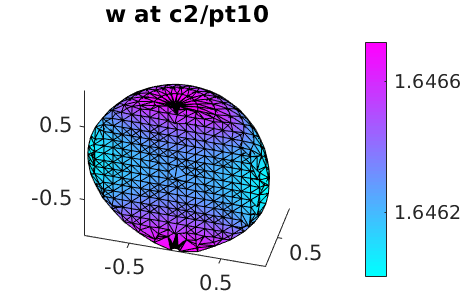}
\end{tabular}
\ece
\vs{-2mm}
   \caption{{\small Results for \reff{cpol}; $(R,k_0,\ga,\eps)=(1,0.1,1,0.1)$. 
(a)--(c) Bifurcation 
diagrams, homogeneous (black), primary (blue) and secondary (magenta) patterns; 
(c) shows the Lagrange multiplier $\lam_m$ for mass conservation. (d) Sample 
solutions, where the cut-away plots ({\tt pstyle=4}) 
of $w$ are for the cutting plane $y=0.01$.  
  \label{cpolf1}}}
\end{figure}

In Fig.~\ref{cpolf1} we show a simple bifurcation diagram for \reff{cpol} 
with \reff{cpolm}, base parameters 
\huga{\label{cpolbp}
(R,k_0,\ga,\eps)=(1,0.1,1,0.1), 
}
using $m$ as the primary bifurcation parameter, and $\lam_m$ as the Lagrange  
parameter for $n_q=1$ due to \reff{cpolm}. Table \ref{cptab} comments on 
the files for the implementation. Concerning the spatial discretizations, 
we proceed similarly to  \cite{MC16, CEM19}: like  
in \S\ref{accylsec} for the case of two surfaces, 
we generate two compatible meshes and FEM spaces, 
one for $\Om$ and one for $\Ga$. As we restrict to $\Om=B_1(0)$ and 
hence $\Ga=S^1$, we reuse the {\tt sppdeo} for $\Ga$ such that 
we have a natural Laplace Beltrami operator $\Del_\Ga$ on the preimage 
of $\Ga$. Then we generate the mesh for $\Om$ such that the boundary points 
for $\Om$ are the mesh points for $\Ga$ (plus two mesh points for 
the north pole and the south pole, respectively). 
This can be generalized to other domains $\Om$, and is a convenient 
method as long as we have a good parametrization of $\Ga$ (or of patches 
of $\Ga$) and associated LBOs $\Del_\Ga$ on the preimages. 
Moreover, on suitable domains it easily generalizes to models 
as in \cite{RR14, GKRR16, MC16} consisting of RD systems in the bulk 
and/or on the surface. 

Thus, in {\tt cpinit} we start 
with {\tt sppdeo} to generate a 2D pde-object {\tt p.pdeo} with 
mesh $(x_i,y_i)$ of {\tt p.nus} points 
for the preimage 
$$
Q=(-\pi,\pi)\times (-\pi/2+\del,\pi-\del)=\phi^{-1}(\Ga)
$$ 
of the sphere $\Ga$ without the poles $N=(0,0,1)$ and $S=(0,0,-1)$, 
cf.~\S\ref{acs-sec} and Fig.~\ref{acSf1}(a). Then we map 
the grid-points from $Q$ to $\Ga$, extend these by $N$ and $S$, 
then extend these further by points in $B_{1-h_0/2}(0)$ 
using {\tt distmesh}, and thus generate a PDE object {\tt p.p2} for 
$\Om$, where $h_0$ is a parameter for the mesh--width in the 3D mesh.  
The first {\tt p.nus+2} points in the mesh for $\Om$ 
then are $(\phi(x_i,y_i)_{i=1,{\tt p.nus}},N,S)$, and we refer 
to {\tt cpinit.m}, {\tt oosetfemops.m} and {\tt sGcp.m} for 
details and comments how to further set up and use the coupling matrices 
for the Robin BCs \ref{cpol2}(c). We need to 
locally modify some library functions to deal with the two meshes, 
and in {\tt userplot} we essentially shift 
{\tt p.p2} to {\tt p.pdeo} to enable the {\tt cutplots} in 
Fig.~\ref{cpolf1}(e).  
\def\Uhom{{U_{{\rm hom}}}}

\begin{table}[ht]\taskip
\caption{Scripts and functions in {\tt cpol}; at the bottom are 
local modifications of library functions. 
\label{cptab}}
\bce\vs{-4mm}
{\small 
\begin{tabular}{l|p{0.73\tew}}
script/function&purpose, remarks\\
\hline
cpcmds1&main script, essentially yielding Fig.~\ref{cpolf1}. {\tt cpcmds2} runs 
on a coarser mesh, shows the structure of Jacobian, and explains the use 
of the rotational phase condition. \\
cpinit&initialization, setting up {\em two} PDE objects {\tt pde} and 
{\tt p2}.\\ 
sphere2ballpdeo&convenience function to generate a mesh for $\Om$ compatible 
with the one for $\Ga$; based on {\tt hudistmesh}\\
oosetfemops&set two sets of FEM matrices {\tt M,K} etc, and coupling matrices.  \\
cpsG,cpsGjac,cpbra&rhs and Jacobian, branch-output\\
qfm,qfmder&aux.eqn and derivative for mass conservation \reff{cpolm}\\
qf2,qf2der&for augmenting qfm, qfmder by the rotational phase  condition on 
certain branches\\ 
userplot&producing plots as in Fig.~\ref{cpolf1}, i.e., $u$ on $\Ga$ and 
$w\in\Om$ using a cutplot, see also {\tt mapogrid}\\
\hline
box2per, getGupde&small modifications of respective 
library functions to deal with the two meshes present here, 
for instance the special sparsity structure of Jacobians in getGupde \\
\end{tabular}
}
\ece
\end{table}
\teskip

The results in Fig.~\ref{cpolf1} illustrate some main features 
of \reff{cpol}, extending results from time integration from \cite{CEM19}. 
There is a unique homogeneous steady state $\Uhom(m)$ 
(other parameters fixed), which is stable at small and large $m$. 
For sufficiently small $\eps$, 
increasing $m$ from low values we find a subcritical 
pitchfork at $m=m_1\approx 12.23$ to a polarized state oriented 
along the $N$--$S$--axis (blue branch in Fig.~\ref{cpolf1}(a). 
The kernel at bifurcation is three--dimensional 
including two other orientations, but the $N-S$--orientation 
can be continued without further phase condition due to the surface 
mesh.%
\footnote{See {\tt cpcmds2} for using {\tt cswibra} to obtain 
and continue the same branch in different orientations.} 
This polarized state becomes stable after a fold at $m\approx 12.01$, 
looses stability at a second fold at $m=m_1\approx 15.99$, and 
then returns to $\Uhom$ in another subcritical pitchfork at $\wt{m}_1\approx 15.72$. 
For the parameters \reff{cpolbp}, in particular $\eps=0.1$, 
there are two more bifurcation points $m_{2}<m_3$  
on $\Uhom$ close to $m=12.98$: At $m_2$ the kernel is three-dimensional, 
and Fig.~\ref{cpolf1}(f) illustrates the bifurcating branch with $N$--$S$ 
orientation, while at $m_3$ we have a doube bifurcation point with 
a genuine angular dependence, and these bifurcating branches again 
return to $\Uhom$ at $\wt{m}_3<\wt{m}_2$ near $m=14.59$. 

Similarly, for smaller $\eps$ we find several more 
branches bifurcating at some $m_i(\eps)$ with 
$m_1(\eps)<m_i(\eps)<\wt{m}_i(\eps)<\wt{m}_1(\eps)$ with 
higher wave numbers, which again form loops with between 
$m_i(\eps)$ and $\wt{m}_i(\eps)$. However, they are all 
unstable, and hence we content ourselves with Fig.~\ref{cpolf1}.

\section{Demos {\tt bruosc} and {\tt bruosc-tpf}: Oscillating Turing patterns}\label{brusec}
\def\dhome{./pftut/bruosc}\def\dname{bruosc}
So far we restricted to steady patterns. We now consider, 
rather closely following \cite{YZE04}, 
oscillating Turing patterns in the Brusselator model 
\begin{subequations}\label{br1}
\hual{
\pa_t u_1&=a-(b+1)u_1+u_1^2u_2+\al\cos(2\pi\beta t)+D_u\Delta u_1, \\
\pa_t u_2&=bu_1-u_1^2u_2+D_v\Delta u_2, 
}
\end{subequations}
with parameters $a,b,D_u,D_v,\al,\beta$, where $\al\cos(2\pi\beta t)$ 
describes a time--periodic forcing. First, however, we focus on the 
autonomous case $\al=0$, in which \reff{br1} has the spatially homogeneous 
steady solution 
\huga{\label{br2}
u^*=(u_1,u_2)^*\equiv (a,b/a).
}

\begin{figure}[ht]
\bce 
\begin{tabular}{ll}{\small (a) Hopf and Turing lines}&{\small (b) Dispersion relations; left to right corresponding to labels 1 to 4 in (a).}\\
\hs{-0mm}\ig[width=0.24\textwidth]{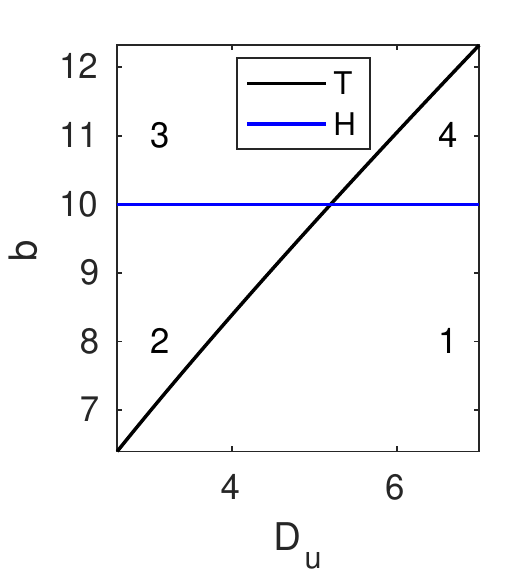}&
\hs{-4mm}\raisebox{23mm}{\begin{tabular}{l}
\ig[height=35mm,width=0.17\textwidth]{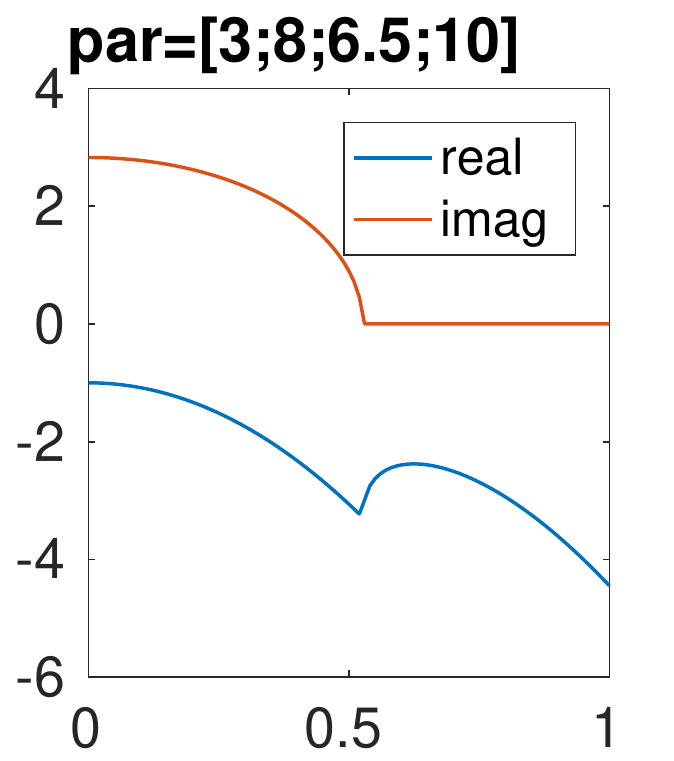}
\ig[height=35mm,width=0.17\textwidth]{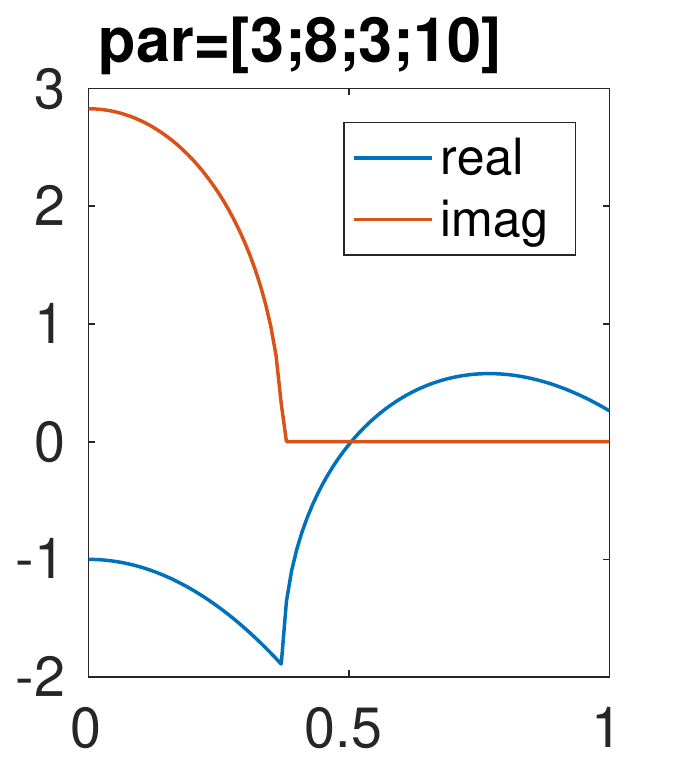}
\ig[height=35mm,width=0.17\textwidth]{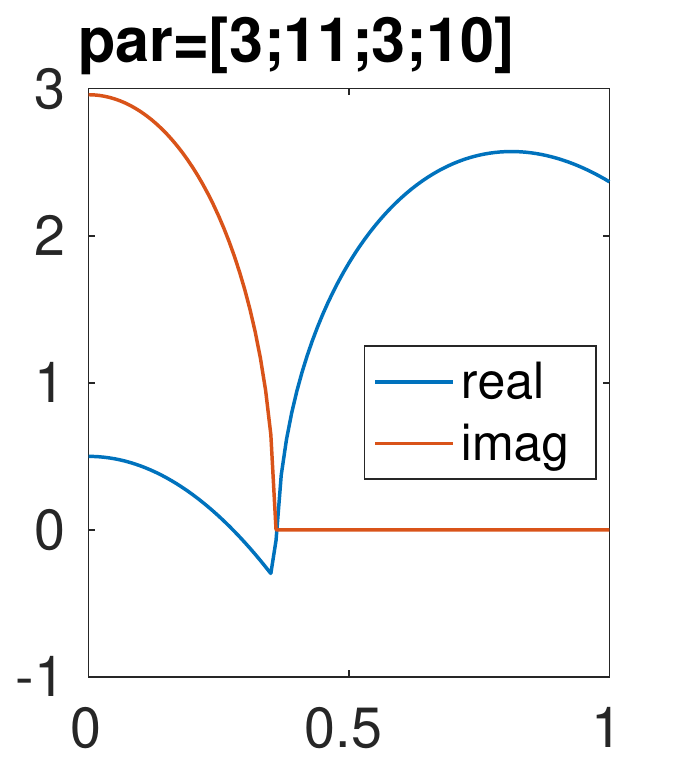}
\ig[height=37mm,width=0.18\textwidth]{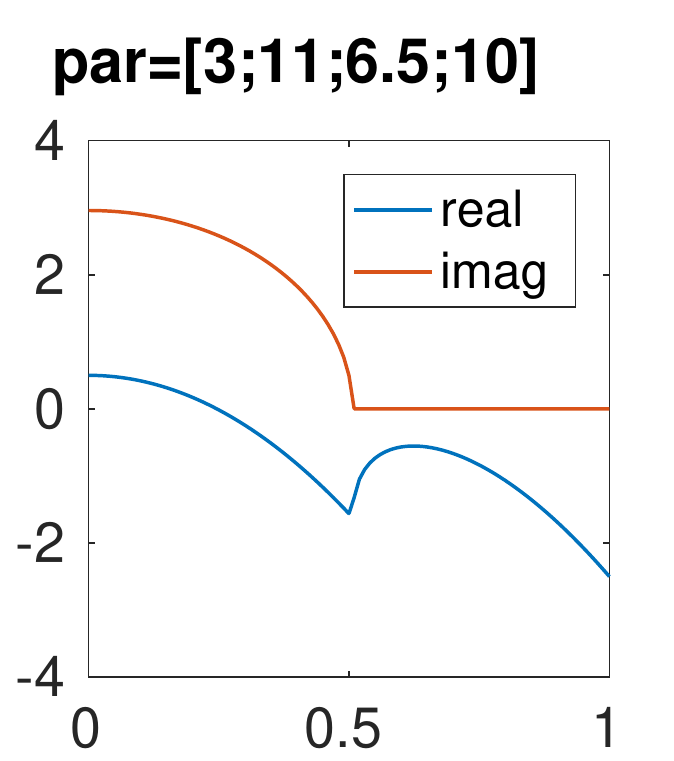}
\end{tabular}}
\end{tabular}
\ece\vs{-5mm}
\caption{{\small Phase diagram for fixed $a=3$ and $D_v=10$, and  
sketches of dispersion relations $\mu(k;{\tt par})$ with parameters ${\tt par}=(a,b,D_u,D_v)$ as indicated and wave number $k$ horizontally. 
  \label{bf0}}}
\end{figure}
The dispersion relation for the linearization of \reff{br1} around 
\reff{br2} can be studied analytically. For instance, 
fixing $a$ and $D_v$ we obtain a phase diagram in the $D_u,b$ plane 
as shown in Fig.~\ref{bf0}(a), where the Hopf and Turing lines are given 
by 
\huga{
b^H=1+a^2, \quad b^T=(1+a\sqrt{D_u/D_v})^2, 
}
respectively. In Fig.~\ref{bf0}(b), the dispersion 
plots are obtained from {\tt spufu} as in 
Fig.~\ref{schnakf00}, and the Turing line in (a) can also be obtained 
from branch point continuation as in \S\ref{intersec}, and the 
Hopf line from a Hopf point continuation which works via 
an analogous extended system, see \cite{hotut}. 
These commands are in {\tt \dname/auxcmds1}, with the pertinent 
extended Jacobians in {\tt bpjac.m} and {\tt hpjac.m}, as usual. 

If for instance we cross the Turing line by decreasing 
$d_u$ from sector 1 to 2, then we find a Turing bifurcation with wave number 
$k_c=\sqrt{a/\sqrt{D_uD_v}}$, resp.~wave vector $k$ with $|k|=k_c$. 
In this sense we have a completely analogous situation as for the 
SH equation in \S\ref{shsec} or the Schnakenberg system (which 
is obviously quite related to \reff{br1}) in \S\ref{schnaknum}, 
and obtain analogous steady patterns. See, e.g., \cite{UW18ab} 
for (localized) BCCs in \reff{br1}, similar to \S\ref{3dshnumsec}. 
Moreover, a three--component version of \reff{br1} (but without 
periodic forcing) was already 
used as a demo problem for Hopf bifurcations and Turing--Hopf bifurcations 
in \cite{hotheo,hotut}, following \cite{yd02}. 

In \S\ref{brasec} we focus on oscillating (time--periodic) solutions 
for \reff{br1} with $\al=0$ as we cross the Hopf line from sector 1 to 4 in Fig.~\ref{bf0}(a). 
The primary bifurcation then is to a spatially homogeneous Hopf orbit, 
and oscillating spatial patterns are afterwards created in (temporal) period 
doubling bifurcations from the homogeneous Hopf orbit. 
In \S\ref{tpfsec} we explain a trick how to add the 
time periodic forcing ($\al>0$) to \reff{br1}, which will yield 
subcritical oscillating Turing patterns. Additionally, we remark that 
\cite{UW18ab} considers (localized) 3D Turing patterns for 
\reff{br1} with $\al=0$, and that \cite{hotut} also contains some examples 
of an alternate setup for non--autonomous problems.


\subsection{The autonomous case}\label{brasec}
  Hopf bifurcation means the bifurcation of time periodic orbits, Hopf 
orbits in short, from steady solutions, due to complex conjugate 
eigenvalues crossing the imaginary axis with non-zero imaginary parts and 
non--zero speeds. The stability of a Hopf--orbit is determined by its 
Floquet multipliers. There always is the trivial multiplier $\ga=1$, 
and if all other multipliers are inside the unit disk, then the orbit is 
(orbitally) stable. See \cite{hotheo,hotut} and the references therein for 
basics of Hopf bifurcations and in particular the implementation in \pdep.

Bifurcations from Hopf orbits can occur if multipliers cross the 
unit circle, and currently \pdep\ can deal with 
\bci 
\item Hopf pitchforks or transcritical bifurcations, associated to 
a multiplier going through 1; 
\item Period doubling (PD) bifurcations, associated to 
a multiplier going through -1. 
\eci 
The case of complex conjugate multipliers crossing the unit circle elsewhere 
is called  Neimark-Sacker case, and numerically expensive even in low dimensional 
ODEs. A multiplier also passes through 1 at non--degenerate folds of Hopf-orbits, but like for folds of steady states we do not take special care of this. 

\begin{table}[ht]\taskip
\caption{Scripts and functions in {\tt \dname}. 
\label{brtab1}}
\bce\vs{-4mm}
{\small 
\begin{tabular}{l|p{0.73\textwidth}}
script/function&purpose, remarks\\
\hline
cmds1d,cmds2d&scripts for 1D and 2D, yielding Figs.~\ref{bf1} and \ref{bf2}. \\%
auxcmds1&script for Fig.~\ref{bf0}, including Hopf and branch point continuation\\\hline
bruinit, oosetfemops&initialization, and setting of 
FEM matrices\\
sG,nodalf,sGjac,njac&encode $G$ and $\pa_u G$\\
bpjac, hpjac&Jacobians for BP and HP continuation\\
spufu, myhoplot&mod of stanufu for the disp.~relations in Fig.~\ref{bf0}, 
and convenience mod of hoplot
\end{tabular}
}
\ece
\end{table}\teskip
Table \ref{brtab1} lists the files in {\tt \dname}. The setup of {\tt sG, sGjac}, etc is as usual, 
and we can concentrate on the main scripts {\tt cmds1d} and {\tt cmds2d}. 
As already said, the basic setups and algorithms to deal with Hopf bifurcations and Hopf orbits 
are described in \cite{hotheo, hotut}. One key point is the use of 
{\tt p.sw.bifcheck=2}, and often also the initialization of spectral 
shifts as guesses for the crossing of Hopf eigenvalues. In \pdep, 
this can be done with {\tt initeig}, see Cell1 of {\tt cmds1d.m} in 
Listing \ref{brl1}. Moreover, if one expects Hopf bifurcations we recommend {\tt hobra} as the standard setting for branch output. For steady solutions, 
this is analogous to {\tt stanbra}, but after a {\tt hoswibra} to a Hopf orbit it takes the data from {\tt p.hopf.y} (the periodic orbits). The default 
norm in {\tt hobra} thus is 
\huga{\label{honorm}
\|u\|:=\left\{\barr{ll}\|u\|_{L^2(\Om\times(0,T))}/\sqrt{T|\Om|}&\text{ if } T>0 
\text{ (genuine Hopf orbit)}, \\
\|u\|_{L^2(\Om)}/\sqrt{|\Om|}&\text{ if } T=0  
\text{ (steady solution)}, \earr\right.
}
where $T$ is the time--period and $|\Om|$ the volume of the domain. 

\begin{figure}[ht]
\bce 
\begin{tabular}{ll}{\small (a) period doubling BD}&{\small (b) solution plots, 
blue branch, red branch, magenta branch (left to right)}\\
\hs{-5mm}\ig[width=0.25\textwidth]{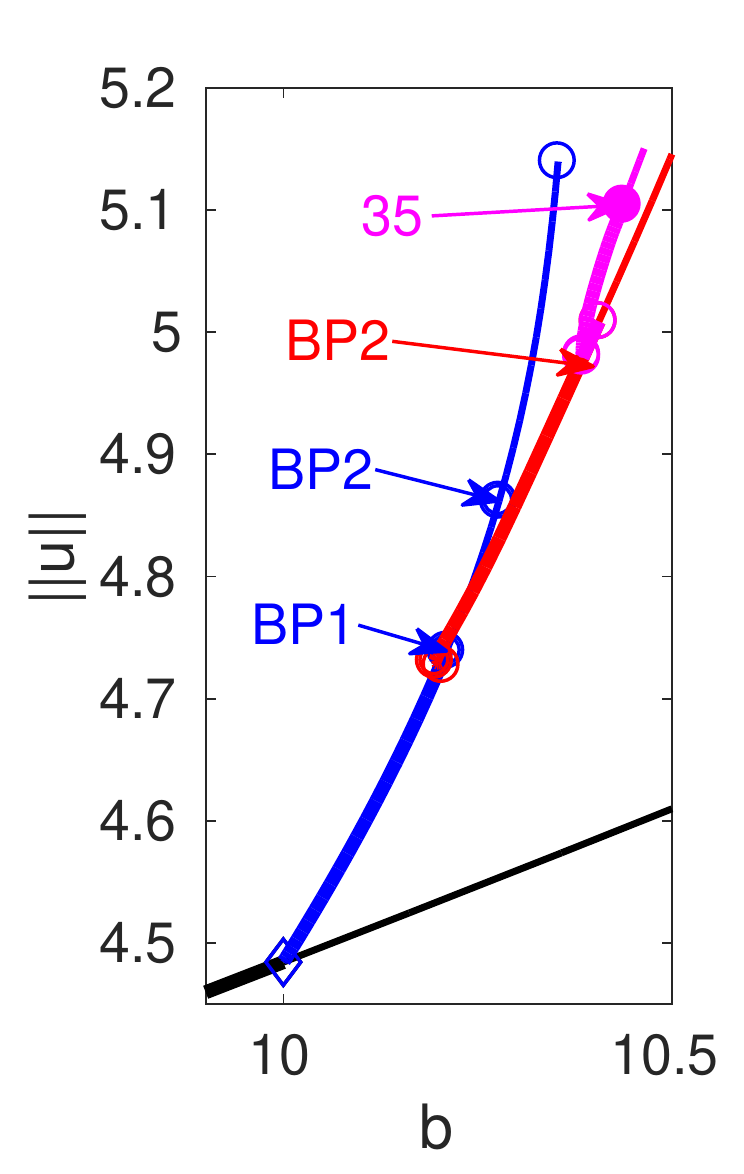}&
\raisebox{29mm}{\begin{tabular}{l}
\ig[width=0.21\textwidth]{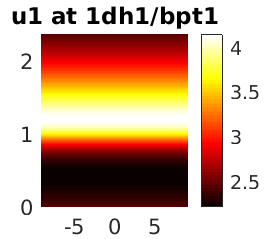}
\ig[width=0.21\textwidth]{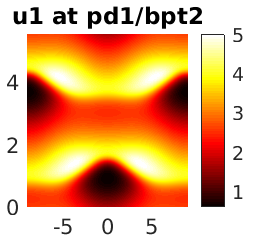}
\ig[width=0.21\textwidth]{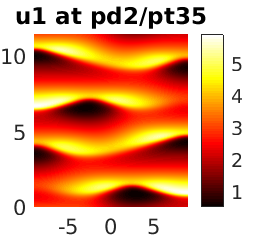}\\
\ig[width=0.2\textwidth]{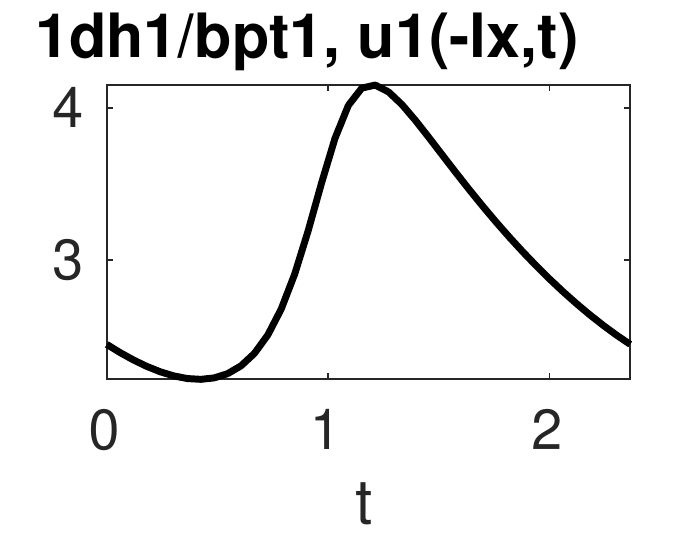}
\ig[width=0.2\textwidth]{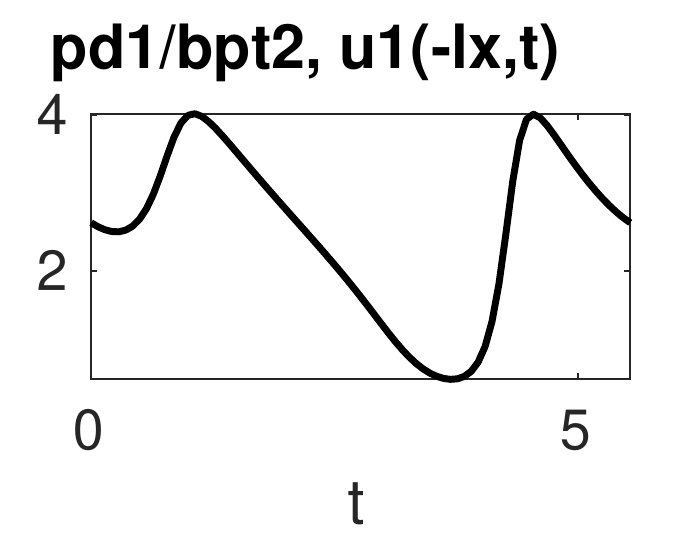}
\ig[width=0.2\textwidth]{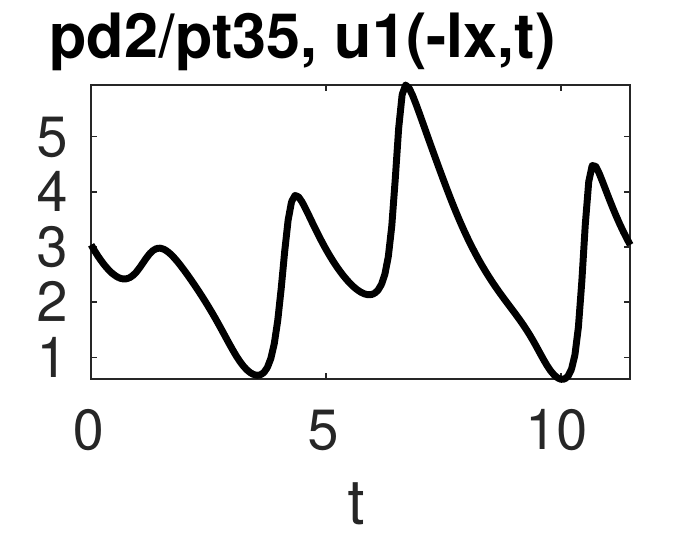}
\end{tabular}}\\
(c) Selected Floquet spectra\\
\end{tabular}\\
\ig[height=31mm]{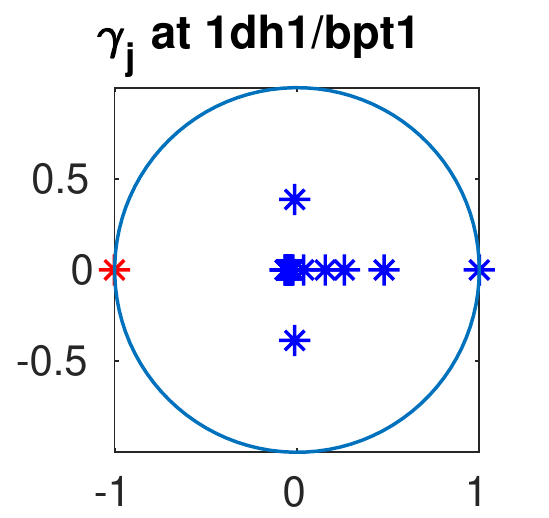}
\raisebox{-4mm}{\ig[height=36mm]{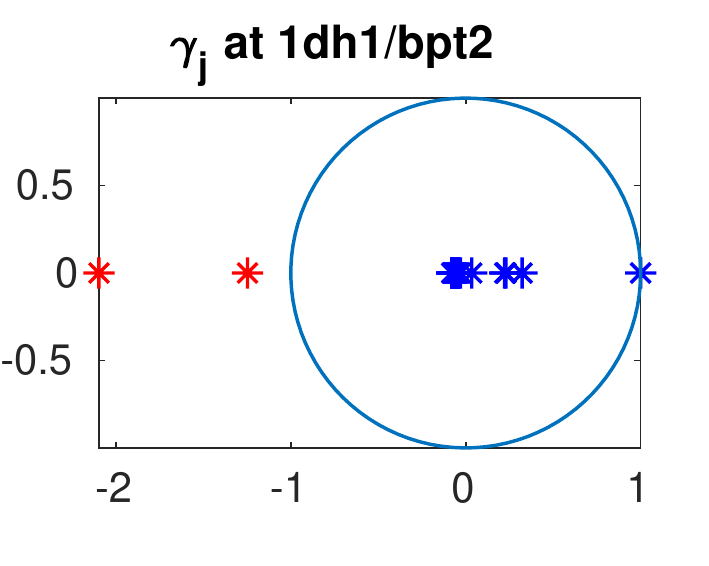}}
\ig[height=31mm]{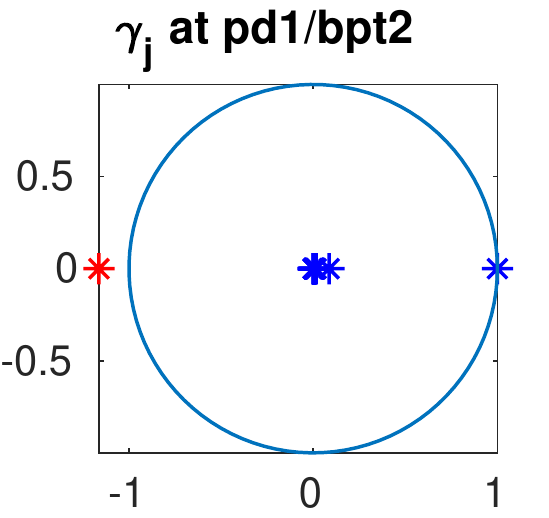}
\ig[height=32mm]{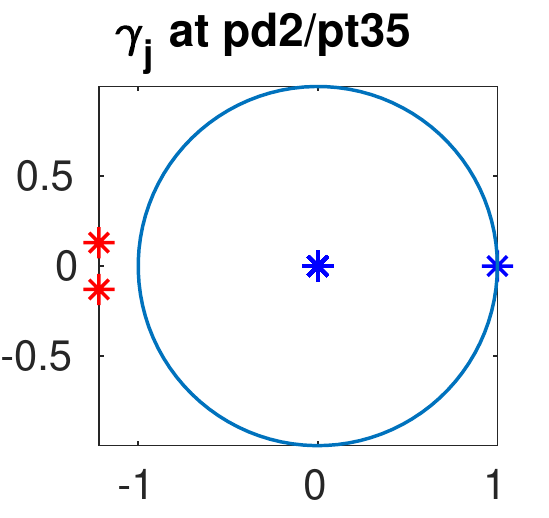}
\ece
\vs{-5mm}
   \caption{{\small Period doubling bifurcations to oscillating Turing patterns 
in \reff{br1} in 1D, $(a,D_u,D_v)=(3,6,10)$, $\Om=(-l_x,l_x)$, $l_x=2\pi/0.7$, 
with Neumann BCs. (a) BD of $u^*=(a,b/a)$ (steady state, black branch), 
primary Hopf branch (blue), first period doubling (red), secondary 
period doubling (magenta). The norm in (a) is \reff{honorm}. (b)  
space (horizontal)--time (vertical) solution plots of $u_1$ (top row), and 
associated time--series at left boundary (bottom row). (c) Leading 20 Floquet 
multipliers at selected solutions. 
  \label{bf1}}}
\end{figure}

In Cell 1 of {\tt cmds1d} we initialize \reff{br1} in sector 1 in Fig.~\ref{bf0}(a), and then increase $b$, 
which yields a Hopf bifurcation point (HBP) at $b=10$. 
The branch switching {\em to} the bifurcating spatially homogeneous 
Hopf branch {\tt 1dh1} works by {\tt hoswibra}, followed by {\tt cont} as usual. 
For the Hopf orbits we use {\tt lssbel} (bordered elimination) 
as the linear system solver, which typically is more efficient than 
the standard solver {\tt lss} (basically \mlab's $\backslash$), see \cite{lsstut, hotheo}. 
On {\tt 1dh1} we find a sequence of PD bifurcations, and the 
branch switching {\em from} Hopf branches is implemented in {\tt poswibra} 
(periodic orbit swibra). Subsequently (Cell 4) we plot the BD and 
example solutions to obtain Fig.~\ref{bf1}. 

\hulst{caption={{\small {\tt \dname/cmds1d}, first 3 cells, the 
remainder dealing with plotting. See text for comments. }},
label=brl1,language=matlab,stepnumber=0, linerange=2-14}{\dhome/cmds1d.m}

The left to right plots in Fig.~\ref{bf1}(b) show example solutions 
from a sequence of period--doublings which occurs for increasing $b$. 
We used $n_p=51$ spacial points (1D), and $n_t=30$ temporal points on 
the primary Hopf orbit, and hence $n_t=59$ and $n_t=117$ after the 
period doublings. 
The first plot in (c) shows the multipliers at the well localized
(see Remark \ref{porem1}) first 
PD point from {\tt 1dh1}. The second and third plots in (c) show 
the multipliers at the rather poorly localized 2nd BPs from {\tt 1dh1} and 
{\tt pd1} (the first BP on {\tt pd1} is a fold). The fourth plot in (c), 
shortly after {\tt pd2} looses stability, indicates that this loss of stability 
is due to a Neimark--Sacker bifurcation. 
\brem\label{porem1}
{\rm The localization of BPs from periodic orbits currently works 
by a simple bisection for the change of the number of unstable multipliers. 
This is expensive, and the multipliers often have a rather 
sensitive dependence on the parameters, but our experience is 
that typically we do not need 
a very accurate localization of the BPs for successful branch-switching, 
and thus we content ourselves with, e.g., the results from Fig.~\ref{bf1}. 
However, see also Cell 3 of Listing \reff{brl1b} for a more careful 
localization of PD points in 2D. }
\eex\erem 

\begin{figure}[ht]
\bce 
\begin{tabular}{ll}{\small (a) period doubling BD}&{\small 
(b) solution plots, from red, dark blue and magenta branch (top--down)}\\
\hs{-0mm}\ig[width=0.25\textwidth,height=74mm]{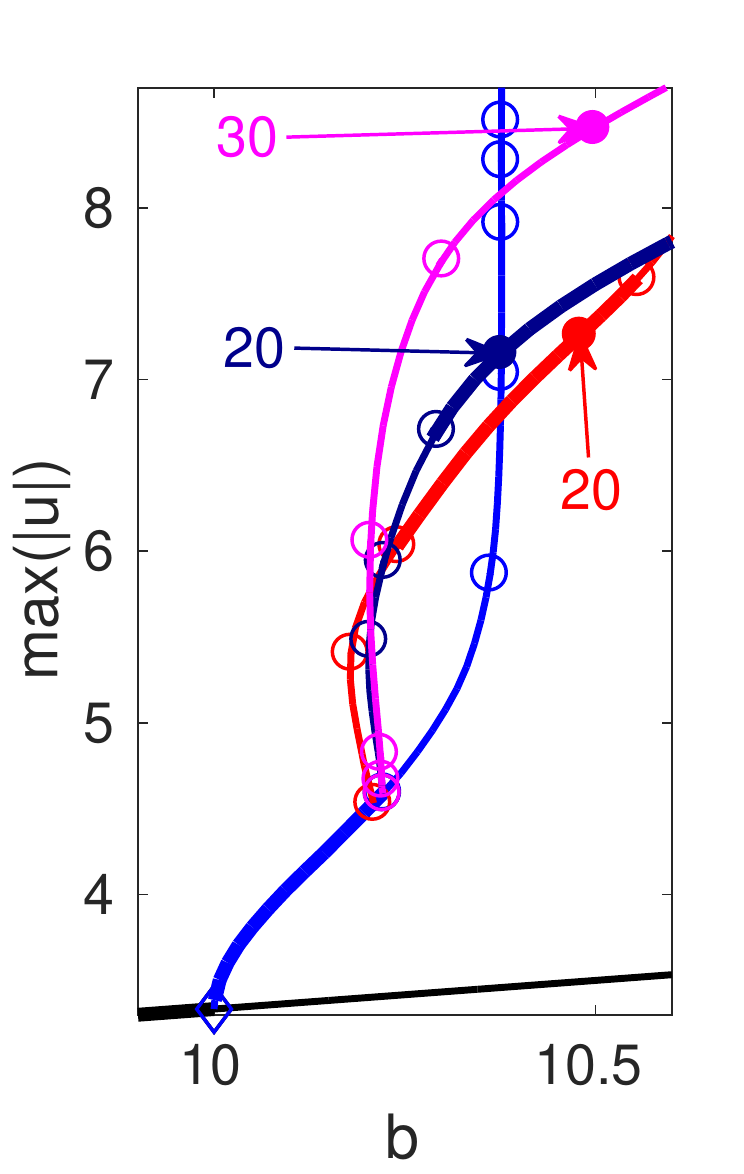}&
\hs{-8mm}\raisebox{34mm}{\begin{tabular}{l}
\raisebox{3mm}{\ig[width=0.5\textwidth]{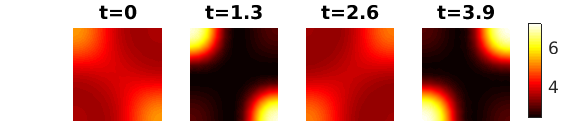}}
\ig[width=0.19\textwidth]{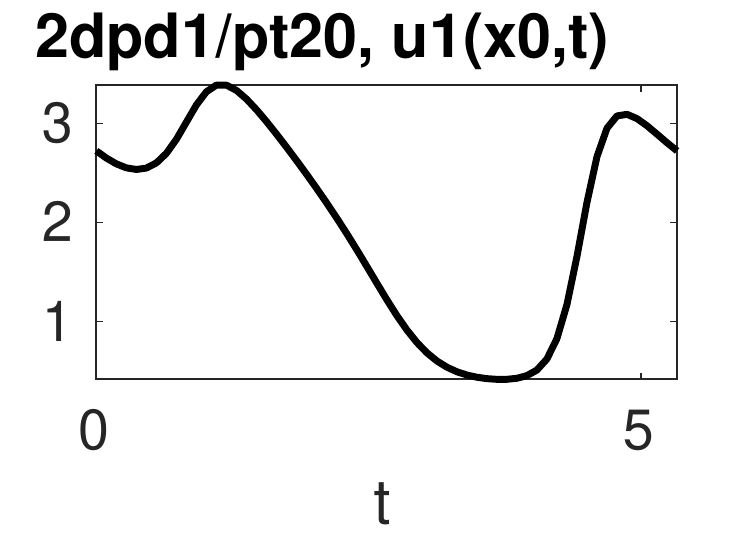}\\
\raisebox{3mm}{\ig[width=0.5\textwidth]{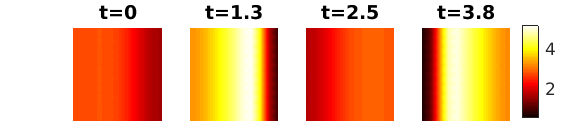}}
\ig[width=0.19\textwidth]{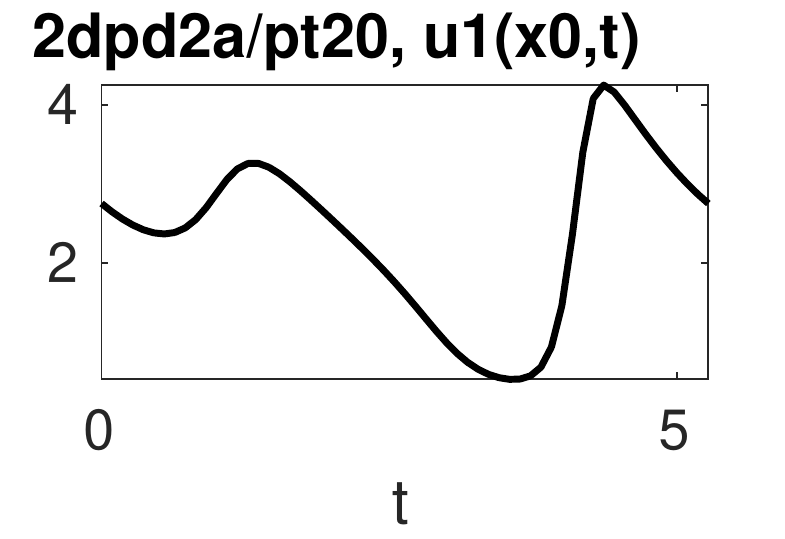}\\
\raisebox{3mm}{\ig[width=0.5\textwidth]{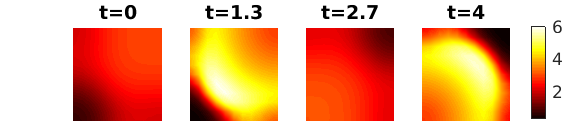}}
\ig[width=0.19\textwidth]{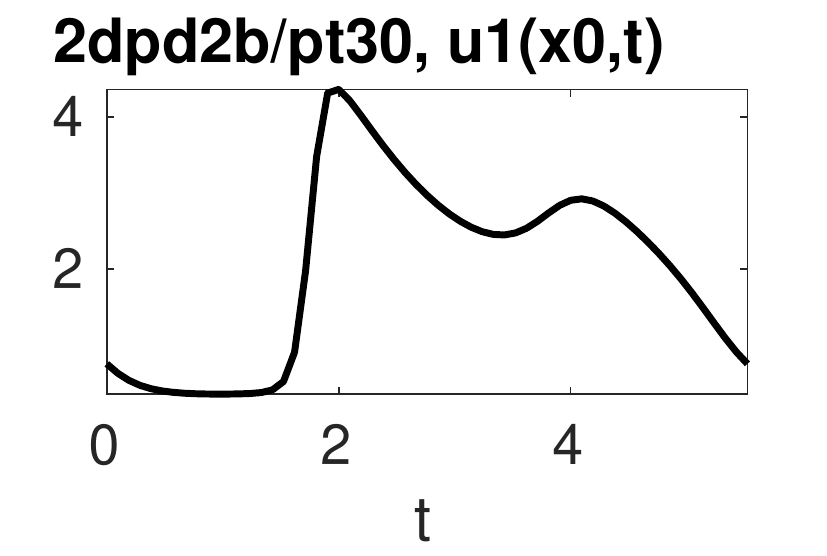}
\end{tabular}}
\end{tabular}
\ece
\vs{-5mm}
   \caption{{\small Period doubling to stable squares and other osc.~patterns for \reff{br1} over $\Om=(-l_x,l_x)^2$, $l_x=\sqrt{2}\pi/0.7$. (a) BD of 
$u^*=(a,b/a)$ (steady state, black branch), 
primary Hopf branch (blue), squares (red), and 
two branches at the 2nd PD from the blue branch. The stripes branch 
(dark blue) corresponds to {\tt pd1} from Fig.~\ref{bf1} but with a 
longer spatial period, while the magenta branch has squares of larger wavelength, which however are unstable. 
  \label{bf2}}}
\end{figure}

Figure \ref{bf2} shows analogous results for \reff{br1} over the 2D 
square domain $\Om=(-l_x,l_x)^2$, $l_x=\sqrt{2}\pi/0.7$, with a 
spatial criss-cross mesh of $n_p=221$ discretization points. 
The factor $\sqrt{2}$ is as in Fig.~\ref{shf12b} to have 
squares as the first oscillating Turing mode. 
The script {\tt cmds2d} works quite analogous to {\tt cmds1d}, 
with the main difference that now some of the PD branch points 
have multiplicity two, defined as in the steady case as the dimension 
of the critical eigenspace. We do not yet have algorithms (like {\tt q(c)swibra} for 
the steady case) for the systematic treatment of HBPs or POBPs 
of higher multiplicity. However, similar to {\tt gentau}, the user 
may select coefficients to choose initial states for the {\tt poswibra} 
predictor (and hence for the predictor as a function of $t$) 'by hand', 
here by passing coefficients for the kernel vectors via {\tt aux.coeff}.  

\hulst{caption={{\small {\tt \dname/cmds2d}, first 5 Cells. 
After following the primary Hopf branch with large {\tt ds} in C2, 
in C3 we reload a point and decrease {\tt ds} to localize 
the first 2 BPs reasonably well. In C4 we deal with the primary PD
 bifurcation to squares, and in C5 with 
PD bifurcations at the second BP, which is of multiplicity two. 
The remainder of {\tt cmds2d.m} deals with plotting. }},
label=brl1b,language=matlab,stepnumber=0, linerange=3-21}{\dhome/cmds2d.m}

\subsection{Time periodic forcing}\label{tpfsec}
\def\dhome{./pftut/bruosc-tpf}\def\dname{bruosc-tpf}
Another interesting result from \cite{YZE04} is that stable 
oscillating Turing patterns can occur under 
(spatially homogeneous) time periodic forcing $\al>0$ 
in the range where both, Turing and Hopf modes are subcritical, i.e., 
the steady state $u^*$ is (exponentially) stable. As an example we 
consider the point $(D_u,b)=(6,9.5)$, choose $\beta=0.42$ as in \cite{YZE04}, 
motivated by ${\rm Im}\mu(0)=2\pi\beta\approx 2.65$ for $(a,b,D_u,D_v)=(3,10,6,10)$, and increase 
$\al$. The first task is to implement the time--periodic forcing. We do this 
by augmenting $G$ by a (nonlinear) oscillator (in complex notation) 
$\dot v=(\del+2\pi\beta\ri)v-|v|^2 v$, which in polar coordinates $v(t)=r(t)\er^{\ri\phi(t)}$ for 
$\del>0$ yields the periodic orbit $r=\sqrt{\del}, \phi=2\pi\beta t$. The full 
system thus reads 
\huga{\label{br10}
\begin{split}
\pa_t u_1&=a-(b+1)u_1+u_1^2u_2+v_1+D_u\Delta u_1, \\
\pa_t u_2&=bu_1-u_1^2u_2+D_v\Delta u_2, \\
\pa_t v_1&=\del v_1-2\pi\beta v_2-(v_1^2+v_2^2)v_1,\\
\pa_t v_2&=\del v_2+2\pi\beta v_1-(v_1^2+v_2^2)v_2, 
\end{split}
}
i.e., we have the forcing amplitude $\al=\sqrt{\del}$ for $\del\ge 0$. 
To put \reff{br10} into \pdep\ we proceed similar to \S\ref{chsec}: 
After creating the FEM mesh and nodal values for $u$ we add two 'virtual 
nodal values', i.e., set 
$n_u=n_u+2$ and append $v=(v_1,v_2)$ to {\tt u}. The pertinent 
commands in {\tt bruinit.m} are 
{\tt p.nu=p.nu+2; p.u=[u;v;0;0;par']}. In {\tt oosetfemops} we then 
extend the left hand side mass matrix {\tt M} accordingly, and we 
append the $v$ equations in {\tt sGpf}, see Listing \ref{brl10}. 
Additionally, we need some minor modifications for generating the branch data and plotting (replacing back $n_u$ by $n_u-2$), and then we can proceed as 
before. 

\hulst{language=matlab,stepnumber=0}{\dhome/oosetfemops.m}
\hulst{language=matlab,stepnumber=0}{\dhome/nodalfpf.m}
\hulst{caption={{\small {\tt \dname/oosetfemops, nodalpf} and 
{\tt sGpf}. With $n_u=n_u+2$ (from {\tt bruinit}), the PDE components $u_1,u_2$ 
are now at {\tt u(1:p.nu-2)}, and the auxiliary oscillator components $v$ at {\tt u(p.nu-1:p.nu)}.}},
label=brl10,language=matlab,stepnumber=0}{\dhome/sGpf.m}. 

\begin{figure}[ht]
\bce 
\begin{tabular}{ll}{\small (a) period doubling BD}&{\small (b) solution plots}\\
\hs{-0mm}\ig[width=0.25\textwidth,height=54mm]{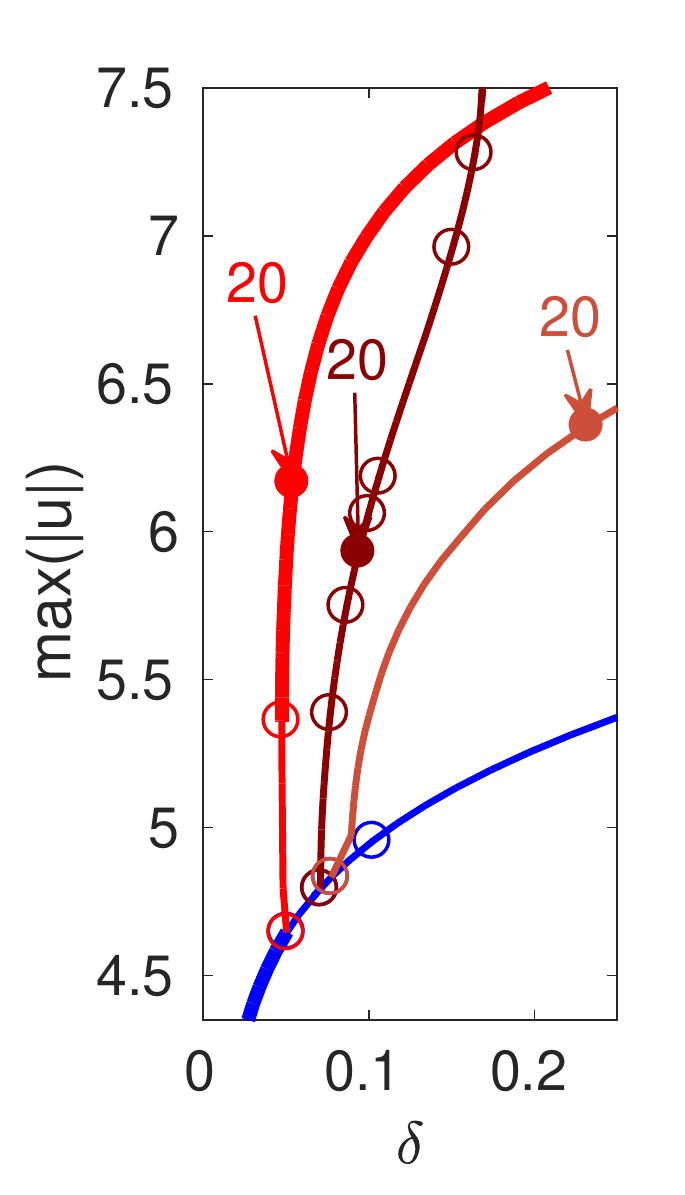}&
\hs{-4mm}\raisebox{26mm}{\begin{tabular}{l}
\ig[width=0.23\textwidth]{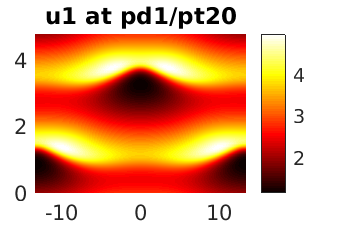}
\ig[width=0.23\textwidth]{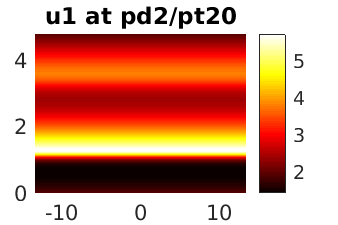}
\ig[width=0.23\textwidth]{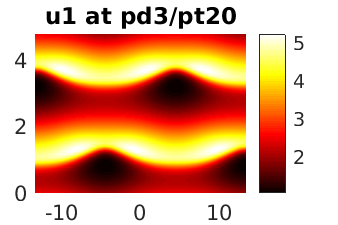}\\
\ig[width=0.22\textwidth]{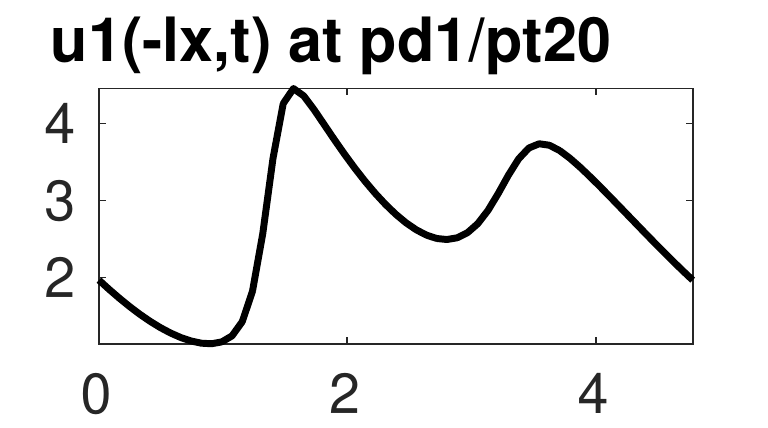}
\ig[width=0.22\textwidth]{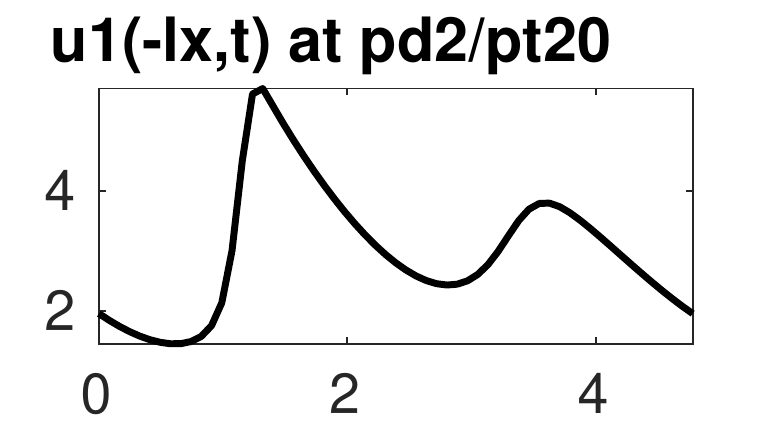}
\ig[width=0.22\textwidth]{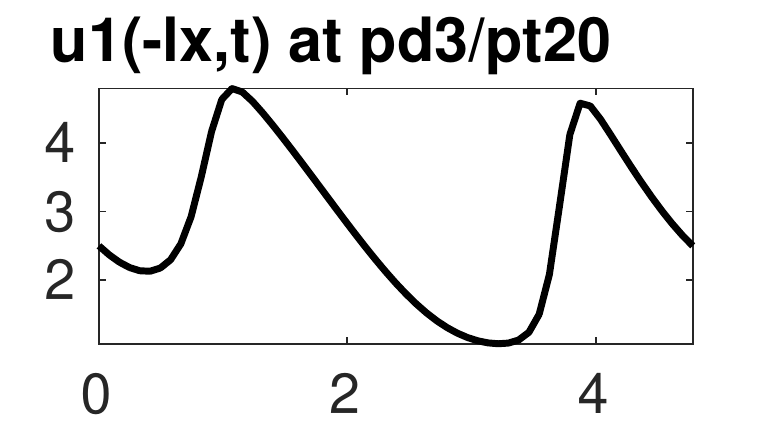}
\end{tabular}}
\end{tabular}
\ece
\vs{-5mm}
   \caption{{\small BD under periodic forcing with amplitude $\al=\sqrt{\del}$, $\beta=0.42$ fixed, $(a,b,D_u,D_v)=(3,9.5,6,10)$. Subcritical oscillating Turing patterns under time-periodic forcing via period doubling bifurcations from the 'natural' 
forced branch. 
  \label{bf3}}}
\end{figure}

Listing \ref{brl11} shows the main commands from the script {\tt cmds1d}, 
and Fig.~\ref{bf3} a selection of results. The first PD bifurcation from 
the blue branch is subcritical to an oscillating Turing pattern, which becomes stable after the 
fold. There are further PD bifurcations, but the bifurcating branches are 
unstable. 
\hulst{caption={{\small {\tt cmds1d} (first 4 cells). In C2/C3 we 
increase $\del$ and follow the oscill.~branch for $\del>0$ to switch on 
the periodic forcing and increase its amplitude. We then find secondary 
period--doubling bifurcations to oscillatory Turing patterns.}},  label=brl11, language=matlab,stepnumber=0,linerange=2-12}{\dhome/cmds1d.m}

\renewcommand{\arraystretch}{0.5}\renewcommand{\baselinestretch}{1}
{\small 
\newcommand{\etalchar}[1]{$^{#1}$}

}
\end{document}